\documentclass[a4paper,11pt]{article}
\usepackage{amsmath}
\DeclareMathOperator*{\argmin}{arg\,min}
\usepackage{amssymb}
\usepackage{amsfonts}
\usepackage{xcolor}
\definecolor{winered}{rgb}{0.5,0,0}
\usepackage[bookmarks=true, bookmarksnumbered=true, allbordercolors={1 1 1}]{hyperref}
\hypersetup{
  colorlinks   = true, 
  urlcolor     = violet, 
  linkcolor    = violet, 
  citecolor    = winered,
}
\usepackage{placeins}
\usepackage[cal=boondoxo]{mathalfa}
\usepackage{tikz}
\usetikzlibrary{fit,positioning}
\usepackage{natbib}
\usepackage{lscape}
\usepackage{longtable}
\usepackage{epstopdf}
\usepackage{float}
\usepackage{setspace}
\usepackage{euscript}
\usepackage{textcomp}
\usepackage{algorithm}
\usepackage[end]{algpseudocode}
\algnewcommand\algorithmicto{\textbf{to }}
\usepackage{makeidx}

\usepackage{subcaption}
\usepackage{arydshln}

\makeindex

\newcommand{\bm}[1]{\boldsymbol{#1}}
\newcommand{\wh}{\widehat}

\usepackage{etoolbox}
\patchcmd{\thebibliography}{\section*{\refname}}{}{}{}

\emergencystretch=\maxdimen
\title{\bf Bayesian Approaches to Shrinkage and Sparse Estimation\\
\begin{large} 
A guide for applied econometricians
\end{large} 
} 
\author{Dimitris Korobilis \\ \emph{University of Glasgow} \and Kenichi Shimizu \\ \emph{University of Glasgow}}              
\date{\today}

\begin{document}

\maketitle                              
\setcounter{page}{1}

\begin{abstract}

In all areas of human knowledge, datasets are increasing in both size and complexity, creating the need for richer statistical models. This trend is also true for economic data, where high-dimensional and nonlinear/nonparametric inference is the norm in several fields of applied econometric work. The purpose of this paper is to introduce the reader to the world of Bayesian model determination, by surveying modern shrinkage and variable selection algorithms and methodologies. Bayesian inference is a natural probabilistic framework for quantifying uncertainty and learning about model parameters, and this feature is particularly important for inference in modern models of high dimensions and increased complexity.

We begin with a linear regression setting in order to introduce various classes of priors that lead to shrinkage/sparse estimators of comparable value to popular penalized likelihood estimators (e.g.\ ridge, lasso). We explore various methods of exact and approximate inference, and discuss their pros and cons. Finally, we explore how priors developed for the simple regression setting can be extended in a straightforward way to various classes of interesting econometric models. In particular, the following case-studies are considered, that demonstrate application of Bayesian shrinkage and variable selection strategies to popular econometric contexts: i) vector autoregressive models; ii) factor models; iii) time-varying parameter regressions; iv) confounder selection in treatment effects models; and v) quantile regression models. A MATLAB package and an accompanying technical manual allow the reader to replicate many of the algorithms described in this review.

\end{abstract}

\newpage
\tableofcontents                        

\newpage
\onehalfspacing
\section[Introduction]{Introduction}

In all areas of human knowledge, datasets are increasing in both size and complexity, creating the need for richer models. This trend is also true for economic data, where high-dimensional and nonlinear/noparametric inference is the norm in several fields of applied econometric work. The purpose of this survey is to introduce the reader to Bayesian inference using shrinkage and variable selection priors. In particular we intend to demonstrate that the benefits of a Bayesian approach to high-dimensional estimation are manifold. Bayesian inference allows for a more accurate quantification of uncertainty. Parameters are treated as random variables that have their own probability density (or mass) functions. The use of a prior distribution provides a natural ground for enhancing possibly weak information in the likelihood.\footnote{Note that our interest here is in ``wide'' data (e.g.\ a linear regression model with more predictors than observations) where unrestricted estimation based only on the likelihood is either unreliable or impossible. In cases with ``tall'' data (many observations) the Bayesian posterior will tend to concentrate towards a point mass, i.e.\ uncertainty is small.} Our first aim is to explore in this review classes of priors that can recover popular penalized regression estimators, such as the lasso of \cite{Tibshirani1996}. Next, we want to demonstrate how the Bayesian paradigm becomes a natural framework for combining prior forms in order to capture more complicated patterns of shrinkage and/or sparsity in the data. For example, \cite{RockovaGeorge2018} extend the lasso with ideas from the Bayesian variable selection literature in order to obtain a ``spike and slab lasso'' estimator that is empirically superior to shrinkage or variable selection alone, and has desirable theoretical guarantees. Finally, we aim to illustrate that the Bayesian framework is ideal for applied economists who want to use shrinkage or sparsity in more complex or unconventional settings. Economists might be interested in combining data-rigorous statistical variable selection with economic restrictions on certain parameters\footnote{For example, instead of the typical statistical shrinkage towards zero that indicates whether an effect is important or not, economists might want to shrink a parameter towards a calibrated value or a sign restriction provided by the solution of an economic model.}, or use a shrinkage estimator in a model with breaks, stochastic volatility, missing data or other complexities. Penalized and constrained maximum likelihood frameworks can deal with such cases, but computation is non-trivial because it relies on optimizing complex functions. We demonstrate emphatically in this survey paper that Bayesian computation provides numerous tools and algorithms for shrinkage and sparsity that can be incorporated in very complex statistical models with the same ease they are used in univariate linear regression settings.

Even though the notions of sparsity and shrinkage estimation are ubiquitous since the explosion of Big Data in all fields of science (e.g. we doubt there are many economists these days who haven't heard about the lasso), we want to clarify these terms before proceeding with our formal definitions. Sparsity refers to finding parameter estimates that have more zeros than not (where zeros in estimation means absence of some effect or relationship). Shrinkage means estimation where many parameter elements are suppressed towards zero, but they are not necessarily zero. While many readers might be familiar with these concepts, interpretation from a Bayesian point of view is slightly different from frequentist approaches. Sparsity is not identical for the simple reason that parameters in the Bayesian paradigm are (continuous, in many cases) random variables. Similarly, shrinkage estimation is embedded in Bayesian inference since any non-diffusing (non-flat) prior will tend to bias the likelihood; the frequentist statistician can only achieve shrinkage if they specify the estimation problem using an explicit penalized likelihood approach. 

We explain these differences, and many more concepts, in this detailed review. We build our discussion gradually by introducing in this section basic components of Bayesian decision theory and estimation, and the principles of Bayesian model determination using the marginal likelihood. In Section 2 we introduce the concept of hierarchical priors and present the basic properties of a large class of hierarchical representations of Bayesian sparsity and shrinkage estimators. In Section 3 we focus on computation using hierarchical priors, and strategies for making inference in high-dimension computationally feasible. Section 4 demonstrates how the hierarchical priors and computational tools discussed in the previous sections, can be readily applied to a wide class of models that are important in economics and finance, as well as other fields of science. Section 5 concludes this review. 

Throughout this review we make the assumption that the reader has a broad understanding of the concept of a prior distribution. If this is not the case, novice readers are advised to begin reading about the basics of Bayesian inference in \autoref{sec:principles} and then move to \autoref{sec:estimators}. More experienced readers, can move directly to \autoref{sec:HP}, skipping the material in this section.

\subsection{Bayesian decision theory and estimation} \label{sec:estimators}
In order to motivate shrinkage and sparsity, we first introduce the concept of loss-based estimation using a Bayesian decision theoretic approach. Detailed introductions can be found in \cite{Shrinkage2018} and \cite{Robert2007}. Assume we have data $X \in \mathcal{X}$ where $\mathcal{X}$ (the sample space) is a measurable set of ${\rm I\!R}^{n}$, and parameters $\theta \in \Theta$ where $\Theta$ (the parameter space) is a measurable set of ${\rm I\!R}^{p}$. We define two probability density functions (p.d.f.) that are measurable on $\mathcal{X}$ and $\Theta$: a the likelihood function $p\left(X \vert \theta \right)$, and a prior function $\pi(\theta)$. Denote with $\wh \theta(X)$ an estimator of $\theta$, that is, a measurable function of data $X$ that maps from ${\rm I\!R}^{n}$ to ${\rm I\!R}^{p}$.

Under these definitions we can now specify what is the loss and risk associated with the estimator $\wh \theta(X)$. First, we can define loss functions of the form $L\left( \wh \theta(X), \theta \right) = \rho\left( \wh \theta(X), \theta \right)$ where $ \rho(\bullet)$ can be a symmetric loss function (the quadratic being the most popular) or any asymmetric loss function that measures how close $\wh \theta(X)$ is to the true $\theta$. The Bayes risk associated with ``decision'' $\wh \theta$ is defined as \citep[see also][]{Shrinkage2018}
\begin{equation}
r\left(\pi, \wh \theta \right) = \int_{\Theta} E_{\theta}\left( L \left(\wh \theta(X), \theta\right) \right) d\pi(\theta). \label{Bayes_risk}
\end{equation}
The quantity $\mathcal{R} (\theta, \hat{\theta}) = E_{\theta} \left( L \left(\wh \theta(X), \theta \right) \right)$ is the frequentist risk of $\wh \theta$, which is defined as the expected value of the loss function over the data realization for a fixed $\theta$. In contrast, the Bayes risk in \autoref{Bayes_risk} is the average of frequentist risk $\mathcal{R}$ with respect to the prior distribution $\pi(\theta)$. Frequentist decision theory aims at making the expected loss $\mathcal{R}(\theta, \hat{\theta})$ small, while Bayesian decision theory aims at finding the minimum of $r\left(\pi, \wh \theta \right)$. In particular, the quantity
\begin{equation}
r(\pi) = \inf_{\wh \theta} r\left(\pi, \wh \theta \right),
\end{equation}
is the Bayes risk of the prior distribution $\pi$.
Given a prior $\pi$, an associated Bayes estimator $\hat{\theta}_\pi$  is a minimizer in the sense that $r(\pi, \hat{\theta}_\pi) =r(\pi)  $.

We can now define the concepts of minimaxity and admissibility. A decision rule (estimator) is \emph{admissible} with respect to the loss function $L$ if and only if no other rule dominates it. 
That is, iff $r\left(\pi, \widetilde{\theta} \right) < r\left(\pi, \wh \theta \right)$ then $\widetilde{\theta}$ is admissible. An estimator is $\hat{\theta}_0$ is \emph{minimax} for a given  loss function $L$ if 
\begin{equation}
\sup_{\theta}\mathcal{R}(\theta,\hat{\theta}_0)
=
\inf_{\hat{\theta} } \sup_{\theta} \mathcal{R}(\theta,\hat{\theta});
\end{equation}
that is, it is the minimizer of the worst-case frequentist risk. For a given prior $\pi$, define an associated Bayes estimator $\hat{\theta}_\pi$.
If $\sup_{\theta}\mathcal{R}(\theta, \hat{\theta}_\pi) = r(\pi,\hat{\theta}_\pi)$,
then $\hat{\theta}_\pi$ can be shown to be minimax. In this case, the prior $\pi$ is least favorable in the sense that $r(\pi', \hat{\theta}_\pi) \leq r(\pi, \hat{\theta}_\pi)$ for all other priors $\pi'$.
That is, $\hat{\theta}_\pi$ is the best with respect to the least favorable prior distribution $\pi(\theta)$. Minimaxity is a desirable feature for comparing estimators but, of course, it can still become a misleading measure of comparison; see a counterexample and further discussion in \cite{Robert2007}. Finally, note that if a minimax estimator is a unique (Bayes) estimator, then this is also admissible.

Why is it important to think in terms of optimality of an estimator with respect to a loss function? To answer this question, consider the expected value of the squared error loss of a \emph{scalar, point} estimator $\wh \theta = \wh \theta(X)$, which is also known as the mean squared error:
\begin{eqnarray}
MSE(\wh \theta) & = & E\left[ L \left(\wh \theta, \theta \right) \right] = E\left[ \left( \wh \theta - \theta \right)^{2} \right], \\
& = & E \left[ \left( \wh \theta - E\left\{  \wh \theta  \right\} + E\left\{  \wh \theta  \right\} - \theta \right)^{2}\right], \\
& = & E\left[ \left(\wh \theta -  E\left\{  \wh \theta  \right\}  \right)^{2} \right] + \left( E\left\{  \wh \theta  \right\} - \theta \right)^{2}. \label{MSE_est}
\end{eqnarray}
The first term in the last equation above is the variance of $\wh \theta$, and the second term is the square of its bias. The least squares estimator, which in many simple linear settings coincides with the maximum likelihood estimator, has zero bias (unbiased) and is the ``best'' meaning that it has narrowest sampling distribution (minimum variance) among all unbiased estimators. Despite these two desirable properties, it is not necessarily the case that OLS will always have the lowest mean squared error. Indeed, in high-dimensional cases with fat data ($p$ large relative to $n$) the sample variance of the OLS will tend to become very large. In cases with more parameters than observations ($p>n$), the OLS estimator has infinite solutions and infinite variance. In such cases, there exist biased estimators that achieve much lower variance compared to the unbiased estimator, to the extend that this reduction in variance compensates for any increase in the square of the bias (making the total MSE of the biased estimator lower). Specifically in the case of out-of-sample prediction the MSE of our modeled variable will be larger if the estimation MSE in \autoref{MSE_est} is high, showing that evaluating estimation loss might be more important than looking only at (minimum variance) unbiasedness.

A well-known illustration of this concept, that changed dramatically the way statisticians think about estimators, is the example of the James-Stein estimator. Assume our likelihood is $X \sim N_{p} \left(\theta, \underline{\sigma}^{2} I_{p} \right)$ where $\theta \in {\rm I\!R}^{p}$ is the unknown parameter and $\underline{\sigma}^{2}$ is assumed to be known. \cite{Stein1956} proved that the maximum likelihood estimator $\wh \theta^{mle} = X$ is the minimum risk equivariant estimator under various loss functions, it is minimax, and it is admissible for $p=1,2$. However, for $p \geq 3$ the maximum likelihood estimator is inadmissible under a square loss function, and the James-Stein estimator
\begin{equation}
\wh \theta^{JS} = \left(1 - \frac{(p-2)\underline{\sigma}^{2}}{\sum_{i=1}^{n} X_{i}}\right)X, \label{Stein}
\end{equation}
has lower risk than the MLE, that is, $\mathcal{R}(\wh \theta^{JS}) < \mathcal{R}(\wh \theta^{mle})$. \cite{EfronMorris1973} showed that the James-Stein estimator is a special case of an empirical Bayes estimator of $\theta$, that is, an estimator that places a Gaussian prior on $\theta$ and sets its prior variance to be a certain function of the data $X$. Stein's estimator minimizes the \emph{total} quadratic risk of $\theta$, but there may be elements $\widehat{\theta}_{i}^{JS}$, $i \in [1,p]$, which have higher risk than the MLE. For that reason, \cite{EfronMorris1973} also propose a \emph{limited translation empirical Bayes estimator}, which offers a compromise between Stein's estimator and the MLE.

Bayesian estimators are by default biased towards the prior expectation, which is a result of doing inference by using the information in both the likelihood and prior functions. Similarly, penalized likelihood estimators, such as the popular lasso of \cite{Tibshirani1996}, constrain the likelihood function with a penalty that intends to introduce a similar bias. The purpose of this subsection is to introduce an alternative view to traditional econometric inference with small parameter space, where unbiasedness is the holy grail. In high-dimensional settings some estimation bias may be desirable, especially when the purpose is prediction in which case richly parameterized specifications are not welcome. In many instances, in-sample parameter estimation accuracy (instead of out-of-sample prediction) is of primary importance, for example, when the quantity of interest is an elasticity or a causal effect that can inform policy decisions. We show later in this survey that even in such cases Bayesian and frequentist penalized regression estimators can be desirable.

\subsection[Principles of Bayesian Model Choice]{Principles of Bayesian Model Choice: A regression perspective} \label{sec:principles}
According to \cite{BDA2013} the process of Bayesian data analysis involves three steps
\begin{enumerate}
\item Setting up a full probability model. This doesn't only involve specifying a likelihood for our data (observables), but we need to specify a joint distribution for both observables and unobservables (parameters, or other unobserved data/variables)
\item Conditioning on the observed data in order to calculate posterior probabilities of all unobservables
\item Assessing model fit, for example, understanding limitations of the chosen likelihood and prior for recovering interpretable and useful parameters estimates, and addressing sensitivity of the results to these choices 
\end{enumerate}
In the first part of this review, we use a simple linear regression setting as the basis for developing shrinkage and sparsity priors (step 1), for discussing posterior computation (step 2) and assessing model fit (step 3). By doing so we aim to offer the same level playing field for presenting various hierarchical prior formulations. The final section presents several extensions of shrinkage and sparsity priors in more complex settings, such as factor models, time-varying parameter regression, and cofounder selection in treatment effect estimation.

The regression model we build upon has the form
\begin{equation}
y_{i} = \bm X_{i} \bm \beta + \varepsilon_{i},   \text{ \ \ \ } i=1,...,n, \label{regression}
\end{equation}
where $n$ is the number of observations, $y_{i}$ is a scalar dependent variable, $\bm X_{i}$ is a $1 \times p$ vector of covariates (or \emph{regressors} or \emph{predictors}) that can possibly include an intercept, dummies, exogenous variables or other effects (e.g.\ trend in a time-series setting), $\bm \beta$ is a $p \times 1$ vector of regression coefficients, and $\varepsilon_{i} \sim N(0,\sigma^2)$ is a Gaussian disturbance term with zero mean and scalar variance parameter $\sigma^2$. Within this setting our interest lies in obtaining ``good'' estimates of $\bm \beta$ and $\sigma^{2}$, specifically in settings with many covariates (``large $p$, small $n$'' regression). 

The linear regression formulation implies a certain Gaussian likelihood function $\mathcal{L}(\bm \beta,\sigma^{2} \vert \bm y, \bm X)$ that is proportional to the sampling density $p(\bm y \vert \bm \beta,\sigma^2)$. These two quantities are not identical because the likelihood is not a true density function.\footnote{The likelihood is a product of densities that lacks a normalizing constant. } The Bayesian needs to specify a joint prior distribution of the parameters, in the form $p(\bm \beta,\sigma^2)$. Bayes Theorem postulates that 
\begin{equation}
p(\bm \beta,\sigma^2 \vert \bm y) = \frac{p(\bm y \vert \bm \beta,\sigma^2) \times p(\bm \beta,\sigma^2)}{p(\bm y)}, \label{Bayes1}
\end{equation}
but for the purpose of parameter estimation, in particular, it is easier to ignore $p(\bm y)$ since it is a normalizing constant (i.e.\ not a function of the parameters of interest $\bm \beta$, $\sigma^2$) and work instead with the formula
\begin{equation}
p(\bm \beta,\sigma^2 \vert \bm y) \propto p(\bm y \vert \bm \beta,\sigma^2) \times p(\bm \beta,\sigma^2). \label{Bayes2}
\end{equation}

A default prior setting in Bayesian inference is the natural conjugate prior which is defined as
\begin{eqnarray}
p(\bm \beta, \sigma^{2}) & = & p(\bm \beta \vert \sigma^{2})p(\sigma^{2}) \\
                         & = & N(\bm 0, \sigma^{2} \bm D) \times Inv-Gamma\left( \frac{v_0}{2},\frac{s_{0}^{2}}{2} \right) , \\
                         & \propto & \left( \sigma^{2} \right)^{-\frac{p}{2}} \exp\left\lbrace -\frac{1}{2\sigma^{2}} \bm \beta^{\prime} \bm D^{-1} \bm \beta   \right \rbrace \\
                         & &\times \left( \sigma^{2} \right)^{-v_0/2-1} \exp \left \lbrace - \frac{s_{0}^{2}/2}{\sigma^{2}}  \right \rbrace,
\end{eqnarray}
where $(\bm D, v_0, s_0)$ are prior hyperparameters chosen by the researcher. Due to the fact that the likelihood has a similar structure to this prior, it is trivial to prove (see the accompanying Technical Document) that the posterior is of the form
{\small \begin{eqnarray}
p(\bm \beta, \sigma^{2} \vert \bm y) & = & N \left( \bm V \left( \bm X^{\prime} \bm y\right),\sigma^{2} \bm V \right) \times Inv-Gamma \left( \frac{v}{2},\frac{s^{2}}{2} \right), \label{CNP_post}
\end{eqnarray}
}where $\bm V = \left( \bm X^{\prime} \bm X + \bm D^{-1} \right)^{-1}$, 
$v = v_0 + n+p$,
$s^{2} = s_{0}^{2} +(\bm y - \bm X \bm \beta)'(\bm y - \bm X \bm \beta)+ \bm \beta' \bm D^{-1} \bm \beta$

 $\bm X = \left[\bm X_{1}^{\prime},..., \bm X_{n}^{\prime}\right]^{\prime}$ and 
 $\bm y = \left(y_{1},...,y_{n} \right)^{\prime}$.

\subsubsection{Goodness of fit measures: Marginal likelihood and information criteria}
While \autoref{Bayes2} is required for the derivation of parameter posterior distributions, the quantity $p(\bm y)$ in \autoref{Bayes1} is of paramount importance for Bayesian model determination. This is the \emph{prior predictive likelihood}, more commonly known as the \emph{marginal likelihood}, that is, the evidence in data $\bm y$ after we integrate out the effect of all possible values that the ``random variables'' $\bm \beta,\sigma^2$ can admit through their prior distribution. This can be proven via solving for $p(\bm y)$ in \autoref{Bayes1}:
{\small
\begin{align}
p(\bm y) p(\bm \beta,\sigma^2 \vert \bm y) & =  p(\bm y \vert \bm \beta,\sigma^2)  p(\bm \beta,\sigma^2) \\
 \Rightarrow \int_{-\infty}^{\infty} \int_{0}^{\infty} p(\bm y)  p(\bm \beta,\sigma^2 \vert \bm y) d\bm \beta d\sigma^2 & =  \int_{-\infty}^{\infty} \int_{0}^{\infty} p(\bm y \vert \bm \beta,\sigma^2) p(\bm \beta,\sigma^2) d\bm \beta d\sigma^2  \\
 \Rightarrow p(\bm y)\int_{-\infty}^{\infty} \int_{0}^{\infty}  p(\bm \beta,\sigma^2 \vert y) d\bm \beta d\sigma^2 & =  \int_{-\infty}^{\infty} \int_{0}^{\infty} p(\bm y \vert \bm \beta,\sigma^2) p(\bm \beta,\sigma^2) d\bm \beta d\sigma^2  \\
 \Rightarrow p(\bm y) & =  \int_{-\infty}^{\infty} \int_{0}^{\infty} p(\bm y \vert \bm \beta,\sigma^2) p(\bm \beta,\sigma^2) d\bm \beta d\sigma^2, \label{ML}
\end{align}
}where $\int_{-\infty}^{\infty} \int_{0}^{\infty}  p(\bm \beta,\sigma^2 \vert y) d\bm \beta d\sigma^2= 1$ because this is a proper density. The marginal likelihood is the expected value of the likelihood where the expectation is taken with respect to the prior. Put differently, it is the prior mean of the likelihood function. An important characteristic of the marginal likelihood is that the integral in \autoref{ML} can only be calculated when the prior is a proper density, that is, if $p(\bm \beta,\sigma^2)$ integrates to one. The benchmark Uniform (Jeffrey's) prior on $\bm \beta$ and $\log(\sigma^2)$ is a key example where this condition fails and the marginal likelihood does not exist.

Assume we want to predict a new (future) observation $y_{n+1}$ given $\bm X_{n+1}$ using the prediction (out-of-sample) model $p(y_{n+1} \vert \bm \beta,\sigma^2, \bm y)$ which, in turn, is based on the in-sample estimated model $p(\bm y \vert \bm \beta,\sigma^2)$. We can then define the \emph{posterior predictive likelihood} 
\begin{equation}
p(y_{n+1} \vert \bm y)  =  \int_{-\infty}^{\infty} \int_{0}^{\infty} p( y_{n+1} \vert \bm \beta,\sigma^2,\bm y) p(\bm \beta,\sigma^2 \vert \bm y) d\bm \beta d\sigma^2, \label{PPL}
\end{equation}
which is the distribution of the out-of-sample data point marginalized over the posterior distribution of the model parameters.

Both quantities -- prior and posterior predictive distributions -- are fundamental for model assessment in Bayesian inference. In the benchmark case of the linear regression with the natural conjugate prior, the marginal likelihood can be derived analytically and is of the form
\begin{eqnarray}
p\left( \bm y \right) & = & \frac{p(y \vert \bm \beta,\sigma^2) \times p(\bm \beta,\sigma^2)}{p(\bm \beta,\sigma^2 \vert \bm y)} \\
& = & 
\frac{\Gamma \left( \frac{v_0}{2} \right)^{-1}  \left(s_{0}/2\right)^{\frac{v_{0}}{2}}   }{(2\pi)^{\frac{n}{2}}\Gamma\left( \frac{v}{2} \right)^{-1}\left(s/2\right)^{\frac{v}{2}}}   \frac{\vert \bm D \vert^{-\frac{1}{2}}}{\vert \bm V \vert^{-\frac{1}{2}} }\\
&\times &
\bigg[\frac{1}{2} \left( s_0 + \bm y' \bm y - \bm \mu^* \bm V^{-1} \bm \mu^*\right) \bigg]\\
\label{ML_NCP}
\end{eqnarray}
where $v_0,s_0,\bm D$ are parameters of the prior distribution (chosen by the researcher), and
 $v,s,\bm V$ are parameters of the posterior distribution whose values are provided in \autoref{CNP_post} and
 $\bm \mu^* = \bm V (\bm X' \bm y)$.

The predictive likelihood is also available analytically and it is of the form
\begin{eqnarray}
y_{n+1} \vert \bm y 
& \sim &
t_1\left( y_{n+1};  \bm X_{n+1} \bm V \left( \bm X^{\prime} \bm y\right), \frac{s}{v}\left( 1+ \bm X_{n+1} \bm V \bm X_{n+1}'\right),  v  \right)
\end{eqnarray}

where we define the $p$-dimensional t-density with location $\bm \mu$, scale matrix $\bm \Sigma$, and degrees of freedom $d$ as 
\begin{eqnarray}
t_p\left( \bm x;  \bm \mu, \bm \Sigma,  d \right)
=
\frac{ \Gamma\left( \frac{d+p}{2} \right)  }{\Gamma \left( \frac{d}{2} \right) d^{p/2} \pi^{p/2} | \bm \Sigma |^{1/2} }
\left[1 + \frac{1}{d} \left( \bm x - \bm \mu \right)' \bm \Sigma^{-1} \left( \bm x - \bm \mu \right) \right]
\end{eqnarray}

The marginal likelihood is rarely available analytically, and in most cases the integral in \autoref{ML} has to be approximated using Monte Carlo or numerical methods.\footnote{Two early examples are \cite{GelfandDey1994} and \cite{Chib1995}; see also \cite{ChibJeliazkov2001} for a review.} In cases of either a complex model or a complex prior structure, or both, evaluating the marginal likelihood can become challenging, if not impossible. In such cases it might be easier to calculate the posterior predictive likelihood in \autoref{PPL} using a procedure called \emph{leave one out cross-validation} (LOO-CV). This would involve fitting the model in training data and then using a hold-out sample to evaluate the posterior predictive likelihood. Notice that if MCMC samples from the parameter posterior are available, evaluation of \autoref{PPL} is straightforward using Monte Carlo integration.\footnote{Recognizing the numerical and computational shortcomings of model choice based on marginal likelihoods, there are several early studies that propose model choice criteria that are based on variants of the posterior predictive distribution, see \cite{Davison1986}, \cite{GelfandGhosh1998}, \cite{Gelmanetal1996}, \cite{LaudIbrahim1995}, \cite{IbrahimLaud1994} and \cite{SanMartiniSpezzaferri1984}.}

When marginal or posterior predictive likelihoods are difficult to obtain, a (computationally) straightforward alternative strategy is to rely on information criteria. For example, the Bayesian information criterion (BIC), is a first-order approximation to the marginal likelihood. Performing a Taylor expansion around the posterior mode\footnote{The posterior mode is chosen such that the first derivative of the posterior is zero, which simplifies terms when taking the Taylor expansion; see \cite{Raftery1995} for a detailed proof.} $(\widetilde{\bm \beta},\widetilde{\sigma}^{2})$ for the logarithm of the term $p \left( \bm y \vert \bm \beta, \sigma^{2} \right) p \left(\bm \beta,\sigma^{2} \right)$ in \autoref{ML}, we can write the log-marginal likelihood as
\begin{equation}
\begin{array}{ccl}
\log p \left(\bm y \right) & = & \log p\left( \bm y \vert \widetilde{\bm \beta}, \widetilde{\sigma}^2 \right) + \log p \left( \widetilde{\bm \beta},\widetilde{\sigma}^{2} \right) + \frac{p}{2} \log(2\pi) \\ && - \frac{p}{2}\log n - \frac{1}{2} \log \left\vert J_{n}\left(\widetilde{\bm \beta},\widetilde{\sigma}^{2} \right) \right\vert + O\left(n^{-1} \right),
\end{array} \label{BIC_temp}
\end{equation}
where $J_{n}\left(\widetilde{\bm \beta},\widetilde{\sigma}^{2} \right)$ is the expected Fisher information matrix of $p\left( \bm y \vert \bm \beta, \sigma^{2} \right) p \left(\bm \beta,\sigma^{2} \right)$ evaluated at the posterior mode $(\widetilde{\bm \beta},\widetilde{\sigma}^{2})$. In large samples, the posterior mode coincides with the MLE $(\widehat{\bm \beta},\widehat{\sigma}^{2})$. Considering this approximation and removing from \autoref{BIC_temp} any terms of order $O\left(1\right)$ or less, we obtain
\begin{equation}
\log p \left(\bm y \right) = \log p\left( \bm y \vert \widehat{\bm \beta}, \widehat{\sigma}^2 \right)  - \frac{p}{2}\log n + O\left(1\right).
\end{equation}
The approximation above provides the basis for defining the Bayesian information criterion
\begin{equation}
BIC = -2 \log \mathcal{L} \left( \widehat{\bm \beta}, \widehat{\sigma}^2 \vert \bm y, \bm X \right) + p \log n,
\end{equation}
where $\mathcal{L} \left( \widehat{\bm \beta}, \widehat{\sigma}^2 \vert \bm y, \bm X \right)$ is the likelihood function evaluated at the MLE.

The BIC is only a crude approximation to the marginal likelihood and it is based on a point estimate. An alternative popular criterion is the deviance information criterion (DIC) proposed by \cite{Spiegelhalteretal2002} which is of the form
\begin{equation}
DIC = -4 E_{p(\bm \beta, \sigma^2 \vert \bm y)} \left[ \log p\left( \bm y \vert \bm \beta, \sigma^2 \right)\right] + 2  \log p\left( \bm y \vert \widetilde{\bm \beta}, \widetilde{\sigma}^2 \right).
\end{equation}
The first term is the expectation of the data density with respect to the posterior\footnote{For that reason, the DIC is related to the posterior predictive likelihood, i.e.\ the integral in \autoref{PPL}, rather than the marginal likelihood.} which can be evaluated numerically from the MCMC output by taking the mean of $p(\bm \beta, \sigma^2 \vert \bm y)$ over all MCMC samples of the parameters. The second term is the value of the data density evaluated at the posterior mode $(\widetilde{\bm \beta},\widetilde{\sigma}^{2})$. For more information on the DIC see also \cite{ChanGrant2016}, \cite{Spiegelhalteretal2014} and \cite{vanderLinde2005}.

\cite{ChenChen2008} propose a modification to the Bayesian information criterion for high-dimensional spaces, which they call the extended Bayesian information criterion (EBIC). In the context of a proportional hazards model, \cite{VolinskyRaftery2000} propose a modification of the BIC penalty term that is consistent with a conjugate unit-information prior under this model. \cite{FosterGeorge1994} propose the risk inflation criterion (RIC) while \cite{FosterGeorge2000} present empirical Bayes selection criteria. \cite{Watanabe2010,Watanabe2013} derives the widely applicable information criterion (WAIC), also known as the Watanabe-Akaike information criterion since this criterion can be considered to be a Bayesian variant of the popular Akaike information criterion. \cite{Gelmanetal2014} and \cite{Vehtarietal2017} perform informative comparisons of the properties of BIC, DIC, WAIC and LOO-CV in a Bayesian context.

\subsubsection{Testing hypotheses: Bayes factors}
Consider now the case of two competing models, model one (denoted as $M_{1}$) and model two (denoted as $M_{2}$). For example, a key scenario that fits this setting, is that of testing hypotheses of the form $H_0: \beta_{j} = 0$ vs $H_1: \beta_{j} \neq 0$, for some $j=1,...,p$. Evidence in favor of either $H_{0}$ or $H_{1}$, corresponds to how good is the fit of two corresponding nested regression models ($M_{1}$ is unrestricted, and $M_{2}$ has the restriction $\beta_{j} = 0$ imposed). In this setting it is convenient to condition parameter posteriors and marginal likelihoods for each model on the random variable $M_{i}$, $i=1,2$, that indexes each of the two models. For example, $p(\bm \beta, \sigma^{2} \vert \bm y, M_{1})$ and $p(\bm y \vert M_{1})$ denote the parameter posterior and marginal likelihood, respectively, of regression model $1$. Consequently, the quantity
\begin{equation}
BF_{12} = \frac{p(\bm y \vert M_{1})}{p(\bm y \vert M_{2})},
\end{equation}
is the \emph{Bayes Factor} between models $1$ and $2$. The quantity
\begin{equation}
PO_{12} \equiv \frac{p\left(M_1 \vert \bm y \right)}{p\left(M_2 \vert \bm y \right)}  = \frac{p(\bm y \vert M_{1})}{p(\bm y \vert M_{2})} \times \frac{p\left(M_1 \right)}{p\left(M_2 \right)}
\end{equation}
is the \emph{posterior odds} between models $1$ and $2$. It is defined as the product of the Bayes factor and the prior odds. If we assign equal model probabilities a-priori, then $p\left(M_1 \right)=p\left(M_2 \right)=\frac{1}{2}$ and the Bayes factor is identical to the posterior odds ratio. The Bayes factor above is a primary tool for assessing evidence in favor of a statistical model versus a competing model.
 
\cite{KassRaftery1995} provide a rule-of-thumb on how to interpret the statistical evidence against model $2$ based on ranges of values of $BF_{12}$: for values higher than three the evidence is substantial, for values higher than 10 it is strong, and for values higher than 100 it is decisive. Given that marginal likelihoods are not available with improper priors (even if the posterior is proper), there has been plenty of interest in calculating Bayes factors when such priors are used. \cite{Aitkin1991} proposes to calculate Bayes factors based on integrating the likelihood with the posterior -- this is equivalent to replacing $p(\bm \beta,\sigma^2)$ with $p(\bm \beta,\sigma^2 \vert \bm y)$ in \autoref{ML}. This formulation allows to calculate ``posterior'' Bayes factors regardless of the prior structure of each model, and at the same time it avoids Lindley's paradox \citep{Aitkin1991}. \cite{BergerPericchi1996,BergerPericchi1998} suggest the use of the \emph{intrinsic} Bayes factor. Their suggestion involves splitting the data into $n$ subsets, such that one can obtain the marginal likelihood of the $i^{th}$ subset conditional on all other subsets. Subsequently, either the arithmetic or geometric average of the Bayes factors estimated in all $n$ subsets of the data can be used as the final estimate.

For nested model comparisons, \cite{VerdinelliWasserman1995} show that Bayes factors can be calculated using the Savage-Dickey density ratio (SDDR) approach. Consider two regression models as in \autoref{regression} but for notational simplicity set $p=1$, that is, only a single covariate is available. The first model, $M_{1}$, is an unrestricted model while model $M_{2}$ imposes the restriction $\beta = \beta^{\star}$ for some scalar value $\beta^{\star}$ (the previous example of testing of $H_0:\beta = 0$ vs $H_1:\beta \neq 0$ fits this setting). In this case the Bayes factor can be written as
\begin{eqnarray}
BF_{12} & = & \frac{p(\bm y \vert M_{1})}{p(\bm y \vert M_{2})} \\
        & = & \frac{\int_{-\infty}^{\infty} \int_{0}^{\infty} p(\bm y \vert \beta,\sigma^2, M_{1}) p(\beta,\sigma^2 \vert M_{1}) d\beta d\sigma^2}{\int_{0}^{\infty} p(\bm y \vert \beta^{\star},\sigma^2, M_{2}) p(\beta^{\star},\sigma^2 \vert M_{2})  d\sigma^2 } \\
        & = & \frac{\int_{0}^{\infty} p \left(\beta^{\star}, \sigma^{2} \vert \bm y, M_{2} \right) d\sigma^2}{\int_{0}^{\infty}  p\left( \beta^{\star},\sigma^{2} \vert M_{2} \right)d\sigma^2},
\end{eqnarray}
that is, SSDR is the ratio of the marginal posterior and prior of $\beta$ under model $M_{2}$, evaluated at the point $\beta = \beta^{\star}$. In general it will be easy to evaluate these two distributions, especially when the Gibbs sampler is used for approximating the posterior distribution. This is because evaluation of the numerator using Monte Carlo integration would be fairly straightforward. Additionally, in the case of an independent prior of the form $p(\beta,\sigma^{2}) = p(\beta)p(\sigma^{2})$ the denominator above becomes $\int_{0}^{\infty}  p\left( \beta^{\star},\sigma^{2} \vert M_{2} \right)d\sigma^2 = p\left( \beta^{\star} \vert M_{2} \right) \int_{0}^{\infty}  p\left( \sigma^{2} \vert M_{2} \right)d\sigma^2 = p\left( \beta^{\star} \vert M_{2} \right)$, i.e.\ we only need to evaluate the (Gaussian) prior of $\beta$ at the point $\beta^{\star}$.

There are of course numerous other ways of obtaining approximations to the Bayes factors that do not explicitly involve calculating ratios of marginal likelihoods. \cite{GoutisRobert1998} propose an alternative procedure for testing nested models based on the Kullback-Leibler divergence. The idea is to compute the projection of the unrestricted model to the restricted parameter space, and use the corresponding minimum distance to judge whether or not the restricted model is appropriate. The same way we used the BIC to obtain a first-order approximation to the marginal likelihood, we can also use the BIC to obtain approximations to Bayes factors -- this approach is illustrated in \cite{Raftery1995}. Notable early studies on the topic of Bayes factors include \cite{KassWasserman1995}, \cite{DeSantisSpezzaferri1997}, \cite{OHagan1995}, \cite{BergerPericchi2001}, \cite{BergerMortera1999}, \cite{LewisRaftery1997}, \cite{Raftery1996} and \cite{DiCiccioetal1997}. A systematic review of methods for calculating Bayes factors can be found in \cite{KadaneLazar2004}.

Finally, it is worth noting that in the case of nested hypothesis testing we can derive an optimal Bayesian point estimate by minimizing expected loss averaged over the two hypotheses, using posterior model probabilities as weights. That is, considering again the simple case with $p=1$ and ignoring the variance parameter $\sigma^2$ for simplicity, we aim to find point estimate $\widehat{\beta}$ such that the joint expected loss under the two models/hypotheses
\begin{eqnarray}
E\left( L \left( \beta,\widehat{\beta} \right) \right) & = & \left[ p(M_{1} \vert \bm y) E\left( L \left( \beta,\widehat{\beta}\right) \vert M_1 \right) \right. + \\ && \left. p(M_{2} \vert \bm y)E\left( L \left( \beta,\widehat{\beta}\right) \vert M_2 \right) \right],
\end{eqnarray}
achieves a minimum. Under a quadratic loss function $L \left( \beta,\widehat{\beta}\right)$, the posterior means are optimal meaning that the optimal estimator is
\begin{equation}
\widehat{\beta}^{BPE} = p(M_{1} \vert \bm y) E\left( p\left(\beta \vert \bm y, M_{1} \right) \right) + p(M_{2} \vert \bm y)E\left( p\left(\beta \vert \bm y, M_{1} \right) \right).
\end{equation}
This estimator can be considered a \emph{Bayesian pre-test estimator}, hence the acronym BPE in the equation above; see \cite{judgeetal1985} for a detailed discussion. In the next section we will generalize this result to the case of $K$ models, in order to motivate model choice in the presence of many models.

\subsubsection{Model choice with many models: Bayesian model averaging}
Model choice can have many forms, but the benchmark scenario that will motivate later in this paper to focus on shrinkage and sparse estimation, is that of model determination among many nested models. In particular, consider the problem of deciding which of $p$ variables in the covariate matrix $\bm X$ should be in the ``optimal'' regression model. Each covariate can have two outcomes, either it is included in a model or it is excluded, meaning that the model space in the presence of $p$ covariates is $2^{p}$. We denote the model set as $\mathcal{M}=\left\{ M_r: r=1,\ldots, 2^p \right\}$. The covariates that pertain to model $M_{r}$ are denoted in this subsection as $\bm X_{r}$ and their associated coefficients as $\bm \beta_{r}$. That is, $\bm X_{r}$ is a matrix that is constructed using only a subset of the columns in $\bm X$. Therefore, we denote regression model $M_r$ as\footnote{For simplicity we do not explicitly allow for an intercept. If an intercept is present in all competing models, then it is important to remove the sample mean from all covariates $\bm X$ (and, as a result, in all subsets $\bm X_{r}$) in order to ensure that the estimated intercept has exactly the same interpretation in all models. With demeaned covariates and the use of a flat prior, the intercept term becomes identical to the sample mean of $\bm y$ in all $2^{p}$ competing models.} 
\begin{equation}
M_{r} : \bm y = \bm X_{r} \bm \beta_{r} + \bm \varepsilon.
\end{equation}
where $\bm X_{r}$ is $n \times p_{r}$ and $\bm \beta_{r}$ is $p_{r} \times 1$ with $p_{r} \in \{ 1,...,p\}$. Now with $2^p$ models, even for small $p$, pairwise model comparison based on Bayes factors is impractical and alternative computational methods are needed. Most importantly, in the presence of many models the researcher might not want to give the same weight to each and every model. For example, she might want to give more weight on parsimonious models or models that include a certain predictor suggested by some theory or common sense. For that reason we define prior model probabilities $p(M_r)$ with $\sum_{r=1}^{2^p} p(M_r) = 1$. Based on Bayes theorem, prior model probabilities combined with marginal likelihoods $p(\bm y \vert M_r)$ give posterior model probabilities
\begin{equation}
p(M_r \vert \bm y) \propto p(\bm y \vert M_r)p(M_r).
\end{equation}
\emph{Bayesian model selection (BMS)} corresponds to selecting the best model, that is, the model $M_r$ with the highest $p(M_r \vert \bm y)$. \emph{Bayesian model averaging (BMA)} involves averaging over many models using $p(M_r \vert \bm y)$ as weights. That is, for a quantity of interest $\Delta$ (e.g.\ an out-of-sample observation $y_{n+1}$ of $\bm y$) BMA is constructed as the following weighted average
\begin{eqnarray}
p(\Delta \vert \bm y ) = \sum_{r=1}^{2^p} p(\Delta \vert \bm y, M_r ) p(M_r \vert \bm y).
\end{eqnarray}
For small model spaces, typically when $p<30$ posterior model probabilities can be calculated analytically such that we can enumerate and estimate all $2^{p}$ available models. For $p>30$ it is impossible to enumerate and estimate all models in a deterministic way. In such cases, one can rely on Markov chain Monte Carlo algorithms which are able to ``visit'' in each iteration, in a stochastic way, the most probable models. \cite{Hoetingetal1999} and \cite{Fragosoetal2018} provide two systematic reviews on the topic.

While model selection and model averaging with an arbitrary number of models are straightforward extensions of the case with only two models, prior elicitation in multi-parameter and multi-model settings is anything but straightforward. In order to explain the intuition behind why this is the case, consider the natural conjugate prior defined previously, which in the case of model $M_r$ can be written as
\begin{equation}
p(\bm \beta_{r}, \sigma^{2} \vert M_{r} ) = N_{p_{r}}(\bm 0_{p_{r}}, \sigma^{2} \bm D_{r}) \times Inv-Gamma\left( \frac{v_0}{2},\frac{s_{0}^{2}}{2} \right).
\end{equation}
Prior elicitation involves choice of $\bm D_{r}, v_0, s_0$. The hyperparameters $v_0, s_0$ are scalar in all regression models can be simply set to a small value close to zero, implying a Jeffrey's (diffuse) prior on $\sigma^2$. However, $\bm D_{r}$ is a matrix that changes size based on the number of predictors in model $M_{r}$. Assume for simplicity we define $\bm D_{r} = \tau \bm I_{p_{r}}$, with $\bm I_{p_{r}}$ the $p_{r} \times p_{r}$ identity matrix. In this case, prior elicitation breaks down to choosing a single hyperparameter $\tau$. We can't use the diffuse choice $\tau \rightarrow \infty$ because the marginal likelihood in \autoref{ML_NCP} will become infinite, hence, $\tau$ should be finite in the multi-model case. However, using the same finite value of $\tau$ in all models, doesn't mean that the effect of this prior is identical (that is, ``objective'') for each model. Consider for instance two models, one with two predictors $\bm X_{2} = (\bm x_{1}, \bm x_{2})$ and a restricted model with only the first predictor $\bm X_{1} = \bm x_{1}$. The posterior variance is $\bm V_{r} = \sigma^{2} \left( \bm X_{r}^{\prime}\bm X_{r} + \left(\tau \bm I_{p_{r}}\right)^{-1} \right)^{-1}$ for each model $r=1,2$, so that the impact of $\tau$ on the common predictor in the two models will be identical only if $\bm x_{1}$ is not correlated with $\bm x_{2}$ and $\bm X_{2}^{\prime}\bm X_{2}$ becomes diagonal. If this is not the case, the correlation between the two predictors will imply that the effect of $\tau$ on the regression coefficient of $\bm x_{1}$ will not be the same in the two models. This issue complicates prior elicitation further when considering $p \gg 2$ correlated covariates, that also potentially have different units of measurement.\footnote{The scaling issue in $\bm X$ can be dealt with by standardizing the data, that is, dividing each column with its sample standard deviation. High correlation in columns of $\bm X$ can also be dealt with by orthogonalizing this matrix. While standardization is easy to apply and is recommended in all model averaging and variable selection algorithms, orthogonalization of the columns of $\bm X$ is only feasible when $n>p$. Therefore this latter procedure is not available in the high-dimensional case ($p>n$), which is exactly where there is higher chance of encountering many correlated predictors!}.

For that reason, many researchers have proposed empirical Bayes priors, in the spirit of the empirical Bayes formulation of Stein's estimation rule; see equation \autoref{Stein} and discussion of \cite{EfronMorris1973}. Empirical Bayes procedures allow to choose prior hyperparameters as a function of the data observations, sometimes also chosen to optimize some criterion (e.g.\ maximum marginal likelihood). A default prior for multi-model settings is the \emph{g-prior} due to \cite{Zellner1986}. The $g$-prior for model $M_{r}$ takes the form
\begin{equation}
\bm \beta_{r} \vert \sigma^{2}, M_{r} \sim N_{p_{r}} \left(\bm 0_{p_{r}}, \frac{1}{g} \sigma^{2} \left(\bm X_{r}^{\prime} \bm X_{r} \right)^{-1}\right),
\end{equation}
where $\sigma^{2} \left(\bm X_{r}^{\prime} \bm X_{r} \right)^{-1}$ is essentially the covariance matrix associated with the OLS estimator $\widehat{\bm \beta}_{r}$ and $g$ a scalar tuning parameter. Under this prior, the posterior variance of $\bm \beta$ conditional on $\sigma^2$ becomes $\bm V_{r} = \frac{1}{1 + g} \times \sigma^{2}\left( \bm X_{r}^{\prime}\bm X_{r}\right)^{-1}$, such that the posterior variance is uniformly affected by selection of $g$. Consequently, the posterior mean/mode is
\begin{equation}
\bm \beta_{r}^{\star} = \frac{1}{1 + g} \widehat{\bm \beta}_{r}.
\end{equation}
When $g \rightarrow 0$ the posterior mean tends to the OLS estimate of model $M_r$ ($\widehat{\bm \beta}_{r}$) while when $g \rightarrow \infty$ the posterior contracts towards zero. While the effect of the prior now depends in a straightforward, transparent way\footnote{We avoid using the term ``objective'', first, because as \cite{GelmanHenning2017} argue it is counterproductive to do so and, second, because the $g$-prior is not in any way an objective prior.} on a single hyperparameter, choice of this hypeparameter is very important for determining marginal likelihoods and model probabilities.

\cite{Fernandezetal2001a,Fernandezetal2001b} propose default values of $g$ in the context of Bayesian model averaging and \cite{Eicheretal2011} expand this discussion by considering further values of $g$. A benchmark suggestion of \cite{Fernandezetal2001b} is to set $g \equiv g_{r} = p_{r}/n$, that is, a value of $g$ that is the ratio of the number of coefficients in each model $r$ over the total number of observations. Wide models with many covariates models will have larger $g$, thus, tending to shrink their posterior towards zero more aggressively. Put differently, the prior variance is getting smaller meaning that the information in the prior increases relative to the information in the likelihood. This is a basic principle of shrinkage and variable selection estimators: when $p$ is large and especially when $p>n$, the information in the likelihood is not sufficient to estimate all $p$ coefficients and the prior becomes increasingly important for determining posterior outcomes. That is, for both Bayesian and non-Bayesian approaches, the concepts of shrinkage and sparsity amount to the prior expectation that increasingly many coefficients a priori will be zero or close to zero.

Of course, there are more rigorous ways of selecting $g$. A key contribution is that of \cite{Liangetal2008} who put hyper-priors on $g$, treating it as a random variable. Such hierarchical approaches are the topic of close examination of the next section, so we won't expand on it here. \cite{Krishnaetal2009} extend the $g$-prior into an \emph{adaptive powered correlation prior} of the form
\begin{equation}
\bm \beta_{r} \vert \sigma^{2}, M_{r} \sim N_{p_{r}} \left(\bm 0_{p_{r}}, \frac{1}{g} \sigma^{2} \left(\bm X_{r}^{\prime} \bm X_{r} \right)^{\lambda}\right),
\end{equation}
where $\lambda \in \mathbb{R}$ controls the prior's response to collinearity in predictors. $\lambda=-1$ gives the original prior proposed by Arnold Zellner, while $\lambda=0$ gives the ridge regression prior.

While the $g$-prior addresses the issue of setting a prior on different regression models that might be nested and have correlated covariates, another important issue is how to define a prior on model space. For both conceptual and computational reasons Bayesians prefer to index all possible $2^p$ models using dummy variables $\bm \gamma = (\gamma_{1},...,\gamma_p)^{\prime}$. When $\gamma_j=0$ a covariate is excluded from a model and when $\gamma_{j}=1$ it is included. Therefore, the model with no predictors is indexed as $\bm \gamma = (0,...,0)^{\prime}$ and the model with all predictors is indexed as $\bm \gamma = (1,...,1)^{\prime}$. All intermediate models are indexed by vectors $\bm \gamma$ that are sequences of zeros and ones. Instead of placing priors on the model space, we can now explicitly consider priors on $\bm \gamma$, and the binomial distribution is a good candidate for a parameter that takes $0/1$ values. The binomial prior can become both uniform but also more informative when this is desirable (e.g.\ in high-dimensional spaces, where our prior is that only a small number of predictors will be important).

This setting that combines the $g$-prior on regression coefficients with a binomial prior on model space, is  the major workhorse model for implementing Bayesian variable selection. While its theoretical underpinnings are well-understood (see \citealp{Hoetingetal1999} for a thorough description), it provides the ground for some of the most interesting Bayesian work on computation in high-dimensional settings.\footnote{See for example, \cite{BottoloRichardson2010}, \cite{Clydeetal2011}, \cite{Dellaportasetal2002}, \cite{Hansetal2007}, \cite{JiSchmidler2013}, \cite{MadiganYork1995}, \cite{NottKohn2005} and \cite{Peltolaetal2012b}.} At the same time this setting possesses implicitly the benefits of a hierarchical prior approach. Therefore, we use this brief discussion of BMA as a stepping stone for introducing in the next the concept of full-Bayes/hierarchical Bayes priors that result in shrinkage and sparse estimators.

\section[Hierarchical Priors]{Hierarchical (full Bayes) priors} \label{sec:HP}

When interest lies in models with many parameters, simple priors such as the benchmark natural conjugate prior presented in the previous section, are inadequate for learning interesting features about our parameters and for quantifying uncertainty. In statistics, the concept of hierarchical or multi-level modeling refers to the process of enhancing a simpler model with a richer specification that allows for learning interesting features of a multi-parameter vector, such as groupings or sparsity and shrinkage towards zero, where the latter being the main focus of this review. The Bayesian interpretation of hierarchical modeling involves specifying prior distributions for the prior hyperparameters of regression coefficients, especially when $p$ is large. A simple hierarchical specification for the regression coefficients $\bm \beta$,\footnote{Ignore estimation uncertainty of $\sigma^2$ for the moment, e.g.\ assume it is known and fixed.} takes the form
\begin{eqnarray}
p \left( y_{i} \vert \bm \beta,\sigma^{2} \right) & \sim & N \left( x_{i} \bm \beta, \sigma^{2} \right), \text{ \ \ } i=1,...,n, \label{hier1} \\
p \left(\beta_{j} \vert \mu,\tau^{2} \right) & \sim & N \left( 0,\tau^{2} \right), \text{ \ \ } j=1,...,p, \label{hier2} \\
p \left( \tau^{2} \right) & \sim & F(a,b), \label{hier3} 
\end{eqnarray}
where $F(a,b)$ denotes some distribution function with hyperparameters $(a,b)$. Due to the fact that choice of $\tau^{2}$ is so crucial for the posterior outcome of $\beta_{j}$, the idea behind this hierarchical specification is to treat the hyperparameter $\tau^{2}$ as a random variable and learn about it from the data, via Bayes Theorem. For that reason, a prior such as the one in equations \eqref{hier2} - \eqref{hier3} is many times referred to as a \emph{full-Bayes prior}, as it allows for full quantification of uncertainty around parameters of interest. While the example above pertains to linear regressions with Gaussian likelihood and prior distributions, Section 4 demonstrates that the concept of hierarchical priors is much more powerful and can be applied to numerous multivariate, non-Gaussian, nonlinear or other settings. Additionally, adaptive hierarchies can be defined in which $\beta_{j}$ depends on hyperparameters specific to this $j$-th element ($\tau^{2}_{j}$) that have their individual hyperprior distributions. Finally, if needed, further layers of the hierarchy can be defined: for instance, if choice of the hyperperameter $a$ of $\tau^{2}$ is not straightforward, we can define another level for the prior distribution of $a$, or we could introduce two variance parameters for $\beta_j$ in \autoref{hier2}.\footnote{For example, a powerful class of hierarchical priors called \emph{global-local shrinkage priors} \citep{PolsonScott2010} provides an excellent benchmark for specifying appropriate hierarchical priors. Such priors are of the form
\begin{eqnarray}
p \left(\beta_{j} \vert \tau^{2}, \lambda^{2}_{j} \right) & \sim & N \left( 0,\tau^{2}\lambda^{2}_{j} \right), \text{ \ \ } j=1,...,p, \\
p \left( \tau^{2} \right) & \sim & F_{\tau}(a,b), \\
p \left( \lambda^{2}_{j} \right) & \sim & F_{\lambda}(c,d),
\end{eqnarray}
where $\tau^{2}$ is a global shrinkage parameter (applying the same shrinkage to the whole parameter vector $\bm \beta$) and $\lambda_{j}$ is a local shrinkage parameter (applying shrinkage only to $\beta_j$). As we see next, such priors will typically have at least three hierarchical layers, but in practical situations they tend to have many more (e.g.\ by putting priors on some or all of the hyperparameters $a, b, c, d$). \label{glob_loc}} 

An important feature of the hierarchical prior in equations \eqref{hier2} - \eqref{hier3} is that, while the conditional prior $p \left(\beta_{j} \vert \tau^{2} \right)$ is Gaussian, unconditionally the prior for $\beta_{j}$ is non-Gaussian. Indeed the marginal prior for $\beta_j$ becomes
\begin{equation}
p\left( \beta_j \right)  = \int N\left(\beta_j;0,\tau^{2}\right)p(\tau^{2})d\tau^{2},
\end{equation}
that is, a \emph{scale mixture of normals} representation that allows to approximate very complex prior shapes for $\beta_{j}$.\footnote{It is trivial to show that if $\tau^{2}$ is not a fixed parameter, then unconditionally the prior for $\beta_{j}$ always has excess kurtosis higher than zero, thus, being a leptokurtic distribution with tails thicker than the normal distribution.} Mixtures have the benefit of allowing for classification and grouping of parameters. In the case of identifying sparsity and shrinkage, we can think of the mixture prior as grouping parameters into ``important'' and ``non-important''. Therefore, it is this implied mixture representation of hierarchical modeling with prior distributions that allows to extract interesting features in a multi-parameter setting. Finally, the posterior mode of $\bm \beta$ under a hierarchical prior has a penalized likelihood representation. For the linear regression model, penalized likelihood problems admit the following regularized least squares form
\begin{equation}
\min_{\bm \beta \in \mathbb{R}^{p}} \frac{1}{n}\vert \vert \bm y - \bm X \bm \beta \vert \vert_{2} + g\left( \bm \beta, \lambda \right), \label{penalized}
\end{equation}
where the first term gives the solution to the usual least squares problem and the term $g\left( \bm \beta, \lambda \right)$ defines the penalty as a function of the regression parameters $\bm \beta$ and a scalar (or possibly vector) tuning parameter $\lambda$. Numerous penalized estimators, such as ridge (Tikhonov regularization), lasso, and elastic net fall under the general form in \autoref{penalized}, and Bayesian modal estimators under suitable hierarchical priors can fully recover all of them.

\begin{figure}[H]
\includegraphics[width = \textwidth, trim = {5cm 5cm 4cm 0cm}]{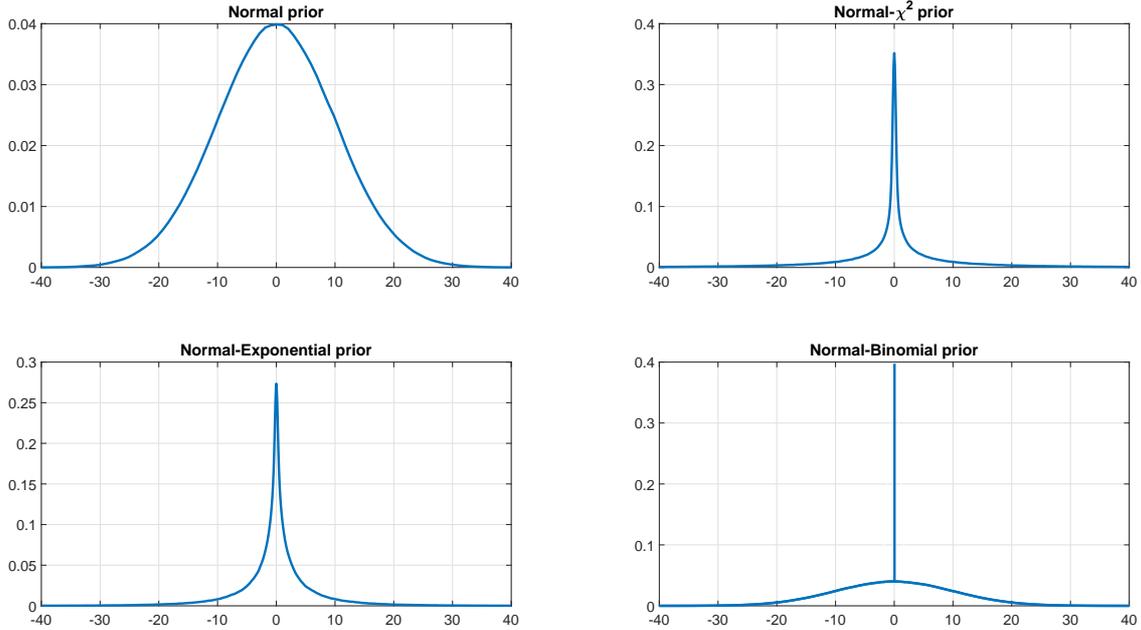}
\quad 
\caption{\emph{Hierarchical priors for a scalar parameter. In all four panels the base distribution is $\beta \sim N(0,100 \times \tau^{2})$. In the top left panel $\tau^{2}=1$ is a fixed hyperparameter, while in the remaining panels it follows a $\chi^{2}_{(1)}$, $Exp(0.5)$ and $Bernoulli(0.9)$ priors.}}
\label{fig:hier_priors}
\end{figure}

In order to understand the ability of hierarchical priors to classify parameters as important and non-important (or non-penalized and penalized), we plot in \autoref{fig:hier_priors} a normal prior with fixed variance vs three cases of a normal prior with variance parameter distributed as $\chi^{2}$ with one degree of freedom, exponential with rate parameter $\lambda=0.5$, and binomial with one trial and probability $\pi = 0.9$ (that is, a Bernoulli distribution). The simple normal prior provides more probability at the origin (zero) relative to its tails, however, it is fairly flat (diffuse) in a small area around zero. What the three mixture priors are introducing, is a more pronounced peak at zero such that when a parameter is in the region of zero it can be shrunk at a faster rate. At the same time, all three mixture distributions have fat tails, providing positive probability to parameter values that are far from zero. That is, these shapes allow for a clearer separation and classification of a parameter as being zero or non-zero. The extreme case of the Bernoulli prior on $\tau^{2}$ (bottom right panel of \autoref{fig:hier_priors}) creates a distribution that looks normal but also has a point mass at zero with high probability. Therefore, all three examples of hierarchical priors provide sharper inference in favor or against the groups of interest (important and non-important parameters).

Computation with hierarchical priors is reviewed in detail in the next section. For now it suffices to note that because of the conditional structure of hierarchical priors, conditional posteriors are typically easy to derive even if the joint parameter posterior is intractable. Sampling from these conditional posteriors using Markov chain Monte Carlo (the \emph{Gibbs sampler}, in particular) is equivalent to taking samples from the intractable joint posterior. Additionally, several approximate methodologies such as variational Bayes and maximum a-posterior (MAP) estimation rely on similar conditional distributions. Therefore, in our discussion in this section we present various hierarchical priors, explain their properties and focus on deriving conditional posteriors. In the next section we discuss in more detail how to use these conditional posteriors to estimate the desired parameters.\footnote{Additional derivations and computational details can be found in the accompanying Technical Document.}

\subsection{Diffusing hierarchical prior}
A natural choice for the variance parameter $\tau^{2}$ in the hierarchical model of equations \eqref{hier1} - \eqref{hier3} is a prior distribution that is diffuse. Similar to Jeffrey's prior for the regression variance $\sigma^{2}$, the choice $\tau^{2} \sim U(0,\infty) $ equivalently $p(\tau^{2}) \propto \tau^{-2}$ can be thought as a default prior choice that reflects our lack of information about sparsity patterns in the data. We might want to also allow for each $\beta_j$ to be determined adaptively, in which case a Jeffrey's prior on hyperparameters $\tau_j^{2}$, $j=1,...,p$ can be defined. Therefore, the full hierarchical prior specification for the regression model is of the form
\begin{eqnarray}
\bm \beta \vert \lbrace \tau_{j}^{2} \rbrace_{j=1}^{p} & \sim & N_{p}(\bm 0,   \bm D_{\tau}), \\
\tau_{j}^{2} & \sim & \frac{1}{\tau_{j}^{2}}, \text{ \ \ for } j=1,...,p, \\
\sigma^{2} & \sim &  \frac{1}{\sigma^{2}},
\end{eqnarray}
where $\bm D_{\tau} =  diag(\tau_{1}^{2},...,\tau_{p}^{2})$. While a Jeffrey's prior on $\tau_j^2$ is a first natural attempt towards hierarchical prior modeling, as \cite{Lindley1983} notes, ``a prior for $\tau^{2}$ that behaves like $\tau^{-2}$ will cause trouble" meaning it will lead to an improper posterior. \cite{gelman2006} examines this issue in more detail and explains why a $Uniform(-\infty,\infty)$ prior on $\log\left( \tau^{2}\right)$ would also not work. However, as \cite{KahnRaftery1992} and \cite{gelman2006} note, under certain conditions, Jeffrey's prior on $\tau^{2}$ yields a limiting proper posterior density. Note that the same improper density can be obtained from the prior $\tau^{2} \sim Inv-Gamma(\epsilon,\epsilon)$ for $\epsilon \rightarrow 0$ (see also \autoref{sec:student} below). \cite{gelman2006} argues that the $Inv-Gamma(\epsilon,\epsilon)$ prior does not have any proper limiting posterior distribution, such that inference becomes sensitive to the choice of $\epsilon$ -- simply setting $\epsilon$ to any ``small'' value is not a reliable solution.

\cite{Figueiredo2003} and \cite{BaeMallick2004} are examples of empirical studies that rely on shrinkage using a uniform hyperprior distribution. \cite{Tipping2001} specifies an inverse gamma prior on $\tau^{2}$ (and calls the resulting hierarchical structure a \emph{sparse Bayesian learning} prior) and adopts the limiting case $\epsilon = 10^{-4}$ as the default hyperparmeter choice. Diffusing priors should not be the first choice in empirical settings especially in high-dimensional and ultra-high-dimensional settings. There are numerous other hyperprior distributions that are interpretable and have better theoretical guarantees \citep{gelman2006}.

\subsection{Student-t shrinkage}  \label{sec:student}
\begin{figure}[H]
\includegraphics[width=\textwidth]{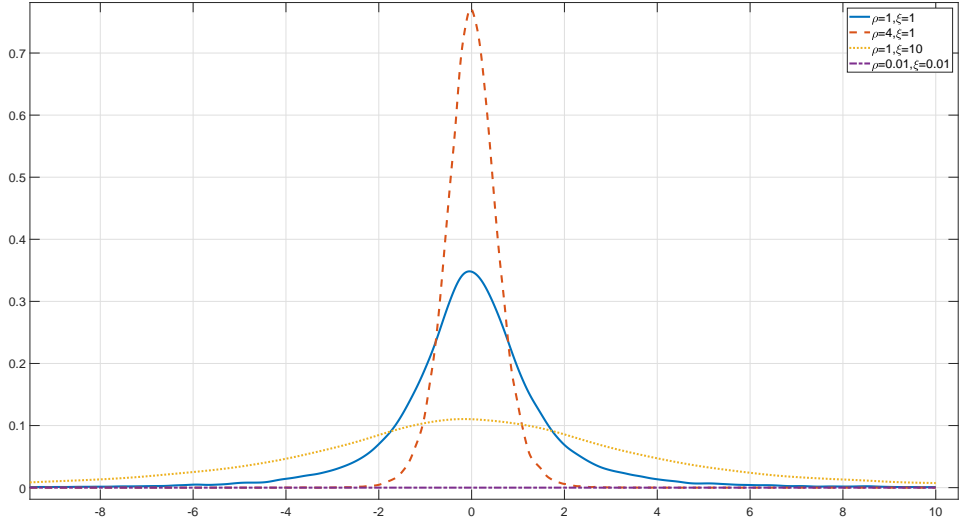}
\caption{\emph{Marginal distribution of $\beta_j$ for the Student-t prior.}}\label{fig:student_t}
\end{figure}

While we just argued that it is not desirable to use the inverse gamma distribution as a way of imposing a diffusing prior on $\tau^{2}$, informative inverse gamma priors provide flexible parametric shrinkage. Following the specification of the normal-inverse gamma prior in \cite{ArmaganZaretzki2010}, we write this prior using the following form
\begin{eqnarray}
\bm \beta \vert \lbrace \tau_{j}^{2} \rbrace_{j=1}^{p} & \sim & N_{p}(\bm 0,   \bm D_{\tau}), \\
\frac{1}{\tau_{j}^{2}} & \sim & Gamma\left(\rho,\xi \right), \text{ \ \ for } j=1,...,p, \\
\sigma^{2} & \sim &  \frac{1}{\sigma^{2}},
\end{eqnarray}
where $\bm D_{\tau} =  diag(\tau_{1}^{2},...,\tau_{p}^{2})$. This is a scale mixture of normals representation of the fat-tailed and leptokurtic Student-t distribution. Similar to our arguments in \autoref{fig:hier_priors} the excess kurtosis of the Student-t results in shrinkage towards zero at a faster rate than the simple normal distribution. At the same time the fatter tails accommodate values of $\tau^2$ that can be far from zero.
 In \autoref{fig:student_t} we illustrate the shape of the marginal distribution of $\beta_j$ for various values of the parameters $\rho,\xi$.

Similar to an inverse gamma prior for the variance parameter $\sigma^2$, the conjugacy of this distribution allows for numerous methods of inference using this prior. For example, \cite{Tipping2001} uses \emph{type-II maximum likelihood} methods \citep{Berger1985}, but (as we discuss in the following section) variational Bayes and other approximate algorithms are also trivial to derive. \cite{ArmaganZaretzki2010} show that conditional posteriors are of the form
\begin{eqnarray}
\bm \beta \vert \lbrace \tau_{j}^{2} \rbrace_{j=1}^{p} , \sigma^{2}, \bm y & \sim & N_p \left( \bm V \times \left(\bm X^{\prime} \bm y \right), \bm V \right), \\
\frac{1}{\tau_{j}^{2}} \bigg\vert \beta_{j}, \bm y & \sim & Gamma\left(\rho +  \frac{1}{2},\xi + \frac{\beta_{j}^{2}}{2} \right), \text{  \ \ } j=1,...,p, \label{student_t_tau}\\
\frac{1}{\sigma^{2}} \bigg\vert \bm \beta, \bm y  & \sim & Gamma \left( \frac{n}{2},\frac{\Psi}{2} \right)
\end{eqnarray}
where $\bm V = \left( \sigma^{-2} \bm X^{\prime} \bm X + \bm D_{\tau}^{-1} \right)^{-1}$ and $\Psi = (\bm y - \bm X \bm \beta)^{\prime}(\bm y - \bm X \bm \beta)$. The Gibbs sampler can be used to sample sequentially from these conditional posteriors, as these samples are guaranteed to be samples from the desired joint parameter posterior.

For the conditional posterior \eqref{student_t_tau} of the prior precisions, $\frac{1}{\tau_{j}^{2}} $, we have 
\begin{align*}
p\left( \frac{1}{\tau_{1}^{2}},\ldots,\frac{1}{\tau_{p}^{2}} \bigg\vert \bm \beta,  \bm y \right)
\propto& \prod_{j=1}^p (2\pi \tau_{j}^{2} )^{-1/2} \exp\left[ -\frac{1}{2\tau_j^2} \beta_j^2 \right] \left( \frac{1}{\tau_j^2} \right)^{\rho-1} \exp\left[ -\frac{\xi}{\tau^2_j}  \right]\\
\propto& \prod_{j=1}^p \left( \frac{1}{\tau_j^2} \right)^{\left( \rho+\frac{1}{2} \right)-1} \exp\left[ -\frac{1}{\tau_j^2} \left( \xi + \frac{\beta_j^2}{2}  \right) \right] 
\end{align*}
where the proportional sign is with respect to $\frac{1}{\tau_{j}^{2}}$'s. It can be seen that the conditional posterior of $\frac{1}{\tau_{j}^{2}}$'s is independent across $j$ and that it has the form in \autoref{student_t_tau}.

\subsection{Normal-gamma priors}
\cite{CaronDoucet2008} proposed the normal-gamma family of hierarchical priors, and \cite{GriffinBrown2010,GriffinBrown2017} established further results and their excellent properties. This prior takes the following hierarchical form
\begin{eqnarray}
\bm \beta & \sim & N_p(\bm 0,\bm D_{\tau}),\\
\tau_{j}^{2} & \sim & Gamma(\lambda,\gamma^{2}/2), \\
\sigma^{2} & \sim & \frac{1}{\sigma^{2}},
\end{eqnarray}
where again $\bm D_{\tau} =  diag(\tau_{1}^{2},...,\tau_{p}^{2})$. The pdf of $\tau_j$ is
\begin{equation}
p\left( \tau^{2} \right) = \frac{\left(\frac{\gamma^2}{2}\right)^{\lambda}}{\Gamma\left( \lambda \right)} \left(\tau_{j}^{2}\right)^{\lambda-1}\exp \left(-\frac{\gamma^2}{2} \tau_{j}^{2} \right),
\end{equation}
such that the marginal pdf of $\beta_j$ is
\begin{equation}
p(\beta_j)  = \frac{\gamma^{(\lambda + 1/2)}}{2^{(\lambda - 1/2)}\sqrt{\pi}\Gamma\left( \lambda \right)} \vert \beta_j \vert^{(\lambda - 1/2)}\mathcal{K}_{(\lambda - 1/2)}\left( \gamma \vert \beta_j \vert \right)
\end{equation}
where $\mathcal{K}_v$ is the modified Bessel function of the second kind, and the tails of this distribution decrease in $\vert \beta_j \vert^{(\lambda - 1)}\exp\left( \gamma \vert \beta_j \vert \right)$. 

Due to the connection of the gamma distribution with a wide array of other distributions (e.g.\ inverse gamma, inverse Gaussian, $\chi^{2}$, etc) choice of the hyperparameters $\lambda$ and $\gamma^{2}$ can result in various shapes for the unconditional distribution of $\bm \beta$ that have different shrinkage properties. This prior becomes diffusing when $\lambda,\gamma^2 \rightarrow 0$, however, this choice falls under the same critique of \cite{gelman2006} for the diffusing inverse gamma prior. This is due to the fact that when $\lambda < 1/2$ the normal-gamma prior places infinite mass in the vicinity of zero, that is, $\lim_{\beta_j \rightarrow 0} p(\beta_j) = \infty$.

\subsection{LASSO prior and extensions}
The least absolute shrinkage and selection operator (lasso) of \cite{Tibshirani1996} has been established as a key workhorse of scientists in all fields working with high-dimensional settings. The estimator takes the form
\begin{equation}
\arg\min_{\bm \beta} \sum_{i=1}^{n} \left( y_i - \bm X_i \bm \beta \right)^2  \text{subject to} \sum_{j=1}^{p} \vert \beta_{j} \vert < t,
\end{equation}
where  $t$ is a prespecified free parameter that determines the degree of regularization. The Lagrangian form of this program is
\begin{equation}
\arg\min_{\bm \beta \in \mathbb{R}^{p}} \vert \vert \bm y - \bm X \bm \beta \vert \vert_{2} + \lambda \vert\vert \bm \beta \vert\vert_{1},
\end{equation} 
where $\vert\vert x \vert\vert_{1} = \sum \vert x_{i} \vert $ is the $\mathcal{l}_{1}$ norm and $\vert\vert x \vert\vert_{2} = \sqrt{\sum x_{i} ^{2} }$ is the $\mathcal{l}_{2}$ norm. $\lambda$ is a tuning parameter related to $t$, controlling for how strongly shrinkage is exercised. As $\lambda \rightarrow 0$ the penalty term vanishes and the lasso becomes indistinguishable from the least squares problem. This optimization formula is related to \emph{basis pursuit denoising}, which is the preferred term for the lasso among researchers in computer science and signal processing.

\citet[][Section 5]{Tibshirani1996} first noted that the lasso estimate can be derived as a Bayes posterior mode under the following Laplace prior distribution
\begin{equation}
p( \bm \beta )  = \prod_{j=1}^{p} \frac{\lambda}{2} \exp \left( - \lambda\vert \beta_j \vert\right) =  \left(\frac{\lambda}{2}\right)^{p} \exp \left( - \lambda \vert \vert \bm \beta \vert \vert_{1} \right).
\end{equation}
However, as \cite{castillo2015} note the full posterior distribution under a Laplace prior does not contract at the same rate as its mode, making uncertainty quantification using the Bayesian lasso unreliable. The intuition behind this is that the $\lambda$ coefficient above needs to be large enough to penalize coefficients $\beta_j$ to zero, but not too large such that nonzero coefficients can be modeled. This issue is addressed by modifications such as the adaptive lasso (\cite{AlhamzawiAli2018}; see end of this section) and the spike and slab lasso (\cite{RockovaGeorge2018}; see section on spike and slab priors) and is related to the motivating arguments of \cite{JohnsonRossell2010} for proposing the non-local priors (see relevant section below).

The first application of the lasso prior stems from computing science and is due to \cite{Girolami2001}. While the joint parameter posterior under a Laplace prior is not of standard form, \cite{Girolami2001} used variational Bayes inference (which at the time was not popular in mainstream statistics) to approximate the posterior mean and variance. \cite{Figueiredo2003} used the fact that the Laplace prior admits a hierarchical representation in the form of a normal-exponential (double exponential) mixture. The hierarchical representation of this prior is of the form
\begin{eqnarray}
\beta_{j} \vert \tau_j &\sim & N(0,\tau^{2}_{j}), \\
\tau_{j}^{2} \vert \lambda^{2} & \sim & Exponential \left( \frac{\lambda^{2}}{2}\right), 
\end{eqnarray}
where the exponential distribution has the functional form $p(\tau^{2} \vert \lambda^2) = \left( \frac{\lambda^{2}}{2} \right) \exp\left( \frac{\lambda^{2}}{2} \tau_{j}^{2} \right)$. The marginal distribution for $\beta$ conditional on $\lambda^2$ is of the form
\begin{equation}
p\left( \beta_j \vert \lambda^{2} \right) = \frac{\sqrt{\lambda^2}}{2} \exp\left(\sqrt{\lambda^2} \vert \beta_j \vert \right) \equiv \frac{\lambda}{2} \exp\left(\lambda \vert \beta_j \vert \right), \label{Laplace}
\end{equation}
which is the desired Laplace distribution for $\beta_j$. \cite{Figueiredo2003} derived an EM algorithm for obtaining the posterior mode (MAP estimator). 

A formal Bayesian treatment of the Bayesian lasso using MCMC can be found in \cite{ParkCasella2008}. These authors choose to specify the Bayesian lasso as a normal-exponential mixture but conditional on the regression variance $\sigma^2$. This is because a hierarchical prior on $\beta_{j}$ that is independent of $\sigma^{2}$ results in a multimodal posterior for $\beta_{j}$. The \cite{ParkCasella2008} Laplace prior takes the form
\begin{eqnarray}
\bm \beta \vert \lbrace \tau_{j}^{2} \rbrace_{j=1}^{p}, \sigma^{2} & \sim & N_{p}( \bm 0, \sigma^{2} \bm D_{\tau}), \\
\tau_{j}^{2} \vert \lambda^{2} & \sim & Exponential\left( \frac{\lambda^{2}}{2} \right), \text{ \ \ for } j=1,...,p, \\
\lambda^{2} & \sim & Gamma(r,\delta) \\
\sigma^{2} & \sim &  \frac{1}{\sigma^{2}},
\end{eqnarray}
where $\bm D_{\tau} =  diag(\tau_{1}^{2},...,\tau_{p}^{2})$. Conditional posteriors under this hierarchical representation are trivial to derive and more details can be found in the accompanying Technical Document. 

The approach in \cite{ParkCasella2008} is probably the most widely used but it is not the only one available. \cite{Hans2009} specified the lasso in terms of the normal orthant distribution.  Let $\mathcal{Z} = \lbrace -1, 1 \rbrace^{p}$ represent the set of all $2^p$ possible vectors of length $p$ whose elements are $\pm 1$. For any realization $z \in \mathcal{Z}$ define the orthant $\mathcal{O}_{z} \subset {\rm I\!R}^p$. If $\bm \beta \in \mathcal{O}_{z}$, then $\beta_{j} \geq 0$ if $z=1$ and $\beta_{j} <0$ if $z=-1$. Then $\bm \beta$ follows the normal-orthant distribution with mean $m$ and covariance $S$, which is of the form
\begin{equation}
\bm \beta \sim N^{[z]} \left( \bm m,\bm S \right) \equiv \Phi\left( \bm m, \bm S\right) N_{p} \left( \bm m,\bm S \right) I\left( \bm \in \mathcal{O}_{z} \right).
\end{equation}
The \cite{Hans2009} prior takes the form
\begin{eqnarray}
\bm \beta \vert \lambda, \sigma & \sim & \left( \frac{\lambda}{2\sqrt{\sigma^{2}}}\right)^{p} \exp\left( - \lambda \sum_{j=1}^{p} \vert \beta_{j} \vert /\sqrt{\sigma^{2}} \right), \\
\lambda & \sim & Gamma(r,\delta), \\
\sigma^{2} & \sim &  \frac{1}{\sigma^{2}},
\end{eqnarray}
and, using the definition of the normal orthant distribution, conditional posteriors are of the form
\begin{eqnarray}
\beta_{j} \vert  \beta_{-j}, \lambda, \sigma^{2}, \bm y & \sim & \phi_{j}N^{[+]} \left(\mu_{j}^{+}, \omega_{jj}^{-1} \right) + (1-\phi_{j})N^{[-]} \left(\mu_{j}^{-}, \omega_{jj}^{-1} \right), \\
\lambda \vert \bm y & \sim & Gamma\left(p+r, \frac{\sum_{j=1}^{p} \vert \beta \vert}{\sqrt{\sigma^{2}}} + \delta \right), \\
\sigma \vert \bm \beta, \bm y & \propto & (\sigma^{2})^{-(\frac{n+p}{2} + 1)} \exp \left(  \frac{\Psi}{2\sigma^2}  - \frac{\lambda \sum_{j=1}^{p} \vert \beta \vert}{\sqrt{\sigma^{2}}} \right),
\end{eqnarray}
where:
\begin{itemize}
\item $N^{[-]}$ and $N^{[+]}$ correspond to the $N^{[z]}$ distribution for $z=-1$ and $z=1$, respectively;
\item $\mu_{j}^{+} = \widehat{\beta}_{j}^{OLS} + \left \lbrace \sum_{i=1,i \neq j}^{p} \left(\widehat{\beta}_{i}^{OLS} - \beta_{i} \right)\left(\omega_{ij}/\omega_{jj}\right)  \right \rbrace + \left(- \frac{\lambda}{\sqrt{\sigma^{2}}\omega_{jj}}\right) $;
\item $\omega_{ij}$ is the $ij$ element of the matrix $\Omega  = \Sigma^{-1} = \left( \sigma^{2}(\bm X^{\prime} \bm X)^{-1} \right)^{-1}$;
\item $\phi_{j} = \frac{\Phi \left( \frac{\mu_{j}^{+}}{\sqrt{\omega_{jj}}} \right)/N\left(0 \vert \mu_{j}^{+}, \omega_{jj}^{-1} \right)}{ \Phi \left( \frac{\mu_{j}^{+}}{\sqrt{\omega_{jj}}} \right)/N\left(0 \vert \mu_{j}^{+}, \omega_{jj}^{-1} \right) + \Phi \left( -\frac{\mu_{j}^{-}}{\sqrt{\omega_{jj}}} \right)/N\left(0 \vert \mu_{j}^{-}, \omega_{jj}^{-1} \right) }$; 
\item $\Psi = (\bm y - \bm X \bm \beta)^{\prime}(\bm y - \bm X \bm \beta)$.
\end{itemize}
The conditional posterior of $\sigma^2$ is not of a standard form and, therefore, cannot be sampled directly. \cite{Hans2009} suggests a simple accept/reject step within the Gibbs sampler that allows to obtain approximate samples from the posterior of $\sigma^2$. Finally, \cite{MallickYi2014} propose a third hierarchical representation of the Laplace prior, this time as a mixture of Uniform distributions (see our Technical Document for details of this algorithm).

There are numerous extensions to the basic lasso that come in various forms. For example, the elastic net combines the benefits of ridge regression ($\mathcal{l}_{2}$ penalization) and the lasso ($\mathcal{l}_{1}$ penalization) by solving the problem
\begin{equation}
\arg\min_{\bm \beta \in \mathbb{R}^{p}} \vert \vert \bm y - \bm X \bm \beta\vert \vert_{2} + \lambda_{1} \vert\vert \bm \beta \vert\vert_{1} + \lambda_{2} \vert\vert \bm \beta \vert\vert_{2},
\end{equation} 
where now $\lambda_{1}$ and $\lambda_{2}$ are tuning parameters. The Bayesian prior that provides the solution to the elastic net estimation problem is of the form
\begin{equation}
\bm \beta \vert \sigma^{2} \sim \exp \left\lbrace -\frac{1}{2\sigma^{2}} \left(\lambda_{1} \sum_{j=1}^{p} \vert\beta_{j} \vert +  \lambda_{2} \sum_{j=1}^{p}\beta_{j}^{2} \right) \right\rbrace.
\end{equation}
\cite{LiLin2010} start from this prior and derive a mixture approximation and a Gibbs sampler that has the minor disadvantage that requires an accept-reject step for obtaining samples from the conditional posterior of $\sigma^{2}$ (similar to the sampler of \cite{Hans2009} for the lasso). The formulation of the elastic net prior in \cite{Kyungetal2010} is slightly different to the one above, but they manage to derive a slightly different mixture representation and a slightly more straightforward Gibbs sampler.

Other popular extensions to the lasso include the group lasso that allows for group shrinkage; the fused lasso that allows for spatial or temporal relationships between neighbouring parameters; and the adaptive lasso that fixes some variable selection consistency issues with the regular lasso. All these extensions have straightforward hierarchical forms, and we refer the reader to discussions in \cite{Kyungetal2010}, \cite{GriffinBrown2011}, \cite{Lengetal2014} and \cite{AlhamzawiAli2018}, among several other studies. Our Technical document provides details of posterior inference using the elastic net, group lasso, fused lasso and adaptive lasso.

\subsection{Generalized double Pareto shrinkage}
\cite{Armaganetal2013a} propose the following generalized double Pareto (GDP) prior on $\bm \beta$
\begin{equation}
\bm \beta \vert \sigma \sim \prod_{j=1}^{p} \frac{1}{2\sigma \delta/r} \left(1 + \frac{1}{r} \frac{\vert \beta_{j} \vert}{\sigma \delta/r} \right) ^{-(r+1)}.
\end{equation}
This distribution can be represented using the familiar, from the Bayesian lasso, normal-exponential-gamma mixture. The only difference is that, while the Exponential component has the same rate parameter for all $j=1,...,p$, in the representation of the GDP mixture this parameter is adaptive. The generalized double Pareto distribution has a spike at zero with Student’s t-like heavy tails.

The generalized double Pareto prior takes the form
\begin{eqnarray}
\bm \beta \vert \lbrace \tau_{j} \rbrace_{j=1}^{p}, \sigma^{2} & \sim &  N_{p} \left( \bm0, \sigma^{2} \bm D_{\tau} \right), \\
\tau_{j}^{2} \vert \lambda_{j} & \sim & Exponential \left( \frac{\lambda_{j}^2}{2} \right), \text{ \ \ for } j=1,...,p, \\
\lambda_{j} & \sim & Gamma(r,\delta), \text{ \ \ for } j=1,...,p, \\
\sigma^{2} & \sim &  \frac{1}{\sigma^{2}},
\end{eqnarray}
where $\bm D_{\tau} =  diag(\tau_{1}^{2},...,\tau_{p}^{2})$.

The conditional posteriors are of the form
\begin{eqnarray}
\bm \beta \vert \lbrace \tau_{j}^{2} \rbrace_{j=1}^{p} , \sigma^{2}, \bm y & \sim & N_p \left( \bm V \times \left(\bm X^{\prime} \bm y \right), \sigma^{2} \bm V \right), \\
\frac{1}{\tau_{j}^{2}} \vert \beta_{j}, \lambda_{j}^2, \bm y & \sim & IG\left(\sqrt{\frac{\lambda_{j}^2\sigma^2}{\beta_{j}^{2}}}, \lambda^2 \right), \text{ \ \ for } j=1,...,p,\\
\lambda_{j}^{2} \vert \bm y &\sim & Gamma\left( r + 1, \sqrt{\frac{ \beta_{j}^{2} }{\sigma^{2}}} + \delta \right), \\
\frac{1}{\sigma^{2}} \vert \bm \beta, \bm y  & \sim & Gamma \left( \frac{n-1+p}{2},\frac{\Psi}{2} + \frac{\bm \beta^{\prime} \bm D_{\tau}^{-1} \bm \beta}{2} \right),
\end{eqnarray}
where $\bm V = \left( \bm X^{\prime} \bm X + \bm D_{\tau}^{-1} \right)^{-1}$, $\bm D_{\tau}^{-1} =  diag(\tau_{1}^{-2},...,\tau_{p}^{-2})$ and $\Psi = (\bm y - \bm X \bm \beta)^{\prime}(\bm y - \bm X \bm \beta)$. \cite{Paletal2017} show,  both theoretically and numerically, that the above ``three-block'' Gibbs sampler is less efficient than a modified two-block Gibbs sampler they propose.

\subsection{Dirichlet-Laplace}
The Dirichlet-Laplace prior was introduced in \cite{Bhattacharyaetal2015}, and \cite{ZhangBondell2018} studied its posterior consistency as well as consistency in variable selection  in the context of a linear regression model. The Dirichlet-Laplace hierarchical prior, which is a generalization of the Laplace prior, takes the form
\begin{eqnarray}
\bm \beta \vert \lbrace \tau_{j} \rbrace_{j=1}^{p},  \lbrace \psi_{j}\rbrace_{j=1}^{p}, \lambda, \sigma^{2} & \sim &  N_{p} \left( \bm 0, \sigma^{2} \bm D_{\lambda,\tau,\psi} \right), \\
\tau_{j}^{2} & \sim & Exponential(1/2), \text{ \ \  } j=1,...,p, \\
\psi_{j} & \sim & Dirichlet(\alpha), \text{ \ \  } j=1,...,p, \\
\lambda & \sim & Gamma(n \alpha, 1/2), \\
\sigma^{2} & \sim &  \frac{1}{\sigma^{2}},
\end{eqnarray}
where $\bm D_{\lambda,\tau,\psi} = diag(\lambda^{2} \tau_{1}^{2}\psi_{1}^{2},...,\lambda^{2} \tau_{p}^{2}\psi_{p}^{2})$.

The conditional posteriors are of the form
\begin{eqnarray}
\bm \beta \vert \lbrace \tau_{j}^{2} \rbrace_{j=1}^{p}, \lbrace \psi_{j}\rbrace_{j=1}^{p}, \lambda,  \sigma^{2} , \bm y & \sim & N_p \left( \bm V \times \left(\bm X^{\prime} \bm y \right), \sigma^{2} \bm V \right), \\
\frac{1}{\tau_{j}^{2}} \vert \lambda^{2}, \sigma^{2} , \bm y & \sim & IG \left(c^*,1\right), \text{ \ } j=1,...,p, \\
\lambda \vert \bm \beta, \bm y & \sim & GIG\left( 2 \frac{\sum_{j=1}^{p}\vert \beta_{j}\vert}{\psi_j \sigma},1 ,p(\alpha - 1) \right), \\
\psi_{j} \vert \bm \beta, \bm y & = & \frac{T_j}{\sum_{j=1}^{p} T_j},  \text{ \  } j=1,...,p, \\ 
\text{where } T_j  &\sim &  GIG \left( 2 \sqrt{\frac{\beta_{j}^{2}}{\sigma^2}},1, \alpha-1 \right)\\
\frac{1}{\sigma^{2}} \vert \bm \beta, \bm y  & \sim & Gamma \left( a^*,b^*\right),
\end{eqnarray}
where 
$a^*=(n+p)/2$,
$b^*=( \Psi +\bm \beta^{\prime} \bm D_{\tau,\lambda,\psi}^{-1} \bm \beta )/2 $, 
$c^*=\sqrt{\lambda^{2} \psi_{j}^{2} \sigma^2 / \beta_{j}^{2}}$,
$\bm V = \left( \bm X^{\prime} \bm X + \bm D_{\tau,\lambda,\psi}^{-1} \right)^{-1}$,  and 
$\Psi = (\bm y - \bm X \bm \beta)^{\prime}(\bm y - \bm X \bm \beta)$. $IG$ is the two-parameter inverse Gaussian distribution, and $GIG$ is the three-parameter generalized inverse Gaussian.

\subsection{Horseshoe prior} \label{sec:horseshoe}
The horseshoe prior was first introduced by \cite{Carvalhoetal2010} and it since its inception has been the most popular and influential hierarchical prior in Bayesian inference. The survey paper by \cite{Bhadraetal2020} provides a thorough review of the applications of this prior in numerous inference problems in statistics and machine learning, including nonlinear models and neural networks. The Horseshoe is a prime representative of the class of global-local shrinkage priors (see Footnote \ref{glob_loc}) and it can be represented as a scale mixture of normals with half-Cauchy mixing distributions. That is, the prior has the following formulation
\begin{eqnarray}
\bm \beta \vert \lbrace \lambda_{j} \rbrace_{j=1}^{p}, \tau & \sim & N_p\left(\bm 0, \sigma^{2} \tau^{2} \bm \Lambda \right), \\
\lambda_{j} \vert \tau & \sim &  C^{+} (0,1), \text{ \ \ for } j=1,...,p,  \\
\tau & \sim &  C^{+} (0,1),
\end{eqnarray}
where $\bm \Lambda = diag(\lambda_{1}^{2},...,\lambda_{p}^{2})$, and $C^{+}(0,\alpha)$ is the half-Cauchy distribution on the positive reals with scale parameter $\alpha$. That is, $\lambda_{j}$ has conditional prior density
\begin{equation}
\lambda_{j} \vert \tau = \frac{2}{\pi \tau \left( 1 + (\lambda_j/\tau)^2 \right)}.
\end{equation}
Under this hierarchical specification, the marginal prior for each $\beta_{j}$ is unbounded at the origin and has tails that decay polynomially. 

There are numerous theoretical results established for this prior, most notably \cite{DattaGhosh2013} and \cite{vanderPasetal2014}, and the reader is referred to \cite{Bhadraetal2020} for a more detailed discussion. There are also various computational approaches to the Horseshoe (see the accompanying Technical Document for details), but the most straightforward is the one proposed by \cite{MakalicSchmidt2016}. These authors note that the half-Cauchy distribution can be written as a mixture of inverse-gamma distributions. In particular, if
\begin{equation}
x^{2} \vert z \sim Inv-Gamma(1/2,1/z), \text{ \ \ \ } z \sim Inv-Gamma(1/2, 1/\alpha^{2}),
\end{equation}
then $x \sim C^{+}(0,\alpha)$. Therefore, the \cite{MakalicSchmidt2016} prior takes the form
\begin{eqnarray}
\bm \beta \vert \lbrace \lambda_{j} \rbrace_{j=1}^{p}, \tau, \sigma^{2} & \sim & N\left( \bm 0, \sigma^{2} \tau^{2} \bm \Lambda \right), \\
\lambda_{j}^{2} \vert v_{j} & \sim & Inv-Gamma(1/2,1/v_{j}), \text{ \ \  } j=1,...,p,   \\
v_{j} & \sim & Inv-Gamma(1/2,1), \text{ \ \  } j=1,...,p,   \\
\tau^{2} \vert \xi & \sim & Inv-Gamma(1/2,1/\xi), \\
\xi & \sim & Inv-Gamma(1/2,1), \\
\sigma^{2} & \sim &  \frac{1}{\sigma^{2}},
\end{eqnarray}
where $\bm \Lambda = diag(\lambda_{1}^{2},...,\lambda_{p}^{2})$.

The conditional posteriors are of the form
\begin{eqnarray}
\bm \beta \vert \lbrace \lbrace \lambda_{j} \rbrace_{j=1}^{p}, \tau^{2} , \sigma^{2}, \bm y & \sim & N_p \left( \bm V \times \left(\bm X^{\prime} \bm y \right), \sigma^{2} \bm V \right), \\
\lambda_{j}^{2} \vert \bm \beta, v_{j}, \tau^{2}, \sigma^{2} \bm y & \sim & Inv-Gamma\left(1, \frac{1}{v_{j}} + \frac{\beta_{j}^{2}}{2\tau^{2}\sigma^{2}} \right), \text{ \ \  } j=1,...,p,  \\
v_{j} \vert \lambda_{j}, \bm y & \sim & Inv-Gamma \left(1, 1+ \frac{1}{\lambda_{j}^{2}} \right), \text{ \ \  } j=1,...,p, \\
\tau^{2} \vert \bm \beta, \xi, \lbrace \lambda_{j} \rbrace_{j=1}^{p}, \sigma^{2}, \bm y & \sim & Inv-Gamma \left( \frac{p+1}{2}, \frac{1}{\xi} + \frac{1}{2\sigma^{2}} \sum_{j=1}^{p} \frac{\beta_{j}^{2}}{\lambda_{j}^{2}} \right)\\
\xi \vert \tau^2, \bm y & \sim & Inv-Gamma \left(1, 1+ \frac{1}{\tau^{2}} \right), \\
\sigma^{2} \vert \bm \beta, \bm y  & \sim & Inv-Gamma \left( \frac{n+p}{2},\frac{\Psi}{2} + \frac{\bm \beta^{\prime} \bm D_{\tau,\lambda}^{-1} \bm \beta}{2} \right),
\end{eqnarray}
where $\bm V = \left( \bm X^{\prime} \bm X + \bm D_{\tau,\lambda}^{-1} \right)^{-1}$, 
$\bm D_{\tau,\lambda}=  diag(\tau^{2}\lambda_{1}^{2},...,\tau^{2}\lambda_{p}^{2}) = \tau^{2} \bm \Lambda$ and $\Psi = (\bm y - \bm X \bm \beta)^{\prime}(\bm y - \bm X \bm \beta)$.

\subsection{Generalized Beta mixtures of Gaussians}
\cite{Armaganetal2011} motivate the use of a three-parameter beta (TPB) distribution for the prior variance parameter, as a flexible class of shrinkage priors. The TPB distribution takes the form
\begin{equation}
p(x \vert a,b,\varphi) = \frac{\Gamma\left(a+b\right)}{\Gamma\left(a\right)\Gamma\left(b\right)} \varphi^{b} x^{b-1} (1-x)^{a-1} \left[ 1 + (\varphi - 1)x\right]^{-(a+b)},
\end{equation}
for $0<x<1$, $a,b,\varphi>0$. The TPB normal scale mixture representation for the distribution of random variable $\beta_{j}$ is given by
\begin{equation}
\beta_{j} \sim N\left( 0, 1/\rho_j - 1 \right), \text{ \ \ \ } \rho_j \sim TBP(a,b,\varphi).
\end{equation}
Proposition 1 in \cite{Armaganetal2011} shows that this distribution can either be written as normal-inverted beta mixture, or a normal-gamma-gamma mixture. The second choice gives a very straightforward Gibbs sampler scheme, and it can be seen as a special case of the normal-gamma class of priors \citep{GriffinBrown2017}.

The Generalized Beta mixtures of Gaussians prior takes the form
\begin{eqnarray}
\bm \beta \vert \lbrace \tau_{j}^2 \rbrace_{j=1}^{p} & \sim &  N_{p} \left( \bm 0,  \bm D_{\tau} \right), \\
\tau_{j}^{2} \vert \lambda_{j} & \sim & Gamma \left( a, \lambda_{j} \right), \text{ \ \ for } j=1,...,p, \\
\lambda_{j} \vert \varphi & \sim & Gamma(b,\varphi ), \text{ \ \ for } j=1,...,p, \\
\varphi & \sim & Gamma\left(\frac{1}{2},\omega \right), \\
\omega & \sim & Gamma\left(\frac{1}{2}, 1 \right), \\
\sigma^{2} & \sim &  \frac{1}{\sigma^{2}},
\end{eqnarray}
where $\bm D_{\tau} =  diag(\tau_{1}^{2},...,\tau_{p}^{2})$. Note that setting $a=b=1/2$ we can obtain the horseshoe prior of \cite{Carvalhoetal2010}. For other choices we can recover popular cases of shrinkage priors.

The conditional posteriors are of the form
\begin{eqnarray}
\bm \beta \vert \lbrace \tau_{j}^{2} \rbrace_{j=1}^{p} , \sigma^{2}, \bm y & \sim & N_p \left( \bm V \times \left(\bm X^{\prime} \bm y \right), \bm V \right), \\
\tau_{j}^{2} \vert \beta_{j}, \lambda_{j}^2, \bm y & \sim & GIG\left(a-\frac{1}{2}, 2\lambda_{j},\beta_{j}^{2} \right), \text{ \ \ for } j=1,...,p,\\
\lambda_{j} \vert \bm y & \sim & Gamma( a+b,\tau_{j}^{2} + \varphi ), \text{ \ \ for } j=1,...,p, \\
\varphi \vert \lbrace \lambda_{j} \rbrace_{j=1}^{p}, \omega, \bm y & \sim & Gamma\left( pb + \frac{1}{2}, \sum_{j=1}^{p} \lambda_{j} + \omega \right), \\
\omega \vert \varphi, \bm y & \sim & Gamma(1, \varphi + 1), \\
\frac{1}{\sigma^{2}} \vert \bm \beta, \bm y  & \sim & Gamma \left( \frac{n+p}{2},\frac{\Psi}{2}  \right),
\end{eqnarray}
where $\bm V = \left( \sigma^{-2} \bm X^{\prime} \bm X + \bm D_{\tau}^{-1} \right)^{-1}$, 
$\bm D_{\tau} =  diag(\tau_{1}^{2},...,\tau_{p}^{2})$ and $\Psi = (\bm y - \bm X \bm \beta)^{\prime}(\bm y - \bm X \bm \beta)$.

The TPB normal mixture includes as special cases Strawderman-Berger and horseshoe priors.

\subsection{Non-local priors}
Non-local priors have been proposed by \cite{JohnsonRossell2010} in the context of hypothesis testing of the form $H_{0}: \beta_j = 0$ vs $H_{1}: \beta_j \neq 0$. From a frequentist perspective, such testing procedures are used in order to find out how likely it would be for a set of observations to occur under the null hypothesis. However, in a Bayesian setting the data are assumed to be observed once, and parameters are continuous random variables. Traditional (local) priors put significant probability in both the null and alternative hypotheses, thus, making it harder for the (continuous) posterior distribution to detect-non zero coefficients asymptotically. Non-local densities place zero probability at zero, and this feature allows such priors to separate more clearly between the null and alternative hypotheses. That is, such priors do not place any prior probability under the null.\footnote{As \cite{JohnsonRossell2010} note: \begin{quote}
[...] to a large extent, we have ignored philosophical issues regarding the logical necessity to specify an alternative hypothesis that is distinct from the null hypothesis. In general, it is our view that one hypothesis (and a test statistic) is enough to obtain a p-value, but that two hypotheses are required to obtain a Bayes factor.
\end{quote}}

\begin{figure}[H]
\centering
\includegraphics[width = \textwidth, trim = {5cm 5cm 4cm 0cm}]{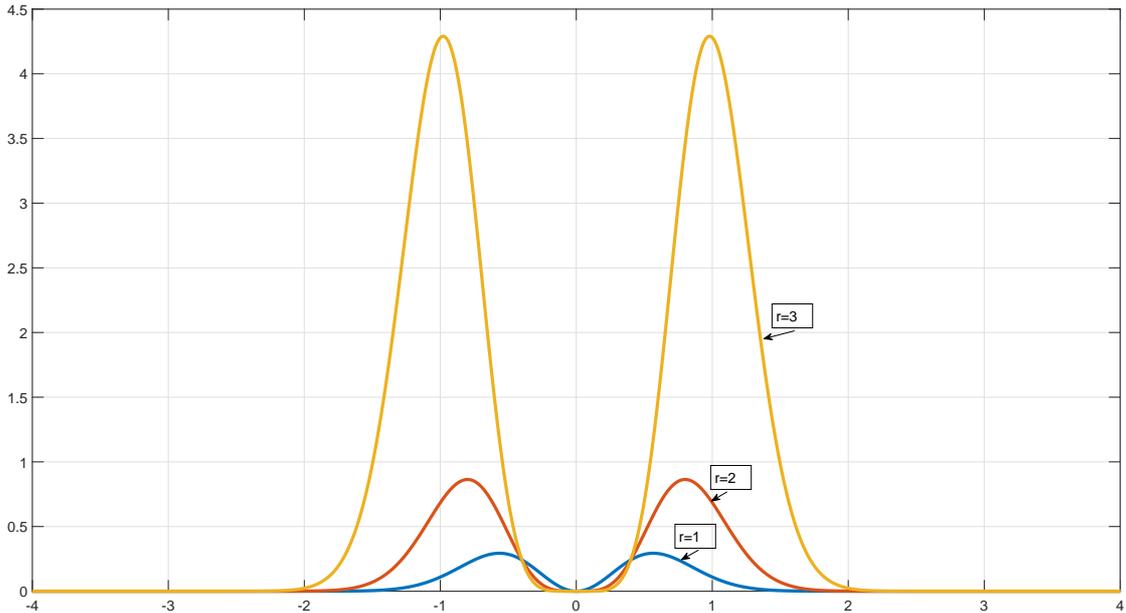}
\quad 
\caption{\emph{Plot of pMOM density for $r=1,2,3$.}}
\label{fig:nonlocal}
\end{figure}

Any distribution that ``decreases to 0 near the boundaries between disjoint null and alternative parameter spaces might be considered'' \citep{JohnsonRossell2010} to be a non-local prior density. Within the context of a linear regression setting similar to the one defined in \autoref{regression}, \cite{JohnsonRossell2012} propose two specific classes of priors.  The first class of prior densities for $\bm \beta$ consists of product moment (pMOM) densities, which are defined as
\begin{equation}
p\left( \bm \beta \vert \tau^2, \sigma^{2}, r \right) \propto \left( 2\pi \right)^{-p/2} \left(\tau^2 \sigma^2 \right)^{-p(r+1/2)} \exp \left\lbrace -\frac{1}{2\tau^2\sigma^2} \bm\beta^{\prime} \bm\beta \right\rbrace \prod_{j=1}^{p} \beta_{j}^{2r}.
\end{equation}
\autoref{fig:nonlocal} plots the pMOM density for $\tau^{2}=\sigma^{2}=1$ and for three values of $r$ ($r=1,2,3$). This graph clearly shows the shapes that this prior can achieve, especially with regards to the rate at which this prior decreases in the region of zero. The second class of prior densities consists of the product inverse moment (piMOM) densities, which are defined as
\begin{equation}
p\left(\bm \beta \vert \tau^2, \sigma^2, r \right) = \frac{\left(\tau^{2} \sigma^{2} \right)^{rp/2}}{\Gamma\left(r/2\right)^p}  \exp \left \lbrace - \tau^{2}\sigma^{2} \left(\bm\beta^{\prime} \bm\beta \right)^{-1} \right \rbrace \prod_{j=1}^{p} \vert \beta_{j} \vert^{-(r+1)}.
\end{equation}

In both of these two priors, $\tau^{2}$ is a scale parameter that determines dispersion of the prior around zero. Therefore, this parameter determines the size of the regression coefficients that will be shrunk to zero, and it is of prime importance. \cite{JohnsonRossell2012} and \cite{Shinetal2018} treat $\tau^2$ to be fixed and show that high-dimensional model selection consistency is achieved under the pMOM prior, as long as $\tau^{2}$ is of a larger order than $\log p$ and it increases subexponentially in $n$. However, fixing this parameter might not be desirable in most applied high-dimensional problems\footnote{For example, \cite{JohnsonRossell2012} note that if the covariate matrix $\bm X$ is not standardized, then it would be important to define an adaptive shrinkage parameter $\tau^2_j$ for each $j=1,...,p$. In such a case, choice of each individual $\tau^2_j$ for large $p$ becomes inconvenient, if not infeasible.}, and a hierarchical approach might be desirable. \cite{Caoetal2020} propose a hyperprior density for $\tau^{2}$ of the form
\begin{equation}
p(\tau^2) = \frac{\left(\frac{n}{2}\right)^{1/2}}{\Gamma\left( \frac{1}{2}\right)} \tau^{-3} \exp\left( -\frac{n}{2\tau^{2}} \right).
\end{equation}
The hierarchical pMOM (or ``hyper-pMOM'') prior they propose achieves strong model selection consistency when $p$ increases at a polynomial rate with $n$. Unfortunately, neither the pMOM, hyper-pMOM or piMOM priors allows for a closed-form computation of joint, marginal or conditional posteriors. Therefore, \cite{Caoetal2020} rely on Laplace approximations.

\subsection{Spike and slab priors} \label{sec:sns}

Similar to non-local priors, spike and slab priors allow for variable selection and testing of the hypotheses $H_{0}: \beta_j = 0$ vs $H_{1}: \beta_j \neq 0$. Unlike non-local priors, spike and slab prior densities place significant probability into both hypotheses. In a regression context, the spike and slab prior \citep{MitchellBeauchamp1988} takes the form
\begin{eqnarray}
\beta_{j} \vert \gamma_{j} & \sim & (1-\gamma_j)\delta_{0}(\beta_j) + \gamma_j N(0,\tau^2), \label{sns} \\ 
\gamma_j & \sim & Bernoulli(\pi_{0}), \label{gamma_prior}
\end{eqnarray}
for each $j=1,...,p$, where $\delta_{0}(\beta_j)$ is the Dirac delta function placing point mass at zero and $\gamma_j$ are 0/1 (dummy) variables indicating whether column $j$ of $\bm X$ is included in the regression or not. The mechanism with which this prior classifies predictors as important or not, is simple: when $\gamma_j=1$ the prior for $\beta_j$ is $N(0,\tau^2)$, that is, estimation is not restricted by the prior for reasonably large values of $\tau^{2}$; when $\gamma_j=0$ the prior becomes a point mass function concentrated at zero and it dominates the likelihood such that the posterior is also concentrates its mass at zero. The concept of variable selection is fully determined by the indicator random variables $\gamma_j$'s. Samples from the posterior of each $\gamma_j$ will be sequences of zeros and ones, and the posterior mean denotes the \emph{posterior inclusion probability} of each predictor in the best model. For example, if we sample MCMC 10,000 draws and find that 2,000 times $\gamma_j=1$, then the posterior mean is simply $2000/10000=0.2$ which translates into $20\%$ posterior inclusion probability of predictor $j$. \cite{BarbieriBerger2004} show that the median probability model, that is, the model where only variables with probabilities larger than 0.5 are selected/retained, is optimal for prediction. \cite{OHaraSilanpaa2009} suggest that as a variable selection mechanism such variable selection priors should work well up to cases where $p$ is 10-15 times larger than $n$, but of course this proportion is only a rule of thumb that is heavily determined by the informativeness of the data and modeling choices.
 
The spike and slab prior belongs to the general class of hierarchical full-Bayes priors introduced earlier in this section, since it can be written in the form 
\begin{eqnarray}
\beta_j \vert \gamma_j & \sim & N\left( 0, \tau^2 \gamma_j \right). \label{hier_sns} 
\end{eqnarray}
If, in addition, we introduce a hyperprior distribution on $\tau^2$ (e.g.\ inverse-gamma, see \citealt{IshwaranRao2003}), then the spike and slab prior is not only a hierarchical prior, but also belongs to the class of local-global shrinkage priors with global shrinkage parameter $\tau^2$ and local shrinkage parameters $\gamma_j$. In signal processing and similar fields, the spike and slab is known as a ``normal-Bernoulli'' or ``Gaussian-Bernoulli'' prior.

A third parametric formulation of this particular spike and slab prior is due to \cite{KuoMallick1998}. In their formulation the regression model with variable selection prior is written as
\begin{equation}
\bm y = \sum_{j=1}^{p} \bm X_{j} \gamma_{j} \beta_{j} + \bm \varepsilon,
\end{equation}
where $\beta_{j}$ is the coefficient on predictor $j$ and $\gamma_j$ is a 0/1 variable indicating whether predictor $j$ is included in the model. This formulation is equivalent to the previous two, but it implies that the vector of indicators $\bm \gamma$ enters only via the likelihood and not through the (hierarchical) prior for $\bm \beta$. In the \cite{KuoMallick1998} formulation each $\beta_{j}$ will simply have a typical Gaussian prior with variance $\tau^2$. Notice that when $\gamma_j=1$, $\beta_j$ will be sampled from its posterior, but when $\gamma_j=0$, $\beta_j$ is not identified. In this case what happens is -- as is the case with any unidentified parameter in a Bayesian setting (e.g.\ mutlicollinearity) -- that $\beta_j$ is sampled from its prior. This lack of identification of $\beta_j$ is not a problem, as what we care about is the joint effect $\gamma_j \times \beta_j$ and the fact that predictor $j$ simply has to be removed whenever $\gamma_j=0$. This detail means that in variable selection a-la \cite{KuoMallick1998} the posterior of $ \beta_j$ with $\gamma_j=0$ will be equal to its normal prior, while the posterior of the same parameter under the spike and slab prior of equation \eqref{sns} is a point mass at zero. Other than this (possibly minor) difference, Bayesian variable selection using all three forms presented above is conceptually and empirically comparable.

The class of spike and slab priors and its theoretical properties have been studied extensively in the literature; see \cite{JohnstoneSilverman2004}, \cite{IshwaranRao2005b}, \cite{Jiang2006}, \cite{Bogdanetal2011} and \cite{castillo2012}. From an applied scientist's point of view, the spike and slab prior is very versatile and can take numerous useful forms.\footnote{For example, \cite{KoopKorobilis2016} specify a spike and slab prior that is able to search for homogeneities in panel data. That is, the spike and slab prior is modified in order to test the hypothesis of the form $H_{0}: \beta_i = \beta_j$ vs $H_{1}: \beta_i \neq \beta_j$.} We next briefly review possible formulations of the spike and slab prior, and their implications for modeling coefficients and selecting variables in a linear regression. We finish this section with a discussion of some key computational aspects of this class of priors.

\subsubsection*{Tuning of parameters in the spike and slab prior}
In the formulation in \autoref{sns} one only has to choose the variance parameter $\tau^2$. This cannot be zero because the slab will become identical to the spike component, and it cannot become infinity because it would also be impossible to separate the spike from the slab component (remember from the previous section that Bayes factors with diffuse priors do not exist). Therefore, $\tau^2$ has to be quite different from zero and not too large (e.g.\ $\tau^2=4$ is a reasonable choice). Of course one can use any of the hyperprior distributions already explored in the previous sections, e.g.\ the choice $\tau^2 \sim exponential(\lambda^{2}/2)$ will convert the slab into a Laplace prior. However, one should be careful not to overshrink the slab (e.g.\ by setting $\lambda$ too large in the Laplace prior) because then the spike and slab will be indistinguishable and posterior inclusion probabilities will be meaningless.

A computationally more efficient formulation of the spike and slab prior (at least within an MCMC setting) is the one proposed by \cite{GeorgeMcCulloch1993,GeorgeMcCulloch1997}, where both the spike and slab distributions are continuous
\begin{equation}
\beta_{j} \vert \gamma_{j} \sim  (1-\gamma_j)N(0,\tau_{0}^2) + \gamma_j N(0,\tau_{1}^2), \\ \label{ssvs}
\end{equation}
where $\tau_0^{2}$ is a ``small'' variance parameter (corresponding to the spike) and $\tau_1^{2}$ is a ``large'' variance parameter (corresponding to the slab). In the limit, when $\tau_{0}^{2} = 0$, the spike becomes the Dirac delta at zero, but for any other values of $\tau_{0}^{2}$ close but different from zero the spike distribution is unable to shrink $\beta_j$ exactly to zero. That is, this version of the spike and slab is appropriate for testing $H_{0}: \beta_j \approx 0$ vs $H_{1}: \beta_j \neq 0$, that is, it provides a soft thresholding rule. \cite{Chipmanetal2001} provide the threshold value above (below) which a regression coefficient is classified as belonging to the slab (spike) component and is not shrunk (shrunk) to zero:
\begin{equation}
\Delta = \sqrt{\frac{\log\left( \frac{\tau_1^2}{\tau_0^2}\right)}{\frac{1}{\tau_{1}^{2}} - \frac{1}{\tau_{0}^{2}}}}.
\end{equation}
Therefore, elicitation of $\tau_{0}^{2},\tau_{1}^{2}$ becomes very important for variable selection in the \cite{GeorgeMcCulloch1993} prior. \cite{NarisettyHe2014} show that fixing these two variance hyperparameters may result in variable selection inconsistency, and propose values that are functions of $n$ and $p$ that ensure good performance of the prior when the data dimensions increase. \cite{IshwaranRao2005b} set $\tau_{0}^2 = \tau^{2}$ and
$\tau_{1}^2 = c \tau^{2}$ where $c>>1$ and $\tau^{2} \sim Inv-Gamma$, although $\tau^2$ could also follow any of the hierarchical distributions defined previously, e.g.\ Horseshoe or Laplace. \cite{SylviaHelga2010} go one step further by motivating a mix-and-match strategy where $\tau_{0}^{2}$ has a Laplace prior, while $\tau_{1}^{2}$ has an inverse-gamma prior. More recently, \cite{RockovaGeorge2018} showed that, under mild conditions, a spike and slab lasso prior produces posterior distributions that concentrate asymptotically around the true regression coefficients at nearly the minimax rate. In their formulation both the spike and the slab are based on Laplace distributions (represented as normal-exponential mixtures), with the spike distribution shrunk more aggressively than the slab distribution.

An important feature of variable selection priors is the prior on $\gamma_j$. As in \autoref{gamma_prior} this is typically Bernoulli with prior probability $\pi_0$, or equivalently a binomial prior for the full vector $\bm \gamma = \left( \gamma_1,...,\gamma_p \right)^{\prime}$. Unfortunately, the choice $\pi_0=0.5$ in a binomial prior is not uniform as it implies a prior expectation that half of the $p$ predictors will be included in the final model. Therefore, in high-dimensional settings it is customary to set this parameter to a value closer to zero, e.g.\ $\pi_0=0.1$. If desired, a prior can be placed on this parameter and a conjugate choice is the beta distribution, that is, $\pi_0 \sim Beta(1,\alpha_0)$. The choice $\alpha_0=1$ makes this prior uniform, but in high-dimensional cases it will be preferable to set $\alpha_0$ to become proportional to the number of predictors $p$. Note that in the presence of a beta hyperprior on $\pi_0$, it is not necessary to use indicator variables $\gamma_j$. For example, following \cite{Dunsonetal2008} we can specify a spike and slab of the form\footnote{See also \cite{Korobilis2013a,Korobilis2013b,Korobilis2016} for related priors applied to econometric contexts such as dynamic regressions and vector autoregresions.}
\begin{eqnarray}
\beta_{j} \vert \pi_0 & \sim & (1-\pi_0)\delta_{0}(\beta_j) + \pi_0 N(0,\tau^2), \\
\pi_0 & \sim & Beta(1,\alpha_0), \label{pi_0}
\end{eqnarray}
that provides a smoother mixture of the two components. (We can, of course, specify an equivalent formulation for the \cite{GeorgeMcCulloch1993} continuous spike and slab formulation.) Finally, \cite{Carvalhoetal2008} turn this latter formulation into a sparsity inducing variable selection prior by replacing \autoref{pi_0} with
\begin{equation}
\pi_0 \vert \rho \sim (1-\rho) \delta_0(\pi_0) + \rho Beta(1,\alpha_0),
\end{equation}
that is, a spike and slab prior for $\pi_0$. Finally, \cite{YuanLin2005} propose a prior for $ \bm \gamma$ that accounts for correlation in predictors, such that if two predictors are highly correlated only one is included in the selected model. In their formulation they multiply the standard binomial prior for $\bm \gamma$ with the determinant of the Gram matrix of predictors, that is, $\vert \bm X^{\prime} \bm X \vert$. High-correlated predictors have small $\vert \bm X^{\prime} \bm X \vert$ and are discouraged from being selected. Such enhancements of the base spike and slab prior are important for variable selection, because marginal inclusion probabilities may be poor under high correlation. In particular, highly correlated predictors may be jointly selected often but each predictor only a small number of times.

\subsubsection*{Computation with spike and slab priors}
Computation with spike and slab priors is as straightforward as is the case with most other hierarchical priors. Conditional on $\gamma_j$ being either zero or one, the prior for $\beta_j$ is either a point mass at zero or normal \citep[in the representation of][]{MitchellBeauchamp1988} or it is one of two normal components \citep[in the representation of][]{GeorgeMcCulloch1993}. Therefore, conditional on $\gamma_j$, results for the normal linear model can be used. The same holds in the case where the components of the spike and slab are non-normal, rather they are Student-t, Laplace etc: as long a hierarchical prior structure is used and the prior can be written in conditionally normal form, derivation of conditional posteriors is straightforward.

Regarding posterior computation of $\gamma_j$'s this usually has to be done element-by-element, that is, we need to derive $\gamma_j$ conditional on $\bm \gamma_{-j}$ (the set $\bm \gamma$ with the $j$-th element removed).\footnote{For that reason, when the Gibbs sampler is used to sample from the conditional posterior of $\gamma_j$ given $\bm \gamma_{-j}$, it is advisable in each Gibbs iteration to sample in random order $j$ to avoid high autocorrelation of samples.} However, in the case of the \cite{MitchellBeauchamp1988} prior of \autoref{sns}, the conditional posterior $p(\gamma_j \vert \bm \gamma_{-j}, \bm \beta, \sigma^{2}, \bm y)$ cannot be used to obtain samples from the posterior of $\gamma_j$. Intuitively, this is because when we sample $\gamma_j = 0$ then the prior for $\beta_j$ is the Dirac delta function that puts infinite mass at zero. Therefore, in the next iteration $p(\gamma_j \vert \bm \gamma_{-j}, \bm \beta, \sigma^{2}, data)$ will give $\gamma_j=0$ with probability one, meaning that the sampler will get stuck in a loop where the only possible outcome is $\beta_j=\gamma_j=0$. This is not an issue in the continuous spike and slab prior of \cite{GeorgeMcCulloch1993}, since the spike is a continuous normal distribution and allows samples of $\beta_j$ to be slightly different from zero.

To see this, let's derive $p(\gamma_j \vert \bm \gamma_{-j}, \bm \beta, \sigma^{2}, \bm y)$ in the case of the spike and slab prior of equations \eqref{sns} - \eqref{hier_sns}, which we rewrite for convenience
\begin{eqnarray}
\beta_{j} \vert \gamma_{j} & \sim & (1-\gamma_j)\delta_{0}(\beta_j) + \gamma_j N(0,\tau^2), \\
\gamma_j & \sim & Bernoulli(\pi_{0}).
\end{eqnarray}
For simplicity, we do not introduce prior distributions on $\tau^{2}$ and $\pi_0$, so we assume these are fixed and chosen by the researcher. Using Bayes theorem, the posterior of $\gamma_j=0$ is
\begin{equation}
p(\gamma_j = 0 \vert \bm \gamma_{-j}, \bm \beta, \sigma^{2}, \bm y) \propto p( \bm y \vert \gamma_j=0,\bm \gamma_{-j},\bm \beta, \sigma^{2}) p(\beta_j \vert \gamma_j = 0) p(\gamma_j = 0).
\end{equation}
In this decomposition, the first term is provided by the likelihood where we set the $j$-th element of $\bm \beta$ equal to zero (since $\gamma_j=0$), regardless of what  the sampled value $\beta_j$ is in the previous iteration of the Gibbs sampler. This is a normal distribution with mean $\bm X \bm \beta^{\star}$, where $\bm \beta^{\star}$ is equal to $\beta$ with the $j$-th element equal to zero, and variance $\sigma^{2}$. The second term is the prior for $\beta_j$ under the restriction $\gamma_j=0$, that is, the Dirac delta density $\delta_{0}(\beta_j)$. The last term is given simply by the Bernoulli prior for $\gamma_j$ and it is equal to $(1-\pi_0)$. Therefore, this posterior is:
\begin{equation}
p(\gamma_j = 0 \vert \bm \gamma_{-j}, \bm \beta, \sigma^{2}, \bm y) \propto N_n(\bm X \bm \beta^{\star},\sigma^{2} \bm I_{n}) \delta_{0}(\beta_j)(1-\pi_0).
\end{equation}
Using similar arguments, we have that
\begin{equation}
p(\gamma_j = 1 \vert \bm \gamma_{-j}, \bm \beta, \sigma^{2}, \bm y) \propto N_n(\bm X \bm \beta,\sigma^{2} \bm I_{n}) N(0,\tau^{2}) \pi_0.
\end{equation}
Therefore, the conditional posterior of $\gamma_j$ is
{\tiny
\begin{equation}
p(\gamma_j \vert \bm \gamma_{-j}, \bm \beta, \sigma^{2}, \bm y) \sim Bernoulli \left( \frac{N_n(\bm X \bm \beta,\sigma^{2} \bm I_{n}) N(0,\tau^{2}) \pi_0}{N_n(\bm X \bm \beta,\sigma^{2} \bm I_{n}) N(0,\tau^{2}) \pi_0 + N_n(\bm X \bm \beta^{\star},\sigma^{2} \bm I_{n}) \delta_{0}(\beta_j)(1-\pi_0)}\right). \label{gamma_post}
\end{equation}
}Notice how the Dirac delta $\delta_{0}(\beta_j)$ enters the denominator term. If in the sampling process it happens to sample $\gamma_j=0$, then $\beta_j=0$ and any subsequent $\gamma_j$'s will also be zero for ever. This is because once a $\beta_j=0$ is observed, $\delta_{0}(\beta_j)$ becomes infinite and the ratio in the Bernoulli posterior is zero.

The solution to this problem is integration. That is, we need to remove dependence to $\beta_j$, and instead of the posterior $p(\gamma_j \vert \bm \gamma_{-j}, \bm \beta, \sigma^{2}, \bm y)$ we compute $p(\gamma_j \vert \bm \gamma_{-j}, \bm \beta_{-j}, \sigma^{2}, \bm y)$, that is, we integrate out $\beta_j$ and condition only on $\bm \beta_{-j}$. Intuitively, because $\gamma_j$ depends only to $\beta_j$ through the spike and slab prior (i.e.\ it is independent to $\bm \beta_{-j}$), the ratio in the Bernoulli posterior of \autoref{gamma_post} will only involve the densities $p( \bm y \vert \gamma_j=0,\bm \gamma_{-j},\bm \beta, \sigma^{2})$, $p( \bm y \vert \gamma_j=1,\bm \gamma_{-j},\bm \beta, \sigma^{2})$ and $p(\gamma_j = 0)$, $p(\gamma_j = 1)$. The accompanying Technical document provides details of conditional posteriors under various forms of spike and slab prior distributions, including cases with more complex hiearchical layers such as the spike and slab lasso of \cite{RockovaGeorge2018}.

\subsection[Monte Carlo study]{Monte Carlo study: Specification of spike and slab priors for variable selection}
Consider a  \cite{GeorgeMcCulloch1993,GeorgeMcCulloch1997} type spike and slab prior
\begin{align*}
\bm y \vert \sigma^2, \bm \beta & \sim  N_n\left(\bm X \bm \beta, \sigma^2 \bm I_n \right), \nonumber \\
\beta_j \vert \tau_{0j}^2, \sigma^2, \gamma_j=0 & \sim   N\left(0, \sigma^2 \tau_{0j}^2 \right), \text{ for } j=1,\ldots, p,\\
\beta_j \vert \tau_{1j}^2, \sigma^2, \gamma_j=1 & \sim   N\left(0, \sigma^2 \tau_{1j}^2 \right), \text{ for } j=1,\ldots, p,\\
\sigma^2 &\sim Inv-Gamma (a,b), \nonumber \\
P(\gamma_j =1 ) & =  \pi_0, \text{ for } j=1,\ldots, p, \nonumber \\
\pi_0 &\sim Beta(c,d) \nonumber 
\end{align*}
%
The conditional posteriors  of $\bm \beta , \sigma^2, \bm \gamma$, and $ \pi_0$ are
\begin{align*}
\bm \beta \vert \bullet &\sim N_p\left( \bm V \bm X^{'} \bm y , \sigma^2 \bm V  \right), \text{ where } \bm V = (\bm D^{-1} +    \bm X^{'} \bm X)^{-1},\\
\sigma^2 \vert \bullet &\sim Inv-Gamma \left( a+\frac{n}{2}+\frac{p}{2}, b + \frac{1}{2} \left[ \left( \bm y - \bm X \bm \beta \right)' \left( \bm y - \bm X \bm \beta \right) +\bm \beta' \bm Q^{-1} \bm \beta \right] \right),\\
\gamma_j \vert \bullet &\sim Bern\left( \frac{\phi \left( \beta_j \vert 0, \sigma^2 \tau_{1j}^2 \right) \pi_0}{\phi \left( \beta_j \vert 0, \sigma^2 \tau_{1j}^2 \right) \pi_0+  \phi \left( \beta_j \vert 0, \sigma^2 \tau_{0j}^2 \right) (1- \pi_0)} \right), \text{ for } j=1,\ldots, p,\\
 \pi_0 \vert \bullet &\sim Beta\left(c+\sum_{j=1}^p \gamma_j, d +\sum_{j=1}^p (1-\gamma_j)  \right)
\end{align*}
where $\phi(\cdot \vert m, v)$ is the normal density with mean $m$ and variance $v$
and 
$\bm D$ is a diagonal matrix with diagonal elements  $\{ (1-\gamma_j)\tau_{0j}^2 + \gamma_j \tau_{1j}^2 \}_{j=1}^p$.

\subsubsection{SSVS-Lasso}
Suppose we employ a Laplace density for the slab component 
\begin{align*}
\tau_{1j}^{2} \vert \lambda_1^{2} & \sim  Exponential\left( \frac{\lambda_1^{2}}{2} \right), \text{ \ \ for } j=1,...,p
\end{align*}
and consider three different ways of defining priors for the spike component that are commonly used in practice, which we define as SSVS-Lasso 1-3.

In SSVS-Lasso-1, $\tau^2_{0j}$ is fixed i.e.\ $\tau^2_{0j}=c_1$ for some  small $c_1>0$ and
in SSVS-Lasso-2, it is  proportional to the prior variance for the slab component i.e.\ $\tau^2_{0j} =c_2 \tau^2_{1j}$ for some small $c_2>0$.
In both SSVS-Lasso-1 and 2, with the prior
$\lambda_1^{2}  \sim  Gamma(r_1,\delta_1) $, 
 the prior variance for the slab is updated according to 
\begin{align*}
\lambda_1^2 \vert \bullet &\sim Gamma\left( \sum_{j=1}^p \gamma_j + r_1, \sum_{j=1}^p \tau_{1j}^2 \gamma_j /2 + \delta_1  \right)\\
1/ \tau_{1j}^{2}  \vert \bullet &\sim IG\left( \sqrt{\lambda_1^2\sigma^2/\beta^2_j} , \lambda_1^2\right), \text{ \ \ for } j=1,...,p
\end{align*}

In SSVS-Lasso-3,  we place two separate Laplace densities on the components  i.e.\
\begin{align*}
\tau_{0j}^{2} \vert \lambda_0^{2} & \sim  Exponential\left( \frac{\lambda_0^{2}}{2} \right), \text{ \ \ for } j=1,...,p, \\
\tau_{1j}^{2} \vert \lambda_1^{2} & \sim  Exponential\left( \frac{\lambda_1^{2}}{2} \right), \text{ \ \ for } j=1,...,p
\end{align*}
with $\lambda_0 \gg \lambda_1$ so that the density for $N(0,\sigma^2 \tau^2_{0j})$ is the ``spike'' and $N(0,\sigma^2 \tau^2_{1j})$ is the ``slab''.
This is similar to the spike-and-slab Lasso in \cite{RockovaGeorge2014} and \cite{BaiRockovaGeorge2021} \footnote{They propose an EM algorithm for estimation.}.
The prior variances are updated according to 
\begin{align*}
1/ \tau_{0j}^{2}  \vert \bullet &\sim IG\left( \sqrt{\lambda_0^2\sigma^2/\beta^2_j} , \lambda_0^2\right),  \text{ \ \ for } j=1,...,p,   \\
1/ \tau_{1j}^{2}  \vert \bullet &\sim IG\left( \sqrt{\lambda_1^2\sigma^2/\beta^2_j} , \lambda_1^2\right),  \text{ \ \ for } j=1,...,p
\end{align*}

Which specifications are appropriate in applications? 
In order to investigate this question, we consider simulation by generating data from the regression model $\bm y = \bm X \bm \beta + \bm \epsilon$ with $\bm \epsilon \sim N_n(\bm 0,\sigma^2 \bm I_n)$. 
We let $n=100$ and $\sigma^2=3$.
We construct the true vector of slope parameters $\bm \beta =c  \tilde{\bm \beta}$ by assigning values $\{1.5, -1.5, 2, -2, 2.5, -2.5  \}$ to the first 6 elements of $ \tilde{\bm \beta}$ and setting others to zero. 
We choose a constant $c>0$ to achieve a desired level of signal-to-noise ratio
\footnote{
In a general linear regression $\bm y = \bm X \bm \beta+ \bm \epsilon$, the signal-to-noise ratio (SNR) is defined as 
$SNR=\frac{|| \bm \Sigma_X^{1/2} \bm \beta ||^2 }{ \sigma^2 }$ 
where $\sigma^2$ is the error variance and $\bm \Sigma_X$ is a $p \times p$ covariance matrix of $\bm X$.
$||\bm \Sigma_X^{1/2} \bm \beta||^2 = \bm \beta' \bm \Sigma_X\bm  \beta$ measures the overall signal strength.
A related quantity is $R^2_{pop}$, the population value of $R^2$, defined as $\frac{SNR}{1+SNR}$.
}.
The data matrix $\bm X$ is generated from the multivariate normal distribution with mean zero and covariance matrix being an identity matrix. 
The covariates are standardized for estimation. 
We examine different values of the number of covariates $p\in \{50, 100,300 \}$ and the signal-to-noise ratio $R^2_{pop}\in \{ 0.4,0.8 \}$.

The analysis was repeated 100 times with new covariates and responses generated each time. 
For each, the metrics recorded were: 
the bias and MSE of the first 6 elements of the coefficients vectors, 
the number of false negatives (FN), 
the number of false positives (FP), and 
the number of true positives (TP).
Posterior means were used as point estimates of the slope coefficients and the error variance. 
We utilize the post-processing approach of \cite{LiPati2017} in order to categorize the covariates into signals and noises.

We compare performance of SSVS-Lasso 1-3 under different data generating processes. 
We fix $c=d=1$ so that the prior on the inclusion probability $\theta$ is uniform. We also let $a=b=0.1$.
For SSVS-Lasso 1 and 2, the hyperparameters for the prior on $\lambda^2_1$ are fixed as $r_1=1$ and $\delta_1=1$.
We let $c_1=10^{-4}$ for SSVS-Lasso-1, $c_2=10^{-4}$ for SSVS-Lasso-2, and $\lambda_0=20, \lambda_1=1$ for SSVS-Lasso-3.
We also show results of \cite{NarisettyHe2014} which is a two component mixture of normals prior with fixed prior variances and 
\cite{KuoMallick1998} which can be seen as a spike-and-slab prior with the spike component being a point mass at zero. 
\autoref{ssvs_table} summarizes results. 

Under a relatively strong signal i.e.\ $R^2_{pop}=0.8$, generally speaking, SSVS-Lasso-3 and Narisetty-He outperform others in all measures, and this tendency becomes more apparent in high-dimensional case i.e.\ $p=300$.
Kuo-Mallick tends to have larger FPs than others. 
When the signal-to-ratio is lower,  i.e.\ $R^2_{pop}=0.4$, 
SSVS-Lasso-3 outperforms others in terms of bias/MSE of the signals and shows reasonable performance in other metrics.

\begin{table}[h]
    \begin{subtable}[h]{\textwidth}
        \centering
        \scriptsize{
      \begin{tabular}{ c c  c c c c} 
 &Bias & MSE & FN & FP & TP \\
   \hline \hline 
\multicolumn{6}{c}{$R^2_{pop}=0.8, p=50 $}\\         
  \hline  
 SSVS-Lasso-1 & 0.14 &0.02& 0.63&0&5.4 \\ 
 SSVS-Lasso-2 & 0.16 &0.03& 0.70&0&5.3 \\ 
 SSVS-Lasso-3 & 0.11 &0.01& 0.53&0&5.4 \\ 
 Narisetty-He & 0.09 &0.01& 0.31&0&5.6 \\ 
Kuo-Mallick & 0.11 &0.02& 0.49&0.04&5.5\\ 
  \hline  
\multicolumn{6}{c}{$R^2_{pop}=0.8, p=100 $}\\         
   \hline  
 SSVS-Lasso-1 & 0.23 &0.06& 0.96&0&5.1 \\ 
 SSVS-Lasso-2 & 0.28 &0.09& 1.15&0&4.8 \\ 
 SSVS-Lasso-3 & 0.13 &0.02& 0.53&0.05&5.4 \\ 
 Narisetty-He & 0.10 &0.01& 0.39&0&5.6 \\ 
Kuo-Mallick & 0.29 &0.13& 1.07&20.8&4.9 \\ 
  \hline
\multicolumn{6}{c}{$R^2_{pop}=0.8, p=300 $}\\         
     \hline 
 SSVS-Lasso-1 & 0.52 &0.29& 1.91&21.0&4.0 \\ 
 SSVS-Lasso-2 & 0.56 &0.33& 0.98&39.6&5.0 \\ 
 SSVS-Lasso-3 & 0.32&0.12& 1.44&9.1&4.5 \\ 
 Narisetty-He & 0.22 &0.08& 1.95&0&4.0 \\ 
Kuo-Mallick & 0.53 &0.30& 0.09&91.1&5.9 \\ 
     \hline \hline
      \end{tabular}
      }
       \caption{ $R_{pop}^2=0.8.$}
    \end{subtable}
    \hfill  \\
    
    \begin{subtable}[h]{\textwidth}
        \centering
        \scriptsize{
      \begin{tabular}{ c c  c c c c} 
&Bias & MSE & FN & FP & TP \\
  \hline \hline 
\multicolumn{6}{c}{$R^2_{pop}=0.4, p=50 $}\\         
  \hline  
 SSVS-Lasso-1 & 0.12 &0.02& 1.1&3.4&4.8 \\ 
 SSVS-Lasso-2 & 0.12 &0.02& 1.1&4.9&4.8 \\ 
 SSVS-Lasso-3 & 0.10 &0.01& 1.1&5.4&4.8 \\ 
 Narisetty-He & 0.16 &0.03& 3.4&2.6&2.6 \\ 
Kuo-Mallick & 0.11 &0.02& 1.2&8.8&4.8 \\ 
\hline 
\multicolumn{6}{c}{$R^2_{pop}=0.4, p=100 $}\\         
  \hline 
 SSVS-Lasso-1 & 0.15 &0.03& 0.8&11.5&5.1 \\ 
 SSVS-Lasso-2 & 0.17 &0.03&0.7&15.9&5.2 \\ 
 SSVS-Lasso-3 & 0.12 &0.02&1.1&17.6&4.8 \\ 
 Narisetty-He & 0.18 &0.04& 2.2&14.9&3.7 \\ 
Kuo-Mallick & 0.28 &0.13& 2.9&29.4&3.0 \\ 
   \hline 
\multicolumn{6}{c}{$R^2_{pop}=0.4, p=300 $}\\         
    \hline 
 SSVS-Lasso-1 & 0.24 &0.06& 0.78&50.3&5.22 \\ 
 SSVS-Lasso-2 & 0.26 &0.07& 0.71&60.4&5.29 \\ 
 SSVS-Lasso-3 & 0.19 &0.04& 0.88&56.9&5.12 \\ 
 Narisetty-He & 0.24 &0.06& 0.91&82.6&5.09 \\ 
Kuo-Mallick & 0.21 &0.05&0.76 & 93.1&5.24 \\ 
     \hline \hline 
      \end{tabular}
      }
       \caption{ $R_{pop}^2=0.4.$}
    \end{subtable}
     \caption{Average metrics over 100 repetitions for each of the approaches. 
Estimated error variance, and bias and MSE of the first 6 elements of the slope vector, and the numbers of False Negatives (FN), False Positives (FP), and True Positives (TP).
The posterior means were used as point estimates. 
The post-processing method of \cite{LiPati2017} was used to distinguish signals from noises. $n=100$. }
\label{ssvs_table}
\end{table}

\FloatBarrier

\section[Bayesian Computation]{Bayesian Computation with hierarchical priors}

We have established that hierarchical priors obey conditional structures that make derivation of conditional posteriors a straightforward business. As a consequence, the Gibbs sampler is the primary computational tool for variable selection and shrinkage problems using hierarchical priors. However, exactly because such full Bayes shrinkage estimators are mostly needed in high and ultra-high dimensions, the Gibbs sampler and related Monte Carlo-based methods become computational costly. In such cases there are numerous other strategies that allow for faster computation. These strategies include approximate methods for computing marginal posterior distributions, or iterative, non-sampling methods that approximate the posterior mode or mean. Many of these algorithms originate in computing science, where data dimensions have always been larger than traditional economic data sets. Currently, Bayesian computation in high-dimensional spaces -- especially in the presence of hierarchical priors -- is the topic of an expanding research agenda in mainstream statistics as well as in the field of machine learning. In this section, we summarize this vibrant research, focusing on both MCMC and fast approximate algorithms.

\subsection{Brute-force/analytical algorithms}
Analytical algorithms for hierarchical priors, in general, do not exist apart from a few special cases that can be fairly restrictive. In the context of estimating a normal mean $\theta$ (see our discussion of \citealp{EfronMorris1973} in \autoref{sec:estimators}), \cite{KahnRaftery1992} put uniform hyperpriors on the mean and variance hyperparameters of a normal prior distribution of $\theta$. In order to obtain the posteriors of these hyperparameters they need to integrate $\theta$, something they are able to do numerically since in their case only univariate integrals are involved on the support $[0,1]$. In the context of a regression with spike and slab prior, \cite{Clyde1999} shows that if the design is orthogonal (that is if the Gram matrix $\bm X^{\prime} \bm X \propto I$) and the regression variance $\sigma^2$ is known, variable selection indicators $\bm \gamma$ can be obtained without resorting to Monte Carlo methods (either Gibbs sampler or Monte Carlo). \cite{PapaspiliopoulosRossell2017} also derive an efficient non-sampling algorithm for Bayesian model averaging in regressions with a block diagonal design.\footnote{Examples of modeling scenarios with block-diagonal matrix $\bm X^{\prime} \bm X$ include time-series regressions with time-varying parameters, and vector autoregressions written in ``seemingly unrelated regressions'' form; see \autoref{sec:VAR} for more details.} Their methods involve calculation of model probabilities and parameter estimates using one-dimensional numerical integration.

An interesting case that allows for approximately analytical posterior inference using hierarchical priors is provided in \cite{vandenBoometal2015a} and \cite{vandenBoometal2015b}. These authors use rotation matrices to partition the regression model into a component explained by predictor $X_{j}$ and all remaining predictors $\bm X_{(-j)}$. In particular, they split the regression into two components
\begin{itemize}
\item one partition that is a regression of $n-1$ observations of a rotation of $\bm y$ on $\bm X_{(-j)}$ (i.e.\ dependence on $X_{j}$ is removed), and 
\item one partition that is a regression of the remaining one observation of a separate rotation of $\bm y$ on $X_{j}$, conditional on $\bm X_{(-j)}$.
\end{itemize}
These authors use a non-shrinking natural conjugate prior in the first part of the rotated regression in order to obtain analytically an estimate of $\bm \beta_{(-j)}$ and $\sigma^2$. Then conditional on these estimates, they introduce in the second part a hierarchical shrinkage prior on $\beta_j$ and derive analytically its posterior, since the regression variance is known using its estimate from the first partition. This procedure requires to repeat the rotation and partition of the regression model for each predictor $j$, $j=1,...,p$. The outcome is an analytical derivation of the posterior of each element $\beta_j$ of $\bm \beta$ under a hierarchical prior that would otherwise require posterior simulation. This algorithm is of course approximate because it requires to obtain the posterior of $\beta_j$ by integrating out the influence of each $\bm \beta_{(-j)}$ using a natural conjugate prior, rather than the same hierarchical prior used on $\beta_j$. \cite{KorobilisPettenuzzo2019} extend this idea to various several hierarchical priors, including normal-Jeffrey's, spike and slab, and normal-gamma.

\subsection{Gibbs sampler}
We have already established that complex distributions (e.g.\ Student-t, Laplace, normal-half Cauchy) can be written in a conditionally conjugate form by using hierarchical representations. Depending on whether we also condition on the regression variance parameter, or not, we obtain the following two hierarchical prior formulations
\begin{equation}
\begin{array}{ccccccc}
\multicolumn{3}{c}{\text{\underline{Natural conjugate prior}}} & & \multicolumn{3}{c}{\text{\underline{Independent prior}}} \\
\bm \beta \vert \sigma^{2}, \bm \tau^{2} & \sim & N_{p} \left(\bm 0, \sigma^{2} \bm D_{\tau} \right), & & \bm \beta \vert \bm \tau^{2} & \sim & N_{p} \left(\bm 0, \bm D_{\tau} \right), \label{cond_vs_indep} \\
\bm \tau^{2} & \sim & p(\bm \tau^2), & & \bm \tau^{2} & \sim & p(\bm \tau^2), \\
\sigma^{2} & \sim & \frac{1}{\sigma^{2}}, & & \sigma^{2} & \sim & \frac{1}{\sigma^{2}},
\end{array}
\end{equation}
where $\bm D_{\tau} = diag(\tau_{1}^2,...,\tau_{p}^2)$ and depending on the structure of the distribution $p(\bm \tau^2)$ (which itself can be a hierarchical mixture of several distributions) we obtained the various interesting cases we explored so far.

Because of the conditional structure of the prior, posterior conditionals are easy to derive. For example, consider the case of the independent prior, then the joint posterior is of the form
\begin{equation}
p \left( \bm \beta, \bm \tau^{2} , \sigma^{2} \vert \bm y \right) \propto p \left(\bm y \vert \bm \beta, \sigma^{2} \right) p(\bm \beta \vert \bm \tau^{2}) p(\bm \tau^{2}) p(\sigma^{2}). 
\end{equation}
We can derive the conditional posterior of $\bm \beta$ as
\begin{equation}
p \left( \bm \beta \vert \bm \tau^{2} , \sigma^{2},  \bm y \right) \propto p \left(\bm y \vert \bm \beta, \sigma^{2} \right) p(\bm \beta \vert \bm \tau^{2}), \label{cond_beta}
\end{equation}
because $p(\bm \tau^{2})$ and $p(\sigma^{2})$ are constants when conditioning on $\bm \tau^{2} , \sigma^{2}$ (because they do not involve the random variable $\bm \beta$). The prior distribution $p(\bm \beta \vert \tau^{2})$ is normal and, due to the modeling assumptions, $p \left(\bm y \vert \bm \beta, \sigma^{2} \right)$ is also normal. As a result, the conditional posterior for $p \left( \bm \beta \vert \bm \tau^{2} , \sigma^{2},  \bm y \right)$ is identical to the conditional posterior under the non-hierarchical version of the same prior \citep[see for example][]{BDA2013}.  Similarly, the conditional posterior for $\sigma^{2}$ becomes
\begin{equation}
p \left( \sigma^{2} \vert \bm \beta, \bm \tau^{2} ,  \bm y \right) \propto p \left(\bm y \vert \bm \beta, \sigma^{2} \right) p(\sigma^{2}),
\end{equation}
which is also identical to the case of the non-hierarchical independent normal-inverse gamma prior. Finally, the conditional posterior for $\tau$ becomes
\begin{equation}
p \left( \bm \tau^{2} \vert \bm \beta,  \sigma^{2} ,  \bm y \right) \propto p(\underline{\bm \beta} \vert \bm \tau^{2}) p(\bm \tau^{2}).
\end{equation}
The data density $p \left(\bm y \vert \bm \beta, \sigma^{2} \right)$ does not contain information about $\bm \tau$ so it becomes a constant. Instead the ``model'' for $\bm \tau^2$ is provided by the density $p(\bm \beta \vert \bm \tau^{2})$ where $\bm \beta$ are observed ``data'' (fixed to their sampled values), since in this conditional posterior the only random variable is $\bm \tau^2$.

It becomes apparent that because of the conditional structure of hierarchical priors, for the vast majority of hierarchical priors we have common formulas for the conditional posteriors of $\bm \beta$ and $\sigma^{2}$, while the formulation of the conditional posterior of $\bm \tau^{2}$ will depend on how complicate its prior is. Under the natural conjugate prior the conditional posteriors are of the form
\begin{equation}
\begin{array}{ccc}
\multicolumn{3}{c}{\text{\underline{Conditional Posteriors (Natural conjugate prior)}}} \\
\bm \beta \vert \sigma^{2}, \bm \tau^{2}, \sigma^{2},\bm y & \sim & N_{p} \left(\bm V \bm X^{\prime} y, \sigma^{2} \bm V \right), \\
\bm \tau^{2} \vert \bm \beta, \sigma^{2}, \bm y & \sim & p(\bm \tau^{2} \vert \bm \beta, \sigma^{2}, \bm y), \\
\sigma^{2} \vert \bm \beta, \bm y & \sim & Inv-Gamma \left( \frac{n+p}{2}, \frac{1}{2} ( \bm \Psi + \bm \beta^{\prime} \bm D_{\tau}^{-1} \bm \beta) \right),
\end{array} \label{Gibbs_conj}
\end{equation}
where $\bm V = \left( \bm X^{\prime} \bm X +  \bm D_{\tau}^{-1} \right)^{-1}$ and $\bm \Psi=\left( \bm y - \bm X \bm \beta\right)'\left( \bm y - \bm X \bm \beta\right)$. Under the independent prior the posteriors are of the form
\begin{equation}
\begin{array}{ccc}
\multicolumn{3}{c}{\text{\underline{Conditional Posteriors (Independent prior)}}} \\
\bm \beta \vert \sigma^{2}, \bm \tau^{2}, \sigma^{2},\bm y & \sim & N_{p} \left(\bm V \bm X^{\prime} y /\sigma^{2},  \bm V \right), \\
\bm \tau^{2} \vert \bm \beta, \sigma^{2}, \bm y & \sim & p(\bm \tau^{2} \vert \bm \beta, \bm y), \\
\sigma^{2} \vert \bm \beta, \bm y & \sim & Inv-Gamma \left( \frac{n}{2}, \frac{1}{2} \bm \Psi \right),
\end{array} \label{Gibbs_ind}
\end{equation}
where $\bm V = \left( \bm X^{\prime} \bm X/\sigma^{2} +  \bm D_{\tau}^{-1} \right)^{-1}$ and $\bm \Psi=\left( \bm y - \bm X \bm \beta\right)'\left( \bm y - \bm X \bm \beta\right)$. Notice how $\sigma^2$ affects the posterior of $\bm \tau^{2}$ in the conjugate prior case, while $\sigma^{2}$ doesn't show up in the posterior of $\bm \tau^{2}$. See the Technical Document for derivations. 

A Gibbs sampler will cycle through equations \eqref{Gibbs_conj} or equations \eqref{Gibbs_ind}, obtaining a sample of parameters conditional on all others, until a large enough sample from the posterior of each parameter is available. Results in \cite{PalKhare2014} and \cite{KhareHobert2013} establish, for a large class of hyper-prior distributions $p(\bm \tau^{2})$, that the above Gibbs sampler is ergodic and has the joint posterior $p\left(\bm \beta,\bm \tau^{2}, \sigma^{2} \vert \bm y \right)$ as its stationary density. Despite its ergodicity, the basic Gibbs sampler for models with hierarchical priors may suffer from slow mixing and convergence to the desired posterior. The conditional structure of a hierarchical prior implies a long chain of dependence of $\bm \beta$ on $\bm \tau^2$ and its hyper-priors. For example, the representation of the Horseshoe prior suggested by \cite{MakalicSchmidt2016} as a hierarchical mixture of a normal distribution and four inverse gamma hyper-prior distributions (see \autoref{sec:horseshoe}) is one example where slow mixing might become a serious issue. For that reason, in the case of the Horseshoe in particular, several authors propose to use more efficient slice sampling schemes, some of which we explore in detail in the accompanying Technical Document. Another disadvantage of the Gibbs sampler is the fact that sampling becomes cumbersome as the dimension $p$ of covariates increases. In the next we explore various approaches for speeding up MCMC and for dealing with convergence issues, particularly in the case where $p$ is large or even $p>>n$.\footnote{All the approaches we explore propose novel ways of sampling from the parameter posterior under the Gibbs sampling scheme. However, given a specific algorithm, the ability of the programming language to handle large matrices is also important. This is illustrated, for example, in \cite{Matusevichetal2016}, where in the context of Bayesian variable selection they combine array database management systems (DBMS) and R processing capabilities, allowing R to handle large matrices without running out of RAM.}

\subsubsection*{Fast sampling from Normal posteriors}
In high-dimensional settings with $p$ large, the most cumbersome step in a Gibbs sampler for the linear regression model with hierarchical priors is sampling from the $p$-variate normal conditional posterior distribution of $\bm \beta$. This step involves an inversion of precision matrix $\bm Q = \bm V^{-1} = \left( \bm X^{\prime} \bm X +  \bm D_{\tau}^{-1} \right)$ (in the case of the natural conjugate prior) or $\bm Q = \bm V^{-1} = \left( \bm X^{\prime} \bm X/ \sigma^{2} +  \bm D_{\tau}^{-1} \right)$ (in the case of the independent prior) in order to obtain the posterior covariance matrix $\bm V$.\footnote{Note that in the case of the natural conjugate prior without hierarchical structure, the matrix $\bm D_{\tau}$ is known (calibrated by the researcher), as is the data information $\bm X^{\prime} \bm X$. In this case, one could calculate and invert $\bm Q$ once, outside the loop of the Gibbs sampler. However, the presence of a hierarchical prior for $\bm \tau^{2}$ means that the matrix $\bm D_{\tau}$ changes values in each Gibbs iteration. Therefore, $\bm V^{-1}$ should be computed and inverted in each iteration (regardless of whether we used the independent prior on $\bm \beta$ or not).} Next, the Cholesky decomposition of $\bm V$ is needed in order to sample from the desired normal distribution. While the inversion step can be sped up (e.g.\ by using Woodbury's identity), standard built-in algorithms (in various programming languages) for obtaining the Cholesky decomposition of a $p \times p$ matrix have a worst asymptotic complexity (measured in flops) of $\mathcal{O}(p^{3})$. Therefore, simple sampling from a normal posterior is deemed to become computationally cumbersome, if not infeasible, as $p$ increases.

\cite{Rue2001} provides a precision-based sampler in order to obtain samples from a normal distribution efficiently when the precision matrix $\bm Q = \left( \bm X^{\prime} \bm X  +  \bm D_{\tau}^{-1} \right)$ is known. Due to the fact that the Bayesian conditional posterior of $\bm \beta$ (ignoring $\sigma^{2}$, e.g.\ assume it is fixed to value 1) is of the form $\bm \beta \vert \bullet \sim N( \bm V \bm X^{\prime} \bm y, \bm V)$, the procedure proposed by \cite{Rue2001} takes the following form
\begin{itemize}
\item Compute the lower Cholesky factorization $\bm Q = \bm L \bm L^{\prime} $
\item Generate $\bm Z \sim N_p(\bm 0, \bm I_p)$
\item Set $\bm v = \bm L^{-1} (\bm X^{\prime} \bm y) $
\item Set $\bm \mu = \bm L^{\prime -1} \bm v$
\item Set $\bm u =  \bm L^{\prime -1} \bm Z $
\item Set $\bm \beta =  \bm \mu + \bm u$
\end{itemize}
It is trivial to show that $E(\bm \beta)  =\bm \mu= (\bm L^{\prime -1}\bm L^{-1}) \bm X^{\prime} \bm y=
( \bm L\bm L^{\prime} )^{-1}\bm X^{\prime} \bm y
=
 \bm V \bm X^{\prime} \bm y$ and $cov(\bm \beta) = cov(\bm \mu + \bm u) =  \bm L^{\prime -1} cov(\bm Z)\bm L^{-1} = \bm L^{\prime -1} \bm L^{-1} = \bm V $, which means that the above procedure provides valid samples from the desired normal distribution. The main feature of this algorithm is that it requires to invert the Cholesky factor of $\bm Q$, instead of inverting $\bm Q$ itself to obtain $\bm V$. While the worst case complexity of this algorithm is also $\mathcal{O}(p^3)$, it provides high efficiency gains in certain classes of models, e.g.\ when the Gram matrix $\bm X^{\prime} \bm X$ is block-diagonal (assuming the prior variance $\bm D_{\tau}$ is diagonal or at most block-diagonal of similar structure).

More recently, \cite{Bhattacharyaetal2016} proposed an efficient algorithm that makes full use of Woodbury matrix inversion lemma in the context of generating normal variates from a distribution of the form $\bm \beta \vert \bullet \sim N( \bm V \bm X^{\prime} \bm y, \bm V)$ (again for simplicity, ignore $\sigma^{2}$). Their algorithm takes the following form
\begin{enumerate}
\item Sample $\bm \eta \sim N_{p}(\bm 0,\bm D_{\tau})$ and $\bm \delta \sim N_{n}(\bm 0,\bm I_{n})$
\item Set $\bm v = \bm X \bm \eta + \bm \delta$
\item Set $\bm w = ( \bm X \bm D_{\tau} \bm X^{\prime} +\bm  I_{n} )^{-1}[\bm y - \bm  v]$
\item Set $\bm \beta = \bm \eta + \bm D \bm X^{\prime} \bm w $  
\end{enumerate}
It is also easy to show that the sample of $\bm \beta$ comes from the desired normal distribution with mean $\bm V \bm X^{\prime} \bm y$ and variance $\bm V$. Note that this algorithm requires inversion of the $n \times n$ matrix $( \bm X \bm D_{\tau} \bm X^{\prime} + \bm I_{n} )^{-1}$, while sampling directly from the normal posterior requires inversion of the $p \times p$ matrix $\bm Q$. Therefore, step 3 in this algorithm only becomes efficient for $p>>n$. The main efficiency gains in this algorithm stem from the fact that it requires to sample from two normal distributions with diagonal covariance matrices (a $p$-variate distribution with covariance $\bm D_{\tau}$ and an $n$-variate distribution with identity covariance). Generating  uncorrelated normal draws is much more efficient than sampling directly from the $p$-variate normal posterior of $\bm \beta$ using the full covariance matrix $\bm V$. In particular, the worst-case complexity (asymptotic upper bound) of this algorithm is $\mathcal{O}(n^{2}p)$, that is, it is only linear in $p$. Therefore, for $n>p$ this algorithm will perform worse than the algorithm of \cite{Rue2001} since the term $n^{2}$ will dominate, but this algorithm shines in the $p > n$ case where it can offer some dramatic improvements in computation times.

Note that for very large $p$, computation of $\bm X \bm D_{\tau} \bm X^{\prime}$ in step 3.\ above will become cumbersome. In such ultra high-dimensional cases, \cite{Johndrowetal2020} provide an approximate version of \cite{Bhattacharyaetal2016}. This involves removing ``irrelevant'' columns of $\bm X$ such that the above product can be computed using a significantly smaller number of algorithmic operations.

\subsubsection*{Scalable Gibbs}
The standard form of the Gibbs sampler for the linear regression model with hierarchical prior contains three blocks as in \autoref{Gibbs_conj}. There is one block for each set of parameters, namely $\bm \beta$, $\bm \tau^{2}$ and $\sigma^{2}$. In the context of hierarchical priors, \cite{Rajaratnam2019} propose to sample $(\bm \beta, \sigma^{2})$ in one block. Their proposed Gibbs sampler is more efficient, as reducing the number of blocks to sample from, also reduces correlation among draws from parameter posteriors. When working with the natural conjugate form of a hierarchical prior, the \emph{scalable Gibbs} algorithm requires only to sample from $p((\bm \beta, \sigma^{2}) \vert \bm \tau^{2}, \bm y)$ and $p(\bm \tau^{2} \vert (\bm \beta, \sigma^{2}), \bm y)$. The first joint distribution for $(\bm \beta, \sigma^{2}) $ can be approximated by first sampling from $p(\sigma^{2} \vert \bm \tau^{2}, \bm y)$ (notice the lack of dependence on $\bm \beta$) and subsequently from $p( \bm \beta \vert \sigma^{2}, \bm \tau^{2} , \bm y)$. The scalable Gibbs algorithm has the following form
\begin{equation}
\begin{array}{ccl}
\left\lbrace
\begin{array}{l}
\sigma^{-2} \vert \bm \tau^{2}, \bm y    \\
\bm \beta \vert \sigma^{2}, \bm \tau^{2}, \bm y
\end{array}  \right. & 
\begin{array}{l}
\sim \\
\sim
\end{array}  &
\begin{array}{l}
Gamma \left( \frac{n-1}{2},\bm y'( \bm I_n - \bm X \bm V \bm X^{\prime}) \bm y /2 \right) \\
N_{p} \left(\bm V \bm X^{\prime} \bm y , \bm V \right) 
\end{array}   , \\
\bm \tau^{2} \vert \bm \beta, \sigma^{2}, \bm y & \sim & p(\bm \tau^{2} \vert \bm \beta, \sigma^{2}, \bm y),
\end{array}
\end{equation}
where $\bm V$ has the same definition as in \autoref{Gibbs_conj}. The proof why the posterior for $\sigma^{2}$, after integrating out $\bm \beta$, has the form shown above, can be found in the Appendix of \cite{Rajaratnam2019}; see also \cite{Paletal2017}.

\subsubsection*{Skinny Gibbs}
In the context of the spike and slab prior with continuous spike and slab distributions, \cite{Narisettyetal2018} propose an efficient sampling scheme that also separates the posterior into a mixture distribution with independent components. We remind that the continuous spike and slab prior \citep{GeorgeMcCulloch1993,NarisettyHe2014} can be written in matrix form as
\begin{equation}
\bm \beta \vert \bm \gamma \sim (\bm I - \bm \Gamma)N_{p}(\bm 0 , \tau_{0}^{2} \bm I) + \Gamma N_{p}(\bm 0 , \tau_{1}^{2} \bm I), 
\end{equation}
where $\bm \Gamma = diag(\bm \gamma)$, or more compactly
\begin{equation}
\bm \beta \vert \bm \gamma \sim N_{p}(\bm 0 ,\bm D_{\gamma}),
\end{equation}
where $\bm D_{\gamma} = diag\left( (1-\gamma_{1})^{2} \tau_{0}^{2} + \gamma_{1}^{2} \tau_{1}^{2},..., (1-\gamma_{p})^{2} \tau_{0}^{2} + \gamma_{p}^{2} \tau_{1}^{2} \right) $. The conditional posterior under this prior is of the form
\begin{equation}
\bm \beta \vert \bullet \sim N_{p}(\bm V \bm X^{\prime} \bm y /\sigma^{2},\bm V), \label{skinny_post}
\end{equation}
where $\bm V = \left( \bm X^{\prime} \bm X/\sigma^2 + \bm D_{\gamma} ^{-1} \right)^{-1}$ and $\vert \bullet$ denotes conditioning on other parameters in the model as well as the data. Computing $\bm V$ requires an inversion as well as obtaining the Cholesky decomposition in order to sample from the $p$-dimensional normal posterior. As we already saw, when $p$ is large these operations can be extremely cumbersome. 

The skinny Gibbs algorithm of \cite{Narisettyetal2018} solves this issue by splitting the conditional posterior into two independent components, an active (A) and an inactive (I) and sample $\bm \beta$ as the union of the following conditionals
\begin{eqnarray}
\bm \beta_{A} \vert  \bullet & \sim & N_{p_{A}}(\bm V_{A} \bm X_{A}^{\prime} \bm y /\sigma^{2},\bm V_{A}), \\
\bm \beta_{I} \vert \bullet & \sim & N_{p_{I}}(\bm 0, \bm V_{I}),
\end{eqnarray}
where $\bm \beta_{A}$ is the $p_{A}$-dimensional vector of elements of $\bm \beta$ corresponding to $\gamma_{j}=1$, and $\bm \beta_{I}$ are the remaining $p_{I} = p - p_{A}$ elements that correspond to $\gamma_{j}=0$. In the first posterior the covariance matrix is $\bm V_{A} = \left( \bm X^{\prime} \bm X/\sigma^2 + \frac{1}{\tau_{1}^{2}}\bm I \right)^{-1} $, while in the second posterior $\bm V_{I} = \left(n + \frac{1}{\tau_{0}^{2}} \right)^{-1} \bm I_{p_I} $. In sparse settings we would expect to find that $p_{I} >> p_{A}$, meaning that the bulk of the elements of $\bm \beta$ would be sampled as restricted elements $\bm \beta_{I}$. This means that we can sample very efficiently $p_{I}$ coefficients from a normal posterior with diagonal convariance matrix, and sample the remaining $p_{A}$ elements from a normal posterior with a full covariance matrix.

Of course, in practical situations it is expected that the two sets of coefficients, $\bm \beta_{A}$ and $\bm \beta_{I}$, will be correlated with each other. As a result, sampling the full vector $\bm \beta$ from two independent conditional posteriors could leave us with a significant approximation error. For that reason, \cite{Narisettyetal2018} add a ``compensation term'' in the conditional posterior of $\bm \gamma$ that accounts for the approximation involved in sampling the coefficients $\bm \beta$. This term ensures that the skinny Gibbs converges to a stationary distribution, while keeping computational complexity minimal. \cite{Narisettyetal2018} show that the skinny Gibbs posterior possesses strong selection consistency property.

\subsubsection*{Orthogonal Data Augmentation}
In the conext of variable selection using spike and slab priors, \cite{GhoshClyde2011} note that when the design matrix $\bm X$ is orthogonal, a stochastic search using MCMC can become very efficient when $p$ is large. Similarly for penalized likelihood problems (whether Bayesian or not), orthogonal designs can be very efficient and result in consistent variable selection regardless of how large $p$ is relative to $n$. The proposal of \cite{GhoshClyde2011} is to augment the $n \times p$ correlated design matrix $\bm X$ with an $n_{a} \times p$ matrix $\bm X_{a}$, such that the $(n+n_{a}) \times p $ ``complete'' design matrix
\begin{equation}
\bm X_{c} = \left[
\begin{array}{c}
\bm X \\
\bm X_{a}
\end{array} 
\right],
\end{equation}
has orthogonal columns, that is, $\bm X_{c}^{\prime} \bm X_{c} = \bm X^{\prime} \bm X + \bm X_{a}^{\prime} \bm X_{a} = \bm W$, where $\bm W = diag(w_{1},...,w_{p})$ is a diagonal matrix with $w_{j}>0$. Furthermore, we have the restriction that the augmented data matrix $\bm X_{a}$ has real entries, and $\bm X_{a}^{\prime} \bm X_{a}$ must be a positive semidefinite symmetric matrix. \cite{GhoshClyde2011} select a diagonal matrix $\bm W$ which then implies the value of $\bm X_{a}$ from the orthogonality condition  $\bm A \equiv \bm X_{a}^{\prime} \bm X_{a} = \bm W - \bm X^{\prime} \bm X $. $\bm X_{a}$ can be obtained as the symmetric matrix square root of $\bm A$, thus, ensuring that its entries are real. \cite{GhoshClyde2011} discuss in detail ways of choosing $\bm W$; for example, since in most variable selection settings the columns of $\bm X$ are typically standardized to have unit norm or variance, one can set $w_{1}=...=w_{p} =w$, such that choice of $\bm W$ collapses into a choice of a scalar $w$.

Once $\bm X_{a}$ has been specified, we can estimate the augmented orthogonal regression model
\begin{equation}
\bm y_{c} = \bm X_{c} \bm \beta + \bm \varepsilon_{c},
\end{equation}
where $y_{c} = \left[ \bm y^{\prime} , \bm y_{a}^{\prime} \right]^{\prime}$ with $\bm y_{a}$ latent data which we need to sample and $\bm \varepsilon_{c} \sim N_{(n+n_a)}(\bm 0, \sigma^{2} \bm I)$. In the most general case, \cite{GhoshClyde2011} consider a hierarchical variable selection prior, which combines a spike at zero with a component that is Student-t, obtained via normal-inverse-gamma mixture (see \autoref{sec:student})
\begin{eqnarray}
\beta_{j} \vert \sigma^{2}, \gamma_{j}, \tau_{j}^{2} & \sim & N(0,\sigma^{2} \gamma_{j} \tau_{j}^{2} ), \\ 
\tau_{j}^{2} & \sim & Gamma(\alpha/2,\alpha/2), \\
\gamma_{j} & \sim & Bernoulli(\pi_{0}),\\
\sigma^{2} & \sim & \frac{1}{\sigma^2},
\end{eqnarray}
for all $j=1,...,p$. For $\alpha=1$ this prior becomes a heavy-tailed Cauchy distribution of the form $\beta_j  \sim C(0,\sigma^{2} \gamma_{j})$.  \cite{GhoshClyde2011} propose a Gibbs sampling scheme the iterates over the following conditional distributions
\begin{enumerate}
\item $p\left( \left( \sigma^{2}, \bm y_{a} \right) \vert \bm \gamma, \bm y \right)$
\item $p(\gamma_{j} \vert \sigma^{2},\bm \tau^{2}, \bm y_{c}) $ for $j=1,...,p$
\item $p(\bm \beta \vert \sigma^{2}, \bm \tau^{2}, \bm \gamma \bm y_{c})$
\item $p(\bm \tau^{2} \vert \sigma^{2}, \bm \beta, \bm \gamma, \bm y_{c})$
\end{enumerate}
All the conditional distributions have standard forms, and details can be found in \cite{GhoshClyde2011}.

\subsection{Approximate computation with hierarchical priors}
Approximate inference methods typically involve optimization algorithms for approximating posterior moments (typically the mean or the mode, and the variance) instead of sampling from the full posterior distribution. Then one can proceed their analysis using only these moments, treating the Bayesian estimator similar to a frequentist point estimator. This approach to Bayesian inference has been popularized in computing science, e.g.\ in estimation of high-dimensional Bayes networks, where large datasets is the norm and MCMC inference is extremely costly. An obvious critique of approximate Bayesian inference of this sort, is that we can't fully take into account parameter uncertainty by characterizing the full parameter posterior distribution. However, it is in high-dimensional and the so-called ultra high-dimensional models \citep[see][]{Shinetal2018}, that shrinkage and sparsity via a hierarchical prior is necessary. Since analytical results for most classes of hierarhical priors are not available, and Monte Carlo sampling is costly in very high dimensions, it is not surprising that approximate methods have become very popular. Additionally, at the conceptual level, we saw that only Bayesian posterior medians/modes correspond to penalized likelihood estimators, but there are not always good theoretical guarantees for the tails of the posterior.\footnote{See for example our discussion of the results of \cite{castillo2015} in the Bayesian lasso prior.} Finally, the concept of sparsity is indeed more interpretable in a setting with point estimates of coefficients, that is, it is more straightforward to test and interpret $H_0: \beta_j = 0$ when $\beta_j$ is approximated with a point estimate rather than when we have thousands of samples from the full posterior of $\beta_j$.

All these reasons lead us to review some of the most popular approximate methods for posterior inference. While many of these methods have been popularized and used extensively in computing science often without  theoretical justifications, investigation of their theoretical/asymptotic properties is currently a topic of  vivid research in mainstream statistics.

\subsubsection{Variational Bayes}
Variational Bayes is probably the most prominent of algorithms, at least when it comes to high-dimensional inference using hierarchical priors. The idea behind this class of algorithms is rather simple, and under certain assumptions (what we will call \emph{mean-field approximation} later) variational Bayes algorithms can be fairly simple to implement by practitioners who are familiar with the Gibbs sampler. For notational simplicity, assume we have a vector of parameters $\bm \theta$ with support $\Theta$ and data $\bm D$, resulting to the posterior distribution 
\begin{equation}
p(\bm \theta \vert \bm D) = \frac{p(\bm D \vert \bm \theta)p(\bm \theta)}{p(\bm D)}.
\end{equation}
Assume that this posterior is intractable because the data density $p(\bm D \vert \bm \theta)$ is complex (e.g.\ it is a highly nonlinear function, or it has unidentified parameters), or because the prior $p(\bm \theta)$ is complex (non-conjugate), or because $\theta$ is high-dimensional (in which case the posterior is a high-dimensional function), or due to combinations of the above cases. In such settings, MCMC is not only computationally costly, but it can also become numerically unstable/unreliable.\footnote{For example, consider the case of a high-dimensional nonlinear regression with highly correlated predictors. In this example, unless modifications are introduced such as adaptive tuning, mixing and convergence of standard Gibbs sampler algorithms with hierarchical priors will tend to be slow.}

The idea behind variational Bayes is to introduce a family of simpler, approximate densities over the parameters $\bm \theta$ which is denoted by the set $\mathcal{Q}$. The objective is to find a member of the family $q(\bm \theta) \in \mathcal{Q}$ that is as close as possible to the true posterior. ``Closeness'' is measured by the Kullback-Leibler (KL) divergence, and the optimal density $q^{\star}(\bm \theta)$, among all densities $q(\bm \theta)$, is the one that minimizes this criterion:
\begin{equation}
q^{\star}(\bm \theta) = \argmin_{q(\bm \theta) \in \mathcal{Q}} 
KL \left( q \left(\bm \theta \right) \vert\vert p\left(\bm \theta \vert \bm  D\right) \right).
\end{equation}
In the above formula the KL measure on the RHS is defined as
{\small \begin{eqnarray}
KL & = &  E_{q(\bm \theta)} \left(\log(q(\bm \theta)) \right)  -  E_{q(\bm \theta)} \left( \log(p(\bm \theta \vert \bm D)) \right)  \\
& = &  E_{q(\bm \theta)} \left(\log(q(\bm \theta)) \right)  -  E_{q(\bm \theta)} \left( \log \left( \frac{p(\bm D \vert \bm \theta)p(\bm \theta)}{p(\bm D)} \right) \right)  \\
& = &   E_{q(\bm \theta)} \left(\log(q(\bm \theta)) \right) +  \log(p(\bm D)) - E_{q(\bm \theta)} \left(\log \left( p(\bm D \vert \bm \theta)p(\bm \theta)\right) \right), \label{finalVB}
\end{eqnarray}
}where all expectations are w.r.t $q(\bm \theta)$, for example, $E_{q(\bm \theta)} \left( \log(p(\bm \theta \vert \bm D)) \right) = \int_{\bm \theta \in \Theta} q(\bm \theta) \log(p(\bm \theta \vert \bm D)) d \bm \theta$. In the last equation we have used the fact that $ E_{q(\bm \theta)} \left(\log(p(\bm D)) \right) = \log(p(\bm D))$ since $p(\bm D)$ does not involve $\bm \theta$. For the same reason the variational Bayes minimization problem is equal to minimizing the difference between the first and the third terms in equation \eqref{finalVB}. We can also solve this equation for the log marginal likelihood $\log(p(\bm D))$ to show that
\begin{eqnarray}
\log(p(\bm D)) & = & KL \left( q \left(\bm \theta \right) \vert\vert p\left(\bm \theta \vert \bm D\right) \right) + ELBO,
\end{eqnarray}
where we define evidence lower bound (ELBO) to be the quantity $ELBO =  + E_{q(\bm \theta)} \left(\log \left( p(\bm D \vert \bm \theta)p(\bm \theta) \right) \right) -  E_{q(\bm \theta)} \left(\log(q(\bm \theta)) \right) $. The ELBO has this name exactly because it is a lower bound for the log evidence (marginal data density). This is because in the equation above the KL divergence term is non-negative, such that $\log(p(\bm D)) \geq ELBO$. Therefore, the optimal $q^{\star}(\bm \theta)$ can be found by equivalently maximizing the ELBO criterion function.

{\large \textbf{The CAVI algorithm}} \\
When latent variables are present, optimizations such as maximizing the ELBO criterion can be implemented using the popular expectation-maximization (EM) algorithm, where the complete log likelihood is computed (E-step) and then it is maximized (M-step). However, in Bayesian inference all parameters are latent (random) variables and as the optimization problem above involves optimizing over the functional $q(\bm \theta)$ and not $\bm \theta$ itself, the EM algorithm is not appropriate. Variational inference instead requires to choose the variational family of distributions $\mathcal{Q}$ and then maximize the ELBO. In most cases this can be done iteratively, with certain schemes that resemble the EM algorithm (but are not identical to EM), and convergence is guaranteed to a local maximum and if the likelihood is log-concave then to a global maximum. The simplest algorithm for maximizing the ELBO is called Coordinate Ascent Variational Inference (CAVI). Its simplicity comes at the cost of certain simplifying assumptions. The first one is that $\mathcal{Q}$ must strictly belong to the exponential family of distributions (e.g.\ the normal satisfies this condition, but the Student's t does not). The second restriction is the use of the mean-field approximation that postulates that the proposed posterior distribution $q(\bm \theta)$ can be decomposed into $M$ independent groups of the form
\begin{equation}
q (\bm \theta) = \prod_{m=1}^{M} q_{m}(\bm \theta_{m}),
\end{equation}
where the groups could either have $\bm \theta_{m}$ being a scalar or a vector. The estimated variational posteriors will be independent, meaning that the mean-field approximation/factorization implies that $\bm \theta_{m}$ will be a-posteriori uncorrelated with $\bm \theta_{k}$, for $k \neq m$ and $k,m = 1,...,M$. This assumption in several modeling settings can be harmless, but in several others it can become harmful -- we discuss this issue in detail later when we examine variational Bayes inference in a linear regression with variable selection prior.

Under the assumption of the mean field approximation it can be shown \citep{Bleietal17} that the optimal densities $q_{m}(\bm \theta_{m})$ satisfy
\begin{equation}
\log \left(q_{m}(\bm \theta_{m})\right) \propto E_{q_{(-m)}(\bm \theta_{(-m)})} \left( \log \left( p(\bm D\vert \bm \theta)p(\bm \theta) \right) \right), \label{vb_it}
\end{equation}
where $E_{q_{(-m)}(\bm \theta_{(-m)})}()$ means that the expectation is w.r.t all variational densities except $q_{m}(\bm \theta_{m})$. Broadly speaking this formula says that in order to optimize w.r.t. $q_{m}(\bm \theta_{m})$ we need to evaluate the posterior under the assumption that all other parameters $\bm \theta_{(-m)}$ are fixed to their posterior expectation (posterior mean). We would obviously need to iterate through \autoref{vb_it} for each $m=1,...,M$ keeping all other parameters fixed to their posterior means, but it can be shown that such iteration results in increasing the ELBO criterion. If the ELBO hasn't changed from one iteration to the next, the algorithm has converged. This criterion resembles the EM algorithm that converges when the value of the likelihood in subsequent iterations is approximately similar. The fact that $\bm \theta_{m}$ is updated conditional on fixing all other parameters $\bm \theta_{(-m)}$ makes variational Bayes resemble Gibbs sampling inference -- despite the fact that there is no sampling involved. We next derive a CAVI algorithm for a linear regression model with variable selection prior, in order to clearly demonstrate how the mean-field approximation is applied and how the functions in \autoref{vb_it} look like.

{\large \textbf{A variational Bayes approach to variable selection}} \\
This subsection follows closely the analysis of \cite{Ormerodetal2017}, and the reader should consult this paper for extensive discussion and proofs. Consider the \cite{KuoMallick1998} regression we explored in \autoref{sec:sns} and is of the form
\begin{eqnarray}
\bm y & = & \bm X \bm \Gamma \bm \beta + \bm \varepsilon, \\
\bm \varepsilon & \sim & N(\bm 0_{n \times 1}, \sigma^{2} \bm I_{n}), \\
\bm \beta & \sim & N_{p}(\bm 0_{p \times 1}, \bm D), \\
\sigma^{2} & \sim & Inv-Gamma(a_0,b_0), \\
\gamma_{j} & \sim & Bernoulli(\pi_{0}), \text{ \ \ } j=1,...,p,
\end{eqnarray}
where $\bm \Gamma = diag(\gamma_{1},...,\gamma_{p})$ and $\bm D$ is a diagonal prior covariance matrix, e.g.\ $\bm D  = c \times \bm I_{p}$ for some constant $c$. Therefore, according to the information above the joint prior is decomposed into $p \left( \bm \beta, \left\lbrace \gamma_{j} \right \rbrace_{j=1}^{p}, \sigma^{2} \right) = p(\sigma^2)\prod_{j=1}^{p}p(\beta_{j}) \times p(\gamma_{j}) $ meaning that all parameters are a-priori uncorrelated. Such choices are both conceptually and practically fine, first because we don't have prior information on how parameters are correlated, and second because we can construct very powerful shrinkage and variable selection algorithms based on these forms. It is important  for the posterior to allow the parameters to be correlated, as this posterior correlation will come from information in the data likelihood. We discussed previously that the mean-field factorization implies that some groups of parameters will be uncorrelated a-posteriori. \cite{Ormerodetal2017} look into three different ways of applying the mean field factorization, based on how we want to define the groups $m=1,...,M$, and their implications for posterior inference. These factorizations are the following
\begin{eqnarray}
\text{(A): \ \ } q(\bm \beta, \sigma^{2}, \bm \gamma) & = & q(\bm \beta,\bm \gamma) q(\sigma^{2}), \\
\text{(B): \ \ } q(\bm \beta, \sigma^{2}, \bm \gamma) & = & q(\sigma^{2}) \prod_{j=1}^{p} q(\beta_{j}, \gamma_{j}), \\
\text{(C): \ \ } q(\bm \beta, \sigma^{2}, \bm \gamma) & = & q(\bm \beta)q(\sigma^{2}) \prod_{j=1}^{p}q( \gamma_{j}).
\end{eqnarray}

The factorization (A) means that application of formula \autoref{vb_it} to the set of parameters $(\bm \beta, \bm \gamma)$ given $\sigma^{2}$ gives
\begin{equation}
\log( q(\bm \beta, \bm \gamma) ) \propto \lambda \bm \gamma -\frac{1}{2}\bm \beta^{\prime} \left( \kappa \bm \Gamma \bm X^{\prime} \bm X \bm \Gamma \bm \beta + \bm D^{-1} \right)\bm \beta + \kappa \bm \beta^{\prime} \bm \Gamma \bm X^{\prime} \bm y,
\end{equation}
or, similarly, that
\begin{equation}
 q(\bm \beta, \bm \gamma) \propto \exp \left\lbrace \lambda \bm \gamma -\frac{1}{2}\bm \beta^{\prime} \left( \kappa \bm \Gamma \bm X^{\prime} \bm X \bm \Gamma \bm \beta + \bm D^{-1} \right)\bm \beta + \kappa \bm \beta^{\prime} \bm \Gamma \bm X^{\prime} \bm y \right\rbrace,
\end{equation}
where $\lambda = \log\left(\frac{\pi_0}{1-\pi_0}\right)$ and $\kappa=E_{q} \left(1/\sigma^2 \right)$. Therefore, we can easily obtain the conditional variational density of $\bm \beta$ and the (marginal) variational density of $\bm \gamma$ as
\begin{eqnarray}
q\left(\bm \beta \vert \bm \gamma \right) & \sim & N\left(\bm \mu_{\bm \gamma}, \bm V_{\bm \gamma}\right), \\
q \left( \bm \gamma\right) & \propto & \int q(\bm \beta, \bm \gamma) d\bm \beta = \vert \bm V_{\bm \gamma} \vert^{1/2} \exp\left[ \lambda \bm \gamma + \frac{1}{2}\bm \mu_{\bm \gamma}^{\prime} \bm V_{\bm \gamma}^{-1} \bm \mu_{\bm \gamma} \right]
\end{eqnarray}
where $\bm V_{\bm \gamma}=\left( \kappa \bm \Gamma \bm X^{\prime} \bm X \bm \Gamma \bm \beta + \bm D^{-1} \right)^{-1}$ and $\bm \mu_{\bm \gamma} = \bm V_{\bm \gamma} \bm \Gamma \bm X^{\prime} \bm y$. In the second equation we only have the kernel of $q \left( \bm \gamma\right)$, but we can easily normalize this to integrate to one by dividing with the sum of the density of all possible combinations of $\bm \gamma$ (which is $2^{p}$ in the regression with $p$ covariates). In order to derive the marginal variational posterior density of $\bm \beta$ we need to integrate out the $\bm \gamma$, which is easily done as these are binary indicators. This marginal density is of the form
\begin{equation}
q(\bm \beta)  = \sum_{\bm \gamma \in {0,1}^{p}} q(\bm \gamma) N\left(\bm \mu_{\bm \gamma}, \bm V_{\bm \gamma}\right),
\end{equation}
which is a combinatorial sum over all $2^{p}$ outcomes for the vector $\bm \gamma$. This sum can be evaluated in finite time only for small $p$. However, for small $p$ there are other numerous analytical algorithms that can be used, for example, we can use a $g$-prior and obtain marginal likelihoods analytically for all $2^{p}$ models, in which case there is no point in using variational Bayes. Therefore, this mean-field factorization is not useful.

The mean-field factorization/approximation (B), which was used in \cite{CarbonettoStephens12}, provides a scalable variational Bayes algorithm where the $q(\beta_{j})$ can be estimated independently and efficiently (by means of parallelization) for each $j=1,...,p$. However, posterior variances of these regression coefficients will tend to be underestimated exactly because of this assumption of posterior independence. Unless the predictors in $\bm X$ are uncorrelated (which is not  realistic  for economic data, and for large-$p$ settings), the bias in posterior variances can be substantial.

The most reasonable case, which is the choice of \cite{Ormerodetal2017}, is case (C). Under this factorization, one full implementation of the iterations in \autoref{vb_it} looks like this:
\begin{enumerate}
\item $q(\bm \beta) =  N\left(\bm \mu, \bm V \right)$ \\
where $\bm V = \left( \kappa (\bm X^{\prime} \bm X) \odot \bm \Omega + \bm D^{-1} \right)^{-1}$ and $\bm \mu =  \kappa\bm V \left( \bm \Pi \bm X^{\prime} \bm y \right)$.
\item $q(\sigma^{2}) = Inv-Gamma \left(a, b \right)$ \\
where $b = b_0 + \frac{1}{2} \left[ \vert\vert\bm y \vert \vert^{2} - 2 \bm y^{\prime} \bm X \bm \Pi \bm \mu + tr\left\lbrace \left(\bm X^{\prime}\bm X \odot \bm \Omega \right) \left( \bm \mu \bm \mu^{\prime} + \bm V \right)  \right\rbrace \right]$ and $a = a_0 + n/2$. The posterior mean of $\sigma^{-2}$ is, thus, $\kappa =\frac{a}{b}$.
\item $q(\gamma_{j})  = Bernoulli(\pi_{j})$, \\
where $\pi_{j} = \frac{exp(\eta_{j})}{1+exp(\eta_{j})}$ with $\eta_{j} = \log\left(\frac{\pi_0}{1-\pi_0}\right) - \frac{\kappa}{2}\left(\mu_{j}^{2} + V_{j,j} \right) \vert \vert \bm X_{j} \vert \vert^{2} + \kappa \left[\mu_{j}\bm X_{j}^{\prime} \bm y - \bm X_{j}^{\prime} \bm X_{(-j)} \bm \Pi_{(-j)} \left( \bm \mu_{(-j)} \mu_j + \bm V_{(-j),j} \right) \right] $.
\end{enumerate}
In the equations above we have used some matrices vectors/matrices that are based on $\pi_{j}$, namely $\bm \pi= (\pi_{1},....,\pi_{p})^{\prime}$, $\bm \Pi = diag(\bm \pi)$ and $\bm \Omega = \bm \pi \bm \pi^{\prime} + \bm \Pi(\bm I - \bm \Pi)$. The symbol $\odot $ denotes the Hadamard product. We used the notations that for a general matrix $\bm A$, $\bm A_j$ is the $j$th column of $\bm A$, $\bm A_{(-j)}$ is $\bm A$ with the $j$th column removed, $A_{i,j}$ is the $(i,j)$th entry of $\bm A$, $\bm A_{(-i),j}$ is the vector corresponding to the $j$th column of $\bm A$ with the $i$th component removed.

The formulas look quite similar to the conditional posteriors in the \cite{KuoMallick1998} Gibbs samplers, but there is no sampling involved. Instead, when the variational posterior mean and variance of $\bm \beta$ are calculated, $\sigma^{-2}$ is fixed to its posterior mean $\kappa$ and the same for $\gamma$ (the vector $\bm \pi$ and its variants, i.e.\ $\bm \Pi$ and $\bm \Omega$). Given that in the very first iteration $\kappa$ and $\bm \pi$ will be initialized to some random values, a convergence period is required until we end up with final estimates of the posterior moments of all parameters.

{\large \textbf{Further readings}} \\
There are numerous papers on variational Bayes inference in computing science problems, for example, natural language processing (text analytics) and Bayesian networks. In statistics and machine learning there has been a consistent effort to establish consistency and other properties of variational Bayes estimates; see for example \cite{Giordanoetal2018} and\cite{WangBlei2019}. With regards to high-dimensional regression and hierarchical priors, the contributions of \cite{CarbonettoStephens12}, \cite{Ormerodetal2017} and \cite{Nevilleetal2014} are an excellent starting point. \cite{KoopKorobilis2018} provide a variational Bayes algorithm for a dynamic spike and slab prior for models featuring time-varying parameters and stochastic volatility (see also next section).

\subsubsection{EM algorithm}
We discussed previously how the expectation-maximization (EM) algorithm is not appropriate for variational Bayes inference, since there we are looking to find the ``best'' density function of $\bm \theta$, rather than a point estimate of our parameters $\bm \beta$. However, the EM algorithm can be used to find the posterior mode of $p(\bm \theta \vert \bm D)$, an inference method known as maximum a-posteriori (MAP) inference. The mode of the posterior under a diffusing (flat) prior distribution is identical to the maximum likelihood estimate, while the MAP estimate under a hierarchical prior corresponds to a penalized likelihood estimator. Therefore, MAP inference -- which was also popularized in computing science -- can be thought of as a bridge between Bayesian and maximum/penalized likelihood inferences where it combines the strengths of both approaches. 

There are numerous implementations of MAP inference using the EM algorithm but, unlike the Gibbs sampler, in many cases algorithms are model-specific and cannot generalize easily. With regards to variable selection and shrinkage we indicatively mention the key contributions of \cite{CaronDoucet2008}, \cite{Figueiredo2003} and \cite{GriffinBrown2011}. A notable recent contribution is the EM variable selection (EMVS) of \cite{RockovaGeorge2014}. These authors adopt a setting (likelihood and prior) that is identical to \cite{GeorgeMcCulloch1993} but they use the EM algorithm as a means of lowering the computational burden of Markov-Chain Monte Carlo methods when estimating posterior distributions over subsets of potential predictors. 
 
\subsubsection{Other approximate algorithms}
There are several other algorithms for approximate high-dimensional inference. These include parallel MCMC, Hamiltonian Monte Carlo, Approximate Bayesian Computation (ABC), Expectation propagation, and Message Passing. A review of all these classes of algorithms can be found in \cite{KorobilisPettenuzzo2020}. A few representative works relying on such algorithms are \cite{DehaeneBarthelme2018}, \cite{KimWand2016}, \cite{Korobilis2021}, \cite{Liuetal2019}, \cite{WainwrightJordan2008} and \cite{Zouetal2016}.

\subsection[Monte Carlo study]{Monte Carlo exercise: Conjugate vs independent hierarchical priors}
Should we be using a conditional or unconditional hierarchical prior (see \autoref{cond_vs_indep})? 
In order to investigate this question, we consider simulation by generating data from the regression model $\bm y = \bm X \bm \beta + \bm \epsilon$ with $\bm \epsilon \sim N_n(\bm 0,\sigma^2 \bm I_n)$. 
We let $n=100$ and $\sigma^2=3$.
We construct the true vector of slope parameters $\bm \beta =c  \tilde{\bm \beta}$ by assigning values $\{1.5, -1.5, 2, -2, 2.5, -2.5  \}$ to the first 6 elements of $ \tilde{\bm \beta}$ and setting others to zero. 
We choose a constant $c>0$ to achieve a desired level of signal-to-noise ratio
\footnote{
In a general linear regression $\bm y = \bm X \bm \beta+ \bm \epsilon$, the signal-to-noise ratio (SNR) is defined as 
$SNR=\frac{|| \bm \Sigma_X^{1/2} \bm \beta ||^2 }{ \sigma^2 }$ 
where $\sigma^2$ is the error variance and $\bm \Sigma_X$ is a $p \times p$ covariance matrix of $\bm X$.
$||\bm \Sigma_X^{1/2} \bm \beta||^2 = \bm \beta' \bm \Sigma_X\bm  \beta$ measures the overall signal strength.
A related quantity is $R^2_{pop}$, the population value of $R^2$, defined as $\frac{SNR}{1+SNR}$.
}.
The data matrix $\bm X$ is generated from the multivariate normal distribution with mean zero and covariance matrix being an identity matrix. 
The covariates are standardized for estimation. 
We examine different values of the number of covariates $p\in \{50, 100,300 \}$ and the signal-to-noise ratio $R^2_{pop}\in \{ 0.4,0.8 \}$.

We consider three shrinkage priors (1) student-t, (2) lasso, and (3) horseshoe and compare performance under the conditional  and independent priors.
The analysis was repeated 100 times with new covariates and responses generated each time. 
For each, the metrics recorded were: 
the estimated value of $\sigma^2$, 
the bias and MSE of the first 6 elements of the coefficients vectors, 
the number of false negatives (FN), 
the number of false positives (FP), and 
the number of true positives (TP).
Posterior means were used as point estimates of the slope coefficients and the error variance. 
We utilize the post-processing approach of \cite{LiPati2017} in order to categorize the covariates into signals and noises.

\autoref{conj_vs_ind_table} summarizes the results. Panel (a) shows results when the signal is relatively strong i.e.\ $R^2_{pop} =0.8$.
Both conjugate and independent priors do well in terms of TPs and FNs. 
However, there are some notable differences. 
First, the error variance tends to be underestimated when conjugate priors are used. This was in fact pointed out by \cite{Moranetal2019}.
Intuitively, conjugate priors implicitly add $p$ ``pseudo-observations'' to the posterior (compare \autoref{Gibbs_conj} with  \autoref{Gibbs_ind}) which can result in underestimations of $\sigma^2$ when $\bm \beta$ is sparse. 
Second, the independent priors tend to have larger bias and MSE of the signals under the high-dimensional case (i.e.\ $p=300$).
\cite{ParkCasella2008} point out that independent priors can induce bi-modality of the posterior on the slope coefficients. This can make the posterior distributions for $\bm \beta$ more spread than in the conjugate case.
We also see that independent priors have larger FPs when $p=300$, which could be a result of this. 
Panel (b) shows results under relatively weak signal i.e.\ $R^2_{pop} = 0.4$. 
We see that all methods face difficulty with distinguishing signals with noise (see FNs, FPs, and TPs) and have large bias and MSEs, compared to the case with   $R^2_{pop} = 0.8$. 
However, the general findings on the difference between conjugate and independent priors are the same: conjugate priors tend to underestimate $\sigma^2$ while independent priors tend to have higher bias and MSE of the signals when $p$ is large compared to $n$. 
We encourage researchers to be aware of these issues when choosing priors and to conduct sensitivity checks.

\begin{table}[h]
    \begin{subtable}[h]{\textwidth}
        \centering
        \scriptsize{
      \begin{tabular}{ c c  c c c c c} 
  &$\hat{\sigma}^2$ &Bias & MSE & FN & FP & TP \\
  \hline \hline 
 \multicolumn{7}{c}{$R^2_{pop}=0.8, p=50 $}\\         
  \hline  
  Student-t (conjugate) & 1.77 &0.19& 0.05&0.10&0&5.9 \\ 
  Bayesian Lasso (conjugate) &1.96 & 0.18&0.05&0.06&0&5.9\\ 
  Horseshoe (conjugate) & 2.35 & 0.19& 0.05&0.07&0&5.9 \\ 
        \hdashline 
  Student-t (independent) & 2.99 &0.19& 0.05&0.09&0&5.9 \\ 
  Bayesian Lasso (independent) &2.86 & 0.18& 0.05&0.05&0&6.0\\ 
  Horseshoe (independent) & 2.93 & 0.19& 0.05&0.08&0&5.9 \\ \hline 
 \multicolumn{7}{c}{$R^2_{pop}=0.8, p=100 $}\\         
  \hline 
  Student-t (conjugate) & 0.42 &0.34& 0.18&0.43&0.01&5.5 \\ 
  Bayesian Lasso (conjugate) &0.69 & 0.28&0.12&0.18&0&5.8\\ 
  Horseshoe (conjugate) & 1.05 & 0.26& 0.11&0.19&0&5.8 \\ 
        \hdashline 
  Student-t (independent) & 2.01 &0.30& 0.15&0.35&0.01&5.6 \\ 
  Bayesian Lasso (independent) &2.02 & 0.25&0.10&0.15&0&5.8\\ 
  Horseshoe (independent) & 2.28 & 0.26& 0.11&0.19&0&5.8 \\
   \hline 
 \multicolumn{7}{c}{$R^2_{pop}=0.8, p=300 $}\\         
    \hline  
    Student-t (conjugate) & 0.35 &0.35& 0.24&0.41&2.79&5.6\\ 
    Bayesian Lasso (conjugate) &0.66 & 0.78&0.71&0.74&2.86&5.2\\ 
    Horseshoe (conjugate) & 0.92 &0.66& 0.63&0.46&12.0&5.5 \\ 
        \hdashline 
    Student-t (independent) & 3.88 &1.52& 2.43&0.03&68.6&5.9 \\ 
    Bayesian Lasso (independent) &2.59 & 1.42&2.12&0.01&68.9&5.9\\ 
    Horseshoe (independent) & 3.27 & 1.43&2.17&0.01&61.1&5.9 \\
     \hline \hline
      \end{tabular}
      }
       \caption{ $R_{pop}^2=0.8.$}
    \end{subtable}
    \hfill  \\
    
    \begin{subtable}[h]{\textwidth}
        \centering
                \scriptsize{
     \begin{tabular}{ c c  c c c c c} 
 &$\hat{\sigma}^2$ &Bias & MSE & FN & FP & TP \\
    \hline    \hline 
  \multicolumn{7}{c}{$R^2_{pop}=0.4, p=50 $}\\         
  \hline  
  Student-t (conjugate) & 1.65 &0.19& 0.06&0.77&0.82&5.2 \\ 
  Bayesian Lasso (conjugate) & 1.78 &0.19& 0.06&0.72&0.54&5.3 \\  
  Horseshoe (conjugate) & 2.16 &0.19&  0.06&0.72&0.62&5.3 \\ 
        \hdashline 
  Student-t (independent) & 2.99 &0.19&  0.06&0.61&1.22&5.4 \\  
  Bayesian Lasso (independent) & 2.86 &0.19& 0.06&0.67&0.59&5.4 \\ 
  Horseshoe (independent) & 2.96 &0.20&  0.06&0.76&0.49&5.2 \\ 
   \hline 
  \multicolumn{7}{c}{$R^2_{pop}=0.4, p=100 $}\\         
  \hline  
  Student-t (conjugate) & 0.35 &0.35& 0.20&1.41&18.7&4.6 \\ 
  Bayesian Lasso (conjugate) & 0.52 &0.29& 0.14&1.11&15.5&4.7 \\  
  Horseshoe (conjugate) & 0.90 &0.27& 0.12&0.92&14.5&5.1 \\ 
        \hdashline 
  Student-t (independent) & 1.85 &0.30& 0.14&0.81&22.8&5.2 \\  
  Bayesian Lasso (independent) & 1.89 &0.26& 0.11&0.56&17.1&5.4 \\ 
  Horseshoe (independent) & 2.20 &0.27& 0.11&0.42&17.0&5.6 \\ 
   \hline 
  \multicolumn{7}{c}{$R^2_{pop}=0.4, p=300 $}\\         
    \hline  
  Student-t (conjugate) & 0.28 &0.46& 0.26&0.99&35.6&5.0 \\ 
  Bayesian Lasso (conjugate) & 0.47 &0.50& 0.29&0.53&49.6&5.4 \\  
  Horseshoe (conjugate) & 0.75 &0.55&0.34 &0.33&53.6&5.6 \\ 
        \hdashline 
  Student-t (independent) &3.50& 0.65&0.46&0.34&70.6&5.6 \\  
  Bayesian Lasso (independent) & 2.28 &0.65& 0.45&0.21&71.1&5.7 \\ 
  Horseshoe (independent) & 2.79 &0.65& 0.45&0.23&66.1&5.7 \\ 
     \hline \hline
      \end{tabular}
      }
        \caption{$R^2_{pop}=0.4.$}
     \end{subtable}
     \caption{Average metrics over 100 repetitions for each of the approaches. 
Estimated error variance, and bias and MSE of the first 6 elements of the slope vector, and the numbers of False Negatives (FN), False Positives (FP), and True Positives (TP).
The posterior means were used as point estimates. 
The post-processing method of \cite{LiPati2017} was used to distinguish signals from noises. $n=100$. }
\label{conj_vs_ind_table}
\end{table}

\FloatBarrier

\section[Beyond linear regression]{Bayesian shrinkage and variable selection beyond linear regression}

So far we explored variable selection in the high-dimensional linear regression, also known as ``large $p$, small $n$'' regression. This setting is already flexible enough, as there are various cases of generalized linear models (GLMs) that have conditionally linear forms. The purpose of this section is to demonstrate that there is an even larger list of models where hierarchical priors have immediate applicability. In particular, we explore key applications of hierarchical shrinkage and variable selection priors in vector autoregressions, factor models, time-varying parameter models, high-dimensional confounder selection in models for treatment effects, and Bayesian quantile regression. This list is far from exhaustive\footnote{For example, one application of Bayesian shrinkage and selection that we do not cover in this section, but is of importance in statistics and in finance, is high-dimensional covariance matrix estimation and selection, see \cite{WangPillai2013} as an indicative example. Another topic we won't cover here, but is becoming increasingly very important in statistics and econometrics, is Bayesian additive regression trees (BART). For an up-to-date review of the topic see \cite{Hilletal2020}.} and its only purpose is to illustrate how Bayesian computation simplifies high-dimensional inference in unconventional settings.

\subsection{Vector autoregressions}  \label{sec:VAR}
The most popular working model for economists using time series variables is the vector autoregression (VAR). VARs are used for the joint modeling of the dynamics of many macroeconomic and financial time series, $\bm Y$, allowing analysts and policy-makers to answer questions regarding dynamic responses of variable $\bm Y_{i}$ to a shock in some other variable $\bm Y_{j}$, $i \neq j$. This is a very important tool especially when variable $\bm Y_{j}$ is controlled by the policy-maker. For example, the central bank controls the short-term interest rate as well as other quantities related to monetary policy, while government controls taxes and fiscal policy in general. Due to the fact that availability of time series observations for macroeconomic and (low-frequency) financial data is limited\footnote{Especially in countries other than the US, where statistical agencies might have available only a handful of decades of data; e.g.\ euro area time series typically begin in 1995 or 1999.}, estimation of macroeconomic VARs by and large relies on Bayesian shrinkage priors. Additionally, parameters in VAR models proliferate at a polynomial rate as the dimensions of the model increases. In univariate linear regression settings, a model with twice as many exogenous predictors has twice as many parameters to estimate. There is not such an analogy in VARs where all variables are endogenous and each variable (and its lagged terms) affects all other variables in the system. 

Unlike our previous notation, consider time series observations $t=1,...,T$ and an $n$-dimensional vector of variables $\bm Y_{t}$, that is, $n$ denotes the number of variables of interest (and not the number of observations anymore) with $\bm Y = \left[ \bm Y_{1}^{\prime},...,\bm Y_{T}^{\prime} \right]^{\prime}$ is a $T \times n$ data matrix. The VAR model for $\bm Y_{t}$ with $p$ lags, also denoted as VAR($p$), is of the form
\begin{equation}
\bm Y_{t} = \bm c + \bm A_{1} \bm Y_{t-1}  + \bm A_{2} \bm Y_{t-2}  + ... + \bm A_{p} \bm Y_{t-p}  + \bm E_{t}, \label{VAR_basic}
\end{equation}
where $\bm c$ is an $n \times 1$ vector of intercepts, $\bm A_{i}$ are $n \times n$ matrices of lagged terms for each $i=1,...,p$, and $\bm E_{t} \sim N(\bm0,\bm \Sigma)$ with $\bm \Sigma$ an $n \times n$ symmetric, semi-positive definite covariance matrix. The VAR is a heavily parametrized model: it has $(1 + np)n$ coefficients $\bm B = \left[\bm c, \bm A_{1}, ..., \bm A_{p} \right]$, plus another $n(n+1)/2$ unique elements in the covariance matrix $\bm \Sigma$. For example, the largest VAR model specified in \cite{Koopetal2019} has $n=129$ and $p=13$ which implies that the total number of parameters is in excess of $200,000$.

Vector autoregressions are effectively linear regression models with parameter matrix $\bm B = \left[\bm c, \bm A_{1}, ..., \bm A_{p} \right]$  and data matrix $\bm X_{t} = \left[ \bm 1, \bm Y_{t-1},..., \bm Y_{t-p} \right]$, such that application of hierarchical priors for shrinkage and variable selection is fairly straightforward. The task of sampling from the conditional posterior of the regression coefficients $\bm B$ can be further simplified if the VAR is written in \emph{seemingly unrelated regressions (SUR)}\footnote{For a thorough and accessible introduction to Bayesian inference in VARs see \cite{KoopKorobilis2010}.} form
\begin{eqnarray}
vec \left( \bm Y \right)  & = & \left(\bm I \otimes \bm X\right) vec\left( \bm B\right) +  vec \left( \bm \varepsilon \right), \\
\bm y & = & \bm Z \bm b + \bm \varepsilon,   \label{SUR_form}
\end{eqnarray}
where $vec(\bullet)$ is the operator that stacks the columns of a matrix into a single column vector. That way, $ \bm y = vec \left( \bm Y \right)$ is a $Tn \times 1$ vector where the first $T$ elements are the observations of the first variable, the next $T$ rows correspond to observations of the second variable, and so on up to variable $n$. The measurement matrix $\bm Z = \left(\bm I \otimes \bm X\right)$ is a block-diagonal matrix with the $T \times (1 + np)$ matrix $\bm X$ repeating on its diagonal $n$ times. The formulation above is observationally identical to the one in \autoref{VAR_basic}, since there are no new parameters or data introduced, but it has the benefit that VAR parameters show up as the $(1 + np)n \times 1$ vector $\bm b = vec\left( \bm B\right)$. Therefore, the SUR form in \autoref{SUR_form} is identical to a univariate regression model, even though this model has both many covariates but also many observations ($\bm y$ and $\bm Z$ both have $Tn$ rows, instead of $T$ rows in a univariate regression). Therefore, it is straightforward to define any hierarchical prior we desire for the vector of VAR parameters $\bm b = vec\left( \bm B\right)$ and derive conditional posteriors, despite the fact that Gibbs sampling might become quite cumbersome as the dimension $n$ of the VAR increases.\footnote{See for example, \cite{Korobilis2013b,Korobilis2016} and \cite{KoopKorobilis2016}.}

In large $n$ cases and when shrinkage on the covariance matrix $\bm \Sigma$ is needed, \cite{Carrieroetal} and \cite{Koopetal2019} propose to estimate the VAR equation-by-equation. For example, in the formulation of \cite{Koopetal2019} one can write \autoref{VAR_basic} as
\begin{equation}
\bm Y_{t} = \bm B \bm X_{t}  + \bm P \bm V_{t},
\end{equation}
where $\bm P$ is a lower triangular matrix with ones on its main diagonal (unitriangular) that satisfies the LDL-decomposition $\bm \Sigma = \bm P \bm D \bm P^{\prime}$. In this case $V_{t} \sim N(\bm0,\bm D)$ where $\bm D$ is a diagonal matrix with variance elements $d_{ii}^{2}$ on its main diagonal, $i=1,...,n$. This formulation is equivalent to \autoref{VAR_basic} because $ \bm P \bm V_{t} \sim N(\bm 0, \bm P \bm D \bm P^{\prime}) = N(\bm 0,\bm \Sigma) \overset{d}{=}  \bm E_{t}$. Since by construction $\bm P$ is invertible, with $\bm P^{-1}$ also a unitriangular matrix, we can write
\begin{eqnarray}
\bm P^{-1} \bm Y_{t} & = & \bm P^{-1}\bm B \bm X_{t} +  \bm V_{t} \Rightarrow \\
(\bm I +  \widetilde{\bm P}^{-1}) \bm Y_{t} & = & \bm \Gamma \bm X_{t}  +  \bm V_{t} \Rightarrow \\
\bm Y_{t} & = & \bm \Gamma \bm X_{t} - \widetilde{\bm P}^{-1} \bm Y_{t} + \bm V_{t}, \label{triang_VAR}
\end{eqnarray}
where $\bm \Gamma = \bm P^{-1}\bm B$, and we have split $\bm P^{-1}$ into an identity matrix and a lower triangular matrix  $\widetilde{\bm P}^{-1}$ by means of the equation $ \bm P^{-1} = \bm I + \widetilde{\bm P}^{-1}$. In \autoref{triang_VAR} we have a VAR on $\bm Y_{t}$ where the covariance matrix elements in the original covariance matrix show up as contemporaneous elements of $\bm Y_{t}$ itself on the right-hand side in  the term $ - \widetilde{\bm P}^{-1} \bm Y_{t}$. Notice that in matrix form this is a nonlinear system as $\bm Y_{t}$ shows up both on the left hand side and the right hand side of \autoref{triang_VAR}. However, exactly because $\widetilde{\bm P}^{-1}$ is lower triangular and $\bm V_{t}$ has a diagonal covariance matrix $\bm D$, equation-by-equation estimation is feasible. In particular, each VAR equation $i$ depends on lags of all other equations and contemporaneous terms in the previous $i-1$ equations. This means that estimation of the VAR collapses to estimation of $n$ independent univariate models, such that specification of hierarchical priors and MCMC estimation are also straightforward. The additional benefit from this approach is that hierarchical priors can be specified to the elements of $\widetilde{\bm P}^{-1}$, thus leading to shrinkage or sparse estimation of the original VAR covariance matrix $\bm \Sigma$, similar to the methodology of \cite{SmithKohn2002}. More details of this approach can be found in \cite{Koopetal2019}, while variants of this approach have been proposed in \cite{Baumeisteretal2020}, \cite{Carrieroetal}, and \cite{KorobilisPettenuzzo2019}.

\subsection{Factor model shrinkage and selection} \label{sec:factor}
Factor models have a long history in econometrics, and an even longer history in statistics and psychology/psychometrics. For that reason, while there are some popular formulations across different literatures, there are also different variations based on the data and applications. We first establish some key results for the specific case of the so-called static factor model, and subsequently we review some of the most popular factor models used in economics and finance. We end our discussion with strategies for Bayesian shrinkage and variable selection in this class of models.

Consider an $n \times 1$ vector of economic variables $\bm X_{t}$ observed over $t=1,...,T$ (without loss of generality $t$ can measure time series, but it can also be observations on individuals or other cross-sectional units). The dimension $n$ can be inconveniently high\footnote{This description includes both the ultra high-dimensional case where $n$ can be in the order of thousands, or larger, but also the case where $n$ is small but much larger than the number of available observations $T$.}, such that unrestricted estimation of models (e.g.\ linear regressions) using the data $\bm X_{t}$ is infeasible. Our target is to estimate a lower-dimensional $k \times 1$ vector ($k<<n$) of latent variables (factors), that summarizes as much as possible the information contained in $\bm X_{t}$. For that reason we define the following multivariate model
\begin{equation}
\bm X_{t}  = \bm \Lambda \bm F_{t}  + \bm \varepsilon_{t}, \label{SF}
\end{equation}
where $\bm \Lambda$ is an $n \times k$ matrix of parameters, $\bm F_{t}$ are the latent variables and $\bm \varepsilon$ is a disturbance term. This is not a regression model, as both $\bm \Lambda$ and $\bm F_{t}$ are latent. For simplicity, we follow \cite{LopesWest2004} and make the assumption that $\bm F_{t} \sim N_{k}(\bm 0, \bm I)$, although we can allow the factors to have a more general covariance matrix such that they are correlated with each other. For the disturbance term we assume $\bm \varepsilon_{t} \sim N_{n}(\bm 0, \bm \Sigma)$. 

While the model in \autoref{SF} looks like a linear regression -- in which case application of hierarchical priors would be straightforward --  this is not the case due to the fact that all terms on the right-hand side of the equation are latent. In many instances, researchers in economics, finance and other disciplines replace $\bm F$ with the first $k$ principal components (PCs). PCs are nonparametric estimates of the factors, that is, they are only approximate estimators of the true factors implied by the likelihood of the model in \autoref{SF}. Plugging in the place of the latent factors (parameters) $\bm F$ the PC estimates turns the factor model into a regression model and inference is simplified. Conditional on the principal component estimates, $\bm \Lambda$ and $\bm \Sigma$ can be estimated simply via least squares, but also standard Bayesian methods for multivariate regression can be used. The benefit of this two-step approach is that it is simple, both conceptually and computationally, and that principal component estimates always provide a sensible fit. However, in many more complex settings (e.g.\ macroeconomic dynamic factor models or financial factor models with stochastic volatility), the PC only provide a rough approximation, and it might be preferable to use the likelihood function to estimate $\bm F$. In such cases, it is imperative to make sure the factor estimates are unique. Therefore, we discuss first how to uniquely identify the factors, loadings and other parameters, before discussing Bayesian inference using hierarchical priors.

\subsubsection*{Identification of the factor model}
The factor model implies that the conditional covariance matrix of $\bm X$ can be decomposed as
\begin{equation}
cov\left( \bm X \vert \bm \Lambda, \bm F, \bm \Sigma \right) \equiv \bm \Omega = \bm \Lambda \bm \Lambda^{\prime}  + \bm \Sigma. \label{fdecomp}
\end{equation}
This decomposition illustrates the fact that likelihood-based (maximum likelihood or Bayesian) estimation of the factor model suffers from lack of identification of a unique set of parameter estimates. For example, consider the case where $\bm \Sigma$ is a full matrix, then there are infinite ways to construct the decomposition in \autoref{fdecomp}. In order to deal with this issue, it is common in factor analysis to set $\bm \Sigma$ to be a diagonal matrix.\footnote{This is known as the \emph{exact factor model} assumption, as opposed to the class of \emph{approximate factor models} that allow for ``some'' weak correlation among the variables in $\bm X$ and a $\bm \Sigma$ covariance that has some non-zero off-diagonal elements. Approximate factor models are typically not estimated with likelihood-based methods, so we don't consider this class of models here.} A consequence of this assumption is that the disturbances $\bm \varepsilon_{j}$ become idiosyncratic to each variable $\bm X_{j}$, $j=1,...,n$, that is, they capture measurement errors and other idiosyncrasies of each variable that are not attributed to its covariation with the remaining $n-1$ variables. Instead, any comovements/commonalities in the $n$ variables $\bm X$ are solely captured by the \emph{common component} $\bm \Lambda \bm F$.

Having $\bm \Sigma$ diagonal is a big step towards identification in the factor model. As \cite{LopesWest2004} mention, $\bm \Omega$ has $n(n+1)/2$ unique elements, therefore, the number of elements in the decomposition of \autoref{fdecomp} should not exceed that threshold. The matrix $\bm \Lambda$ has  $nk$ elements, and $\bm \Sigma$ being diagonal has $n$ elements, therefore we obtain the inequality $n(n+1)/2 \geq nk + n$, which provides an upper bound on the number of factors one can extract: with $n=5$ variables we can extract $k=2$ factors, and when $n=20$ the maximum number of factors that can be extracted is $k=9$. However, there is a further problem impairing identification of the factor model, and this pertains to separating $\bm \Lambda$ from $\bm F$. Without further restrictions, there are infinite ways of finding such matrices that provide exactly the same values for the common component. Put more formally, if $\bm P$ is an $k \times k$ orthogonal matrix such that $\bm P \bm P^{\prime} = \bm I_{k}$, then the factor model can be rewritten as
\begin{eqnarray}
\bm X_{t} &  = & \bm \Lambda \bm F_{t}  + \bm \varepsilon_{t} \Rightarrow \\
\bm X_{t} &  = & \bm \Lambda \bm P \bm P^{\prime} \bm F_{t}  + \bm \varepsilon_{t} \Rightarrow \\
\bm X_{t} &  = & \bm \Lambda^{\star} \bm F^{\star}_{t}  + \bm \varepsilon_{t},
\end{eqnarray}
where $\bm \Lambda^{\star}$ and $\bm F^{\star}_{t}$ are alternative estimates to $\bm \Lambda$ and $\bm F_{t}$ that provide exactly the same likelihood value (they are observationally equivalent). Given that the variance of the factors is normalized to be one, unique identification of the loadings and the factors requires an additional $k(k-1)/2$ restrictions on the loadings matrix $\bm \Lambda$. A standard restriction that is imposed in this case \citep{LopesWest2004} is to restrict $\bm \Lambda$ to be lower triangular, that is, the top $k \times k$ block of this matrix has its $k(k-1)/2$ upper triangular elements equal to zero. This restriction provides local identification up to a rotation of the sign, meaning that we could multiply any column of $\bm \Lambda$ with $-1$ and do the same to the respective column of the factors $\bm F$, and arrive to an observationally equivalent solution. For that reason, \cite{GewekeZhou1996} suggest to further assume the $k$ diagonal elements of $\bm \Lambda$ to be restricted to be positive.

When modeling comovements between financial time series, the assumption $\bm F_{t} \sim N_{k}(\bm 0, \bm I)$ is often not empirically relevant, and instead it is assumed that $\bm F_{t} \sim N_{k}(\bm 0, \bm \Sigma^{F})$ with $\bm \Sigma^{F}$ a diagonal matrix, with possibly heteroskedastic elements that capture changing (over time) volatility of financial variables \citep[see for example][]{Chibetal2006}. In this case further restrictions on $\bm \Lambda$ are needed, and \cite{Chibetal2006} choose to fix the diagonal elements of the loadings matrix to be one. \cite{BBE2005} extract factors from a large macroeconomic dataset and in their methodology it is imperative for $\bm \Sigma^{F}$ to be a full covariance matrix (it is the covariance matrix of a VAR from which they want to identify shocks and estimate impulse response functions). Therefore, with $\bm \Sigma^{F}$ a full matrix, these authors further restrict the upper $k \times k$ block of the loadings matrix to be the identity matrix. 

All these identification restrictions in various applications of the factor model do not come at no cost. Imposing zeros in the loadings matrix means that certain variables are excluded from determining the factors. In the case of \cite{BBE2005}, in particular, the identity restriction means that the first variable exclusively loads on the first factor, the second variable exclusively on the second factor, and so on. Therefore, the ordering of the variables in $\bm X$ ends up affecting the estimates of $\bm F$, and in their case this restriction turns out to be empirically detrimental.\footnote{While their factors are statistically identified, they do not carry any economic content (i.e.\ they are not ``structurally identified''). \cite{BBE2005} is one of the few papers that estimates a factor model both with principal components and least squares (the ``plug-in'' approach explained previously) and with Bayesian inference. Comparing the impulse response functions for some key variables using the two estimation methods, there are marked differences. Whenever PCA has been used to estimate the unknown factors, impulse response functions have the signs and shapes expected by economic theory. When likelihood-based factors have been used, the impulse responses of variables such as inflation degenerate to zero for all horizons; \cite[compare][Figures II and IV]{BBE2005}.}

\subsubsection*{Bayesian shrinkage and variable selection in the factor model}
Despite the fact that statistical identification of the factor model using zero and sign restrictions on the loadings might contradict evidence in the data, once the factor model is fully identified Bayesian inference becomes straightforward. To see this, we re-write for convenience the factor model including now all relevant identification restrictions that are imposed prior to estimation
\begin{eqnarray}
\bm X_{t} & = & \bm \Lambda \bm F_{t}  + \bm \varepsilon_{t}, \text{ \ \ } t=1,...,T, \\
\bm F_{t} & \sim  & N_{k}(\bm 0, \bm I), \\
\bm \varepsilon_{t} & \sim & N_{n} (\bm 0 ,\bm \Sigma), \\
\Sigma_{ij} & = & 0,  \text{ \ \ } i\neq j, \text{ \ \ } i,j=1,...,n \\
\Lambda_{ij} & = & 0, \text{ \ \ } i<j, \text{ \ \ } i=1,...,n, \text{ \ \ } j=1,...,k , \\
\Lambda_{ij} & > & 0, \text{ \ \ } i=j, \text{ \ \ } i=1,...,n, \text{ \ \ } j=1,...,k.
\end{eqnarray}
\cite{LopesWest2004} show that this model is conveniently estimated sequentially via the Gibbs sampler, by sampling from conditional posteriors. The priors are of the form
\begin{eqnarray}
\Lambda_{ij} & \sim & \left \lbrace 
\begin{array}{c}
N(0, \tau^{2}), \text{ \ \ }, i>j \\
N(0, \tau^{2})I(\Lambda_{ij}>0), \text{ \ \ }, i=j \\
\delta_0(\Lambda_{ij}), \text{ \ \ }, i<j
\end{array}
\right. \label{lambda_prior} \\
\Sigma_{ii} & \sim & Inv-Gamma(a_0,b_0),
\end{eqnarray}
where $\delta_0(\Lambda_{ij})$ is the Dirac delta function, that is, a point mass function for $\Lambda_{ij}$ that is concentrated at zero.
The conditional posteriors are
\begin{eqnarray}
\bm F_{i} \vert \bullet & \sim & N\left( \bm V_{F} \left(\bm \Lambda^{\prime}\bm \Sigma^{-1} \bm X \right) ,\bm V_{F}  \right), \\
\bm \Lambda_{i} \vert \bullet & \sim & N\left( \bm V_{L,i}\left(\frac{1}{\Sigma_{ii}}\bm F \bm X_{i}\right), \bm V_{L,i}\right), \text{ \ \ } i=1,...,n, \\
\Sigma_{ii} \vert \bullet & \sim & Inv-Gamma\left( a_0 + n/2, b_0 + SSE_{i} \right), \text{ \ \ } i=1,...,n,
\end{eqnarray}
where $\bm V_{F} = (\bm I + \bm \Lambda^{\prime}\bm \Sigma^{-1} \bm \Lambda)^{-1}$, $\bm V_{L,i} = (\bm D^{-1} + \frac{1}{\Sigma_{ii}}\bm F^{\prime}\bm F)^{-1}$ and $SSE_{i} = (\bm X_{i} -  \bm F\bm \Lambda_{i})^{\prime}(\bm X_{i} -  \bm F\bm \Lambda_{i})$. The $\vert \bullet$ notation above is used to denote conditioning on other parameters and data. 

By updating $\bm \Lambda$ conditional on $\bm F$ and vice-versa, the Gibbs sampler works around the issue that the product of these two parameters shows up in the likelihood function. Of course, this sequential updating of the common component by updating each of $\bm \Lambda$ and $\bm F$ conditional on the other, will inevitably generate unwanted correlation in the Gibbs chain. In order to deal with the sampling inefficiency associated with correlated MCMC draws, \cite{Chibetal2006} proposed an alternative Metropolis-Hasting step for updating $\bm \Lambda$ conditional on the factors, while \cite{GhoshDunson2009} proposed a parameter-expanded Gibbs sampler \citep[see also][for a similar parameter expanded factor model estimated with an EM algorithm]{RockovaGeorge2016}. Notice that the fact that $\bm \Lambda$ has the required zero and sign restrictions imposed prior to estimation, means that every time we sample $\bm F$ conditional on $\bm \Lambda$ the factors will be sampled from a unique, identified posterior distribution. Had we not imposed these restrictions, the Gibbs sampler would still work numerically, but lack of identification means that each sample could correspond to different pairs of solutions for $\bm \Lambda$ and $\bm F$. That is, in the case where $\bm \Lambda$ and $\bm F$ are not separately identified, their product (the common component $\bm \Lambda \bm F$) is always identified. There are certain inference exercises, such as prediction, where it might be the case that identification and interpretation of the factors is not required; see for example the arguments in favor of this approach in \cite{BhattacharyaDunson2011} and in \cite{Korobilis2020}.

An early attempt to full-Bayes inference in factor models is \cite{West2003} who developed a variable selection prior in the loadings of the static factor model. Under the \cite{LopesWest2004} identification scheme, the extension proposed by  \cite{West2003} simply involves replacing the prior in \autoref{lambda_prior} with a spike and slab prior. As long as the identification restrictions are maintained, the presence of the variable selection prior can be used to find further data-based restrictions in the loadings matrix. \cite{Carvalhoetal2008} extend the variable ideas in \cite{West2003} to create a very sparse static factor model for genome data; see our discussion of their prior in \autoref{sec:sns}. \cite{KnowlesGhahramani2011} further extend these ideas to a spike and slab prior that is semiparametric, utilizing the ability of an Indian Buffet Process prior to allow for infinitely many factors. These authors use a Metropolis-within-Gibbs algorithm for inference. \cite{RockovaGeorge2016} propose a similar spike and slab formulation based on the Indian Buffet Process, but unlike \cite{KnowlesGhahramani2011} they propose maximum a posteriori (MAP) inference by means of approximating the posterior mode using an EM algorithm. The Bayesian asymptotic theory and posterior contraction rates for the sparse static factor model with continuous spike and slab priors is explored in detail in \cite{Patietal2014}.

\cite{GhoshDunson2009} proposed a heavy-tailed prior on $\bm \Lambda$ (using a normal/inverse gamma mixture prior) which they argue performs better than the default normal/truncated-normal prior in \autoref{lambda_prior}.  \cite{BhattacharyaDunson2011} proposed a novel multiplicative gamma process prior on the factor loadings that shrinks more aggressively columns of $\bm \Lambda$ that correspond to a higher number of factors. They call their approach a sparse infinite factor model, as it allows to specify a maximum number of factors and the prior is able to determine zero and non-zero loadings, as well as the number of factors. The gamma process prior for the loadings matrix is of the following ``global-local shrinkage'' form
\begin{eqnarray}
\Lambda_{ij} \vert \phi_{ij}, \tau_{j} & \sim & N(0,\phi_{ij}^{-1} \tau_{j}^{-1}), \\
\phi_{ij} & \sim & Gamma(v/2,v/2), \\
\tau_{j} & = & \prod_{l=1}^{j} \delta_{l}, \text{ \ \ } j=1,...,k, \\
\delta_{1} & = & Gamma(a_{1},1), \\
 \delta_{l} & = & Gamma(a_{2},1),  \text{ \ \ } l \geq 2.
\end{eqnarray}
While the local shrinkage parameter is the same for each element of $\bm \Lambda$, the global shrinkage parameter $\tau_{j} $ is shrinking more aggressively as the index $j$ increases, where $j=1,...,k$ indexes the number of factors. This is because $\tau_{j} $ is a $j$-dimensional product of gamma-distributed random variables. \cite{Legramantietal2020} propose a cumulative shrinkage process prior and \cite{Srivastavaetal2017} propose a multi-scale generalized double Pareto prior; both these priors are similar in spirit to the \cite{BhattacharyaDunson2011} prior in terms of shrinking the loadings towards zero and selecting the appropriate number of factors at the same time.

We close this section by mentioning ongoing research on alternative solutions to the identification problem (rotational interdeterminacy) in factor models, that do not rely on preimposing zero restrictions on the loadings matrix. The expanded parametric forms proposed in papers such as  \cite{BhattacharyaDunson2011} and \cite{Legramantietal2020} discussed above, deal with this issue efficiently. Other approaches include the ex-post processing approaches of \cite{Assmann2016} and \cite{KaufmannSchumacher2019}. \cite{FruewirthLopes2018} introduce a generalized lower triangular representation of the factor model and propose a sparsity-inducing prior that overshrinks. While papers like \cite{West2003} apply sparsity after imposing zero identifying restrictions, the parameterization of \cite{FruewirthLopes2018} allows the prior to impose zeros that are sufficient for identification and inference, thus, not suffering from the rotation problem. Finally, \cite{Chanetal2018} propose an invariant parameterization of the static factor model that is based on the singular value decomposition.

\subsection{Dynamic sparsity and shrinkage}
When working in a time series setting the concepts of shrinkage, variable selection, and model averaging need not be static. This is true for economics where there has always been evidence that predictors can be unstable. There is significant theoretical and empirical evidence that when forecasting oil prices, stock prices, consumer prices, exchange rates, and numerous other economic/financial variables there is hardly a single exogenous predictor that can be claimed to be important over a substantial time sample. In practice, we observe ``pockets of predictability'', that is, short periods where a specific variable might have predictive information for another variable of interest. This concept of unstable predictors has been popularized since the global financial crisis of 2007-2009, a period when it was obvious that all constant parameter relationships between economic variables completely broke down. Combined with the availability of new Bayesian tools for high-dimensional inference, a large literature has emerged since then that uses terms such as ``time-varying sparsity'' or ``dynamic model averaging'' or ``time-varying dimension models''. \cite{KoopKorobilis2018} provide a detailed discussion of this literature.\footnote{At the same time, in the field of signal processing there is a related literature on ``dynamic compressive sensing'' for streaming signals (e.g.\ video); see \cite{ZinielSchniter13}. } 

The starting point for imposing dynamic sparsity and dynamic shrinkage is a regression with time-varying parameters (TVPs) and stochastic volatility (henceforth, abbreviated as \emph{TVP regression}) of the form
\begin{eqnarray}
y_{t} & = & \bm X_{t} \bm \beta_{t}  + h_{t} z_{t},  \label{TVP1}\\
\bm \beta_{t} & = & \bm \beta_{t-1} + \bm u_{t}, \text{ \ \ } \bm \beta_{0} \sim N_p(\bm 0,\bm P), \label{TVP2} \\
\log h_{t} &  = & \log h_{t-1} + v_{t}, \text{ \ \ } \log h_{0} \sim N(0,\delta), \label{TVP3}
\end{eqnarray}
where $y_{t}$ is a scalar time series observation for $t=1,...,T$, $\bm X_{t}$ is a $p$-dimensional vector of covariates (that can include an intercept, own lags of $y$ and exogenous predictors), $\bm \beta_{t}$ is a vector of time-varying (or drifting) regression coefficients, and $h_{t}$ is the scalar time-varying (or stochastic) variance/volatility parameter. Additionally, we assume $z_{t} \sim N(0,1)$, $\bm u_{t} \sim N_p( \bm 0, \bm Q)$ with $\bm Q$ a $p \times p$ covariance matrix, and $v_{t} \sim N(0,\delta^{2})$ with $\delta^{2}$ a scalar variance parameter.

As is the case with the constant parameter regression, shrinkage is mainly desirable in the TVPs $\bm \beta_{t}$, but this can take now a dual form: shrinkage towards time-invariance ($\bm \beta_{t}$ becomes a constant parameter) and ``traditional'' shrinkage towards zero.\footnote{Shrinkage of $h_{t}$ towards a constant variance specification is feasible, but it is not desirable for economic and financial time series data, since we know that economic shocks are very volatile and constant parameter specifications are always inferior (both using in-sample and out-of-sample measures of fit).} Notice that the TVP regression as is specified in equations \eqref{TVP1} - \eqref{TVP3} is a hiearchical model, and \autoref{TVP2} in particular can be though of as a hiearchical prior for $\bm \beta_{t}$ of the form $\bm \beta_{t}\vert \bm \beta_{t-1}, \bm Q \sim N(\bm \beta_{t-1}, \bm Q)$. Seen like this, it is straightforward to assume that $\bm Q$ is diagonal and allow its elements to follow of the hyperpriors we examined previously (e.g.\ Student-t, Laplace etc). However, doing so would only regularize the evolution of $\bm \beta_{t}$ around $\bm \beta_{t-1}$, where in the limit of $\bm Q = \bm 0$ then $\bm \beta_{t}$ becomes a constant parameter. Shrinking $\bm \beta_{t}$ towards zero requires different treatment, and there are numerous ways one can deal with this problem.

Dynamic variable selection or dynamic model averaging can be implemented in this setting by simply placing appropriate hierarchical priors that will allow to test the hypothesis $H_{0}:\beta_{jt}= 0$ vs $H_{1}: \beta_{jt} \neq 0$ for all $j=1,...,p$ and for all $t=1,...,T$. Recall that in ``static'' Bayesian model averaging the challenge is to average over $2^{p}$ regressions. Therefore, the dynamic version of model averaging implies that one has to average over $2^{p}$ regression models for each $t=1,...,T$. It is not surprising then that many proposed approaches in the literature for dealing with this problem are not based on computationally intensive MCMC algorithms. For example, \cite{KoopKorobilis2012} and \cite{DanglHalling2012} use variance discounting methods \citep[see for example][]{West1997} in order to provide plug-in estimators of $h_{t}$ and $Q$ and estimate a single time-varying parameter regression very quickly. Subsequently, dynamic variable selection and model averaging can be implemented by enumerating and estimating all $2^{p}$ possible models -- as long as $p$ is fairly small (e.g.\ 20 predictors).

In terms of directly introducing shrinkage and sparsity via hierarchical priors, there are numerous ways of doing so in a TVP regression model. \cite{Belmonteetal2014} and \cite{BittoFruewirthSchnatter2019} place hierarchical priors in an equivalent ``non-centered'' parameterization of the TVP regression that takes the form
\begin{eqnarray}
y_{t} & = & \bm X_{t} \bm \theta + \bm X_{t} \bm W \bm \theta_{t}  + h_{t} z_{t}, \\
\bm \theta_{t} & = & \bm \theta_{t-1} + \bm d_{t},\text{ \ \ } \bm \theta_{0} = \bm 0, \\
\log h_{t} &  = & \log h_{t-1} + v_{t},\text{ \ \ } \log h_{0} \sim N(0,\delta)
\end{eqnarray}
where $\bm d_{t} \sim N_p(\bm 0,\bm I_p)$ and $\bm W$ is a diagonal matrix with elements $w_{j}$, $j=1,...,p$. This formulation is observationally equivalent to the TVP regression of equations \eqref{TVP1} - \eqref{TVP3}. As long as the initial condition is $\bm \theta_{0}$ it holds that $\bm \theta + \bm \theta_{t} = \bm \beta_{t}$. This allows to split the time-varying parameter into a constant parameter level $\bm \theta$ (determined by data $\bm X_{t}$) and the additive time-variation around the constant level. Additionally, notice that the state equation is now standardized ($\bm d_{t}$ has unit variance) which can be done by setting $\bm W^{\prime -1}\bm W^{-1} = \bm Q$. \cite{Belmonteetal2014} place a Bayesian lasso (Laplace) prior on $\bm \theta$ and on the diagonal elements of $\bm W$. By doing so, they can shrink the total coefficient $\beta_{j,t}$ into a constant parameter $\theta_{j}$ (if $w_{j}\rightarrow 0$), or shrink it to zero (when both $\beta_{j,t}$ and $w_{j}$ are shrunk towards zero). Alternatively, the model can become an unrestricted TVP regression when both $\beta_{j,t}$ and $w_{j}$ are not shrunk towards zero by the Laplace prior.

\cite{NakajimaWest2013} convert the TVP regression into a latent threshold dynamic regression of the form
\begin{eqnarray}
y_{t} & = & \bm X_{t} \bm b_{t}  + h_{t} z_{t},\\
\bm b_{t} & = & \bm \beta_{t} \bm S_{t}, \\
\bm \beta_{t} & = & \bm \beta_{t-1} + \bm u_{t}, \text{ \ \ } \bm \beta_{0} \sim N(\bm 0,\bm P), \\
\log h_{t} &  = & \log h_{t-1} + v_{t}, \text{ \ \ } \log h_{0} \sim N(0,\delta),
\end{eqnarray}
where $\bm S_{t}$ is a $p \times p$ diagonal matrix with element $s_{j,t} = I(\beta_{j,t}\geq d_{j})$. That is, the $s_{j,t}$ are 0/1 indicators that can shrink $b_{j,t}$ either towards zero or towards the unrestricted TVP $\beta_{j,t}$. The threshold value $d_{j}$ can be estimated endogenously such that the data decide which coefficients are zero (or not) at each point in time. Of course, similar to interpretation of spike and slab priors, the condition $I(\beta_{j,t}\geq d_{j})$ is a soft, rather than a hard, thresholding rule, due to the fact that $s_{j,t}$ (in a Bayesian setting) is a random variable. This means that once considering the full uncertainty in the posterior the approach of \cite{NakajimaWest2013} provides a class of smooth thresholding models; see also \cite{NakajimaWest2013JFE,NakajimaWest2015DSP,NakajimaWest2017BJPS}.

\cite{RockovaMcAlinn2017} specify a dynamic spike and slab prior of the form
\begin{eqnarray}
\bm \beta_{t} \vert \bm \mu_{t}, \bm \gamma, \lambda_0, \lambda_{1}  & \sim & (\bm I - \bm \Gamma) N(\bm 0,\lambda_0 \bm I_{p}) + \bm \Gamma N(\bm \mu_{t}, \lambda_{1} \bm I_{p}), \\
\bm \mu_{t} & \sim & N(\bm \mu_{t-1}, \bm Q),
\end{eqnarray}  
where $\bm \Gamma = diag(\bm \gamma)=diag(\gamma_{1},...,\gamma_{p})$ and $\lambda_0$ and $\lambda_1$ can also have further exponentional prior distributions, converting this prior into a dynamic version of the spike and slab lasso of \cite{RockovaGeorge2018}. This prior is a spike and slab prior for $\bm \beta_{t}$ but it is only the slab component that incorporates the random walk evolution via the prior mean for $\bm \mu_{t}$. In contrast, \cite{KoopKorobilis2018} propose a similar but non-hierarchical prior of the form
\begin{eqnarray}
\bm \beta_{t}\vert \bm \beta_{t-1}, \bm Q & \sim & N_{p}(\bm \beta_{t-1}, \bm Q), \\
\bm \beta_{t}\vert \bm \gamma, \bm \tau^{2} & \sim&  (\bm I - \bm \Gamma)N_{p}(\bm 0, c \bm D_{\tau}) + \bm \Gamma N_{p}(\bm 0, \bm D_{\tau}),
\end{eqnarray}
where $\bm D_{\tau} = diag(\bm \tau^{2}) = diag(\tau^{2}_1,...,\tau^{2}_p)$ and $c$ is a small constant (set to $c=0.0001$ in \citealp{KoopKorobilis2018}). Again it is trivial to allow $\bm \tau^{2}$ to have its own hyperprior, such that we can combine shrinkage with sparsity in one setting. Finally, \cite{KalliGriffin2014} modify \autoref{TVP2} and introduce a normal-gamma mixture evolution for $\bm \beta_{t}$, which can be written in the following hyperprior form
\begin{eqnarray}
\beta_{j,t} \vert \beta_{j,t-1}, \psi_{j,t}, \psi_{j,t-1} & \sim & N\left( \frac{\psi_{j,t}}{\psi_{j,t-1}} \phi_{j} \beta_{j,t-1} , \left( 1 - \phi_j^{2}\right) \psi_{j,t} \right), \\
\psi_{j,t} \vert \kappa_{j,t-1} & \sim & Gamma\left( \lambda_{j} + \kappa_{j,t-1},  \frac{\lambda_{j}}{\mu_{j}(1-\rho_{j})} \right), \\
\kappa_{j,t}  & \sim & Poisson\left( \frac{\rho_j\lambda_{j}\psi_{j,t-1} }{\mu_{j}(1-\rho_{j})} \right),
\end{eqnarray}
which makes this a normal-gamma-Poisson mixture distribution. While this mixture having a very flexible distributional form, implying interesting shapes for $\bm \beta_{t}$, there might be sensitivity to the choice of the key hyperparameters $(\lambda_{j},\rho_{j}, \mu_{j})$.

All the examples above use the state-space form of the TVP regression and rely on recursive estimation methods, either in the form of the simple Kalman filter or (within the context of simulation methods) forward filtering backward sampling (FFBS) algorithms. However, as noted by \cite{Korobilis2021} one can simply discard the prior $\bm \beta_{t}\vert \bm \beta_{t-1}, \bm Q \sim N(\bm \beta_{t-1}, \bm Q)$ and treat the TVP regression as a constant parameter regression. This can be seen if we stack all observations in \autoref{TVP1} and write it as
\begin{eqnarray}
\left[ \begin{array}{c}
y_{1} \\
y_{2} \\
... \\
y_{T-1} \\
y_{T}
\end{array} \right]
 & =  & \left[ 
\begin{array}{cccc}
\bm X_{1} &  \bm 0    & ...         & \bm 0 \\
\bm 0     & \bm X_{2} & \ddots      & \bm 0 \\
\vdots    & \ddots    &             & \vdots \\
\bm 0     &   ...     & \bm X_{T-1} & \bm 0 \\
\bm 0     &   ...     & \bm0        & \bm X_{T} \\
\end{array}
 \right]
 \left[
\begin{array}{c}
\bm \beta_{1} \\
\bm \beta_{2} \\
... \\
\bm \beta_{T-1} \\
\bm \beta_{T} \\
\end{array} 
  \right]
  + 
 \left[
\begin{array}{c}
\varepsilon_{1} \\
\varepsilon_{2} \\
... \\
\varepsilon_{T-1} \\
\varepsilon_{T} \\
\end{array} 
  \right], \label{TVP_general} \\ 
  \bm y \text{ \ \ \ \ }  &  & \text{ \ \ \ \ \ \ \ \ \ \ \ \ \ \ \ \ }  \mathcal{X}  \text{ \ \ \ \ \ \ \ \ \ \ \ \ \ \ \ \ \ \ \ \ \ \ }\bm B \text{ \ \ \ \ \ \ \ \ \ \ \ \ \ \ } \bm \varepsilon, \label{TVP_general_mat}
\end{eqnarray}
where $\varepsilon_{t} \sim N(0,h_{t}^2)$. In this form, the TVP regression is a model with $T$ observations and $Tp$ covariates and it can be estimated as a high-dimensional ``static'' regression with data matrices $\bm y$ and $\mathcal{X}$ as defined in \autoref{TVP_general_mat}. \cite{Korobilis2021} shows that a large class of hierarchical shrinkage priors can be placed on the $Tp \times 1$ parameter vector $\bm B$, and inference can proceed using the regression form in \autoref{TVP_general_mat} without the need to rely on state-space methods. Since $\bm B$ has $T$ time copies of parameters on the $p$ predictors in $\bm X_{t}$, more structured shrinkage can be placed by using a group lasso or other similar grouping prior.

Applying a shrinkage or variable selection prior directly to the vector of parameters $\bm \beta_{t}$ means that a certain $\beta_{j,t}$ might be unrestricted in period $s$, then restricted to zero in period $s+1$, then switch back to being unrestricted in $s+2$, and so on, for $s \in \{1,...,T\}$. This is a noisy approach to dynamic shrinkage/sparsity, and more persistent estimates over time might be desirable such that we prevent an important coefficient from becoming sparse just for a period or two, and vice-versa for a sparse coefficient. In many economic data, there is evidence of prolonged regimes where coefficients are either important or not important (e.g.\ macroeconomic recessions vs expansions, or bull vs bear stock markets). In this case, it might be desirable to incorporate the information in $\bm \beta_{t}\vert \bm \beta_{t-1}, \bm Q \sim N(\bm \beta_{t-1}, \bm Q)$ alongside a hierarchical shrinkage prior. A simple way to do this is to write the TVP regression as a static regression for the parameters $\Delta \bm \beta_{t} = \bm \beta_{t} - \bm \beta_{t-1}$. This takes the form{\small
\begin{eqnarray}
\left[ \begin{array}{c}
y_{1} \\
y_{2} \\
... \\
y_{T-1} \\
y_{T}
\end{array} \right]
 & =  & \left[ 
\begin{array}{cccc}
\bm X_{1} &  \bm 0    & ...         & \bm 0 \\
\bm X_{2} & \bm X_{2} &  \ddots     & \bm 0 \\
\vdots    & \ddots    &    \ddots          & \vdots \\
\bm X_{T-1}  &   ...     & \bm X_{T-1} & \bm 0 \\
\bm X_{T} &   ...     & \bm X_{T}        & \bm X_{T} \\
\end{array}
 \right]
 \left[
\begin{array}{c}
\bm \beta_{1} \\
\Delta \bm \beta_{2} \\
... \\
\Delta \bm \beta_{T-1} \\
\Delta \bm \beta_{T} \\
\end{array} 
  \right]
  + 
 \left[
\begin{array}{c}
\varepsilon_{1} \\
\varepsilon_{2} \\
... \\
\varepsilon_{T-1} \\
\varepsilon_{T} \\
\end{array} 
  \right], \label{TVP_general2} \\ 
  \bm y \text{ \ \ \ \ }  &  & \text{ \ \ \ \ \ \ \ \ \ \ \ \ \ \ \ \ \ \ }  \mathcal{Z}  \text{ \ \ \ \ \ \ \ \ \ \ \ \ \ \ \ \ \ \ \ \ \ \ \ \ \ }\bm B^{\Delta } \text{ \ \ \ \ \ \ \ \ \ \ \ \ \ \ \ } \bm \varepsilon, \label{TVP_general_mat2}
\end{eqnarray}}
where we implicitly assume that $\bm \beta_{0}=0$ such that $\Delta \bm \beta_{1} = \bm \beta_{1}$. The $t$-th equation of the system above can be written as:
\begin{eqnarray}
y_{t} & = &   \bm X_{t}\Delta \bm \beta_{t}  + \bm X_{t} \Delta \bm \beta_{t-1}  + ... + \bm X_{t} \Delta \bm \beta_{2}  +   \bm X_{t} \bm \beta_{1} + \varepsilon_{t}, \\
  & = &  \bm X_{t} (\Delta \bm \beta_{t} + \Delta \bm \beta_{t-1} + ... + \bm \beta_{1}) + \varepsilon_{t}, \\
  & = & \bm X_{t} \bm \beta_{t} + \varepsilon_{t},
\end{eqnarray}
that is, equations \eqref{TVP1} and  \eqref{TVP_general2} are observationally equivalent. Under this specification the prior implied by \autoref{TVP2} becomes (in matrix form)
\begin{eqnarray}
\bm B^{\Delta}  & \sim & N_{[Tp]}(\bm 0, \bm Q),
\end{eqnarray}
and this prior can now be converted into a hierarchical prior by placing appropriate hyper-prior distributions on $\bm Q$.

Dynamic shrinkage and sparsity is a very active area of research, and there are several other important contributions that we don't explore here due to space constraints. For further readings we direct the reader to \cite{Chanetal2012}, \cite{Irie2019}, \cite{Kowaletal2019} and \cite{UribeLopes2017}, among others.

\subsection{High-dimensional causal inference}
Let $y_i$ denote an outcome variable and $T_i$ be some treatment variable. 
Suppose that the $p$-dimensional vector of cofounders $\bm x_i$ is high-dimensional. 
The parameter of interest is the treatment effect $\alpha$ in the model below: 
\begin{align}
y_i &= \beta_0+\alpha T_i + \bm x_i' \bm \beta + \epsilon_i \label{outcome_model}
\end{align}
A \textit{naive post selection}  approach would be to
apply the lasso to the equation above, excluding $\alpha$ from the $\mathcal{l}_{1}$-penalty and then regress $y_i$ on $T_i$ as well as on the selected covariates to estimate and conduct inference about the treatment effect. 
Any control variable that is highly correlated with $T_i$  but weakly with $y_i$ tends to drop out of the selection in the first stage, 
and could lead to omitted variable bias in estimating $\alpha$ in the second stage.
\cite{Bellonietal2014} propose post-double selection to overcome such bias. 
\cite{Hahnetal2018} and  \cite{Antonellietal2019} offer Bayesian counterparts in linear models, using shrinkage priors.

Using the model \eqref{outcome_model} as a benchmark, \cite{Hahnetal2018} consider the following system of equations: 
\begin{alignat*}{3}
T_i &= \bm x_i' \bm \gamma + \epsilon_i, &&\epsilon_i \sim N(0,\sigma^2_\epsilon) \quad \text{(Selection eq.)} \\
y_i &= \alpha T_i +\bm x_i' \bm \beta + v_i, \quad &&v_i \sim N(0,\sigma^2_v) \quad \text{(Response eq.)}
\end{alignat*}
The likelihood can be factorized: 
\begin{align*}
f\left( Y,T | \bm X, \bm \gamma, \alpha, \bm \beta,\sigma_\epsilon, \sigma_v \right) &=
f\left( Y |\bm X,T, \alpha, \bm \beta,\sigma_\epsilon \right) f\left(  T | \bm X,\bm \gamma, \sigma_v \right)
\end{align*}
With re-parameterization $\left(\alpha, \bm \beta+ \alpha\bm \gamma, \bm \gamma \right)' \to (\alpha, \bm \beta_d,\bm \beta_c)'$, the system can be written as
\begin{alignat*}{3}
T_i &=\bm  x_i' \bm \beta_c + \epsilon_i, &&\epsilon_i \sim N(0,\sigma^2_\epsilon) \quad \text{(Selection eq.)} \\
y_i &= \alpha \left(T_i - \bm x_i' \bm \beta_c \right) +\bm x_i' \bm  \beta_d + v_i, \quad &&v_i \sim N(0,\sigma^2_v) \quad \text{(Response eq.)}
\end{alignat*}
The authors place independent shrinkage priors over $\bm \beta_c$ and $\bm \beta_d$:
\begin{align*}
\pi(\beta_{ij}) &\propto \frac{1}{v} \log \left( 1 + \frac{4}{(\beta_j / v)^2} \right) \quad j=1,...,p, \quad i=c,d \\
v &\sim C^+(0,1) 
\end{align*}
where $C^+(0,1)$ denotes a folded standard Cauchy distribution. This prior is a proxy of the horseshoe prior. 
Non-informative priors are used for other parameters $\alpha \propto 1$, $\sigma_\epsilon \propto \frac{1}{\sigma_\epsilon}$, and $\sigma_v \propto \frac{1}{\sigma_v}$.  
They use a slice sampler for posterior sampling. 
\cite{Hahnetal2020} extends the approach to nonparametric case using regression trees.

 \cite{Antonellietal2019} propose a spike-and-slab lasso prior approach. Their proposed framework is
\begin{align*}
y_i | T_i, \bm x_i, \alpha,\bm  \beta, \sigma^2 &\sim N\left( \beta_0+ \alpha T_i + \bm x_i' \bm  \beta, \sigma^2 \right)  \\
p(\bm \beta |\bm \gamma, \sigma^2) &=\prod_{j=1}^p \gamma_j \psi_1(\beta_j; \lambda_1, \sigma^2)+(1-\gamma_j) \psi_0(\beta_j; \lambda_0, \sigma^2)\\
p(\bm \gamma | \theta) &=\prod_{j=1}^p \theta^{ \omega_j \gamma_j} \left( 1 - \theta^{\omega_j } \right)^{1-\gamma_j}\\
p(\theta|a,b) &\sim Beta(1,0.1p) \\
p(\sigma^2|c,d) &\sim Inv-Gamma(c,d) \\
\beta_0, \alpha &\sim N(0,K)
\end{align*}
where 
$\psi_0(\beta_j; \lambda_0, \sigma^2) =\frac{\lambda_0}{2\sigma}e^{-\lambda_0|\beta_j| / \sigma}$
and 
$\psi_1(\beta_j; \lambda_1, \sigma^2) =\frac{\lambda_1}{2\sigma}e^{-\lambda_1|\beta_j| / \sigma}$.
$\lambda_1$ is fixed to be a small value, say 0.1, so that the prior variance in the slab component is high enough to be uninformative.

The hyperparameter $\lambda_0$ is chosen via empirical Bayes. 
A new feature that they introduce is the weights $\omega_j$ which are tuning parameters that they use to prioritize variables to be included (i.e.\ $\gamma_j=1$) if they are associated with the treatment. 
Specifically, they first fit the standard lasso on the model for predicting $T$ given $X$. If  the $j$th covariate $x_j$ has non-zero coefficient from the lasso, they set $\omega_j=\delta$ for some $\delta \in (0,1)$. For other variables, $\omega_j=1$.
On the one hand, a smaller value of $\delta$ leads to higher inclusion probability and hence more protection against the omitted variable bias. 
On the other hand, one needs to ensure a small enough inclusion probability for an unimportant variable in the outcome model.
To balance the trade off, the authors choose $\delta \in(0,1)$ as the smallest value of $\omega_j$ such that the inclusion probability of $\beta_j=0$  is less than 0.1. 
See also \cite{Antonellietal2020}, who introduce how posterior distributions of treatment and outcome models can be used together with doubly robust estimators.

\subsection{Bayesian quantile regression}
A regression specification can be represented more generally using the formulation
\begin{equation}
y_{i}  = f(y_i \vert \bm X_{i} )  + \varepsilon, \label{quantile_general}
\end{equation}
where $f(y_i \vert \bm X_{i} )$ is a conditional mean function for $y$ (conditional on covariates $\bm X$). Using this notation, the linear regression model can be recovered if we set $f(y_i \vert \bm X_{i} ) = E(y_i \vert \bm X_{i} ) = \bm X_{i} \bm \beta$, that is, the linear regression only models the (conditional) mean of $y$. The distribution of $y$ is fully determined by the assumptions we make about the disturbance term $\varepsilon$. In many cases it is desirable to use the information in covariates in such a way that the full distribution of $y$ is determined by $\bm X$. While such feature can also be incorporated implicitly in a traditional linear regression setting\footnote{For example, we can assume $\varepsilon_{i} \sim N(0,\sigma_{i}^{2})$, where $\sigma_{i}^{2}$ can be some function of $\bm X_{i}$.}, a structured (and popular) way is to model the conditional quantiles of $y$, $\mathcal{Q}_{r}(y_i \vert \bm X_{i} )$, where $r \in (0,1)$ denotes the quantile of $y$. While the conditional quantile can be modeled using either linear or nonlinear functional forms, the linear form is by far the most widely used.

In this case, we replace in \autoref{quantile_general} $f(y_i \vert \bm X_{i} ) = \mathcal{Q}_{r}(y_i \vert \bm X_{i} ) = \bm X_{i} \bm \beta_{r}$ and obtain the following \emph{quantile regression specification}
\begin{equation}
y_{i}  = \bm X_{i} \bm \beta_{r}  + \varepsilon_{i}.
\end{equation}
The model above is a linear regression for each quantile level $r$. \cite{KoenkerBassett1978} show that an estimator of this quantile regression model can be obtained as
\begin{equation}
\widehat{\bm \beta}_{r} = \min_{\bm \beta_{r}} \mathbb{E} \left( \sum_{i=1}^{n} \rho_{r}(\varepsilon_{i}) \right),
\end{equation}
where $\rho_{r}(u) = (r - \mathbb{I}(u<r))$ is a loss function. \cite{YuMoyeed2001} show that the same estimator $\widehat{\bm \beta}_{r}$ can be obtaining by obtaining the maximum likelihood estimator under the assumption that $\varepsilon_{i}$ is distributed as asymmetric Laplace, i.e.\ if it has the density
\begin{equation}
p(\varepsilon_{i}; r,\sigma^{2}) \propto  \frac{\tau(1 - \tau)}{\sigma_{r}^{2}} \left[ e^{(1-r) \frac{\varepsilon_{i}}{\sigma_{r}^{2}}} \mathbb{I}(\varepsilon_{i} \leq 0) + e^{- r \frac{\varepsilon_{i}}{\sigma_{r}^{2}}}\mathbb{I}(\varepsilon_{i} > 0) \right], \label{AL_likelihood}
\end{equation}
where $\sigma_{r}^{2}$ is a scale parameter. Therefore, the contribution of \cite{YuMoyeed2001} provides a parametric framework for implementing Bayesian inference.\footnote{Of course here we have similar conceptual issues as with the Bayesian representation of the lasso estimator: while \cite{Tibshirani1996} showed that the $\mathcal{l}_{1}$ optimization problem for the lasso is equivalent to the posterior mode of Bayesian regression estimator under a Laplace prior, \cite{castillo2015} show that the posterior distribution does not contract at the same rate as the posterior mode. Similarly here, there is an equivalence between quantile regression and maximizing the likelihood under an asymmetric Laplace likelihood as both problems provide unique point solutions. Bayesian inference, in contrast, assumes that coefficients are random variables and (unless one focuses on MAP or MMSE estimation) cannot be obtained as the solution to an optimization problem. In practice, however, it turns out that Bayesian quantile regression estimation using the asymmetric Laplace likelihood is a very flexible model, even if it is not identical to the model introduced by \cite{KoenkerBassett1978}.} In particular, \cite{KozumiKobayashi2011} take advantage of the fact that the asymmetric Laplace likelihood can be written as a normal-exponential mixture of the form
\begin{equation}
\varepsilon_{i} \vert u_{i},z_{i,r} \sim \theta_{r} z_{i,r} + \sqrt{\sigma_{r}^{2} \kappa_{r}^{2} z_{i,r}} u_{t}, \label{mixture_approximation}
\end{equation}
where $z_{i,r} \sim Exp(\sigma^2_{r})$ and $u_{t} \sim N(0,1)$, while $\theta_r,\kappa_{r}^{2}$ are parameters defined as $\theta_{r} = \frac{1-2r}{r(1-r)}$, $\kappa_{r}^{2} = \frac{2}{r(1-r)}$. If we marginalise Equation \eqref{mixture_approximation} over the exponentially distributed $z_{i,r}$ we obtain Equation \eqref{AL_likelihood}; see a derivation in \cite{KhareHobert2012}.

Therefore, the Bayesian quantile regression model has the following representation
\begin{equation}
y_{i} = \bm X_{i} \bm \beta_{r} + \theta_{r} z_{i,r} + \sqrt{\sigma_{r}^{2} \kappa_{r}^{2} z_{i,r}} u_{t},
\end{equation}
where $z_{i,r} \sim Exp(\sigma^2_{r})$ and $u_{t} \sim N(0,1)$ can either be thought of two disturbance terms (similar to frontier models in econometrics), or equivalently $u_{t}$ can be interpreted as the disturbance term and $z_{i,r}$ can be thought of as an unobserved factor (similar to the factors we examined in \autoref{sec:factor}). In any case, conditional posteriors are trivial to derive as conditional on all other parameters, the posterior of each parameter of interest has standard form. This is easier to see for instance for the coefficients $\bm \beta_{r}$ by rewriting the model as
\begin{eqnarray}
(y_{i} - \theta_{r} z_{i,r}) =  \bm X_{i} \bm \beta_{r} + \sqrt{\sigma_{r}^{2} \kappa_{r}^{2} z_{i,r}} u_{t} \Rightarrow \\
y_{i}^{\star} = \bm X_{i} \bm \beta_{r} + \sigma_{r}^{\star}u_{t},
\end{eqnarray}
where $y_{i}^{\star} = (y_{i} - \theta_{r} z_{i,r})$ and $\sigma_{r}^{\star} = \sqrt{\sigma_{r}^{2} \kappa_{r}^{2} z_{i,r}} $. As long as we condition on $z_{i,r},\sigma_{r}^{2},\kappa_{r}^{2},\theta_{r}$, we can obtain a sample of $\bm \beta_{r}$ (assuming a normal prior), from the standard formulas for a linear regression of $y_{i}^{\star}$ on $\bm X_{i}$ with known variance $\sigma_{r}^{\star}$.

In more detail, we assume the following priors
\begin{eqnarray}
\bm \beta_{r} & \sim &  N\left(0,\bm D_{\tau,r}\right), \\
\sigma_{r}^{2} & \sim & Inv-Gamma(n_{0,r},s_{0,r}), \\
z_{i,r}  &  \sim & Exponential(\sigma^{2}_{r}),
\end{eqnarray}
where for simplicity assume that $\bm D_{\tau,r} = \bm D_{\tau} = \tau \times I_p$ where $\tau$ is fixed and known for all $r$. The conditional posteriors \citep{KhareHobert2012} are of the form
\begin{eqnarray}
\bm \beta_{r} \vert \bullet & \sim & N_{p}\left( \bm V \times \left( \bm X^{\prime} \bm U^{-1} \widetilde{\bm y} \right), \bm V \right), \label{post1} \\
\sigma_{r}^{2} \vert  \bullet & \sim & Inv-Gamma \left(a_r,s_{0,r} + \sum_{i=1}^{n} \frac{\left( y^{\star}_{i}\right)^2}{2z_{i,r}\kappa_r^{2}} + \sum_{i=1}^{n} z_{i,r} \right), \label{post2} \\
z_{i,r}^{-1} \vert  \bullet & \sim & IG \left(\frac{\sqrt{\theta_r^2 + 2\kappa_r^2} }{\vert y_{i} - \bm X_{i} \bm \beta_{r} \vert }, \frac{\theta_{r}^2 + 2\kappa_{r}^2}{\sigma_{r}^2 \kappa_{r}^2} \right), \label{post3}
\end{eqnarray}
where the notation $\vert \bullet $ means ``conditioning on other parameters and data'', $\bm V = \left(\bm X^{\prime} \bm U^{-1} \bm X + D_{\tau,r}^{-1} \right)^{-1}$, $\bm U = \left( \sigma_{r}^2 \kappa_r^{2} \right) \times diag \left(z_{1,r},...,z_{n,r} \right)$, $\widetilde{\bm y} = \left(\bm y - \theta_{r} \bm z_{r} \right)$, $y_{i}^{\star} = \left( y_{i} - \bm X_{i} \bm \beta_{r} - \theta_{r} z_{i,r} \right)$, and 
$a_r=n_{0,r} + \frac{3n}{2}$.

We need to note here a few important points regarding the implementation of this model
\begin{enumerate}
\item The Gibbs sampler in equations \eqref{post1} - \eqref{post3} is only one of the many implementations of the Bayesian quantile regression using an asymmetric Laplace likelihood. \cite{KozumiKobayashi2011} first developed a Gibbs sampler were $z_{i,r}$ is sampled from a three-parameter generalized inverse Gaussian (GIG) distribution. \cite{KhareHobert2012} proved that this Gibbs sampler is ergodic, but  proposed to sample $z_{i,r}^{-1}$ from the two-parameter inverse Gaussian (IG) posterior of \eqref{post3}. While the two sampling steps are identical (the transformation utilizes the ability of GIG distribution to be written as an equivalent IG distribution), the consequences in programming might be more important. For example, MATLAB only has a built-in random number generator for the IG distribution but not for the GIG (although contributed packages on the internet do exist), while in R there are libraries that provide reliable generators for both the IG and GIG distributions.
\item The Gibbs sampler  needs to be run for each quantile level $r$. Therefore, we need to choose quantile levels that are reasonable. For most empirical cases the grid $r={0.05,0.10,0.25,0.5,0.75,0.90,0.95}$ covers the most important areas of a distribution of interest, but of course one can consider much finer grids at the cost of increased computation.
\item Estimation of the quantile $r=0.5$ is the most accurate, as $50 \%$ of the data lie on the left/right of the median. As $r$ approaches 0 or 1, estimation accuracy might decrease as for some problems the number of observations in the tails could be too low (e.g.\ short, quarterly macroeconomic time series). If ultra high-frequency financial data are available (e.g.\ 1-min data) then typically the researcher is able to consider values of $r$ closer to 0 or 1 without problems.
\item The conditional posteriors are applied for each quantile level independently. If we consider small grid of quantiles, for example $r={0.05,0.10,0.25,0.5,0.75,0.90,0.95}$ then one can use vectorized operations (for matrix programming languages, such as MATLAB) to obtain $\bm \beta_{r}$ for all $r$ -- despite the fact that these samples of $\bm \beta_{r}$ will be uncorrelated across $r$. If we consider a very fine grid for values of $r$, then further benefits can be achieved if we parallelize and split the sampling equations for each $r$ in different cores.
\item The fact that the regression for each quantile is estimated independently, means that one can obtain estimates of the conditional quantile function $\mathcal{Q}_{r}(y_i \vert \bm X_{i} )$ that are not ordered. That is, solutions of the form $\widehat{\mathcal{Q}}_{r_{1}}(y_{i} \vert \bm X_{i}) > \widehat{\mathcal{Q}}_{r_{2}}(y_{i} \vert \bm X_{i})$ for $r_{1}<r_{2}$ are not compatible with the concept of quantile. Putting this in context, we can't allow the model to predict that the conditional median of inflation is $2\%$ while its first quartile is $2.5\%$! This problem, which is known as quantile crossing, pertains to all quantile models for which every conditional quantile level is estimated independently from the others (regardless of whether estimation is Bayesian or not). However, \cite{RodriguesFan2017}, among numerous others, provide a fully Bayesian algorithm for post-processing MCMC draws of the conditional quantiles in order to ensure monotonicity. The idea is to use a nonparametric smoothing function in order to ensure that this monotinicity exists. The approach of \cite{RodriguesFan2017} is attractive from a Bayesian parametric perspective because it uses the properties of the asymmetric Laplace distribution in order to derive the implied information that a quantile level $r^{\prime}$ conveys for some other quantile level $r$, thus, using an expanded information set when smoothing conditional quantile estimates. 
\end{enumerate}
After taking the modeling points above into consideration, estimation of the Bayesian quantile regression model is not much more challenging than the normal (Gaussian) linear regression model. All the hierarchical priors described previously can be readily applied to the quantile model, with very minor modifications to the conditional posteriors presented in equations \eqref{post1} - \eqref{post3}. The interesting feature is that because a separate regression needs to be run for each $r$, we can also use shrinkage or sparsity priors to allow different covariates in $\bm X$ to affect different quantiles of the distribution of $y$. Exactly because application of hierarchical priors is so trivial, we won't provide detailed examples and derivations here. \cite{KozumiKobayashi2011}, while proposing a Gibbs sampler for the Bayesian quantile model, they also show in a subsection how easy it is to adopt the double-exponential (lasso) prior. Other applications include \cite{AlhamzawiYu2012}, \cite{Yuetal2013}, \cite{Korobilis2017} and \cite{RePEc:ecb:ecbwps:20212600}, and the reader can consult these papers for modeling details. \cite{Limetal2020} is the case of a paper that estimates a Bayesian quantile regression using variational Bayes methods.

\section[Conclusion]{Concluding remarks}
We have reviewed a wide range of concepts and algorithms for Bayesian sparse and shrinkage estimation. Our focus was on recent contributions in the field, covering the mainly academic publications during the decade 2010-2020, that are increasingly focusing on efficient computation and inference in high-dimensional models. A major contribution of our work is to collect in a single document all these recent contributions, as other reviews and surveys we are aware of provide only very focused reviews of certain hierarchical priors and algorithms. While we believe contributions to the field of Bayesian sparse and shrinkage estimation will keep expanding at a polynomial rate, we do hope that this review will become a useful manual for PhD students and researchers who want an accessible introduction to the field. 

\newpage
\bibliographystyle{apa}
\addcontentsline{toc}{section}{References}
\bibliography{BibTex/Bayesian_variable_selection}

\begin{thebibliography}{}

\bibitem[\protect\astroncite{Aitkin}{1991}]{Aitkin1991}
Aitkin, M. (1991).
\newblock Posterior {B}ayes factors.
\newblock {\em Journal of the Royal Statistical Society. Series B
  (Methodological)}, 53(1):111--142.

\bibitem[\protect\astroncite{Alhamzawi and Ali}{2018}]{AlhamzawiAli2018}
Alhamzawi, R. and Ali, H. T.~M. (2018).
\newblock The {B}ayesian adaptive lasso regression.
\newblock {\em Mathematical Biosciences}, 303:75 -- 82.

\bibitem[\protect\astroncite{Alhamzawi and Yu}{2012}]{AlhamzawiYu2012}
Alhamzawi, R. and Yu, K. (2012).
\newblock Variable selection in quantile regression via gibbs sampling.
\newblock {\em Journal of Applied Statistics}, 39(4):799--813.

\bibitem[\protect\astroncite{Antonelli et~al.}{2020}]{Antonellietal2020}
Antonelli, J., Papadogeorgou, G., and Dominici, F. (2020).
\newblock Causal inference in high dimensions: {A} marriage between {B}ayesian
  modeling and good frequentist properties.
\newblock {\em Biometrics}.

\bibitem[\protect\astroncite{Antonelli et~al.}{2019}]{Antonellietal2019}
Antonelli, J., Parmigiani, G., and Dominici, F. (2019).
\newblock High-dimensional confounding adjustment using continuous spike and
  slab priors.
\newblock {\em {B}ayesian Analysis}, 14(3):805 -- 828.

\bibitem[\protect\astroncite{Armagan et~al.}{2011}]{Armaganetal2011}
Armagan, A., Clyde, M., and Dunson, D. (2011).
\newblock Generalized beta mixtures of {G}aussians.
\newblock In Shawe-Taylor, J., Zemel, R., Bartlett, P., Pereira, F., and
  Weinberger, K.~Q., editors, {\em Advances in Neural Information Processing
  Systems}, volume~24. Curran Associates, Inc.

\bibitem[\protect\astroncite{Armagan et~al.}{2013}]{Armaganetal2013a}
Armagan, A., Dunson, D.~B., and Lee, J. (2013).
\newblock Generalized double pareto shrinkage.
\newblock {\em Statistica Sinica}, 23:119--143.

\bibitem[\protect\astroncite{Armagan and Zaretzki}{2010}]{ArmaganZaretzki2010}
Armagan, A. and Zaretzki, R.~L. (2010).
\newblock Model selection via adaptive shrinkage with t priors.
\newblock {\em Computational Statistics}, 25(3):441--461.

\bibitem[\protect\astroncite{Assmann et~al.}{2016}]{Assmann2016}
Assmann, C., Boysen-Hogrefe, J., and Pape, M. (2016).
\newblock {B}ayesian analysis of static and dynamic factor models: An ex-post
  approach towards the rotation problem.
\newblock {\em Journal of Econometrics}, 192(1):190--206.

\bibitem[\protect\astroncite{Bae and Mallick}{2004}]{BaeMallick2004}
Bae, K. and Mallick, B.~K. (2004).
\newblock {Gene selection using a two-level hierarchical {B}ayesian model}.
\newblock {\em Bioinformatics}, 20(18):3423--3430.

\bibitem[\protect\astroncite{Bai et~al.}{2021}]{BaiRockovaGeorge2021}
Bai, R., Rockova, V., and George, E.~I. (2021).
\newblock Spike-and-slab meets lasso: {A} review of the spike-and-slab lasso.
\newblock {\em arXiv preprint arXiv:2010.06451}.

\bibitem[\protect\astroncite{Barbieri and Berger}{2004}]{BarbieriBerger2004}
Barbieri, M.~M. and Berger, J.~O. (2004).
\newblock Optimal predictive model selection.
\newblock {\em The Annals of Statistics}, 32(3):870--897.

\bibitem[\protect\astroncite{Baumeister et~al.}{2020}]{Baumeisteretal2020}
Baumeister, C., Korobilis, D., and Lee, T.~K. (2020).
\newblock {Energy Markets and Global Economic Conditions}.
\newblock {\em The Review of Economics and Statistics}, pages 1--45.

\bibitem[\protect\astroncite{Belloni et~al.}{2014}]{Bellonietal2014}
Belloni, A., Chernozhukov, V., and Hansen, C. (2014).
\newblock Inference on treatment effects after selection among high-dimensional
  controls.
\newblock {\em The Review of Economic Studies}, 81(2):608--650.

\bibitem[\protect\astroncite{Belmonte et~al.}{2014}]{Belmonteetal2014}
Belmonte, M.~A., Koop, G., and Korobilis, D. (2014).
\newblock Hierarchical shrinkage in time-varying parameter models.
\newblock {\em Journal of Forecasting}, 33(1):80--94.

\bibitem[\protect\astroncite{Berger}{1985}]{Berger1985}
Berger, J.~O. (1985).
\newblock {\em Statistical Decision Theory and {B}ayesian Analysis}, volume
  Springer Series in Statistics.
\newblock Springer-Verlag New York.

\bibitem[\protect\astroncite{Berger and Mortera}{1999}]{BergerMortera1999}
Berger, J.~O. and Mortera, J. (1999).
\newblock Default {B}ayes factors for nonnested hypothesis testing.
\newblock {\em Journal of the American Statistical Association},
  94(446):542--554.

\bibitem[\protect\astroncite{Berger and Pericchi}{1996}]{BergerPericchi1996}
Berger, J.~O. and Pericchi, L.~R. (1996).
\newblock The intrinsic {B}ayes factor for model selection and prediction.
\newblock {\em Journal of the American Statistical Association},
  91(433):109--122.

\bibitem[\protect\astroncite{Berger and Pericchi}{1998}]{BergerPericchi1998}
Berger, J.~O. and Pericchi, L.~R. (1998).
\newblock Accurate and stable {B}ayesian model selection: {T}he median
  intrinsic {B}ayes factor.
\newblock {\em Sankhyā: The Indian Journal of Statistics, Series B
  (1960-2002)}, 60(1):1--18.

\bibitem[\protect\astroncite{Berger and Pericchi}{2001}]{BergerPericchi2001}
Berger, J.~O. and Pericchi, L.~R. (2001).
\newblock Objective {B}ayesian methods for model selection: {I}ntroduction and
  comparison.
\newblock In Lahiri, P., editor, {\em Model selection}, volume Volume 38 of
  {\em Lecture Notes--Monograph Series}, pages 135--207. Institute of
  Mathematical Statistics, Beachwood, OH.

\bibitem[\protect\astroncite{Bernanke et~al.}{2005}]{BBE2005}
Bernanke, B.~S., Boivin, J., and Eliasz, P. (2005).
\newblock {Measuring the Effects of Monetary Policy: A Factor-Augmented Vector
  Autoregressive (FAVAR) Approach}.
\newblock {\em The Quarterly Journal of Economics}, 120(1):387--422.

\bibitem[\protect\astroncite{Bhadra et~al.}{2020}]{Bhadraetal2020}
Bhadra, A., Datta, J., Li, Y., and Polson, N. (2020).
\newblock Horseshoe regularisation for machine learning in complex and deep
  models.
\newblock {\em International Statistical Review}, 88(2):302--320.

\bibitem[\protect\astroncite{Bhattacharya et~al.}{2016}]{Bhattacharyaetal2016}
Bhattacharya, A., Chakraborty, A., and Mallick, B.~K. (2016).
\newblock {Fast sampling with {G}aussian scale mixture priors in
  high-dimensional regression}.
\newblock {\em Biometrika}, 103(4):985--991.

\bibitem[\protect\astroncite{Bhattacharya and
  Dunson}{2011}]{BhattacharyaDunson2011}
Bhattacharya, A. and Dunson, D.~B. (2011).
\newblock {Sparse {B}ayesian infinite factor models}.
\newblock {\em Biometrika}, pages 291--306.

\bibitem[\protect\astroncite{Bhattacharya et~al.}{2015}]{Bhattacharyaetal2015}
Bhattacharya, A., Pati, D., Pillai, N.~S., and Dunson, D.~B. (2015).
\newblock {D}irichlet–{L}aplace priors for optimal shrinkage.
\newblock {\em Journal of the American Statistical Association},
  110(512):1479--1490.
\newblock PMID: 27019543.

\bibitem[\protect\astroncite{Bitto and
  Fr\"{u}hwirth-Schnatter}{2019}]{BittoFruewirthSchnatter2019}
Bitto, A. and Fr\"{u}hwirth-Schnatter, S. (2019).
\newblock Achieving shrinkage in a time-varying parameter model framework.
\newblock {\em Journal of Econometrics}, 210(1):75 -- 97.
\newblock Annals Issue in Honor of John Geweke ``Complexity and Big Data in
  Economics and Finance: Recent Developments from a {B}ayesian Perspective''.

\bibitem[\protect\astroncite{Blei et~al.}{2017}]{Bleietal17}
Blei, D.~M., Kucukelbir, A., and McAuliffe, J.~D. (2017).
\newblock Variational inference: {A} review for statisticians.
\newblock {\em Journal of the American Statistical Association},
  112(518):859--877.

\bibitem[\protect\astroncite{Bogdan et~al.}{2011}]{Bogdanetal2011}
Bogdan, M., Chakrabarti, A., Frommlet, F., and Ghosh, J.~K. (2011).
\newblock Asymptotic {B}ayes-optimality under sparsity of some multiple testing
  procedures.
\newblock {\em The Annals of Statistics}, 39(3):1551--1579.

\bibitem[\protect\astroncite{Bottolo and
  Richardson}{2010}]{BottoloRichardson2010}
Bottolo, L. and Richardson, S. (2010).
\newblock Evolutionary stochastic search for {B}ayesian model exploration.
\newblock {\em {B}ayesian Anal.}, 5(3):583--618.

\bibitem[\protect\astroncite{Cao et~al.}{2020}]{Caoetal2020}
Cao, X., Khare, K., and Ghosh, M. (2020).
\newblock High-dimensional posterior consistency for hierarchical non-local
  priors in regression.
\newblock {\em {B}ayesian Analysis}, 15(1):241 -- 262.

\bibitem[\protect\astroncite{Carbonetto and
  Stephens}{2012}]{CarbonettoStephens12}
Carbonetto, P. and Stephens, M. (2012).
\newblock Scalable variational inference for {B}ayesian variable selection in
  regression, and its accuracy in genetic association studies.
\newblock {\em {B}ayesian Analysis}, 7(1):73 -- 108.

\bibitem[\protect\astroncite{Caron and Doucet}{2008}]{CaronDoucet2008}
Caron, F. and Doucet, A. (2008).
\newblock Sparse {B}ayesian nonparametric regression.
\newblock In {\em Proceedings of the 25th International Conference on Machine
  Learning}, ICML '08, pages 88--95, New York, NY, USA. ACM.

\bibitem[\protect\astroncite{Carriero et~al.}{2019}]{Carrieroetal}
Carriero, A., Clark, T.~E., and Marcellino, M. (2019).
\newblock Large bayesian vector autoregressions with stochastic volatility and
  non-conjugate priors.
\newblock {\em Journal of Econometrics}, 212(1):137 -- 154.
\newblock Big Data in Dynamic Predictive Econometric Modeling.

\bibitem[\protect\astroncite{Carvalho et~al.}{2008}]{Carvalhoetal2008}
Carvalho, C.~M., Chang, J., Lucas, J.~E., Nevins, J.~R., Wang, Q., and West, M.
  (2008).
\newblock High-dimensional sparse factor modeling: {A}pplications in gene
  expression genomics.
\newblock {\em Journal of the American Statistical Association},
  103(484):1438--1456.

\bibitem[\protect\astroncite{Carvalho et~al.}{2010}]{Carvalhoetal2010}
Carvalho, C.~M., Polson, N.~G., and Scott, J.~G. (2010).
\newblock {The horseshoe estimator for sparse signals}.
\newblock {\em Biometrika}, 97(2):465--480.

\bibitem[\protect\astroncite{Castillo et~al.}{2015}]{castillo2015}
Castillo, I., Schmidt-Hieber, J., and van~der Vaart, A. (2015).
\newblock {B}ayesian linear regression with sparse priors.
\newblock {\em The Annals of Statistics}, 43(5):1986--2018.

\bibitem[\protect\astroncite{Castillo and van~der Vaart}{2012}]{castillo2012}
Castillo, I. and van~der Vaart, A. (2012).
\newblock Needles and straw in a haystack: Posterior concentration for possibly
  sparse sequences.
\newblock {\em The Annals of Statistics}, 40(4):2069 -- 2101.

\bibitem[\protect\astroncite{Chan et~al.}{2018}]{Chanetal2018}
Chan, J., Leon-Gonzalez, R., and Strachan, R.~W. (2018).
\newblock Invariant inference and efficient computation in the static factor
  model.
\newblock {\em Journal of the American Statistical Association},
  113(522):819--828.

\bibitem[\protect\astroncite{Chan and Grant}{2016}]{ChanGrant2016}
Chan, J.~C. and Grant, A.~L. (2016).
\newblock Fast computation of the deviance information criterion for latent
  variable models.
\newblock {\em Computational Statistics \& Data Analysis}, 100:847 -- 859.

\bibitem[\protect\astroncite{Chan et~al.}{2012}]{Chanetal2012}
Chan, J.~C., Koop, G., Leon-Gonzalez, R., and Strachan, R.~W. (2012).
\newblock Time varying dimension models.
\newblock {\em Journal of Business \& Economic Statistics}, 30(3):358--367.

\bibitem[\protect\astroncite{Chen and Chen}{2008}]{ChenChen2008}
Chen, J. and Chen, Z. (2008).
\newblock Extended {B}ayesian information criteria for model selection with
  large model spaces.
\newblock {\em Biometrika}, 95(3):759--771.

\bibitem[\protect\astroncite{Chib}{1995}]{Chib1995}
Chib, S. (1995).
\newblock Marginal likelihood from the {G}ibbs output.
\newblock {\em Journal of the American Statistical Association},
  90(432):1313--1321.

\bibitem[\protect\astroncite{Chib and Jeliazkov}{2001}]{ChibJeliazkov2001}
Chib, S. and Jeliazkov, I. (2001).
\newblock Marginal likelihood from the metropolis–hastings output.
\newblock {\em Journal of the American Statistical Association},
  96(453):270--281.

\bibitem[\protect\astroncite{Chib et~al.}{2006}]{Chibetal2006}
Chib, S., Nardari, F., and Shephard, N. (2006).
\newblock Analysis of high dimensional multivariate stochastic volatility
  models.
\newblock {\em Journal of Econometrics}, 134(2):341--371.

\bibitem[\protect\astroncite{Chipman et~al.}{2001}]{Chipmanetal2001}
Chipman, H., George, E.~I., and McCulloch, R.~E. (2001).
\newblock The practical implementation of {B}ayesian model selection.
\newblock In Lahiri, P., editor, {\em Model selection}, volume Volume 38 of
  {\em Lecture Notes--Monograph Series}, pages 65--116. Institute of
  Mathematical Statistics, Beachwood, OH.

\bibitem[\protect\astroncite{Clyde}{1999}]{Clyde1999}
Clyde, M.~A. (1999).
\newblock {B}ayesian model averaging and model search strategies.
\newblock In Bernardo, J., Dawid, A., Berger, J., and Smith, A., editors, {\em
  {B}ayesian Statistics 6}. Oxford University Press.

\bibitem[\protect\astroncite{Clyde et~al.}{2011}]{Clydeetal2011}
Clyde, M.~A., Ghosh, J., and Littman, M.~L. (2011).
\newblock {B}ayesian adaptive sampling for variable selection and model
  averaging.
\newblock {\em Journal of Computational and Graphical Statistics},
  20(1):80--101.

\bibitem[\protect\astroncite{Dangl and Halling}{2012}]{DanglHalling2012}
Dangl, T. and Halling, M. (2012).
\newblock Predictive regressions with time-varying coefficients.
\newblock {\em Journal of Financial Economics}, 106(1):157 -- 181.

\bibitem[\protect\astroncite{Datta and Ghosh}{2013}]{DattaGhosh2013}
Datta, J. and Ghosh, J.~K. (2013).
\newblock Asymptotic properties of {B}ayes risk for the horseshoe prior.
\newblock {\em {B}ayesian Anal.}, 8(1):111--132.

\bibitem[\protect\astroncite{Davison}{1986}]{Davison1986}
Davison, A.~C. (1986).
\newblock {Approximate predictive likelihood}.
\newblock {\em Biometrika}, 73(2):323--332.

\bibitem[\protect\astroncite{De~Santis and
  Spezzaferri}{1997}]{DeSantisSpezzaferri1997}
De~Santis, F. and Spezzaferri, F. (1997).
\newblock Alternative {B}ayes factors for model selection.
\newblock {\em Canadian Journal of Statistics}, 25(4):503--515.

\bibitem[\protect\astroncite{Dehaene and
  Barthelm\'{e}}{2018}]{DehaeneBarthelme2018}
Dehaene, G. and Barthelm\'{e}, S. (2018).
\newblock Expectation propagation in the large data limit.
\newblock {\em Journal of the Royal Statistical Society: Series B (Statistical
  Methodology)}, 80(1):199--217.

\bibitem[\protect\astroncite{Dellaportas et~al.}{2002}]{Dellaportasetal2002}
Dellaportas, P., Forster, J.~J., and Ntzoufras, I. (2002).
\newblock On {B}ayesian model and variable selection using mcmc.
\newblock {\em Statistics and Computing}, 12(1):27--36.

\bibitem[\protect\astroncite{DiCiccio et~al.}{1997}]{DiCiccioetal1997}
DiCiccio, T.~J., Kass, R.~E., Raftery, A., and Wasserman, L. (1997).
\newblock Computing {B}ayes factors by combining simulation and asymptotic
  approximations.
\newblock {\em Journal of the American Statistical Association},
  92(439):903--915.

\bibitem[\protect\astroncite{Dunson et~al.}{2008}]{Dunsonetal2008}
Dunson, D.~B., Herring, A.~H., and Engel, S.~M. (2008).
\newblock {B}ayesian selection and clustering of polymorphisms in functionally
  related genes.
\newblock {\em Journal of the American Statistical Association},
  103(482):534--546.

\bibitem[\protect\astroncite{Efron and Morris}{1973}]{EfronMorris1973}
Efron, B. and Morris, C. (1973).
\newblock Stein's estimation rule and its competitors -- an empirical {B}ayes
  approach.
\newblock {\em Journal of the American Statistical Association},
  68(341):117--130.

\bibitem[\protect\astroncite{Eicher et~al.}{2011}]{Eicheretal2011}
Eicher, T.~S., Papageorgiou, C., and Raftery, A.~E. (2011).
\newblock Default priors and predictive performance in {B}ayesian model
  averaging, with application to growth determinants.
\newblock {\em Journal of Applied Econometrics}, 26(1):30--55.

\bibitem[\protect\astroncite{Fern\'{a}ndez et~al.}{2001a}]{Fernandezetal2001a}
Fern\'{a}ndez, C., Ley, E., and Steel, M.~F. (2001a).
\newblock Benchmark priors for {B}ayesian model averaging.
\newblock {\em Journal of Econometrics}, 100(2):381--427.

\bibitem[\protect\astroncite{Fern\'{a}ndez et~al.}{2001b}]{Fernandezetal2001b}
Fern\'{a}ndez, C., Ley, E., and Steel, M. F.~J. (2001b).
\newblock Model uncertainty in cross-country growth regressions.
\newblock {\em Journal of Applied Econometrics}, 16(5):563--576.

\bibitem[\protect\astroncite{Figueiredo}{2003}]{Figueiredo2003}
Figueiredo, M. A.~T. (2003).
\newblock Adaptive sparseness for supervised learning.
\newblock {\em IEEE Trans. Pattern Anal. Mach. Intell.}, 25(9):1150--1159.

\bibitem[\protect\astroncite{Foster and George}{1994}]{FosterGeorge1994}
Foster, D.~P. and George, E.~I. (1994).
\newblock The risk inflation criterion for multiple regression.
\newblock {\em The Annals of Statistics}, 22(4):1947--1975.

\bibitem[\protect\astroncite{Fourdrinier et~al.}{2018}]{Shrinkage2018}
Fourdrinier, D., Strawderman, W.~E., and Wells, M.~T. (2018).
\newblock {\em Shrinkage Estimation}, volume Springer Texts in Statistics.
\newblock Springer International Publishing.

\bibitem[\protect\astroncite{Fragoso et~al.}{2018}]{Fragosoetal2018}
Fragoso, T.~M., Bertoli, W., and Louzada, F. (2018).
\newblock {B}ayesian model averaging: {A} systematic review and conceptual
  classification.
\newblock {\em International Statistical Review}, 86(1):1--28.

\bibitem[\protect\astroncite{Fr\"{u}wirth-Schnatter and
  Lopes}{2018}]{FruewirthLopes2018}
Fr\"{u}wirth-Schnatter, S. and Lopes, H. (2018).
\newblock Sparse {B}ayesian factor analysis when the number of factors is
  unknown.
\newblock Technical Report arXiv:1804.04231v1, ArXiV.

\bibitem[\protect\astroncite{Fr\"{u}wirth-Schnatter and
  Wagner}{2010}]{SylviaHelga2010}
Fr\"{u}wirth-Schnatter, S. and Wagner, H. (2010).
\newblock {B}ayesian variable selection for random intercept modeling of
  {G}aussian and non‐{G}aussian data.
\newblock In Bernardo, J., Bayarri, M., Berger, J., Dawid, A., Heckerman, D.,
  Smith, A., and West, M., editors, {\em {B}ayesian Statistics 9}. Oxford
  University Press.

\bibitem[\protect\astroncite{Gelfand and Dey}{1994}]{GelfandDey1994}
Gelfand, A.~E. and Dey, D.~K. (1994).
\newblock {B}ayesian model choice: {A}symptotics and exact calculations.
\newblock {\em Journal of the Royal Statistical Society. Series B
  (Methodological)}, 56(3):501--514.

\bibitem[\protect\astroncite{Gelfand and Ghosh}{1998}]{GelfandGhosh1998}
Gelfand, A.~E. and Ghosh, S.~K. (1998).
\newblock Model choice: {A} minimum posterior predictive loss approach.
\newblock {\em Biometrika}, 85(1):1--11.

\bibitem[\protect\astroncite{Gelman}{2006}]{gelman2006}
Gelman, A. (2006).
\newblock {Prior distributions for variance parameters in hierarchical models}.
\newblock {\em {B}ayesian Analysis}, 1(3):515 -- 534.

\bibitem[\protect\astroncite{Gelman et~al.}{2013}]{BDA2013}
Gelman, A., Carlin, J.~B., Stern, H.~S., Dunson, D.~B., Vehtari, A., and Rubin,
  D.~B. (2013).
\newblock {\em {B}ayesian Data Analysis}.
\newblock Chapman and Hall/CRC, New York, 3rd ed. edition.

\bibitem[\protect\astroncite{Gelman and Hennig}{2017}]{GelmanHenning2017}
Gelman, A. and Hennig, C. (2017).
\newblock {Beyond subjective and objective in statistics}.
\newblock {\em Journal of the Royal Statistical Society Series A},
  180(4):967--1033.

\bibitem[\protect\astroncite{Gelman et~al.}{2014}]{Gelmanetal2014}
Gelman, A., Hwang, J., and Vehtari, A. (2014).
\newblock Understanding predictive information criteria for {B}ayesian models.
\newblock {\em Statistics and Computing}, 24(6):997--1016.

\bibitem[\protect\astroncite{Gelman et~al.}{1996}]{Gelmanetal1996}
Gelman, A., Meng, X.-L., and Stern, H. (1996).
\newblock Posterior predictive assessment of model fitness via realized
  discrepancies.
\newblock {\em Statistica Sinica}, 6(4):733--760.

\bibitem[\protect\astroncite{George and Foster}{2000}]{FosterGeorge2000}
George, E.~I. and Foster, D.~P. (2000).
\newblock Calibration and empirical {B}ayes variable selection.
\newblock {\em Biometrika}, 87(4):731--747.

\bibitem[\protect\astroncite{George and McCulloch}{1993}]{GeorgeMcCulloch1993}
George, E.~I. and McCulloch, R.~E. (1993).
\newblock Variable selection via {G}ibbs sampling.
\newblock {\em Journal of the American Statistical Association},
  88(423):881--889.

\bibitem[\protect\astroncite{George and McCulloch}{1997}]{GeorgeMcCulloch1997}
George, E.~I. and McCulloch, R.~E. (1997).
\newblock Approaches for {B}ayesian variable selection.
\newblock {\em Statistica Sinica}, 7(2):339--373.

\bibitem[\protect\astroncite{Geweke and Zhou}{1996}]{GewekeZhou1996}
Geweke, J. and Zhou, G. (1996).
\newblock Measuring the price of the arbitrage pricing theory.
\newblock {\em The Review of Financial Studies}, 9(2):557--587.

\bibitem[\protect\astroncite{Ghosh and Clyde}{2011}]{GhoshClyde2011}
Ghosh, J. and Clyde, M.~A. (2011).
\newblock Rao–blackwellization for {B}ayesian variable selection and model
  averaging in linear and binary regression: {A} novel data augmentation
  approach.
\newblock {\em Journal of the American Statistical Association},
  106(495):1041--1052.

\bibitem[\protect\astroncite{Ghosh and Dunson}{2009}]{GhoshDunson2009}
Ghosh, J. and Dunson, D.~B. (2009).
\newblock Default prior distributions and efficient posterior computation in
  {B}ayesian factor analysis.
\newblock {\em Journal of Computational and Graphical Statistics},
  18(2):306--320.

\bibitem[\protect\astroncite{Giordano et~al.}{2018}]{Giordanoetal2018}
Giordano, R., Broderick, T., and Jordan, M.~I. (2018).
\newblock Covariances, robustness, and variational {B}ayes.
\newblock {\em Journal of Machine Learning Research}, 19(51):1--49.

\bibitem[\protect\astroncite{{Girolami}}{2001}]{Girolami2001}
{Girolami}, M. (2001).
\newblock A variational method for learning sparse and overcomplete
  representations.
\newblock {\em Neural Computation}, 13(11):2517--2532.

\bibitem[\protect\astroncite{Goutis and Robert}{1998}]{GoutisRobert1998}
Goutis, C. and Robert, C.~P. (1998).
\newblock Model choice in generalised linear models: {A} {B}ayesian approach
  via {K}ullback-{L}eibler projections.
\newblock {\em Biometrika}, 85(1):29--37.

\bibitem[\protect\astroncite{Griffin and Brown}{2010}]{GriffinBrown2010}
Griffin, J.~E. and Brown, P.~J. (2010).
\newblock Inference with normal-gamma prior distributions in regression
  problems.
\newblock {\em {B}ayesian Analysis}, 5(1):171--188.

\bibitem[\protect\astroncite{Griffin and Brown}{2011}]{GriffinBrown2011}
Griffin, J.~E. and Brown, P.~J. (2011).
\newblock {B}ayesian hyper-lassos with non-convex penalization.
\newblock {\em Australian \& New Zealand Journal of Statistics},
  53(4):423--442.

\bibitem[\protect\astroncite{Griffin and Brown}{2017}]{GriffinBrown2017}
Griffin, J.~E. and Brown, P.~J. (2017).
\newblock Hierarchical shrinkage priors for regression models.
\newblock {\em {B}ayesian Analysis}, 12(1):135--159.

\bibitem[\protect\astroncite{Hahn et~al.}{2018}]{Hahnetal2018}
Hahn, P.~R., Carvalho, C.~M., Puelz, D., and He, J. (2018).
\newblock Regularization and confounding in linear regression for treatment
  effect estimation.
\newblock {\em {B}ayesian Analysis}, 13(1):163--182.

\bibitem[\protect\astroncite{Hahn et~al.}{2020}]{Hahnetal2020}
Hahn, P.~R., Murray, J.~S., and Carvalho, C.~M. (2020).
\newblock {B}ayesian regression tree models for causal inference:
  {R}egularization, confounding, and heterogeneous effects (with discussion).
\newblock {\em {B}ayesian Analysis}, 15(3):965 -- 1056.

\bibitem[\protect\astroncite{Hans}{2009}]{Hans2009}
Hans, C. (2009).
\newblock {B}ayesian lasso regression.
\newblock {\em Biometrika}, 96(4):835--845.

\bibitem[\protect\astroncite{Hans et~al.}{2007}]{Hansetal2007}
Hans, C., Dobra, A., and West, M. (2007).
\newblock Shotgun stochastic search for “large p” regression.
\newblock {\em Journal of the American Statistical Association},
  102(478):507--516.

\bibitem[\protect\astroncite{Hill et~al.}{2020}]{Hilletal2020}
Hill, J., Linero, A., and Murray, J. (2020).
\newblock {B}ayesian additive regression trees: {A} review and look forward.
\newblock {\em Annual Review of Statistics and Its Application}, 7(1):251--278.

\bibitem[\protect\astroncite{Hoeting et~al.}{1999}]{Hoetingetal1999}
Hoeting, J.~A., Madigan, D., Raftery, A.~E., and Volinsky, C.~T. (1999).
\newblock {B}ayesian model averaging: {A} tutorial.
\newblock {\em Statistical Science}, 14(4):382--401.

\bibitem[\protect\astroncite{Ibrahim and Laud}{1994}]{IbrahimLaud1994}
Ibrahim, J.~G. and Laud, P.~W. (1994).
\newblock A predictive approach to the analysis of designed experiments.
\newblock {\em Journal of the American Statistical Association},
  89(425):309--319.

\bibitem[\protect\astroncite{Irie}{2019}]{Irie2019}
Irie, K. (2019).
\newblock Bayesian dynamic fused lasso.
\newblock {\em arXiv preprint arXiv:1905.12275}.

\bibitem[\protect\astroncite{Ishwaran and Rao}{2003}]{IshwaranRao2003}
Ishwaran, H. and Rao, J.~S. (2003).
\newblock Detecting differentially expressed genes in microarrays using
  {B}ayesian model selection.
\newblock {\em Journal of the American Statistical Association},
  98(462):438--455.

\bibitem[\protect\astroncite{Ishwaran and Rao}{2005}]{IshwaranRao2005b}
Ishwaran, H. and Rao, J.~S. (2005).
\newblock Spike and slab variable selection: {F}requentist and {B}ayesian
  strategies.
\newblock {\em The Annals of Statistics}, 33(2):730--773.

\bibitem[\protect\astroncite{Ji and Schmidler}{2013}]{JiSchmidler2013}
Ji, C. and Schmidler, S.~C. (2013).
\newblock Adaptive markov chain monte carlo for {B}ayesian variable selection.
\newblock {\em Journal of Computational and Graphical Statistics},
  22(3):708--728.

\bibitem[\protect\astroncite{Jiang}{2006}]{Jiang2006}
Jiang, W. (2006).
\newblock On the consistency of {B}ayesian variable selection for high
  dimensional binary regression and classification.
\newblock {\em Neural Computation}, 18(11):2762--2776.

\bibitem[\protect\astroncite{Johndrow et~al.}{2020}]{Johndrowetal2020}
Johndrow, J., Orenstein, P., and Bhattacharya, A. (2020).
\newblock Scalable approximate mcmc algorithms for the horseshoe prior.
\newblock {\em Journal of Machine Learning Research}, 21(73):1--61.

\bibitem[\protect\astroncite{Johnson and Rossell}{2010}]{JohnsonRossell2010}
Johnson, V.~E. and Rossell, D. (2010).
\newblock On the use of non-local prior densities in {B}ayesian hvoothesis
  tests hypothesis.
\newblock {\em Journal of the Royal Statistical Society. Series B (Statistical
  Methodology)}, 72(2):143--170.

\bibitem[\protect\astroncite{Johnson and Rossell}{2012}]{JohnsonRossell2012}
Johnson, V.~E. and Rossell, D. (2012).
\newblock {B}ayesian model selection in high-dimensional settings.
\newblock {\em Journal of the American Statistical Association},
  107(498):649--660.

\bibitem[\protect\astroncite{Johnstone and
  Silverman}{2004}]{JohnstoneSilverman2004}
Johnstone, I.~M. and Silverman, B.~W. (2004).
\newblock Needles and straw in haystacks: {E}mpirical {B}ayes estimates of
  possibly sparse sequences.
\newblock {\em The Annals of Statistics}, 32(4):1594--1649.

\bibitem[\protect\astroncite{Judge et~al.}{1985}]{judgeetal1985}
Judge, G.~G., Griffith, W.~E., Hill, R.~C., L\"{u}tkepohl, H., and Lee, T.-C.
  (1985).
\newblock {\em The theory and practice of econometrics}.
\newblock Wiley, New York.

\bibitem[\protect\astroncite{Kadane and Lazar}{2004}]{KadaneLazar2004}
Kadane, J.~B. and Lazar, N.~A. (2004).
\newblock Methods and criteria for model selection.
\newblock {\em Journal of the American Statistical Association},
  99(465):279--290.

\bibitem[\protect\astroncite{Kahn and Raftery}{1992}]{KahnRaftery1992}
Kahn, M.~J. and Raftery, A.~E. (1992).
\newblock Fast exact {B}ayesian inference for the hierarchical normal model:
  {S}olving the improper posterior.
\newblock Technical report, University of Washington.

\bibitem[\protect\astroncite{Kalli and Griffin}{2014}]{KalliGriffin2014}
Kalli, M. and Griffin, J.~E. (2014).
\newblock Time-varying sparsity in dynamic regression models.
\newblock {\em Journal of Econometrics}, 178(2):779 -- 793.

\bibitem[\protect\astroncite{Kass and Raftery}{1995}]{KassRaftery1995}
Kass, R.~E. and Raftery, A.~E. (1995).
\newblock {B}ayes factors.
\newblock {\em Journal of the American Statistical Association},
  90(430):773--795.

\bibitem[\protect\astroncite{Kass and Wasserman}{1995}]{KassWasserman1995}
Kass, R.~E. and Wasserman, L. (1995).
\newblock A reference {B}ayesian test for nested hypotheses and its
  relationship to the schwarz criterion.
\newblock {\em Journal of the American Statistical Association},
  90(431):928--934.

\bibitem[\protect\astroncite{Kaufmann and
  Schumacher}{2019}]{KaufmannSchumacher2019}
Kaufmann, S. and Schumacher, C. (2019).
\newblock {B}ayesian estimation of sparse dynamic factor models with
  order-independent and ex-post mode identification.
\newblock {\em Journal of Econometrics}, 210(1):116 -- 134.
\newblock Annals Issue in Honor of John Geweke “Complexity and Big Data in
  Economics and Finance: {R}ecent Developments from a {B}ayesian
  Perspective”.

\bibitem[\protect\astroncite{Khare and Hobert}{2012}]{KhareHobert2012}
Khare, K. and Hobert, J.~P. (2012).
\newblock Geometric ergodicity of the {G}ibbs sampler for {B}ayesian quantile
  regression.
\newblock {\em Journal of Multivariate Analysis}, 112:108 -- 116.

\bibitem[\protect\astroncite{Khare and Hobert}{2013}]{KhareHobert2013}
Khare, K. and Hobert, J.~P. (2013).
\newblock Geometric ergodicity of the {B}ayesian lasso.
\newblock {\em Electron. J. Statist.}, 7:2150--2163.

\bibitem[\protect\astroncite{Kim and Wand}{2016}]{KimWand2016}
Kim, A. S.~I. and Wand, M.~P. (2016).
\newblock The explicit form of expectation propagation for a simple statistical
  model.
\newblock {\em Electronic Journal of Statistics}, 10(1):550--581.

\bibitem[\protect\astroncite{Knowles and
  Ghahramani}{2011}]{KnowlesGhahramani2011}
Knowles, D. and Ghahramani, Z. (2011).
\newblock Nonparametric {B}ayesian sparse factor models with application to
  gene expression modeling.
\newblock {\em The Annals of Applied Statistics}, 5(2B):1534--1552.

\bibitem[\protect\astroncite{Koenker and Bassett}{1978}]{KoenkerBassett1978}
Koenker, R. and Bassett, G. (1978).
\newblock Regression quantiles.
\newblock {\em Econometrica}, 46(1):33--50.

\bibitem[\protect\astroncite{Koop and Korobilis}{2010}]{KoopKorobilis2010}
Koop, G. and Korobilis, D. (2010).
\newblock {B}ayesian multivariate time series methods for empirical
  macroeconomics.
\newblock {\em Foundations and Trends® in Econometrics}, 3(4):267--358.

\bibitem[\protect\astroncite{Koop and Korobilis}{2012}]{KoopKorobilis2012}
Koop, G. and Korobilis, D. (2012).
\newblock Forecasting inflation using dynamic model averaging.
\newblock {\em International Economic Review}, 53(3):867--886.

\bibitem[\protect\astroncite{Koop and Korobilis}{2016}]{KoopKorobilis2016}
Koop, G. and Korobilis, D. (2016).
\newblock Model uncertainty in panel vector autoregressive models.
\newblock {\em European Economic Review}, 81:115--131.

\bibitem[\protect\astroncite{Koop and Korobilis}{2018}]{KoopKorobilis2018}
Koop, G. and Korobilis, D. (2018).
\newblock {B}ayesian dynamic variable selection in high dimensions.
\newblock Technical Report arXiv:1809.03031, ArXiV.

\bibitem[\protect\astroncite{Koop et~al.}{2019}]{Koopetal2019}
Koop, G., Korobilis, D., and Pettenuzzo, D. (2019).
\newblock Bayesian compressed vector autoregressions.
\newblock {\em Journal of Econometrics}, 210(1):135--154.

\bibitem[\protect\astroncite{Korobilis}{2013a}]{Korobilis2013a}
Korobilis, D. (2013a).
\newblock {B}ayesian forecasting with highly correlated predictors.
\newblock {\em Economics Letters}, 118(1):148 -- 150.

\bibitem[\protect\astroncite{Korobilis}{2013b}]{Korobilis2013b}
Korobilis, D. (2013b).
\newblock {VAR} forecasting using {B}ayesian variable selection.
\newblock {\em Journal of Applied Econometrics}, 28(2):204--230.

\bibitem[\protect\astroncite{Korobilis}{2016}]{Korobilis2016}
Korobilis, D. (2016).
\newblock Prior selection for panel vector autoregressions.
\newblock {\em Computational Statistics \& Data Analysis}, 101:110 -- 120.

\bibitem[\protect\astroncite{Korobilis}{2017}]{Korobilis2017}
Korobilis, D. (2017).
\newblock Quantile regression forecasts of inflation under model uncertainty.
\newblock {\em International Journal of Forecasting}, 33(1):11--20.

\bibitem[\protect\astroncite{Korobilis}{2020}]{Korobilis2020}
Korobilis, D. (2020).
\newblock Sign restrictions in high-dimensional vector autoregressions.
\newblock Working Paper series 20-09, Rimini Centre for Economic Analysis.

\bibitem[\protect\astroncite{Korobilis}{2021}]{Korobilis2021}
Korobilis, D. (2021).
\newblock High-dimensional macroeconomic forecasting using message passing
  algorithms.
\newblock {\em Journal of Business \& Economic Statistics}, 39(2):493--504.

\bibitem[\protect\astroncite{Korobilis
  et~al.}{2021}]{RePEc:ecb:ecbwps:20212600}
Korobilis, D., Landau, B., Musso, A., and Phella, A. (2021).
\newblock The time-varying evolution of inflation risks.
\newblock Working Paper Series 2600, European Central Bank.

\bibitem[\protect\astroncite{Korobilis and
  Pettenuzzo}{2019}]{KorobilisPettenuzzo2019}
Korobilis, D. and Pettenuzzo, D. (2019).
\newblock Adaptive hierarchical priors for high-dimensional vector
  autoregressions.
\newblock {\em Journal of Econometrics}, 212(1):241 -- 271.

\bibitem[\protect\astroncite{Korobilis and
  Pettenuzzo}{2020}]{KorobilisPettenuzzo2020}
Korobilis, D. and Pettenuzzo, D. (2020).
\newblock Machine learning econometrics: {B}ayesian algorithms and methods.
\newblock {\em Oxford Research Encyclopedia of Economics and Finance}.

\bibitem[\protect\astroncite{Kowal et~al.}{2019}]{Kowaletal2019}
Kowal, D.~R., Matteson, D.~S., and Ruppert, D. (2019).
\newblock Dynamic shrinkage processes.
\newblock {\em Journal of the Royal Statistical Society: Series B (Statistical
  Methodology)}, 81(4):781--804.

\bibitem[\protect\astroncite{Kozumi and Kobayashi}{2011}]{KozumiKobayashi2011}
Kozumi, H. and Kobayashi, G. (2011).
\newblock {G}ibbs sampling methods for {B}ayesian quantile regression.
\newblock {\em Journal of Statistical Computation and Simulation},
  81(11):1565--1578.

\bibitem[\protect\astroncite{Krishna et~al.}{2009}]{Krishnaetal2009}
Krishna, A., Bondell, H.~D., and Ghosh, S.~K. (2009).
\newblock {B}ayesian variable selection using an adaptive powered correlation
  prior.
\newblock {\em Journal of Statistical Planning and Inference}, 139(8):2665 --
  2674.

\bibitem[\protect\astroncite{Kuo and Mallick}{1998}]{KuoMallick1998}
Kuo, L. and Mallick, B. (1998).
\newblock Variable selection for regression models.
\newblock {\em Sankhy\={a}: The Indian Journal of Statistics, Series B
  (1960-2002)}, 60(1):65--81.

\bibitem[\protect\astroncite{Kyung et~al.}{2010}]{Kyungetal2010}
Kyung, M., Gill, J., Ghosh, M., and Casella, G. (2010).
\newblock Penalized regression, standard errors, and {B}ayesian lassos.
\newblock {\em {B}ayesian Analysis}, 5(2):369--411.

\bibitem[\protect\astroncite{Laud and Ibrahim}{1995}]{LaudIbrahim1995}
Laud, P.~W. and Ibrahim, J.~G. (1995).
\newblock Predictive model selection.
\newblock {\em Journal of the Royal Statistical Society. Series B
  (Methodological)}, 57(1):247--262.

\bibitem[\protect\astroncite{Legramanti et~al.}{2020}]{Legramantietal2020}
Legramanti, S., Durante, D., and Dunson, D.~B. (2020).
\newblock {B}ayesian cumulative shrinkage for infinite factorizations.
\newblock {\em Biometrika}, 107(3):745--752.

\bibitem[\protect\astroncite{Leng et~al.}{2014}]{Lengetal2014}
Leng, C., Tran, M.-N., and Nott, D. (2014).
\newblock {B}ayesian adaptive lasso.
\newblock {\em Annals of the Institute of Statistical Mathematics},
  66(2):221--244.

\bibitem[\protect\astroncite{Lewis and Raftery}{1997}]{LewisRaftery1997}
Lewis, S.~M. and Raftery, A.~E. (1997).
\newblock Estimating {B}ayes factors via posterior simulation with the
  {L}aplace-{M}etropolis estimator.
\newblock {\em Journal of the American Statistical Association},
  92(438):648--655.

\bibitem[\protect\astroncite{Li and Pati}{2017}]{LiPati2017}
Li, H. and Pati, D. (2017).
\newblock Variable selection using shrinkage priors.
\newblock {\em Computational Statistics \& Data Analysis}, 107:107--119.

\bibitem[\protect\astroncite{Li and Lin}{2010}]{LiLin2010}
Li, Q. and Lin, N. (2010).
\newblock The {B}ayesian elastic net.
\newblock {\em {B}ayesian Analysis}, 5(1):151--170.

\bibitem[\protect\astroncite{Liang et~al.}{2008}]{Liangetal2008}
Liang, F., Paulo, R., Molina, G., Clyde, M.~A., and Berger, J.~O. (2008).
\newblock Mixtures of g priors for {B}ayesian variable selection.
\newblock {\em Journal of the American Statistical Association},
  103(481):410--423.

\bibitem[\protect\astroncite{Lim et~al.}{2020}]{Limetal2020}
Lim, D., Park, B., Nott, D., Wang, X., and Choi, T. (2020).
\newblock Sparse signal shrinkage and outlier detection in high-dimensional
  quantile regression with variational {B}ayes.
\newblock {\em Statistics and Its Interface}, 13(2):237--249.

\bibitem[\protect\astroncite{Lindley}{1983}]{Lindley1983}
Lindley, D.~V. (1983).
\newblock Parametric empirical {B}ayes inference: {T}heory and applications:
  Comment.
\newblock {\em Journal of the American Statistical Association},
  78(381):61--62.

\bibitem[\protect\astroncite{Liu et~al.}{2019}]{Liuetal2019}
Liu, Y., Ro\v{c}kov\'{a}, V., and Wang, Y. (2019).
\newblock Variable selection with abc {B}ayesian forests.
\newblock Technical Report arXiv:1806.02304v2, ArXiV.

\bibitem[\protect\astroncite{Lopes and West}{2004}]{LopesWest2004}
Lopes, H.~F. and West, M. (2004).
\newblock {B}ayesian model assessment in factor analysis.
\newblock {\em Statistica Sinica}, 14(1):41--67.

\bibitem[\protect\astroncite{Madigan et~al.}{1995}]{MadiganYork1995}
Madigan, D., York, J., and Allard, D. (1995).
\newblock {B}ayesian graphical models for discrete data.
\newblock {\em International Statistical Review / Revue Internationale de
  Statistique}, 63(2):215--232.

\bibitem[\protect\astroncite{{Makalic} and
  {Schmidt}}{2016}]{MakalicSchmidt2016}
{Makalic}, E. and {Schmidt}, D.~F. (2016).
\newblock A simple sampler for the horseshoe estimator.
\newblock {\em IEEE Signal Processing Letters}, 23(1):179--182.

\bibitem[\protect\astroncite{Mallick and Yi}{2014}]{MallickYi2014}
Mallick, H. and Yi, N. (2014).
\newblock A new {B}ayesian lasso.
\newblock {\em Statistics and its Interface}, 7(4):571–582.

\bibitem[\protect\astroncite{Martini and
  Spezzaferri}{1984}]{SanMartiniSpezzaferri1984}
Martini, A.~S. and Spezzaferri, F. (1984).
\newblock A predictive model selection criterion.
\newblock {\em Journal of the Royal Statistical Society. Series B
  (Methodological)}, 46(2):296--303.

\bibitem[\protect\astroncite{Matusevich et~al.}{2016}]{Matusevichetal2016}
Matusevich, D.~S., Cabrera, W., and Ordonez, C. (2016).
\newblock Accelerating a {G}ibbs sampler for variable selection on genomics
  data with summarization and variable pre-selection combining an array {DBMS}
  and {R}.
\newblock {\em Machine Learning}, 102(3):483--504.

\bibitem[\protect\astroncite{Mitchell and
  Beauchamp}{1988}]{MitchellBeauchamp1988}
Mitchell, T.~J. and Beauchamp, J.~J. (1988).
\newblock {B}ayesian variable selection in linear regression.
\newblock {\em Journal of the American Statistical Association},
  83(404):1023--1032.

\bibitem[\protect\astroncite{Moran et~al.}{2019}]{Moranetal2019}
Moran, G.~E., Ročková, V., and George, E.~I. (2019).
\newblock Variance prior forms for high-dimensional {B}ayesian variable
  selection.
\newblock {\em {B}ayesian Analysis}, 14(4):1091--1119.

\bibitem[\protect\astroncite{Nakajima and West}{2013a}]{NakajimaWest2013}
Nakajima, J. and West, M. (2013a).
\newblock {B}ayesian analysis of latent threshold dynamic models.
\newblock {\em Journal of Business \& Economic Statistics}, 31(2):151--164.

\bibitem[\protect\astroncite{Nakajima and West}{2013b}]{NakajimaWest2013JFE}
Nakajima, J. and West, M. (2013b).
\newblock Bayesian dynamic factor models: {L}atent threshold approach.
\newblock {\em Journal of Financial Econometrics}, 11:116--153.

\bibitem[\protect\astroncite{Nakajima and West}{2015}]{NakajimaWest2015DSP}
Nakajima, J. and West, M. (2015).
\newblock Dynamic network signal processing using latent threshold models.
\newblock {\em Digital Signal Processing}, 47:6--15.
\newblock https://doi.org/10.1016/j.dsp.2015.04.008.

\bibitem[\protect\astroncite{Nakajima and West}{2017}]{NakajimaWest2017BJPS}
Nakajima, J. and West, M. (2017).
\newblock Dynamics and sparsity in latent threshold factor models: {A} study in
  multivariate {EEG} signal processing.
\newblock {\em Brazilian Journal of Probability and Statistics}, 31:701--731.

\bibitem[\protect\astroncite{Narisetty and He}{2014}]{NarisettyHe2014}
Narisetty, N.~N. and He, X. (2014).
\newblock {B}ayesian variable selection with shrinking and diffusing priors.
\newblock {\em The Annals of Statistics}, 42(2):789--817.

\bibitem[\protect\astroncite{Narisetty et~al.}{2018}]{Narisettyetal2018}
Narisetty, N.~N., Shen, J., and He, X. (2018).
\newblock Skinny {G}ibbs: {A} consistent and scalable {G}ibbs sampler for model
  selection.
\newblock {\em Journal of the American Statistical Association}, 0(0):1--13.

\bibitem[\protect\astroncite{Neville et~al.}{2014}]{Nevilleetal2014}
Neville, S.~E., Ormerod, J.~T., and Wand, M.~P. (2014).
\newblock Mean field variational {B}ayes for continuous sparse signal
  shrinkage: {P}itfalls and remedies.
\newblock {\em Electronic Journal of Statistics}, 8(1):1113--1151.

\bibitem[\protect\astroncite{Nott and Kohn}{2005}]{NottKohn2005}
Nott, D.~J. and Kohn, R. (2005).
\newblock Adaptive sampling for {B}ayesian variable selection.
\newblock {\em Biometrika}, 92(4):747--763.

\bibitem[\protect\astroncite{O'Hagan}{1995}]{OHagan1995}
O'Hagan, A. (1995).
\newblock Fractional {B}ayes factors for model comparison.
\newblock {\em Journal of the Royal Statistical Society. Series B
  (Methodological)}, 57(1):99--138.

\bibitem[\protect\astroncite{O'Hara and
  Sillanp\"{a}\"{a}}{2009}]{OHaraSilanpaa2009}
O'Hara, R.~B. and Sillanp\"{a}\"{a}, M.~J. (2009).
\newblock A review of {B}ayesian variable selection methods: {W}hat, how and
  which.
\newblock {\em {B}ayesian Analysis}, 4(1):85--117.

\bibitem[\protect\astroncite{Ormerod et~al.}{2017}]{Ormerodetal2017}
Ormerod, J.~T., You, C., and M\"{u}ller, S. (2017).
\newblock A variational {B}ayes approach to variable selection.
\newblock {\em Electronic Journal of Statistics}, 11(2):3549--3594.

\bibitem[\protect\astroncite{Pal and Khare}{2014}]{PalKhare2014}
Pal, S. and Khare, K. (2014).
\newblock Geometric ergodicity for {B}ayesian shrinkage models.
\newblock {\em Electronic Journal of Statistics}, 8(1):604 -- 645.

\bibitem[\protect\astroncite{Pal et~al.}{2017}]{Paletal2017}
Pal, S., Khare, K., and Hobert, J.~P. (2017).
\newblock Trace class {M}arkov chains for {B}ayesian inference with generalized
  double {P}areto shrinkage priors.
\newblock {\em Scandinavian Journal of Statistics}, 44(2):307--323.

\bibitem[\protect\astroncite{Papaspiliopoulos and
  Rossell}{2017}]{PapaspiliopoulosRossell2017}
Papaspiliopoulos, O. and Rossell, D. (2017).
\newblock {B}ayesian block-diagonal variable selection and model averaging.
\newblock {\em Biometrika}, 104(2):343--359.

\bibitem[\protect\astroncite{Park and Casella}{2008}]{ParkCasella2008}
Park, T. and Casella, G. (2008).
\newblock The {B}ayesian lasso.
\newblock {\em Journal of the American Statistical Association},
  103(482):681--686.

\bibitem[\protect\astroncite{Pati et~al.}{2014}]{Patietal2014}
Pati, D., Bhattacharya, A., Pillai, N.~S., and Dunson, D. (2014).
\newblock Posterior contraction in sparse {B}ayesian factor models for massive
  covariance matrices.
\newblock {\em The Annals of Statistics}, 42(3):1102--1130.

\bibitem[\protect\astroncite{Peltola et~al.}{2012}]{Peltolaetal2012b}
Peltola, T., Marttinen, P., and Vehtari, A. (2012).
\newblock Finite adaptation and multistep moves in the metropolis-hastings
  algorithm for variable selection in genome-wide association analysis.
\newblock {\em PLOS ONE}, 7(11):1--11.

\bibitem[\protect\astroncite{Polson and Scott}{2010}]{PolsonScott2010}
Polson, N.~G. and Scott, J.~G. (2010).
\newblock Shrink globally, act locally: {S}parse {B}ayesian regularization and
  prediction.
\newblock In Bernardo, J., Bayarri, M., Berger, J., Dawid, A., Heckerman, D.,
  Smith, A., and West, M., editors, {\em {B}ayesian Statistics 9}. Oxford
  University Press.

\bibitem[\protect\astroncite{Raftery}{1995}]{Raftery1995}
Raftery, A.~E. (1995).
\newblock {B}ayesian model selection in social research.
\newblock {\em Sociological Methodology}, 25:111--163.

\bibitem[\protect\astroncite{Raftery}{1996}]{Raftery1996}
Raftery, A.~E. (1996).
\newblock {Approximate {B}ayes factors and accounting for model uncertainty in
  generalised linear models}.
\newblock {\em Biometrika}, 83(2):251--266.

\bibitem[\protect\astroncite{Rajaratnam et~al.}{2019}]{Rajaratnam2019}
Rajaratnam, B., Sparks, D., Khare, K., and Zhang, L. (2019).
\newblock Uncertainty quantification for modern high-dimensional regression via
  scalable {B}ayesian methods.
\newblock {\em Journal of Computational and Graphical Statistics},
  28(1):174--184.

\bibitem[\protect\astroncite{Robert}{2007}]{Robert2007}
Robert, C. (2007).
\newblock {\em The {B}ayesian Choice: {F}rom Decision-Theoretic Foundations to
  Computational Implementation}, volume Springer Texts in Statistics.
\newblock Springer-Verlag New York.

\bibitem[\protect\astroncite{Rodrigues and Fan}{2017}]{RodriguesFan2017}
Rodrigues, T. and Fan, Y. (2017).
\newblock Regression adjustment for noncrossing {B}ayesian quantile regression.
\newblock {\em Journal of Computational and Graphical Statistics},
  26(2):275--284.

\bibitem[\protect\astroncite{Ro\v{c}kov\'{a} and
  George}{2014}]{RockovaGeorge2014}
Ro\v{c}kov\'{a}, V. and George, E.~I. (2014).
\newblock {EMVS}: {T}he {EM} approach to {B}ayesian variable selection.
\newblock {\em Journal of the American Statistical Association},
  109(506):828--846.

\bibitem[\protect\astroncite{Ro\v{c}kov\'{a} and
  George}{2016}]{RockovaGeorge2016}
Ro\v{c}kov\'{a}, V. and George, E.~I. (2016).
\newblock Fast {B}ayesian factor analysis via automatic rotations to sparsity.
\newblock {\em Journal of the American Statistical Association},
  111(516):1608--1622.

\bibitem[\protect\astroncite{Ro\v{c}kov\'{a} and
  George}{2018}]{RockovaGeorge2018}
Ro\v{c}kov\'{a}, V. and George, E.~I. (2018).
\newblock The spike-and-slab lasso.
\newblock {\em Journal of the American Statistical Association},
  113(521):431--444.

\bibitem[\protect\astroncite{Ro\v{c}kov\'{a} and
  McAlinn}{2017}]{RockovaMcAlinn2017}
Ro\v{c}kov\'{a}, V. and McAlinn, K. (2017).
\newblock Dynamic variable selection with spike-and-slab process priors.
\newblock Technical Report arXiv:1708.00085v2, ArXiV.

\bibitem[\protect\astroncite{Rue}{2001}]{Rue2001}
Rue, H. (2001).
\newblock Fast sampling of {G}aussian markov random fields.
\newblock {\em Journal of the Royal Statistical Society. Series B (Statistical
  Methodology)}, 63(2):325--338.

\bibitem[\protect\astroncite{Shin et~al.}{2018}]{Shinetal2018}
Shin, M., Bhattacharya, A., and Johnson, V.~E. (2018).
\newblock Scalable {B}ayesian variable selection using nonlocal prior densities
  in ultrahigh-dimensional settings.
\newblock {\em Statistica Sinica}, 28(2):1053--1078.

\bibitem[\protect\astroncite{Smith and Kohn}{2002}]{SmithKohn2002}
Smith, M. and Kohn, R. (2002).
\newblock Parsimonious covariance matrix estimation for longitudinal data.
\newblock {\em Journal of the American Statistical Association},
  97(460):1141--1153.

\bibitem[\protect\astroncite{Spiegelhalter
  et~al.}{2002}]{Spiegelhalteretal2002}
Spiegelhalter, D.~J., Best, N.~G., Carlin, B.~P., and Van Der~Linde, A. (2002).
\newblock {B}ayesian measures of model complexity and fit.
\newblock {\em Journal of the Royal Statistical Society: Series B (Statistical
  Methodology)}, 64(4):583--639.

\bibitem[\protect\astroncite{Spiegelhalter
  et~al.}{2014}]{Spiegelhalteretal2014}
Spiegelhalter, D.~J., Best, N.~G., Carlin, B.~P., and van~der Linde, A. (2014).
\newblock The deviance information criterion: 12 years on.
\newblock {\em Journal of the Royal Statistical Society: Series B (Statistical
  Methodology)}, 76(3):485--493.

\bibitem[\protect\astroncite{Srivastava et~al.}{2017}]{Srivastavaetal2017}
Srivastava, S., Engelhardt, B.~E., and Dunson, D.~B. (2017).
\newblock Expandable factor analysis.
\newblock {\em Biometrika}, 104(3):649--663.

\bibitem[\protect\astroncite{Stein}{1956}]{Stein1956}
Stein, C. (1956).
\newblock Inadmissibility of the usual estimator for the mean of a multivariate
  normal distribution.
\newblock In Neyman, J., editor, {\em Berkeley Symposium on Mathematical
  Statistics and Probability}, pages 197--206. University of California Press.

\bibitem[\protect\astroncite{Tibshirani}{1996}]{Tibshirani1996}
Tibshirani, R. (1996).
\newblock Regression shrinkage and selection via the lasso.
\newblock {\em Journal of the Royal Statistical Society. Series B
  (Methodological)}, 58(1):267--288.

\bibitem[\protect\astroncite{Tipping}{2001}]{Tipping2001}
Tipping, M.~E. (2001).
\newblock Sparse {B}ayesian learning and the relevance vector machine.
\newblock {\em Journal of Machine Learning Research}, 1:211--244.

\bibitem[\protect\astroncite{Uribe and Lopes}{2017}]{UribeLopes2017}
Uribe, P. and Lopes, H. (2017).
\newblock Dynamic sparsity on dynamic regression models.
\newblock Technical report, Available at
  \href{http://hedibert.org/wp-content/uploads/2018/06/uribe-lopes-Sep2017.pdf}{http://hedibert.org/wp-content/uploads/2018/06/uribe-lopes-Sep2017.pdf}.

\bibitem[\protect\astroncite{{van den Boom}
  et~al.}{2015a}]{vandenBoometal2015a}
{van den Boom}, W., {Dunson}, D., and {Reeves}, G. (2015a).
\newblock Quantifying uncertainty in variable selection with arbitrary
  matrices.
\newblock In {\em 2015 IEEE 6th International Workshop on Computational
  Advances in Multi-Sensor Adaptive Processing (CAMSAP)}, pages 385--388.

\bibitem[\protect\astroncite{{van den Boom}
  et~al.}{2015b}]{vandenBoometal2015b}
{van den Boom}, W., {Dunson}, D., and {Reeves}, G. (2015b).
\newblock Scalable approximations of marginal posteriors in variable selection.
\newblock Technical Report arXiv:1506.06629v1, ArXiV.

\bibitem[\protect\astroncite{van~der Linde}{2005}]{vanderLinde2005}
van~der Linde, A. (2005).
\newblock {DIC} in variable selection.
\newblock {\em Statistica Neerlandica}, 59(1):45--56.

\bibitem[\protect\astroncite{van~der Pas et~al.}{2014}]{vanderPasetal2014}
van~der Pas, S.~L., Kleijn, B. J.~K., and van~der Vaart, A.~W. (2014).
\newblock The horseshoe estimator: {P}osterior concentration around nearly
  black vectors.
\newblock {\em Electronic Journal of Statistics}, 8(2):2585--2618.

\bibitem[\protect\astroncite{Vehtari et~al.}{2017}]{Vehtarietal2017}
Vehtari, A., Gelman, A., and Gabry, J. (2017).
\newblock Practical {B}ayesian model evaluation using leave-one-out
  cross-validation and {WAIC}.
\newblock {\em Statistics and Computing}, 27(5):1413--1432.

\bibitem[\protect\astroncite{Verdinelli and
  Wasserman}{1995}]{VerdinelliWasserman1995}
Verdinelli, I. and Wasserman, L. (1995).
\newblock Computing {B}ayes factors using a generalization of the
  {S}avage-{D}ickey density ratio.
\newblock {\em Journal of the American Statistical Association},
  90(430):614--618.

\bibitem[\protect\astroncite{Volinsky and Raftery}{2000}]{VolinskyRaftery2000}
Volinsky, C.~T. and Raftery, A.~E. (2000).
\newblock {B}ayesian information criterion for censored survival models.
\newblock {\em Biometrics}, 56(1):256--262.

\bibitem[\protect\astroncite{Wainwright and
  Jordan}{2008}]{WainwrightJordan2008}
Wainwright, M.~J. and Jordan, M.~I. (2008).
\newblock Graphical models, exponential families, and variational inference.
\newblock {\em Foundations and Trends® in Machine Learning}, 1(1–2):1--305.

\bibitem[\protect\astroncite{Wang and Pillai}{2013}]{WangPillai2013}
Wang, H. and Pillai, N.~S. (2013).
\newblock On a class of shrinkage priors for covariance matrix estimation.
\newblock {\em Journal of Computational and Graphical Statistics},
  22(3):689--707.

\bibitem[\protect\astroncite{Wang and Blei}{2019}]{WangBlei2019}
Wang, Y. and Blei, D.~M. (2019).
\newblock Frequentist consistency of variational {B}ayes.
\newblock {\em Journal of the American Statistical Association},
  114(527):1147--1161.

\bibitem[\protect\astroncite{Watanabe}{2010}]{Watanabe2010}
Watanabe, S. (2010).
\newblock Asymptotic equivalence of {B}ayes cross validation and widely
  applicable information criterion in singular learning theory.
\newblock {\em Journal of Machine Learning Research}, 11:3571--3594.

\bibitem[\protect\astroncite{Watanabe}{2013}]{Watanabe2013}
Watanabe, S. (2013).
\newblock A widely applicable {B}ayesian information criterion.
\newblock {\em Journal of Machine Learning Research}, 14(1):867--897.

\bibitem[\protect\astroncite{West}{2003}]{West2003}
West, M. (2003).
\newblock {B}ayesian factor regression models in the ``large p, small n''
  paradigm.
\newblock In Bernardo, J., Bayarri, M., Berger, J., Dawid, A., Heckerman, D.,
  Smith, A., and West, M., editors, {\em {B}ayesian Statistics 7}, pages
  723--732. Oxford University Press.

\bibitem[\protect\astroncite{West and Harrison}{1997}]{West1997}
West, M. and Harrison, J. (1997).
\newblock {\em {B}ayesian Forecasting and Dynamic Models}, volume Springer
  Series in Statistics.
\newblock Springer-Verlag New York.

\bibitem[\protect\astroncite{Yu et~al.}{2013}]{Yuetal2013}
Yu, K., Chen, C., Reed, C., and Dunson, D. (2013).
\newblock {B}ayesian variable selection in quantile regression.
\newblock {\em Statistics and its Interface}, 6(2):261--274.
\newblock cited By 22.

\bibitem[\protect\astroncite{Yu and Moyeed}{2001}]{YuMoyeed2001}
Yu, K. and Moyeed, R.~A. (2001).
\newblock {B}ayesian quantile regression.
\newblock {\em Statistics \& Probability Letters}, 54(4):437 -- 447.

\bibitem[\protect\astroncite{Yuan and Lin}{2005}]{YuanLin2005}
Yuan, M. and Lin, Y. (2005).
\newblock Efficient empirical {B}ayes variable selection and estimation in
  linear models.
\newblock {\em Journal of the American Statistical Association},
  100(472):1215--1225.

\bibitem[\protect\astroncite{Zellner}{1986}]{Zellner1986}
Zellner, A. (1986).
\newblock On assessing prior distributions and {B}ayesian regression analysis
  with g-prior distributions.
\newblock In Goel, P. and Zellner, A., editors, {\em {B}ayesian Inference and
  Decision Techniques: Essays in Honor of Bruno de Finetti}, pages 233--243,
  New York. Elsevier Science Publishers, Inc.

\bibitem[\protect\astroncite{Zhang and Bondell}{2018}]{ZhangBondell2018}
Zhang, Y. and Bondell, H.~D. (2018).
\newblock Variable selection via penalized credible regions with
  {D}irichlet–{L}aplace global-local shrinkage priors.
\newblock {\em {B}ayesian Analysis}, 13(3):823 -- 844.

\bibitem[\protect\astroncite{Ziniel and Schniter}{2013}]{ZinielSchniter13}
Ziniel, J. and Schniter, P. (2013).
\newblock Dynamic compressive sensing of time-varying signals via approximate
  message passing.
\newblock {\em IEEE Transactions on Signal Processing}, 61(21):5270--5284.

\bibitem[\protect\astroncite{{Zou} et~al.}{2016}]{Zouetal2016}
{Zou}, X., {Li}, F., {Fang}, J., and {Li}, H. (2016).
\newblock Computationally efficient sparse {B}ayesian learning via generalized
  approximate message passing.
\newblock In {\em 2016 IEEE International Conference on Ubiquitous Wireless
  Broadband (ICUWB)}, pages 1--4.

\end{thebibliography}

\newpage

\begin{appendix}
\begin{center}
{\Large Technical Document to accompany ``Bayesian Approaches to Shrinkage and Sparse Estimation: A guide for applied econometricians''}
\newline
\newline
\begin{tabular}{c}
\large{\begin{tabular}{cccc}
Dimitris Korobilis & & & Kenichi Shimizu \\  University of Glasgow & & & University of Glasgow
\end{tabular}}
\end{tabular}
\newline
\newline
\newline
\end{center}

\setcounter{page}{1}
\renewcommand{\theequation}{A.\arabic{equation}} \setcounter{equation}{0} %
\renewcommand{\thetable}{A\arabic{table}} \setcounter{table}{0}
\renewcommand{\thefigure}{A\arabic{figure}} \setcounter{figure}{0}
\setcounter{footnote}{0}

\section{Inference with non-hierarchical natural conjugate and independent priors}
In this section, we review non-hierarchical Bayesian estimation of  simple regression models under natural conjugate and independent priors. 
Most of the shrinkage priors that we review in this paper have forms of either  conjugate or  independent prior, conditional on the parameters such as prior variances of the slope coefficients.
Therefore, it is helpful to first review the conditional posterior distributions under the non-hierarchical priors.

Consider the simple linear regression model of the form
\begin{equation}
y_{i} = x_{i} \bm \beta + \varepsilon_{i}, \text{ \ } \varepsilon_{i} \sim N(0,\sigma^{2}),  \text{ \ } i=1,...,n
\end{equation}
where $\bm \beta$ is a $p \times 1$ vector. We define $\bm y = (y_{1},...,y_{n})^{\prime}$, $\bm X = (x_{1}^{\prime},...,x_{n}^{\prime})^{\prime}$ and $\bm \varepsilon = (\varepsilon_{1},...,\varepsilon_{n})^{\prime}$, such that the stacked form of the regression model is
\begin{equation}
\bm y = \bm X \bm \beta + \bm \varepsilon,
\end{equation}
where $\bm \varepsilon \sim N_{n}(\bm 0_{n \times 1}, \sigma^{2} \bm I_{n} )$.
In this section, we assume that the prior variances on $\bm \beta$ are fixed and will review posterior sampling under generic normal-inverse-gamma priors on 
$\bm \beta$ and $\sigma^2$. The prior of $\bm \beta$ can be defined either dependent or independent on $\sigma^2$. 
In both cases, assume an inverse gamma prior\footnote{The conditional posteriors under the improper prior $\sigma \sim \frac{1}{\sigma^2} d\sigma^2$ are similar.} on $\sigma^2$.
\begin{eqnarray}
\sigma^{2} & \sim &  Inv-Gamma \left(a,b\right)
\end{eqnarray}
where we use the parametrization so that if $x \sim  Inv-Gamma \left(a,b \right)$, then it has density $p(x) = \frac{b^a}{\Gamma(a)} \left( \frac{1}{x} \right)^{a+1} \exp\left( -\frac{b}{x} \right)$.

\subsection{Natural conjugate prior}\label{normal_inv_gamma_dep}
In the first case, the prior on $\bm \beta$ is defined conditional on $\sigma^2$. The hierarchical structure is summarized as follows.
\begin{eqnarray}
\bm \beta \vert  \sigma^{2} & \sim & N_{p}\left( \bm \mu_\beta, \sigma^{2} \bm V_\beta \right)
\end{eqnarray}
The conditional posteriors are of the form
{\small
\begin{eqnarray}
\bm \beta \ \vert \ \bullet & \sim & N_{p}\left( \bm V \times \left[  \bm X' \bm y + \bm V_\beta^{-1}\bm \mu_\beta \right] , \sigma^2 \bm V \right), \label{beta_dep}\\
\sigma^2  \ \vert \ \bullet & \sim &   Inv-Gamma 
\left( 
a+\frac{n+p}{2},
b+\frac{1}{2} \left[
 \left( \bm y- \bm X \bm \beta \right)'  \left( \bm y- \bm X \bm \beta \right)  + \left( \bm \beta - \bm \mu_\beta \right) \bm V_\beta^{-1} \left( \bm \beta - \bm \mu_\beta \right) 
 \right] 
 \right)\label{sigma2_dep}
\end{eqnarray}}
where 
$\bm V =  (\bm X' \bm X + \bm V_\beta^{-1})^{-1}$.
and 
$\bullet$ denotes data and all the parameters except for the parameter that is being updated.

\subsubsection*{Derivation}
The joint prior is 
\begin{align*}
p(\bm \beta, \sigma^2) &=
(2\pi)^{-p/2} |\sigma^2 \bm V_\beta|^{-1/2} 
exp\left[ -\frac{1}{2\sigma^2}(\bm \beta-\bm \mu_\beta)'\bm V_\beta^{-1} (\bm \beta-\bm \mu_\beta) \right] 
\frac{b^a}{\Gamma(a)} \left( \frac{1}{\sigma^2} \right)^{a+1} exp\left(-\frac{b}{\sigma^2}\right)\\
&\propto
\left( \frac{1}{\sigma^2} \right)^{a+p/2+1}
exp\left[ -\frac{1}{\sigma^2} \left\{b+\frac{1}{2}(\bm \beta-\bm \mu_\beta)'\bm V_\beta^{-1} (\bm \beta-\bm \mu_\beta) \right\}\right] 
\end{align*}
where the proportionality sign is with respect to the parameters $(\bm \beta, \sigma^2)$.
The likelihood is 
\begin{align*}
p(\bm y \vert \bm \beta, \sigma^2)
(2\pi)^{-n/2}
\left( \frac{1}{\sigma^2} \right)^{n/2}
exp\left[ -\frac{1}{2\sigma^2}(\bm y - \bm X \bm \beta)'(\bm y - \bm X \bm \beta) \right]
\end{align*}
The posterior is 
\begin{align*}
p(\bm \beta, \sigma^2 \vert \bm y) &\propto 
p(\bm y\vert  \bm \beta, \sigma^2 )
p(\bm \beta, \sigma^2)\\
&\propto 
\left( \frac{1}{\sigma^2} \right)^{a+\frac{p+n}{2}+1}
exp\left[ -\frac{1}{\sigma^2} \left\{b+\frac{1}{2}
\left[
(\bm \beta-\bm \mu_\beta)'\bm V_\beta^{-1} (\bm \beta-\bm \mu_\beta)
+
(\bm y - \bm X \bm \beta)'(\bm y - \bm X \bm \beta)
\right]
 \right\}\right] 
\end{align*}
From the right-hand-side above, it is easy to see that the conditional posterior $p(\sigma^2 \vert \bm \beta, \bm y)$ is  of the form \eqref{sigma2_dep}.

To see \eqref{beta_dep}, note that
{\small
\begin{align*}
(\bm \beta-\bm \mu_\beta)'\bm V_\beta^{-1} (\bm \beta-\bm \mu_\beta)
+
(\bm y - \bm X \bm \beta)'(\bm y - \bm X \bm \beta)
&=
\bm \beta' \bm V_\beta^{-1}  \bm \beta
-2 \bm \beta' \bm V_\beta^{-1} \bm \mu_\beta
+\bm \mu_\beta' \bm V_\beta^{-1} \bm \mu_\beta\\
&+
\bm y' \bm y -2\bm \beta' \bm X' \bm y + \bm \beta'  \bm X' \bm X \bm \beta \\
&=
\bm \beta' \left[\bm V_\beta^{-1}+ \bm X \bm X \right]\bm \beta
-2\bm \beta' \left[\bm V_\beta^{-1} \bm \mu_\beta + \bm X' \bm y \right]
+
\left[\bm \mu_\beta \bm V_\beta^{-1} \bm \mu_\beta + \bm y' \bm y \right]\\
&=
(\bm \beta - \bm \mu_*)' \bm V_*^{-1} (\bm \beta - \bm \mu_*)
-\bm \mu_*' \bm V_*^{-1} \bm \mu_* 
+\left[ \bm \mu_\beta' \bm V_\beta^{-1} \bm \mu_\beta + \bm y' \bm y \right]
\end{align*}}
where we used the identity
\begin{align*}
\bm u'\bm A \bm u-2\bm \alpha' \bm u
= (\bm u-\bm A^{-1} \bm \alpha)' \bm A(\bm u-\bm A^{-1} \bm \alpha) -\bm \alpha' \bm A^{-1} \bm \alpha
\end{align*}
in the last equality with 
$\bm u=\bm \beta, 
\bm A=\bm V_\beta^{-1}+ \bm X \bm X$, and 
$\bm \alpha = \bm V_\beta^{-1} \bm \mu_\beta + \bm X' \bm y$
and defined 
\begin{align*}
\bm \mu_* &=
\bm A^{-1} \bm \alpha =
\left[ \bm V_\beta^{-1}+ \bm X \bm X \right]^{-1}\left[ \bm V_\beta^{-1} \bm \mu_\beta + \bm X' \bm y \right]\\
\bm V_*&= \bm A^{-1} 
=\left[ \bm V_\beta^{-1}+ \bm X \bm X \right]^{-1} 
\end{align*}
Hence the posterior is 
\begin{align*}
p(\bm \beta, \sigma^2 \vert \bm y) &\propto 
\left( \frac{1}{\sigma^2} \right)^{a_*+1}
exp\left[ -\frac{1}{\sigma^2} \left\{b_*+\frac{1}{2}
(\bm \beta - \bm \mu_*)' \bm V_*^{-1} (\bm \beta - \bm \mu_*)
 \right\}\right] 
\end{align*}
where $a_*=a+n/2+p/2$
and 
$b_*=b + \frac{1}{2}\left[ \bm \mu_\beta' \bm V_\beta^{-1} \bm \mu_\beta + \bm y' \bm y
-\bm \mu_*' \bm V_*^{-1} \bm \mu_* 
\right]$.
Therefore, the conditional posterior for $\bm \beta$ is of the form \eqref{beta_dep}.

\subsection{Independent prior}\label{normal_inv_gamma_indep}
In this case, $\bm \beta$ and $\sigma^2$ are \textit{a priori} independent. 
\begin{eqnarray}
\bm \beta  & \sim & N_{p}\left( \bm \mu_\beta,  \bm V_\beta \right),
\end{eqnarray}
The conditional posteriors are of the form
\begin{eqnarray}
\bm \beta \ \vert \ \bullet & \sim & N_{p}\left( \bm V\times \left[  \bm X' \bm y / \sigma^2 + \bm V_\beta^{-1}\bm \mu_\beta \right] ,  \bm V\right),\label{beta_indep}\\
\sigma^2  \ \vert \ \bullet & \sim &  Inv-Gamma \left(a+\frac{n}{2},b+ \frac{1}{2}  \left( \bm y- \bm X \bm \beta \right)'  \left( \bm y- \bm X \bm \beta \right) \right)\label{sigma2_indep}
\end{eqnarray}
where 
$\bm V = ( \bm X' \bm X / \sigma^2 + \bm V_\beta^{-1})^{-1}$.

\subsubsection*{Derivation}
The joint prior is 
\begin{align*}
p(\bm \beta, \sigma^2) &=
(2\pi)^{-p/2} |\bm V_\beta|^{-1/2} 
exp\left[ -\frac{1}{2}(\bm \beta-\bm \mu_\beta)'\bm V_\beta^{-1} (\bm \beta-\bm \mu_\beta) \right] 
\frac{b^a}{\Gamma(a)} \left( \frac{1}{\sigma^2} \right)^{a+1} exp\left(-\frac{b}{\sigma^2}\right)\\
&\propto
\left( \frac{1}{\sigma^2} \right)^{a+1}
exp\left[ -\frac{1}{\sigma^2} \left\{b+\frac{1}{2}(\bm \beta-\bm \mu_\beta)'(\bm V_\beta/\sigma^2)^{-1} (\bm \beta-\bm \mu_\beta) \right\}\right] 
\end{align*}
The posterior is 
\begin{align*}
p(\bm \beta, \sigma^2 \vert \bm y) &\propto 
p(\bm y\vert  \bm \beta, \sigma^2 )
p(\bm \beta, \sigma^2)\\
&\propto 
\left( \frac{1}{\sigma^2} \right)^{a+\frac{n}{2}+1}
exp\left[ -\frac{1}{\sigma^2} \left\{b+\frac{1}{2}
\left[
(\bm \beta-\bm \mu_\beta)'(\bm V_\beta/\sigma^2)^{-1} (\bm \beta-\bm \mu_\beta)
+
(\bm y - \bm X \bm \beta)'(\bm y - \bm X \bm \beta)
\right]
 \right\}\right] 
\end{align*}
To see \eqref{sigma2_indep}, note that 
\begin{align*}
p( \sigma^2 \vert \bm\beta, \bm y) &\propto 
\left( \frac{1}{\sigma^2} \right)^{a+\frac{n}{2}+1}
exp\left[ -\frac{1}{\sigma^2} \left\{b+\frac{1}{2}
(\bm y - \bm X \bm \beta)'(\bm y - \bm X \bm \beta)
 \right\}\right] 
\end{align*}
To see \eqref{beta_indep}, note that 
\begin{align*}
&(\bm \beta-\bm \mu_\beta)'(\bm V_\beta/\sigma^2)^{-1} (\bm \beta-\bm \mu_\beta)
+
(\bm y - \bm X \bm \beta)'(\bm y - \bm X \bm \beta)\\
&=
\bm \beta' (\bm V_\beta/\sigma^2)^{-1}  \bm \beta
-2 \bm \beta'(\bm V_\beta/\sigma^2)^{-1} \bm \mu_\beta
+\bm \mu_\beta' (\bm V_\beta/\sigma^2)^{-1} \bm \mu_\beta
+
\bm y' \bm y -2\bm \beta' \bm X' \bm y + \bm \beta'  \bm X' \bm X \bm \beta \\
&=
\bm \beta' \left[(\bm V_\beta/\sigma^2)^{-1}+ \bm X \bm X \right]\bm \beta
-2\bm \beta' \left[(\bm V_\beta/\sigma^2)^{-1}\bm \mu_\beta + \bm X' \bm y \right]
+
\left[\bm \mu_\beta (\bm V_\beta/\sigma^2)^{-1}\bm \mu_\beta + \bm y' \bm y \right]\\
&=
(\bm \beta - \bm \mu_*)' \bm V_*^{-1} (\bm \beta - \bm \mu_*)
-\bm \mu_*' \bm V_*^{-1} \bm \mu_* 
+\left[ \bm \mu_\beta' (\bm V_\beta/\sigma^2)^{-1}  \bm \mu_\beta + \bm y' \bm y \right]
\end{align*}
where
\begin{align*}
\bm \mu_* &=
\left[ (\bm V_\beta/\sigma^2)^{-1} + \bm X \bm X \right]^{-1}\left[ (\bm V_\beta/\sigma^2)^{-1}  \bm \mu_\beta + \bm X' \bm y \right]
=
\left[ \bm V_\beta^{-1} + \bm X \bm X/\sigma^2 \right]^{-1}\left[ \bm V_\beta^{-1}  \bm \mu_\beta + \bm X' \bm y /\sigma^2\right]
\\
\bm V_*&= 
\left[ (\bm V_\beta/\sigma^2)^{-1} + \bm X \bm X \right]^{-1}
=
\sigma^2
\left[ \bm V_\beta^{-1} + \bm X \bm X/\sigma^2 \right]^{-1} 
\end{align*}
Hence the posterior is 
\begin{align*}
p(\bm \beta, \sigma^2 \vert \bm y) &\propto 
\left( \frac{1}{\sigma^2} \right)^{a_*+1}
exp\left[ -\frac{1}{\sigma^2} \left\{b_*+\frac{1}{2}
(\bm \beta - \bm \mu_*)' \bm V_*^{-1} (\bm \beta - \bm \mu_*)
 \right\}\right] 
\end{align*}
where $a_*=a+n/2$
and 
$b_*=b + \frac{1}{2}\left[ \bm \mu_\beta' (\bm V_\beta/\sigma^2)^{-1} \bm \mu_\beta + \bm y' \bm y
-\bm \mu_*' \bm V_*^{-1} \bm \mu_* 
\right]$.

\newpage

\renewcommand{\theequation}{B.\arabic{equation}} \setcounter{equation}{0} %
\renewcommand{\thetable}{B\arabic{table}} \setcounter{table}{0}
\renewcommand{\thefigure}{B\arabic{figure}} \setcounter{figure}{0}

\section{MCMC inference in linear regression model with hierarchical priors}
We use the simple linear regression model of the form
\begin{equation}
y_{i} = x_{i} \bm \beta + \varepsilon_{i}, \text{ \ } \varepsilon_{i} \sim N(0,\sigma^{2}),  \text{ \ } i=1,...,n
\end{equation}
where $\bm \beta$ is a $p \times 1$ vector. We define $\bm y = (y_{1},...,y_{n})^{\prime}$, $\bm X = (x_{1}^{\prime},...,x_{n}^{\prime})^{\prime}$ and $\bm \varepsilon = (\varepsilon_{1},...,\varepsilon_{n})^{\prime}$, such that the stacked form of the regression model is
\begin{equation}
\bm y = \bm X \bm \beta + \bm \varepsilon,
\end{equation}
where $\bm \varepsilon \sim N_{n}(\bm 0_{n \times 1}, \sigma^{2} \bm I_{n} )$.
\subsection{Normal-Jeffreys}
The Normal-Jeffreys hierarchical prior takes the form
\begin{eqnarray}
\bm \beta \vert \lbrace \tau_{j}^{2} \rbrace_{j=1}^{p}, \sigma^{2} & \sim & N_{p}(\bm 0, \sigma^{2} \bm D), \\
\tau_{j}^{2} & \sim & \frac{1}{\tau_{i}^{2}}, \text{ \ \ for } j=1,...,p, \\
\sigma^{2} & \sim &  \frac{1}{\sigma^{2}}
\end{eqnarray}
where $\bm D =  diag(\tau_{1}^{2},...,\tau_{p}^{2})$.

The conditional posteriors are of the form
\begin{eqnarray}
\bm \beta  \ \vert \ \bullet & \sim &  N_p \left(  \bm V \times \bm X^{\prime} \bm y, \sigma^{2} \bm V \right) , \\
\tau_{j}^{2}   \ \vert \ \bullet & \sim &  Inv-Gamma\left(  \frac{1}{2}, \frac{\beta_{j}^{2}}{2\sigma^2} \right), \text{ for } j=1,...,p, \\
\sigma^{2}   \ \vert \ \bullet & \sim &  Inv-Gamma \left( \frac{n+2}{2},\frac{\Psi+\bm \beta' \bm D^{-1} \bm \beta}{2} \right) 
\end{eqnarray}
where $\bm V =  \left( \bm X^{\prime} \bm X + \bm D^{-1} \right)^{-1} $ and 
$\Psi = (\bm y - \bm X \bm \beta)^{\prime}(\bm y - \bm X \bm \beta)$.

\subsection{Student-t shrinkage}
The Normal-Inv-Gamma prior is the scale mixture of Normals representation of the fat-tailed Student-t distribution. This hierarchical prior, which is also called ``sparse Bayesian Learning'' prior in signal processing, takes the form
\begin{eqnarray}
\bm \beta \vert \lbrace \tau_{j}^{2} \rbrace_{j=1}^{p}, \sigma^{2} & \sim & N_{p}(\bm 0,  \sigma^{2} \bm D), \\
\tau_{j}^{2} & \sim & inv-Gamma\left(\rho,\xi \right), \text{ \ \ for } j=1,...,p, \\
\sigma^{2} & \sim &  \frac{1}{\sigma^{2}}
\end{eqnarray}
where $\bm D=  diag(\tau_{1}^{2},...,\tau_{p}^{2})$.

The conditional posteriors are of the form
\begin{eqnarray}
\bm \beta   \ \vert \ \bullet & \sim &  N_p \left(  \bm V \times \bm X^{\prime} \bm y, \sigma^{2} \bm V \right), \\
\tau_{j}^{2}   \ \vert \ \bullet & \sim &  Inv-Gamma\left(\rho +  \frac{1}{2},\xi + \frac{\beta_{j}^{2}}{2\sigma^2} \right), \text{  for } j=1,...,p, \label{student_t_tau}\\
\sigma^{2}   \ \vert \ \bullet & \sim &  Inv-Gamma \left( \frac{n+p}{2},\frac{\Psi+\bm \beta' \bm D^{-1} \bm \beta}{2} \right)
\end{eqnarray}
where $\bm V =  \left( \bm X^{\prime} \bm X + \bm D^{-1} \right)^{-1} $ and 
$\Psi = (\bm y - \bm X \bm \beta)^{\prime}(\bm y - \bm X \bm \beta)$.

\subsection{Bayesian Lasso}
As noted first by \cite{Tibshirani1996}, the lasso estimator
\begin{equation}
\hat{\bm \beta} = \arg\min_\beta \left( \bm y - \bm X \bm \beta \right)' \left( \bm y - \bm X \bm \beta \right) + \lambda_1 \sum_{j=1}^p |\beta_j| 
\end{equation}
is equivalent to the posterior mode  under the  Laplace prior
\begin{equation}
\bm \beta \vert \sigma \sim \prod_{j=1}^{p}  \frac{\lambda}{2\sqrt{\sigma^2}}  e^{-\lambda \vert \beta_{j} \vert / \sqrt{\sigma^2}}, \label{Laplace}
\end{equation}
which 
can be written as the following Normal-Exponential mixture
\begin{equation}
\bm \beta \vert \sigma \sim \prod_{j=1}^{p} \int_{0}^{\infty} \frac{1}{\sqrt{2 \pi \sigma^2 s_{j}}} e^{\left( -\frac{\beta_{j}^{2}}{2\sigma^{2}s_{j}} \right)} \frac{\lambda^{2}}{2}e^{-\frac{\lambda}{2s_j}} ds_{j}.
\end{equation}
This is the mixture prior analyzed by \cite{ParkCasella2008}, which is by far the most popular form for the Bayesian lasso. \cite{Hans2009} provides an alternative formulation by means of the orthant-truncated Normal distribution. A third possible formulation of the Laplace prior is the scale mixture of uniform distributions proposed by \cite{MallickYi2014}. A related representation is that of a mixture of truncated Normal distributions \citep[see][]{AlhamzawiAli2020}.

\subsubsection{\cite{ParkCasella2008} algorithm} \label{ParkCasella}
The \cite{ParkCasella2008} Laplace prior takes the form
\begin{eqnarray}
\bm \beta \vert \lbrace \tau_{j}^{2} \rbrace_{j=1}^{p}, \sigma^{2} & \sim & N_{p}(\bm 0, \sigma^{2} \bm D), \\
\tau_{j}^{2} \vert \lambda^{2} & \sim & Exponential\left( \frac{\lambda^{2}}{2} \right), \text{ \ \ for } j=1,...,p, \\
\lambda^{2} & \sim & Gamma(r,\delta) \\
\sigma^{2} & \sim &  \frac{1}{\sigma^{2}}
\end{eqnarray}
where $\bm D =  diag(\tau_{1}^{2},...,\tau_{p}^{2})$.

The conditional posteriors are of the form
\begin{eqnarray}
\bm \beta   \ \vert \ \bullet & \sim &  N_p \left( \bm V \times \bm X^{\prime} \bm y , \sigma^{2} \bm V \right), \\
\frac{1}{\tau_{j}^{2}}   \ \vert \ \bullet & \sim &  IG\left(\sqrt{\frac{\lambda^2\sigma^2}{\beta_{j}^{2}}}, \lambda^2 \right), \text{ \ \ for } j=1,...,p,\\
\lambda^{2}   \ \vert \ \bullet & \sim &  Gamma\left( r + p, \frac{\sum_{j=1}^{p} \tau_{j}^{2} }{2} + \delta \right), \\
\sigma^{2}   \ \vert \ \bullet & \sim &  Inv-Gamma \left( \frac{n+p}{2},\frac{\Psi+\bm \beta' \bm D^{-1} \bm \beta}{2}  \right)
\end{eqnarray}
where $\bm V = \left( \bm X^{\prime} \bm X + \bm D^{-1} \right)^{-1}$, $\bm D =  diag(\tau_{1}^{2},...,\tau_{p}^{2})$, 
and $\Psi = (\bm y - \bm X \bm \beta)^{\prime}(\bm y - \bm X \bm \beta)$.

\subsubsection{\cite{Hans2009} algorithm}
Before we proceed we need to define the notion of the Normal orthant distribution, following \cite{Hans2009}. Let $\mathcal{Z} = \lbrace -1, 1 \rbrace^{p}$ represent the set of all $2^p$ possible vectors of length $p$ whose elements are $\pm 1$. For any realization $z \in \mathcal{Z}$ define the orthant $\mathcal{O}_{z} \subset {\rm I\!R}^p$. If $\bm \beta \in \mathcal{O}_{z}$, then $\beta_{j} \geq 0$ if $z=1$ and $\beta_{j} <0$ if $z=-1$. Then $\bm \beta$ follows the Normal-orthant distribution with mean $m$ and covariance $S$, which is of the form
\begin{equation}
\bm \beta \sim N^{[z]} \left( \bm m,\bm S \right) \equiv \Phi\left( \bm m, \bm S\right) N_{p} \left( \bm m,\bm S \right) I\left( \bm \in \mathcal{O}_{z} \right).
\end{equation}
The \cite{Hans2009} prior takes the form
\begin{eqnarray}
\bm \beta \vert \lambda, \sigma & \sim & \left( \frac{\lambda}{2\sqrt{\sigma^{2}}}\right)^{p} \exp\left( - \lambda \sum_{j=1}^{p} \vert \beta_{j} \vert /\sqrt{\sigma^{2}} \right), \\
\lambda & \sim & Gamma(r,\delta), \\
\sigma^{2} & \sim &  \frac{1}{\sigma^{2}},
\end{eqnarray}
where the prior for $\bm \beta$ is an equivalent representation of the Laplace density in \autoref{Laplace}.

The conditional posteriors are of the form
\begin{eqnarray}
\beta_{j} \vert  \beta_{-j}, \lambda, \sigma^{2}, \bm y & \sim & \phi_{j}N^{[+]} \left(\mu_{j}^{+}, \omega_{jj}^{-1} \right) + (1-\phi_{j})N^{[-]} \left(\mu_{j}^{-}, \omega_{jj}^{-1} \right), \\
\lambda \vert \bm y & \sim & Gamma\left(p+r, \frac{\sum_{j=1}^{p} \vert \beta \vert}{\sqrt{\sigma^{2}}} + \delta \right), \\
\sigma \vert \bm \beta, \bm y & \propto & (\sigma^{2})^{-(\frac{n+p}{2} + 1)} \exp \left(  \frac{\Psi}{2\sigma^2}  - \frac{\lambda \sum_{j=1}^{p} \vert \beta \vert}{\sqrt{\sigma^{2}}} \right),
\end{eqnarray}
where:
\begin{itemize}
\item $N^{[-]}$ and $N^{[+]}$ correspond to the $N^{[z]}$ distribution for $z=-1$ and $z=1$, respectively;
\item $\mu_{j}^{+} = \widehat{\beta}_{j}^{OLS} + \left \lbrace \sum_{i=1,i \neq j}^{p} \left(\widehat{\beta}_{i}^{OLS} - \beta_{i} \right)\left(\omega_{ij}/\omega_{jj}\right)  \right \rbrace + \left(- \frac{\lambda}{\sqrt{\sigma^{2}}\omega_{jj}}\right) $;
\item $\omega_{ij}$ is the $ij$ element of the matrix $\Omega  = \Sigma^{-1} = \left( \sigma^{2}(\bm X^{\prime} \bm X)^{-1} \right)^{-1}$;
\item $\phi_{j} = \frac{\Phi \left( \frac{\mu_{j}^{+}}{\sqrt{\omega_{jj}}} \right)/N\left(0 \vert \mu_{j}^{+}, \omega_{jj}^{-1} \right)}{ \Phi \left( \frac{\mu_{j}^{+}}{\sqrt{\omega_{jj}}} \right)/N\left(0 \vert \mu_{j}^{+}, \omega_{jj}^{-1} \right) + \Phi \left( -\frac{\mu_{j}^{-}}{\sqrt{\omega_{jj}}} \right)/N\left(0 \vert \mu_{j}^{-}, \omega_{jj}^{-1} \right) }$; 
\item $\Psi = (\bm y - \bm X \bm \beta)^{\prime}(\bm y - \bm X \bm \beta)$.
\end{itemize}

Notice that the conditional posterior of $\sigma^{2}$ does not belong to a standard form we can sample from. \cite{Hans2009} proposes a simple accept/reject algorithm in order to obtain samples from $\sigma^{2}$. The posterior of $\sigma^{2}$ simplifies to the standard Inv-Gamma form, if we consider a Laplace prior for $\bm \beta$ that is independent of $\sigma$, i.e. the prior $\bm \beta \vert \lambda \sim \left( \frac{\lambda}{2}\right)^{p} \exp\left( - \lambda \sum_{j=1}^{p} \vert \beta_{j} \vert  \right)$. Finally, notice that sampling of $\beta_{j}$ conditional on $\beta_{-j}$ (i.e. all elements of the vector $\bm \beta$ other than the $j$-th) becomes very inefficient when predictors $\bm X$ are correlated. \cite{Hans2009} proposes to use an alternative Gibbs sampler algorithm that orthogonalizes predictors, which comes at the cost of increased computational complexity (due to the rotations of data and parameters involved when orthogonalizing the predictors).

\subsubsection{\cite{MallickYi2014} algorithm}
The \cite{MallickYi2014} Laplace prior takes the form
\begin{eqnarray}
\bm \beta \vert \lbrace \tau_{j}^{2} \rbrace_{j=1}^{p}, \sigma^{2} & \sim & \prod_{j=1}^{p} Uniform\left( -\sqrt{\sigma^{2}}\tau_{j}, \sqrt{\sigma^{2}}\tau_{j}\right), \\
\tau_{j} \vert \lambda & \sim & Gamma \left( 2, \lambda \right), \text{ \ \ for } j=1,...,p, \\
\lambda & \sim & Gamma\left( r, \delta \right), \\
\sigma^{2} & \sim &  \frac{1}{\sigma^{2}}.
\end{eqnarray}

The conditional posteriors are of the form
\begin{eqnarray}
\bm \beta   \ \vert \ \bullet & \sim &  N_p \left(\widehat{\bm \beta}_{OLS}, \sigma^{2} \left(\bm X^{\prime} \bm X\right)^{-1} \right) \prod_{j=1}^{p}I\left( \vert \beta_{j} \vert < \sqrt{\sigma^{2}}\tau_{j} \right), \label{beta_lasso} \\
\tau_{j}   \ \vert \ \bullet & \sim &  Exponential\left(\lambda \right) I\left( \tau_{j} > \frac{\vert \beta_{j} \vert}{\sqrt{\sigma^{2}}} \right), \text{ \ \ for } j=1,...,p, \label{tau_lasso}\\
\lambda  & \sim & Gamma \left(r + 2p, \delta + \sum_{j=1}^{p} \vert \beta_{j} \vert \right), \label{lambda_lasso} \\
\frac{1}{\sigma^{2}}   \ \bigg\vert \ \bullet & \sim &  Gamma \left( \frac{n-1+p}{2},\frac{\Psi}{2}\right) I\left(\sigma^{2} < \frac{1}{\max_{j}\left(\beta_{j}^{2} / \tau_{j}^{2} \right) } \right), \label{sigma_lasso}
\end{eqnarray}
where $I(\bullet)$ is the indicator function and $\Psi = (\bm y - \bm X \bm \beta)^{\prime}(\bm y - \bm X \bm \beta)$. Because of the truncation of the conditional posteriors, \cite{MallickYi2014} suggest the following sampling steps:
\begin{enumerate}
\item Generate first $\tau_{j}$ from the truncated Exponential distribution in \autoref{tau_lasso}: Sample a $\tau_{j}^{\star} \sim Exponential(\lambda)$, and then set $\tau_{j} = \tau_{j}^{\star} + \frac{\vert \beta_{j} \vert}{\sqrt{\sigma^{2}}}$.
\item Sample $\beta$ from the truncated Normal distribution in \autoref{beta_lasso}
\item Sample $\lambda$ from the Gamma distribution in \autoref{lambda_lasso}
\item Generate $\sigma^{2}$ from the right truncated Gamma distribution in \autoref{sigma_lasso}: Use simple accept/reject sampling, that is, sample $\frac{1}{\sigma^{2 \star}}$ from $Gamma \left( \frac{n-1+p}{2},\frac{\Psi}{2}\right)$ until the condition $\sigma^{2 \star} < \frac{1}{\max_{j}\left(\beta_{j}^{2} / \tau_{j}^{2}\right)}$ is met. If it is, set $\sigma = \frac{1}{\sigma^{2 \star}}$.
\end{enumerate}

\subsection{Bayesian Adaptive Lasso}
\cite{FanLi2001} showed that the lasso can perform automatic variable selection
but it produces biased estimates for the larger coefficients. Thus, they argued that the
oracle properties do not hold for the lasso. To obtain the oracle property, \cite{Zou2006}
introduced the adaptive lasso estimator as
\begin{equation}
\hat{\bm \beta} = \arg\min_\beta \left( \bm y - \bm X \bm \beta \right)' \left( \bm y - \bm X \bm \beta \right) + \sum_{j=1}^p  \lambda_j |\beta_j| 
\end{equation}
with the weight vector $ \lambda_j=\lambda | \hat{ \beta}_j |^{-r}$ for $j=1,...,p$ where $ \hat{ \beta}_j$ is a $\sqrt{n}$ consistent estimator such as the least squares estimator. 
The adaptive lasso enjoys the oracle property and it leads to a near-minimax-optimal estimator.

\cite{AlhamzawiAli2018} proposed Bayesian adaptive lasso.
They show that a Laplace density can be written as a exponential scale mixture of truncated normal distribution i.e.\
\begin{align*}
\frac{\lambda_j}{2\sqrt{\sigma^2}}  e^{-\lambda \vert \beta_{j} \vert / \sqrt{\sigma^2}}
&=
\int_{0}^{\infty}
\int_{u_j > \sqrt{\lambda_j^2/ \sigma^2 } |\beta_j|}\frac{1}{\sqrt{2 \pi \sigma^2 s_{j}}} e^{\left( -\frac{\beta_{j}^{2}}{2\sigma^{2}s_{j}} \right)} e^{\left( -\frac{u_j}{2} \right)}\frac{\lambda_j^{2}}{8} 
e^{\left( -\frac{\lambda_j^2s_j }{8} \right) } d u_j ds_{j}\\
&=
\int_{0}^{\infty}
\int_{u_j > \sqrt{\lambda_j^2/ \sigma^2 } |\beta_j|}
N(\beta_j; 0, \sigma^2 s_j)
Exponential \left(u_j; \frac{1}{2} \right)
Exponential \left( s_j; \frac{\lambda_j^2}{8} \right) d u_j ds_{j}
\end{align*}
Based on this fact, they propose  the following conditional prior for Bayesian adaptive lasso
\begin{eqnarray}
\beta_j \vert  \sigma^{2}, \lambda_j^2, s_j & \sim & N(0,\sigma^2 s_j)I \left(  |\beta_j| < \sqrt{\sigma^2 / \lambda_j^2} u_j \right) \text{ \ \ } j=1,...,p, \\
s_j \vert \lambda_j^2 & \sim & Exponential \left( \frac{\lambda_j^2}{8} \right)  \text{ \ \ } j=1,...,p,  \\
u_j & \sim & Exponential \left( \frac{1}{2} \right)  \text{ \ \ } j=1,...,p,  \\
\lambda_j^2 & \sim & Gamma(a,b)\text{ \ \ } j=1,...,p,  \\
\sigma^2 &\sim& \sigma^{-2}d\sigma^2 
\end{eqnarray}
The conditional posteriors are of the form
\begin{eqnarray}
\bm \beta   \ \vert \ \bullet & \sim &  N_p\left( \bm V \times \bm X \bm y, \sigma^2 \bm V \right) \prod_{j=1}^p I \left(  |\beta_j| < \sqrt{\sigma^2 / \lambda_j^2} u_j \right)   \\
\sigma^2  \ \vert \ \bullet & \sim &   Inv-Gamma\left( a^*, b^* \right) I \left( \sigma^2 > max_j \left\{ \frac{\lambda_j^2\beta_j^2}{ u_j^2} \right\} \right) \\
s_j^{-1}  \ \vert \ \bullet & \sim & IG\left( \sqrt{\frac{\sigma^2 \lambda_j^2}{4\beta_j^2}},\frac{\lambda_j^2}{4} \right) \text{ \ \ } j=1,...,p,  \\
p(u_j   \ \vert \ \bullet \ )& \propto & Exponential\left( \frac{1}{2} \right) I \left(u_j >  \sqrt{ \frac{\lambda_j^2}{\sigma^2} } |\beta_j| \right) \text{ \ \ } j=1,...,p,  \\
p(\lambda^2_j   \ \vert \ \bullet \ )& \propto & Gamma\left( a+p,b+\frac{s_j}{8} \right) I \left( \lambda_j^2 < \frac{\sigma^2u_j^2}{\beta_j^2}\right)  \text{ \ \ } j=1,...,p
\end{eqnarray}
where $\bm V =( \bm X' \bm X +\bm S^{-1} )^{-1}$ with $\bm S=diag(s_1,...,s_p)$,
$a^*=\frac{n-1+p}{2}$, and 
$b^*=\frac{1}{2} \left[\left( \bm y- \bm X \bm \beta \right)'  \left( \bm y- \bm X \bm \beta \right)  +\sum_{j=1}^p \frac{\beta_j^2}{s_j} \right] $.

\subsection{Bayesian Fused Lasso}
In some applications, there might be  a meaningful order among the covariates (e.g.\ time). The original lasso ignores such ordering.  To compensate the ordering limitations of the lasso, the fused lasso was introduced. It penalizes the $L_1$-norm of both the coefficients and their differences: 
\begin{equation}
\hat{\bm \beta} = \arg\min_\beta \left( \bm y - \bm X \bm \beta \right)' \left( \bm y - \bm X \bm \beta \right) + \lambda_1 \sum_{j=1}^p |\beta_j| +  \lambda_2 \sum_{j=2}^p |\beta_j - \beta_{j-1}| 
\end{equation}

\cite{Kyungetal2010}  proposed Bayesian group lasso with the following conditional prior.
\begin{eqnarray}
p\left( \bm \beta \vert  \sigma^{2} \right)& \propto & \exp \left( - \frac{\lambda_1}{\sigma} \sum_{j=1}^p |\beta_j|  - \frac{\lambda_2}{\sigma} \sum_{j=2}^p |\beta_j - \beta_{j-1}| \right)\\
 \sigma^2 &\sim& \sigma^{-2}d\sigma^2
\end{eqnarray}
where the conditional prior is equivalent to the following gamma mixture of normals prior.
\begin{eqnarray}
\bm \beta \vert  \lbrace \tau_{j}^{2} \rbrace_{j=1}^{p}, \lbrace \omega_{j}^{2} \rbrace_{j=1}^{p-1}, \sigma^{2} & \sim & N_{p}(\bm 0, \sigma^{2}\bm \Sigma_\beta ), \\
 \tau_{j}^{2}  & \sim & \frac{\lambda^2_1}{2}e^{-\lambda_1\tau_j^2/2} d\tau_j^2 \text{ for }, j=1,...,p,\\
 \omega_{j}^{2}  & \sim & \frac{\lambda^2_2}{2}e^{-\lambda_2\omega_j^2/2} d\omega_j^2 \text{ for }, j=1,...,p-1
\end{eqnarray}
where $\tau_1^2,...,\tau_p^2$ and $\omega_1^2,...,\omega_{p-1}^2$ are mutually independent, and  $\bm \Sigma_\beta$ is a tridiagonal matrix with 
\begin{eqnarray}
\text{Main diagonal }  & = &\left\{\frac{1}{\tau_i^2} + \frac{1}{\omega_{i-1}^2} + \frac{1}{\omega_i^2} , i=1,...,p  \right\},\\
\text{Off diagonals }  & = &\left\{- \frac{1}{\omega_{i}^2} ,i=1,...,p-1  \right\}
\end{eqnarray}
where $1/\omega_0^2 = 1/\omega_p^2 = 0$.

The conditional posteriors are of the form
\begin{eqnarray}
\bm \beta   \ \vert \ \bullet & \sim &  N_p\left( \bm V \times \bm X \bm y, \sigma^2 \bm V \right)\\
1/\tau_j^2   \ \vert \ \bullet & \sim &  IG \left( \left( \frac{\lambda_1^2 \sigma^2}{\beta_j^2}\right)^{1/2}, \lambda_1^2\right)1(1/\tau_j^2 >0), j=1,...,p\\
1/\omega_j^2   \ \vert \ \bullet & \sim &  IG \left( \left( \frac{\lambda_2^2 \sigma^2}{(\beta_{j+1}-\beta_j)^2}\right)^{1/2}, \lambda_2^2\right)1(1/\omega_j^2 >0), j=1,...,p-1\\
\sigma^2  \ \vert \ \bullet & \sim &   Inv-Gamma 
\left( 
a^*,
b^*
 \right)
\end{eqnarray}
where $\bm V =( \bm X' \bm X +\bm \Sigma_\beta^{-1} )^{-1}$, $a^*=\frac{n-1+p}{2}$, and $b^*=\frac{1}{2} \left[
 \left( \bm y- \bm X \bm \beta \right)'  \left( \bm y- \bm X \bm \beta \right)  +\bm \beta \bm \Sigma_\beta^{-1}  \bm \beta 
 \right] $.

When we place $Gamma(r,\delta)$ priors on $\lambda_1$ and $\lambda_2$, the conditional posteriors are
\begin{eqnarray}
\lambda_1^2   \ \vert \ \bullet & \sim &  Gamma\left( p+r, \frac{1}{2}\sum_{j=1}^p \tau_j^2 + \delta \right)\\
\lambda_2^2   \ \vert \ \bullet & \sim &  Gamma\left( p-1+r, \frac{1}{2}\sum_{j=1}^{p-1} \omega_j^2 + \delta \right)
\end{eqnarray}

\subsection{Bayesian Group Lasso}
If there is a group of covariates among which the pairwise correlation is high (e.g.\ dummy variables), the lasso tends to select only individual variables from the group. The group lasso takes   such group structure into account:
\begin{equation}
\hat{\bm \beta} = \arg\min_\beta \left( \bm y - \sum_{k=1}^K \bm X_k \bm \beta_k \right)' \left( \bm y - \sum_{k=1}^K \bm X_k \bm \beta_k  \right) + \lambda \sum_{j=k}^K ||\bm \beta_k||_{G_k}
\end{equation}
where $K$ is the number of groups, $\bm \beta_k$ is the vector of $\beta$'s in the group $k$, and $ ||\bm \beta||_{G_k}= \left( \bm \beta' \bm G_k \bm \beta \right)^{1/2}$ with positive definite matrices $\bm G_k$'s. Typically, $\bm G_k = \bm I_{m_k}$ where $m_k$ is the number of variables in group $k$.  

\cite{Kyungetal2010} proposed Bayesian group lasso with the following conditional prior.
\begin{eqnarray}
p\left( \bm \beta \vert  \sigma^{2} \right)& \propto & \exp \left( - \frac{\lambda}{\sigma} \sum_{j=k}^K ||\bm \beta_k||_{G_k} \right)\\
 \sigma^2 &\sim& \sigma^{-2}d\sigma^2
\end{eqnarray}
where the conditional prior is equivalent to the following gamma mixture of normals prior.
\begin{eqnarray}
\bm \beta_{G_k} \vert  \tau_{k}^{2}, \sigma^{2} & \sim & N_{m_k}(\bm 0, \sigma^{2}\tau^2_k  \bm I_{m_k}), \\
\tau_{k}^{2} \vert \sigma^{2} & \sim & Gamma\left( \frac{m_k +1}{2} , \frac{\lambda^2}{2} \right) \text{ for } k=1,...,K
\end{eqnarray}

The conditional posteriors are of the form
\begin{eqnarray}
\bm \beta_{G_k} \vert  \bm \beta_{-G_k}, \sigma^{2}, \tau_{1}^{2},...,\tau_K^2 ,  \lambda, \bm y & \sim & N_{p}\left(\bm V_k \times \bm X_k' \left( \bm y - \frac{1}{2}\sum_{k'\ne k}\bm X_{k'} \bm \beta_{G_{k'}} \right), \sigma^2\bm V_k  \right), \\
1/\tau_k^2   \ \vert \ \bullet & \sim &  IG \left( \left( \frac{\lambda^2 \sigma^2}{|| \bm \beta_{G_k} ||^2}\right)^{1/2}, \lambda^2\right)1(1/\tau_k^2 >0), \text{ for } k=1,...,K\\
\sigma^2  \ \vert \ \bullet & \sim &  Inv-Gamma \left( \frac{n-1+p}{2}, \frac{1}{2} || \bm y - \bm X  \bm \beta||^2 +\frac{1}{2} \sum_{j=k}^K\frac{1}{\tau_k^2}|| \bm \beta_{G_k} ||^2  \right)
\end{eqnarray}
where 
$\bm \beta_{-G_k}=\left( \bm\beta_{G_1},...,\bm\beta_{G_{k-1}},\bm\beta_{G_{k+1}}, ..., \bm\beta_{G_{K}} \right)$
and 
$\bm V_k = (\bm X_k' \bm X_k +\tau_k^{-2} \bm I_{m_k})^{-1}$.

When we place a $Gamma(r,\delta)$ prior on $\lambda$,  the posterior conditional on $\lambda$ is 
\begin{eqnarray}
\lambda^2  \ \vert \ \bullet & \sim &  Gamma\left( \frac{p+K}{2} +r, \frac{1}{2}\sum_{k=1}^K \tau_k^2 + \delta \right)
\end{eqnarray}

\subsection{Bayesian Elastic Net}
Here again we have various alternative algorithms. We look into the algorithm of \cite{Kyungetal2010} and the algorithm of \cite{LiLin2010}, but we can also mention here the algorithm of \cite{Hans2011} that is based on the algorithm of \cite{Hans2009} we examined for the Bayesian lasso.

The elastic net combines the benefits of ridge regression ($\mathcal{l}_{2}$ penalization) and the lasso ($\mathcal{l}_{1}$ penalization). The Bayesian prior that provides the solution to the elastic net estimation problem is of the form
\begin{equation}
\bm \beta \vert \sigma^{2} \sim \exp \left\lbrace -\frac{1}{2\sigma^{2}} \left(\lambda_{1} \sum_{j=1}^{p} \vert\beta_{j} \vert +  \lambda_{2} \sum_{j=1}^{p}\beta_{j}^{2} \right) \right\rbrace.
\end{equation}
\cite{LiLin2010} start from this prior and derive a mixture approximation and a Gibbs sampler that has the minor disadvantage that requires an accept-reject algorithm for obtaining samples from the conditional posterior of $\sigma^{2}$ (similar to the sampler of \cite{Hans2009} for the lasso). The formulation of the elastic net prior in \cite{Kyungetal2010} is slightly different to the one above, but they manage to derive a slightly different mixture representation and a slightly more straightforward Gibbs sampler.

\subsubsection{\cite{LiLin2010} algorithm}
The \cite{LiLin2010} prior takes the form
\begin{eqnarray}
\bm \beta \vert \lbrace \tau_{j}^{2} \rbrace_{j=1}^{p}, \lambda_{2}, \sigma^{2} & \sim & N_{p} \left( \bm 0, \frac{\sigma^{2}}{\lambda_{2}} \bm D_{\tau} \right), \\
\tau_{j}^{2} \vert \sigma^{2} & \sim & TG_{(1,\infty)} \left( \frac{1}{2}, \frac{8 \lambda_2 \sigma^2}{ \lambda_{1}^{2} } \right), \text{ \ \ for } j=1,...,p, \\
\sigma^{2} & \sim &  \frac{1}{\sigma^{2}}.
\end{eqnarray}
where $TG_{(1,\infty)}$ is the Gamma distribution truncated to the support $(1,\infty)$, and $\bm D_{\tau} =  diag\left( \frac{\tau_{1}^{2}-1}{\tau_{1}^{2}},...,\frac{\tau_{p}^{2}-1}{\tau_{p}^{2}} \right)$. Notice that $\lambda_{1}, \lambda_{2}$ do not have their own prior distributions, that is, they are not considered to be random variables in this algorithm. Instead, \cite{LiLin2010} suggest to use empirical Bayes methods to calibrate these two parameters.

The conditional posteriors are of the form
\begin{eqnarray}
\bm \beta  \ \vert \ \bullet & \sim &  N_p \left( \bm V \times \bm X^{\prime} \bm y, \sigma^{2} \bm V \right), \\
\tau_{j}^{2} - 1   \ \vert \ \bullet & \sim &  GIG\left(\frac{1}{2}, \frac{\lambda_{1}}{4 \lambda_{2} \sigma^{2}}, \frac{\lambda_{2} \beta_{j}^{2}}{\sigma^2} \right), \text{ \ \ for } j=1,...,p,\\
p(\sigma^{2} | \  \bullet \ )  & \propto & \left(\frac{1}{\sigma^{2}} \right)^{\frac{n}{2} + p+1} \left\lbrace \Gamma_{U}\left( \frac{1}{2}, \frac{\lambda_{1}^{2}}{8\lambda_{2}\sigma^{2}}\right) \right\rbrace \\
& & \exp\left[ -\frac{1}{2\sigma^2} \left\lbrace \Psi + \lambda_{2} \sum_{j=1}^{p} \frac{\tau_{j}^{2}}{\tau_{j}^{2} - 1} \beta_{j}^{2} + \frac{\lambda_{1}^{2}}{4\lambda_{2}}\sum_{j=1}^{p} \tau_{j}^{2} \right\rbrace \right]
\end{eqnarray}
where 
$\bm V = \left( \bm X^{\prime} \bm X + \lambda_{2} \bm D_{\tau}^{-1} \right)^{-1}$, $\bm D_{\tau}^{-1} =  diag\left( \frac{\tau_{1}^{2}}{\tau_{1}^{2}-1},...,\frac{\tau_{p}^{2}}{\tau_{p}^{2}-1} \right)$, and $\Psi = (\bm y - \bm X \bm \beta)^{\prime}(\bm y - \bm X \bm \beta)$. $\Gamma_{U}(\bullet)$ is the upper incomplete gamma function. $GIG$ is the three parameter Generalized Inverse Gaussian distribution.\footnote{CRAN has several implementations in R of random number generators that allow sampling from the $GIG$ distribution. As of the time of writing of this document, Mathworks does not provide a built-in function for MATLAB that allows to generate from this distribution, but external contributions do exist.} The conditional posterior distribution of $\sigma^{2}$ does not belong to a known density we can sample from. Therefore, for each Monte Carlo iteration we sample the first two parameters directly from their conditional posteriors but we sample $\sigma^{2}$ indirectly from its conditional posterior using an accept/reject step.

\subsubsection{\cite{Kyungetal2010} algorithm}
The \cite{Kyungetal2010} prior takes the form
\begin{eqnarray}
\bm \beta \vert \lbrace \tau_{j}^{2} \rbrace_{j=1}^{p},\lambda_2, \sigma^{2} & \sim & N_{p}(\bm 0, \sigma^{2} \bm D_{\tau,\lambda_{2}}), \\
\tau_{j}^{2} \vert \lambda^{2} & \sim & Exponential \left(\frac{\lambda_{1}^{2}}{2} \right), \text{ \ \ for } j=1,...,p, \\
\lambda_{1}^{2} & \sim & Gamma(r_{1},\delta_{1}), \\
\lambda_{2} & \sim & Gamma(r_{2},\delta_{2}), \\
\sigma^{2} & \sim &  \frac{1}{\sigma^{2}}
\end{eqnarray}
where $\bm D_{\tau,\lambda_{2}} =  diag((\tau_{1}^{-2} + \lambda_{2})^{-1},...,(\tau_{p}^{-2} + \lambda_{2}))^{-1})$.

The conditional posteriors are of the form
\begin{eqnarray}
\bm \beta   \ \vert \ \bullet & \sim &  N_p \left( \bm V \times \bm X^{\prime} \bm y, \sigma^{2} \bm V \right), \\
\frac{1}{\tau_{j}^{2}}   \ \vert \ \bullet & \sim &  IG\left(\sqrt{\frac{\lambda_{1}^{2}\sigma^2}{\beta_{j}^{2}}}, \lambda_{1}^{2} \right) I(1/\tau_{j}^{2}>0), \text{ \ \ for } j=1,...,p,\\
\lambda_{1}^{2}  \ \vert \ \bullet & \sim &  Gamma\left( r_{1} + p, \frac{\sum_{j=1}^{p} \tau_{j}^{2} }{2} + \delta_{1} \right), \\
\lambda_{2}   \ \vert \ \bullet & \sim &  Gamma\left( r_{2} + \frac{p}{2}, \frac{\sum_{j=1}^{p} \beta_{j}^{2} }{2\sigma^{2}} + \delta_{2} \right), \\
\sigma^{2} \ \vert \ \bullet & \sim &  Inv-Gamma \left( \frac{n-1+p}{2},\frac{\Psi+\bm \beta^{\prime} \bm D_{\tau,\lambda_{2}}^{-1} \bm \beta}{2}  \right)
\end{eqnarray}
where $\bm V = \left( \bm X^{\prime} \bm X + \bm D_{\tau,\lambda_{2}}^{-1} \right)^{-1}$, 
$\bm D_{\tau,\lambda_{2}}^{-1} =  diag((\tau_{1}^{-2} + \lambda_{2}),...,(\tau_{p}^{-2} + \lambda_{2})))$,
 and $\Psi = (\bm y - \bm X \bm \beta)^{\prime}(\bm y - \bm X \bm \beta)$.

\subsection{Generalized Double Pareto}
\cite{Armaganetal2013a} propose the following Generalized Double Pareto (GDP) prior on $\bm \beta$
\begin{equation}
\bm \beta \vert \sigma \sim \prod_{j=1}^{p} \frac{1}{2\sigma \delta/r} \left(1 + \frac{1}{r} \frac{\vert \beta_{j} \vert}{\sigma \delta/r} \right) ^{-(r+1)}.
\end{equation}
This distribution can be represented using the familiar, from the Bayesian lasso, Normal-Exponential-Gamma mixture, see \autoref{ParkCasella}. The only difference is that, while the Exponential component has the same rate parameter for all $j=1,...,p$, in the representation of the GDP mixture this parameter is adaptive.

The Generalized Double Pareto prior takes the form
\begin{eqnarray}
\bm \beta \vert \lbrace \tau_{j} \rbrace_{j=1}^{p}, \sigma^{2} & \sim &  N_{p} \left( \bm 0, \sigma^{2} \bm D\right), \\
\tau_{j}^{2} \vert \lambda_{j} & \sim & Exponential \left( \frac{\lambda_{j}^2}{2} \right), \text{ \ \ for } j=1,...,p, \\
\lambda_{j} & \sim & Gamma(r,\delta), \text{ \ \ for } j=1,...,p, \\
\sigma^{2} & \sim &  \frac{1}{\sigma^{2}},
\end{eqnarray}
where $\bm D =  diag(\tau_{1}^{2},...,\tau_{p}^{2})$.

The conditional posteriors are of the form
\begin{eqnarray}
\bm \beta   \ \vert \ \bullet & \sim &  N_p \left( \bm V \times \bm X^{\prime} \bm y , \sigma^{2} \bm V \right), \\
\frac{1}{\tau_{j}^{2}}   \ \bigg\vert \ \bullet & \sim &  IG\left(\sqrt{\frac{\lambda_{j}^2\sigma^2}{\beta_{j}^{2}}}, \lambda^2 \right), \text{ \ \ for } j=1,...,p,\\
\lambda_{j}^{2}   \ \vert \ \bullet & \sim &  Gamma\left( r + 1, \sqrt{\frac{ \beta_{j}^{2} }{\sigma^{2}}} + \delta \right), \\
\sigma^{2}  \ \vert \ \bullet & \sim &  Inv-Gamma \left( \frac{n-1+p}{2},\frac{\Psi+\bm \beta^{\prime} \bm D^{-1} \bm \beta}{2} \right)
\end{eqnarray}
where $\bm V = \left( \bm X^{\prime} \bm X + \bm D^{-1} \right)^{-1}$, $\bm D =  diag(\tau_{1}^{2},...,\tau_{p}^{2})$, and 
$\Psi = (\bm y - \bm X \bm \beta)^{\prime}(\bm y - \bm X \bm \beta)$.

\subsection{Normal-Gamma}
The Normal-Gamma prior of \cite{GriffinBrown2010} takes the form
\begin{eqnarray}
\bm \beta \vert \lbrace \tau_{j} \rbrace_{j=1}^{p} & \sim & N\left( \bm 0, \bm D \right), \\
\tau \vert \lambda, \gamma^{2} & \sim & Gamma \left( \lambda, \frac{1}{2\gamma^{2}} \right), \\
\sigma^{2} & \sim &  \frac{1}{\sigma^{2}},
\end{eqnarray}
where $\bm D=  diag(\tau_{1}^{2},...,\tau_{p}^{2})$.

The conditional posteriors $\bm \beta$ and $\sigma^2$ are of the usual form
\begin{eqnarray}
\bm \beta   \ \vert \ \bullet & \sim &  N_{p}\left( \bm V \times \bm X' \bm y / \sigma^2   ,  \bm V\right),\label{beta_indep}\\
\sigma^2    \ \vert \ \bullet & \sim &   Inv-Gamma \left( \frac{n}{2}, \frac{1}{2}  \left( \bm y- \bm X \bm \beta \right)'  \left( \bm y- \bm X \bm \beta \right) \right)\label{sigma2_indep}
\end{eqnarray}
where 
$\bm V =  (\bm X' \bm X / \sigma^2 +\bm D^{-1})^{-1}$.

The parameters $\tau_1,...,\tau_p$ can be updated in a block since the full conditional distributions of $\tau_1,...,\tau_p$
are independent. The full conditional distribution of $\tau_j$ follows a generalized inverse Gaussian distribution
\begin{eqnarray}
\tau_j  \ \vert \ \bullet & \sim & GIG( \lambda-0.5,1/\gamma^2,\beta^2_j ), \text{ \ \ } j=1,...,p
\end{eqnarray}


\subsection{Multiplicative Gamma process}

Suppose we have the factor model 
\begin{eqnarray}
\bm X_t & = & \bm \Lambda \bm F_t + \bm \epsilon_t\\
\bm \epsilon_t & \sim & N_n(\bm 0,\bm \Sigma) , t=1,...,T
\end{eqnarray}
where 
$\bm X_t$ is a $n \times 1$ vector,
$ \bm \Lambda$ is a $n \times k$ matrix of factor loadings,
$\bm F_t$ is a $k \times 1$ vector, and 
$\bm \Sigma=diag(\Sigma_{11},...,\Sigma_{nn})$.

\cite{BhattacharyaDunson2011} proposed a novel multiplicative gamma process prior on the factor loadings that shrinks more aggressively columns of $\bm \Lambda$ that correspond to a higher number of factors. They call their approach the sparse infinite factor model, as it allows to specify a maximum number of factors and the prior is able to determine zero and non-zero loadings, as well as the number of factors. The gamma process prior for the loadings matrix is of the following ``global-local shrinkage'' form
\begin{eqnarray}
\Lambda_{ij} \vert \phi_{ij}, \tau_{j} & \sim & N(0,\phi_{ij}^{-1} \tau_{j}^{-1}), \\
\phi_{ij} & \sim & Gamma(v/2,v/2), \\
\tau_{j} & = & \prod_{l=1}^{j} \delta_{l}, \text{ \ \ } j=1,...,k, \\
\delta_{1} & \sim & Gamma(a_{1},1), \\
 \delta_{l} & \sim & Gamma(a_{2},1),  \text{ \ \ } l \geq 2,\\
 \Sigma_{ii} &\sim & Inv-Gamma(a_0,b_0), i=1,...,n
\end{eqnarray}
While the local shrinkage parameter is the same for each element of $\bm \Lambda$, the global shrinkage parameter $\tau_{j} $ is shrinking more aggressively as the index $j$ increases, 
where $j=1,...,k$ indexes the number of factors. This is because $\tau_{j} $ is a $j$-dimensional product of gamma distributions. 

Let $\bm X^{(i)}$ be the $i$th column of the $n \times k$ matrix  $\bm X$
$\bm \Lambda_i'$ be the $i$th row of $\bm \Lambda$.
The conditional posterior distributions are 
\begin{eqnarray}
\bm \Lambda_{i}   \ \vert \ \bullet & \sim &  N_k \left( \bm V_{L_i} \left( \bm F' \Sigma_{ii}^{-1} \bm X^{(i)} \right) ,\bm V_{L_i} \right) \text{ \ \ } i=1,...,n , \\
\bm F_{t}   \ \vert \ \bullet & \sim &  N_k\left( \bm V_{F} \left(\bm \Lambda^{\prime}\bm \Sigma^{-1} \bm X_t \right) ,\bm V_{F}  \right)\text{ \ \ } t=1,...,T , \\
\phi_{ij}    \ \vert \ \bullet & \sim & Gamma \left( \frac{v+1}{2}, \frac{v+\tau_j  \Lambda^2_{ij} }{2} \right) \text{ \ \ } i=1,...,n ,j=1,...,k,\\
\tau_\ell^{(j)} &=& \prod_{t=1, t \ne j}^\ell \delta_t \text{ \ \ } j=1,...,k \\
\delta_{1}   \ \vert \ \bullet & \sim &  Gamma \left( a_{1} +0.5nk,1+0.5\sum_{\ell=1}^k \tau_\ell^{(1)} \sum_{i=1}^n \phi_{i\ell} \Lambda^2_{i\ell}\right), \\
 \delta_{j}   \ \vert \ \bullet & \sim &   Gamma \left(a_{2} +0.5n(k-j+1),1+0.5\sum_{\ell=j}^k \tau_\ell^{(j)} \sum_{i=1}^n \phi_{i\ell} \Lambda^2_{i\ell}\right),   \text{ \ \ } j \geq 2,\\
\Sigma_{ii}   \ \vert \ \bullet & \sim &  Inv-Gamma\left( a_0 + n/2, b_0 + SSE_{i} \right), \text{ \ \ } i=1,...,n,
\end{eqnarray}
where 
$\bm V_{L_i} = (\bm D_i^{-1} + \Sigma_{ii}^{-1} \bm F' \bm F)^{-1}$, 
$ \bm D_i^{-1} = diag( \phi_{i1}\tau_1,...,\phi_{ik}\tau_k )$, 
$\bm V_{F} = (I + \bm \Lambda^{\prime}\bm \Sigma^{-1} \bm \Lambda)^{-1}$,
 and 
$SSE_{i} = (\bm X^{(i)}  -  \bm F\bm \Lambda_{i})^{\prime}(\bm X^{(i)} -  \bm F\bm \Lambda_{i})$.

\subsection{Dirichlet-Laplace}
The Dirichlet-Laplace prior of \cite{Bhattacharyaetal2015}, as analyzed in \cite{ZhangBondell2018}, takes the form
\begin{eqnarray}
\bm \beta \vert \lbrace \tau_{j} \rbrace_{j=1}^{p},  \lbrace \psi_{j}\rbrace_{j=1}^{p}, \lambda, \sigma^{2} & \sim &  N_{p} \left( \bm 0, \sigma^{2} \bm D_{\lambda,\tau,\psi} \right), \\
\tau_{j}^{2} & \sim & Exponential(1/2), \text{ \ \ for } j=1,...,p, \\
\psi_{j} & \sim & Dirichlet(\alpha), \text{ \ \ for } j=1,...,p, \\
\lambda & \sim & Gamma(n \alpha, 1/2), \\
\sigma^{2} & \sim &  \frac{1}{\sigma^{2}},
\end{eqnarray}
where $\bm D_{\lambda,\tau,\psi} = diag(\lambda^{2} \tau_{1}^{2}\psi_{1}^{2},...,\lambda^{2} \tau_{p}^{2}\psi_{p}^{2})$.

The conditional posteriors are of the form
\begin{eqnarray}
\bm \beta   \ \vert \ \bullet & \sim &  N_p \left( \bm V \times \bm X^{\prime} \bm y, \sigma^{2} \bm V \right), \\
\frac{1}{\tau_{j}^{2}}   \ \bigg\vert \ \bullet & \sim &  IG \left(\sqrt{\frac{\lambda^{2} \psi_{j}^{2} \sigma^2}{\beta_{j}^{2}}},1\right), \text{ \ \ for } j=1,...,p, \\
\lambda   \ \vert \ \bullet & \sim &  GIG\left( 2 \frac{\sum_{j=1}^{p}\vert \beta_{j}\vert}{\psi_j \sigma},1 ,p(\alpha - 1) \right), \\
T_j   \ \vert \ \bullet & \sim &  GIG \left( 2 \sqrt{\frac{\beta_{j}^{2}}{\sigma^2}},1, \alpha-1 \right), \text{ \ \ for } j=1,...,p,  \\
\psi_{j}    &= &  \frac{T_j}{\sum_{j=1}^{p} T_j}, \text{ \ \ for } j=1,...,p, \\ 
\sigma^{2}  \ \vert \ \bullet & \sim &  Inv-Gamma \left( \frac{n+p}{2},\frac{\Psi+\bm \beta^{\prime} \bm D_{\tau,\lambda,\psi}^{-1} \bm \beta}{2} \right)
\end{eqnarray}
where 
$\bm V = \left( \bm X^{\prime} \bm X + \bm D_{\tau,\lambda,\psi}^{-1} \right)^{-1}$, 
$\bm D_{\tau,\lambda,\psi} =  diag(\lambda^{2}\tau_{1}^{2}\psi_{1}^{2},...,\lambda^{2}\tau_{p}^{2}\psi_{p}^{2})$,
 and $\Psi = (\bm y - \bm X \bm \beta)^{\prime}(\bm y - \bm X \bm \beta)$.

\subsection{Horseshoe}
The horseshoe prior on a regression coefficient $\bm \beta$ takes the following hierarchical form
\begin{eqnarray}
\bm \beta \vert \lbrace \lambda_{j} \rbrace_{j=1}^{p}, \tau & \sim & N\left( \bm 0, \sigma^{2} \tau^{2} \bm \Lambda \right), \\
\lambda_{j} \vert \tau & \sim &  C^{+} (0,1), \text{ \ \ for } j=1,...,p,  \\
\tau & \sim &  C^{+} (0,1),
\end{eqnarray}
where $\bm \Lambda = diag(\lambda_{1}^{2},...,\lambda_{p}^{2})$, and $C^{+}(0,\alpha)$ is the half-Cauchy distribution on the positive reals with scale parameter $\alpha$. That is, $\lambda_{j}$ has conditional prior density
\begin{equation}
\lambda_{j} \vert \tau = \frac{2}{\pi \tau \left( 1 + (\lambda_j/\tau)^2 \right)}.
\end{equation}
\subsubsection{\cite{MakalicSchmidt2016} algorithm}
\cite{MakalicSchmidt2016} note that the half-Cauchy distribution can be written as a mixture of inverse-Gamma distributions. In particular, if
\begin{equation}
x^{2} \vert z \sim Inv-Gamma(1/2,1/z), \text{ \ \ \ } z \sim Inv-Gamma(1/2, 1/\alpha^{2}),
\end{equation}
then $x \sim C^{+}(0,\alpha)$. Therefore, the \cite{MakalicSchmidt2016} prior takes the form
\begin{eqnarray}
\bm \beta \vert \lbrace \lambda_{j} \rbrace_{j=1}^{p}, \tau, \sigma^{2} & \sim & N\left( \bm 0, \sigma^{2} \tau^{2} \bm \Lambda \right), \\
\lambda_{j}^{2} \vert v_{j} & \sim & Inv-Gamma(1/2,1/v_{j}), \text{ \ \ for } j=1,...,p,   \\
v_{j} & \sim & Inv-Gamma(1/2,1), \text{ \ \ for } j=1,...,p,   \\
\tau^{2} \vert \xi & \sim & Inv-Gamma(1/2,1/\xi), \\
\xi & \sim & Inv-Gamma(1/2,1), \\
\sigma^{2} & \sim &  \frac{1}{\sigma^{2}},
\end{eqnarray}
where $\bm \Lambda = diag(\lambda_{1}^{2},...,\lambda_{p}^{2})$.

The conditional posteriors are of the form
\begin{eqnarray}
\bm \beta   \ \vert \ \bullet & \sim &  N_p \left( \bm V \times \bm X^{\prime} \bm y, \sigma^{2} \bm V \right), \\
\lambda_{j}^{2}   \ \vert \ \bullet & \sim &  Inv-Gamma\left(1, \frac{1}{v_{j}} + \frac{\beta_{j}^{2}}{2\tau^{2}\sigma^{2}} \right), \text{ \ \ for } j=1,...,p,  \\
v_{j}   \ \vert \ \bullet & \sim &  Inv-Gamma \left(1, 1+ \frac{1}{\lambda_{j}^{2}} \right), \text{ \ \ for } j=1,...,p, \\
\tau^{2}   \ \vert \ \bullet & \sim &  Inv-Gamma \left( \frac{p+1}{2}, \frac{1}{\xi} + \frac{1}{2\sigma^{2}} \sum_{j=1}^{p} \frac{\beta_{j}^{2}}{\lambda_{j}^{2}} \right)\\
\xi   \ \vert \ \bullet & \sim &  Inv-Gamma \left(1, 1+ \frac{1}{\tau^{2}} \right), \\
\sigma^{2}   \ \vert \ \bullet & \sim &  Inv-Gamma \left( \frac{n+p}{2},\frac{\Psi+\bm \beta^{\prime} \bm D^{-1} \bm \beta}{2} \right),
\end{eqnarray}
where $\bm V = \left( \bm X^{\prime} \bm X + \bm D^{-1} \right)^{-1}$, $\bm D=  diag(\tau^{2}\lambda_{1}^{2},...,\tau^{2}\lambda_{p}^{2}) = \tau^{2} \bm \Lambda$,
 and $\Psi = (\bm y - \bm X \bm \beta)^{\prime}(\bm y - \bm X \bm \beta)$.

\subsubsection{Slice sampler}
Under the hosrshoe prior, 
\begin{eqnarray}
\bm \beta \vert \lbrace \lambda_{j} \rbrace_{j=1}^{p}, \tau , \sigma^{2} & \sim & N\left( \bm 0, \sigma^{2} \tau^2 diag(\lambda_{1}^{2},...,\lambda_{p}^{2}) \right), \\
\lambda_{j}  & \sim &  C^{+} (0,1), \text{ \ \ for } j=1,...,p,  \\
\tau & \sim &  C^{+} (0,1) \\
\sigma^{2} & \sim &  \frac{1}{\sigma^{2}}
\end{eqnarray}
the conditional posteriors are of the form
\begin{eqnarray}
\bm \beta   \ \vert \ \bullet & \sim &  N_p\left( \bm V \times \bm X^{'} \bm y , \sigma^2 \bm V  \right),\\
\sigma^2   \ \vert \ \bullet & \sim &  Inv-Gamma \left(\frac{n}{2}+\frac{p}{2},  \frac{1}{2} \left[ \left( \bm y - \bm X \bm \beta \right)' \left( \bm y - \bm X \bm \beta \right) +\bm \beta' \bm D^{-1} \bm \beta \right] \right)\\
p( \lambda_{j}   \ \vert \ \bullet \ ) & \propto & \left( \frac{1}{\lambda_j^2}\right)^{1/2} \exp\left[ - \frac{\beta_{j}}{2\sigma^{2} \tau^{2}} \frac{1}{\lambda_{j}^{2}} \right] \frac{1}{1+\lambda_{j}^{2} } d\lambda_j , \text{ \ \ for } j=1,...,p\\
p(\tau   \ \vert \ \bullet \ )& \propto &  \left( \frac{1}{\tau^2} \right)^{p/2} \exp \left[ -\frac{1}{2\sigma^2} \sum_{j=1}^p \frac{\beta_j^2}{\lambda_j^2} \frac{1}{\tau^2} \right] \frac{1}{1+\tau^2}  d\tau
\end{eqnarray}
where $\bm V = (    \bm X' \bm X +\bm D^{-1}  )^{-1}$ with
$\bm D=  diag(\tau^2 \lambda_{1}^{2},..., \tau^2 \lambda_{p}^{2})$. 

%

With a change of variable $\eta_j =\frac{1}{\lambda_j^2}$, it can be seen that 
\begin{eqnarray}
\eta_{j} \vert \bm \beta, \tau^{2}, \sigma^{2}& \propto &\exp\left( - \mu_j \eta_j \right) \frac{1}{1+\eta_j} d\eta_j
\end{eqnarray}
where $\mu_j = \frac{\beta_{j}}{2\sigma^{2} \tau^{2}}$.
The $\lambda_{j}$'s are updated with a slice sampler (see Section \ref{slice_sampler}):
\begin{enumerate}
\item Sample $u_j \sim Unif \left[0,\frac{1}{1 + \eta_j} \right]$,
\item Sample $\eta_j \vert u_j \sim \exp\left( - \mu_j \eta_j \right) I \left( \eta_j <  \frac{1-u_j}{u_j} \right)$ \footnote{This is an exponential density with parameter $\mu_j^{-1}$ truncated on $\left(0, \frac{1-u_j}{u_j} \right)$.},
\item Set $\lambda_j =\eta_j^{-1/2}$.
\end{enumerate} 
Similarly, with a change of variable $\eta = \frac{1}{\tau^2}$, we have 
\begin{eqnarray}
\eta \vert  \bm \beta, \lbrace \lambda_{j} \rbrace_{j=1}^{p}, \sigma^{2}, \bm y& \propto &\eta^{\frac{p+1}{2}-1} \exp\left( - \mu \eta \right) \frac{1}{1+\eta} d\eta
\end{eqnarray}
where $\mu=\frac{1}{2\sigma^2} \sum_{j=1}^p \frac{\beta_j^2}{\lambda_j^2}$.
The $\tau$ can be updated in a similar fashion:
\begin{enumerate}
\item Sample $u \sim Unif \left[0,\frac{1}{1 + \eta} \right]$,
\item Sample $\eta \vert u \sim \eta^{\frac{p+1}{2}-1}  \exp\left( - \mu \eta \right) I \left( \eta <  \frac{1-u}{u} \right)$ \footnote{This is a gamma density with the shape parameter $\frac{p+1}{2}$ and the scale parameter $\mu^{-1}$ truncated on $\left( 0, \frac{1-u}{u}\right)$.},
\item Set $\tau =\eta^{-1/2}$.
\end{enumerate}

\subsubsection{\cite{Johndrowetal2020} algorithm}
The horseshoe prior in \cite{Johndrowetal2020} has its original form
\begin{eqnarray}
\bm \beta \vert \lbrace \lambda_{j} \rbrace_{j=1}^{p}, \tau , \sigma^{2} & \sim & N\left( \bm 0, \sigma^{2} \tau^{2} \bm \Lambda \right), \\
\lambda_{j} \vert \tau & \sim &  C^{+} (0,1), \text{ \ \ for } j=1,...,p,  \\
\tau & \sim &  C^{+} (0,1), \\
\sigma^{2} & \sim &  \frac{1}{\sigma^{2}},
\end{eqnarray}

In order to improve the mixing of the global parameter $\tau^2$, they propose a blocked Metropolis-within-Gibbs sampler 
where  $(\bm \beta, \tau^2, \sigma)$ are updated  in one block. 
The conditional posterior of $\tau^2 $ given $\bm \lambda=( \lambda_1^2,...,\lambda_p^2)$ is 
\begin{eqnarray}
p( \tau^2 \vert \bm \lambda, \bm y)&  \propto & \vert \bm M \vert ^{-1/2} \left(\frac{1}{2} \bm y^{\prime} \bm M^{-1} \bm y  \right)^{-\frac{n}{2}} \times \frac{\tau}{1 + \frac{1}{\tau^{2}}}
\end{eqnarray}
where 
$\bm M = I_{n} + \bm X \bm D \bm X^{\prime} $.
Their Metropolis-within-Gibbs algorithm is as follows
\begin{eqnarray}
p(\lambda_{j}^{2}   \ \vert \tau^2, \bm \beta, \sigma^2 )  &\propto&   \frac{\lambda_{j}^{2}}{\lambda_{j}^{2} + 1} \exp\left( - \frac{\beta_{j}}{2\sigma^{2} \tau^{2}} \frac{1}{\lambda_{j}^{2}} \right), \text{ \ \ for } j=1,...,p,\\
log (\tau^{-2*}) &\sim& N\left( log(\tau^{-2}), s \right), \text{ accept }  \tau^{2*} \text{ w.p. } \frac{p( \tau^{2*} \vert \bm \lambda, \bm y)\tau^{2*}}{p( \tau^2 \vert \bm \lambda, \bm y)\tau^{2}},\\
\sigma^{2}   \ \vert \tau^2, \bm \lambda^2  &\sim&    Inv-Gamma \left( \frac{n}{2},\frac{\bm y^{\prime} \bm M^{-1} \bm y }{2} \right), \\
\bm \beta   \ \vert \tau^2, \bm \lambda^2, \sigma^2 &\sim&    N_p \left( \bm V \times  \bm X^{\prime} \bm y , \sigma^{2} \bm V \right)
\end{eqnarray}
where $\bm V = \left( \bm X^{\prime} \bm X + \bm D^{-1} \right)^{-1}$ and $\bm D =  diag(\tau^{2}\lambda_{1}^{2},...,\tau^{2}\lambda_{p}^{2}) $.

The $\lambda_{j}^2$ can be updated via a slice sampler:
\begin{enumerate}
\item Sample $u \sim Unif \left[0,\frac{\lambda_{j}^{2}}{\lambda_{j}^{2} + 1} \right]$,
\item Sample $\lambda_{j}^{2} \vert u \sim \exp\left( - \frac{\beta_{j}}{2\sigma^{2} \tau^{2}} \frac{1}{\lambda_{j}^{2}} \right) I \left( \frac{1-u}{u} > \frac{1}{\lambda_{j}^{2}} \right)$.
\end{enumerate} 


\subsection{Generalized Beta mixtures of Gaussians}
In their paper, \cite{Armaganetal2011} motivate the use of a three-parameter beta (TPB) distribution as a flexible class of shrinkage priors. The TPB distribution takes the form
\begin{equation}
p(x \vert a,b,\varphi) = \frac{\Gamma\left(a+b\right)}{\Gamma\left(a\right)\Gamma\left(b\right)} \varphi^{b} x^{b-1} (1-x)^{a-1} \left[ 1 + (\varphi - 1)x\right]^{-(a+b)},
\end{equation}
for $0<x<1$, $a,b,\varphi>0$. Proposition 1 in \cite{Armaganetal2011} shows that this distribution can either be written as Normal-inverted beta mixture, or a Normal-Gamma-Gamma mixture. The second choice gives a very straightforward Gibbs sampler scheme so we present an algorithm based on the Normal-Gamma-Gamma representation of TPB.

The Generalized Beta mixtures of Gaussians prior takes the form
\begin{eqnarray}
\bm \beta \vert \lbrace \tau_{j}^2 \rbrace_{j=1}^{p}, \sigma^{2} & \sim &  N_{p} \left(0, \sigma^{2} \bm D_{\tau} \right), \\
\tau_{j}^{2} \vert \lambda_{j} & \sim & Gamma \left( a, \lambda_{j} \right), \text{ \ \ for } j=1,...,p, \\
\lambda_{j} \vert \varphi & \sim & Gamma(b,\varphi ), \text{ \ \ for } j=1,...,p, \\
\varphi & \sim & Gamma\left(\frac{1}{2},\omega \right), \\
\omega & \sim & Gamma\left(\frac{1}{2}, 1 \right), \\
\sigma^{2} & \sim &  \frac{1}{\sigma^{2}},
\end{eqnarray}
where $\bm D_{\tau} =  diag(\tau_{1}^{2},...,\tau_{p}^{2})$. Note that setting $a=b=1/2$ we can obtain the horseshoe prior of \cite{Carvalhoetal2010}. For other choices we can recover popular cases of shrinkage priors.

The conditional posteriors are of the form
\begin{eqnarray}
\bm \beta   \ \vert \ \bullet & \sim &  N_p \left( \bm V \times \bm X^{\prime} \bm y, \sigma^{2} \bm V \right), \\
\tau_{j}^{2}   \ \vert \ \bullet & \sim &  GIG\left(a-\frac{1}{2}, 2\lambda_{j},\frac{\beta_{j}^{2}}{ \sigma^{2}} \right), \text{ \ \ for } j=1,...,p,\\
\lambda_{j}  \ \vert \ \bullet & \sim &  Gamma( a+b,\tau_{j}^{2} + \varphi ), \text{ \ \ for } j=1,...,p, \\
\varphi   \ \vert \ \bullet & \sim & Gamma\left( pb + \frac{1}{2}, \sum_{j=1}^{p} \lambda_{j} + \omega \right), \\
\omega   \ \vert \ \bullet & \sim &  Gamma(1, \varphi + 1), \\
\sigma^{2}   \ \vert \ \bullet & \sim &  Gamma \left( \frac{n+p}{2},\frac{\Psi+\bm \beta^{\prime} \bm D_{\tau}^{-1} \bm \beta}{2} \right),
\end{eqnarray}
where $\bm V = \left( \bm X^{\prime} \bm X + \bm D_{\tau}^{-1} \right)^{-1}$, $\bm D_{\tau}=  diag(\tau_{1}^{2},...,\tau_{p}^{2})$, and 
$\Psi = (\bm y - \bm X \bm \beta)^{\prime}(\bm y - \bm X \bm \beta)$.

\subsection{Spike and slab}

\subsubsection{\cite{KuoMallick1998} algorithm}

\cite{KuoMallick1998} consider the following modified formulation of the regression problem. 
\begin{eqnarray}
\bm y \vert \bm \beta, \bm \gamma, \sigma^2 & \sim & N_{p} \left(\bm X \bm \theta, \sigma^{2} \bm I \right)
\end{eqnarray}
where $\bm X = \left( \bm x_1,\ldots, \bm x_p \right)$ and $\bm \theta =\left( \beta_1 \gamma_1, \ldots,   \beta_p \gamma_p \right)'$ with $\gamma_j = 1$ if $\bm x_j$ is included in the model and $0$ otherwise. 

The authors consider the following independent prior. 
\begin{eqnarray}
\bm \beta & \sim &  N_{p} \left(\bm 0, \bm D \right), \\
\gamma_j &\sim& Bernoulli(p_j), \text{ for } j=1,\ldots,p, \\
\sigma^2 &\sim& Inv-Gamma \left( a,b \right)
\end{eqnarray}

With $\bm X^*=(\gamma_1 \bm x_1,...,\gamma_p \bm x_p)$, the conditional posteriors can be written as follows. 
\begin{eqnarray}
\bm \beta   \ \vert \ \bullet & \sim &  N_p\left( \bm V \times  \bm X^{*'} \bm y / \sigma^2 , \bm V  \right),\\
\sigma^2  \ \vert \ \bullet & \sim &  Inv-Gamma \left( a+\frac{n}{2}, b + \frac{1}{2} \left( \bm y - \bm X^* \bm \beta \right)' \left( \bm y - \bm X^* \bm \beta \right)  \right),\\
\gamma_j   \ \vert \ \bullet & \sim &  Bernoulli \left( \frac{c_j}{c_j + d_j} \right)
\end{eqnarray}
where 
$\bm \gamma_{-j}=\left(\gamma_1,\ldots, \gamma_{j-1},\gamma_{j+1},\ldots,\gamma_p\right)$
and 
$\bm V = \left(  \bm X^{*'} \bm X^* / \sigma^2 + \bm D^{-1} \right)^{-1}$ and 
\begin{eqnarray}
c_j &=& p_j \exp\left[ -\frac{1}{2\sigma^2}\left( \bm y - \bm X \bm \theta^*_j  \right)'\left(\bm y - \bm X \bm \theta^*_j   \right) \right],  \\
d_j &=& (1-p_j) \exp\left[ -\frac{1}{2\sigma^2}\left( \bm y - \bm X \bm \theta^{**}_j  \right)'\left(\bm y - \bm X \bm \theta^{**}_j   \right) \right]
\end{eqnarray}
where 
$\bm \theta^*_j $ is $\bm \theta$ with the $j$-component replaced by $\beta_j$ and 
$\bm \theta^{**}_j $ is $\bm \theta$ with the $j$-component replaced by $0$.
Note that the conditional posterior of $\gamma_j$ depends on $\bm \gamma_{-j}$. 
In order to facilitate the mixing, it is preferred to update $\gamma_j$ for $j=1,\ldots, p$ in random order.

%
Note that although the formulation above holds for a generic prior variance $\bm V_\beta$, but an important special case is when it is a diagonal matrix $\bm V_\beta =diag\left( \tau^2_1,\ldots,\tau^2_p \right)$.
This is equivalent to assume a spike and slab prior on $\theta_j$, which is 
a mixture of a point mass at $0$ with probability $1-p_j$ and 
a normal density $N\left( \mu_{\beta,j}, \tau_j^2 \right)$ with probability $p_j$.

\subsection{Stochastic search variable selection}

Consider the following stochastic search variable selection prior with fixed values of the prior variances. 
\begin{eqnarray}
\beta_j \vert \sigma^2, \gamma_j=0 & \sim &  N\left(0, \sigma^2 \tau_{0j}^2 \right), \label{ssvs_spike}\\
\beta_j \vert \sigma^2, \gamma_j=1 & \sim &  N\left(0, \sigma^2 \tau_{1j}^2 \right),  \label{ssvs_slab}\\
P(\gamma_j =1 ) & = & \theta \text{ for } j=1,\ldots, p, \label{ssvs_gamma}\\
\theta &\sim& Beta(c,d)  \label{ssvs_theta}\\
\sigma^2 &\sim& Inv-Gamma (a,b) \label{ssvs_sigma2}
\end{eqnarray}
\cite{GeorgeMcCulloch1993} use non-conjugate prior in \eqref{ssvs_spike} and \eqref{ssvs_slab}.

\eqref{ssvs_spike} and \eqref{ssvs_slab} can be equivalently
written as 
\begin{eqnarray}
\bm \beta \vert \sigma^2, \bm \gamma, \{ \tau_{0j}^2, \tau_{1j}^2 \}_{j=1}^p & \sim &  N_{p} \left(\bm 0, \sigma^2 \bm D \right)
\end{eqnarray}
where $\bm D$ is a diagonal matrix with diagonal elements with $\{ (1-\gamma_j)\tau_{0j}^2 + \gamma_j \tau_{1j}^2 \}_{j=1}^p$

The conditional posteriors are of the form
\begin{eqnarray}
\bm \beta  \ \vert \ \bullet & \sim &  N_p\left( \bm V \times \bm X' \bm y , \sigma^2 \bm V \right), \text{ where } \bm V = (   \bm X^{'} \bm X + \bm D^{-1} )^{-1},\\
\sigma^2   \ \vert \ \bullet & \sim &  Inv-Gamma \left( a+\frac{n}{2}+\frac{p}{2}, b + \frac{1}{2} \left[ \left( \bm y - \bm X \bm \beta \right)' \left( \bm y - \bm X \bm \beta \right) +\bm \beta' \bm D^{-1} \bm \beta \right] \right),\\
\gamma_j \vert \ \bullet &\sim &  Bernoulli \left( \frac{\phi \left( \beta_j \vert 0, \sigma^2 \tau_{1j}^2 \right) \theta}{\phi \left( \beta_j \vert 0, \sigma^2 \tau_{1j}^2 \right) \theta +  \phi \left( \beta_j \vert 0, \sigma^2 \tau_{0j}^2 \right) (1-\theta)} \right), \text{ for  } j=1,...,p,\\
\theta \vert \ \bullet &\sim & Beta\left(c+\sum_{j=1}^p \gamma_j, d +\sum_{j=1}^p (1-\gamma_j)  \right), \text{ for  } j=1,...,p
\end{eqnarray}
where 
 $\phi(x\vert m, v)$ is the normal density with mean $m$ and variance $v$.

\cite{Narisettyetal2018} propose to fix the value of the prior variance parameters  as 
$\tau_{0j}^2 = \frac{ \hat{\sigma}^2 }{ 10 n}$ and
$ \tau_{1j}^2 = \hat{\sigma}^2 \max \left( \frac{ p^{2.1} }{ 100 n}, \log (n) \right)$
where $\hat{\sigma}^2 $ is the sample variance of $y_i$.
The prior inclusion probability $\theta$ is chosen so that $Pr\left( \sum_{j=1}^p \gamma_j > K \right)=0.1$
for $K=\max \left( 10, \log (n) \right)$.

\subsection{Spike and slab lasso}
Consider the generic SSVS prior \eqref{ssvs_spike}-\eqref{ssvs_sigma2}.
Instead of fixing the prior variances $\tau_{0j}$ and $\tau_{1j}$, one could place priors on them.
A hierarchical Bayes version of the spike and slab lasso prior in \cite{RockovaGeorge2014} and \cite{BaiRockovaGeorge2021}\footnote{They propose an EM algorithm for estimation.}
would correspond to placing two separate Laplace densities on the components  i.e.\
\begin{eqnarray}
\tau_{0j}^{2} \vert \lambda_0^{2} & \sim  & Exponential\left( \frac{\lambda_0^{2}}{2} \right), \text{ \ \ for } j=1,...,p, \\
\tau_{1j}^{2} \vert \lambda_1^{2} & \sim  & Exponential\left( \frac{\lambda_1^{2}}{2} \right), \text{ \ \ for } j=1,...,p
\end{eqnarray}
with $\lambda_0 \gg \lambda_1$ so that the density for $N(0,\sigma^2 \tau^2_{0j})$ is the ``spike'' and $N(0,\sigma^2 \tau^2_{1j})$ is the ``slab''.

The prior variances are updated according to 
\begin{eqnarray}
1/ \tau_{0j}^{2}  \vert \ \bullet &\sim & IG\left( \sqrt{\lambda_0^2\sigma^2/\beta^2_j} , \lambda_0^2\right),  \text{ \ \ for } j=1,...,p,   \\
1/ \tau_{1j}^{2}  \vert \ \bullet &\sim & IG\left( \sqrt{\lambda_1^2\sigma^2/\beta^2_j} , \lambda_1^2\right),  \text{ \ \ for } j=1,...,p
\end{eqnarray}

\subsection{Semiparametric spike and slab}
\cite{Dunsonetal2008}
allows for simultaneous selection of important predictors and soft clustering
of predictors having similar impact on the variable of interest. This prior is a generalization of the typical “spike and slab” priors used for Bayesian variable selection
and model averaging in the statistics literature.
The coefficient $\bm \beta$ admit a prior of the form
\begin{align*}
\beta_j & \sim \pi \delta_0(\beta) + (1-\pi)G\\
G&\sim DP(\alpha G_0)\\
G_0 &\sim N(0,\tau^2) 
\end{align*}
$G$ is a nonparametric density which follows a Dirichlet process
with base measure $G_0$ and concentration parameter $\alpha$.
In this case the base measure $G_0$ is Gaussian with zero mean and variance $\tau^2$,
which is the typical conjugate prior distribution used on linear regression coefficients. Hence, this prior implies that each coefficient $\beta_j$ will either be restricted to 0
with probability $\pi$, or with probability $1-\pi$ will come from a mixture of Gaussian
densities. If it comes from a mixture of Gaussian densities, then 
due to a property of the Dirichlet process, 
$\beta_j$'s in the same mixture component will share the same mean and the variance. 

As an example, consider coefficients $\beta_j$, $j=1,...,6$ with 
$(\beta_1,\beta_3) \sim N(0,10^6)$,
$(\beta_2,\beta_4) \sim N(0,0.1)$, and 
$(\beta_5,\beta_6) \sim \delta_0$. In this case, 
$(\beta_1,\beta_3)$ are clustered together and have a Gaussian prior with variance $10^6$
which means that their posterior mean/median will be close to the least squares estimator.
The second cluster consists of $(\beta_2,\beta_4)$ which have prior variance $0.1$, hence their posterior median will be equivalent to a ridge regression estimator.
Finally, $(\beta_5,\beta_6)$ are restricted to be zero.

Inference using the Gibbs sampler is straightforward, once we write the Dirichlet process prior using its stick-breaking representation, that is, an infinite sum of point mass functions. The general form of the semiparametric spike and slab prior we use is of the form
\begin{eqnarray}
\beta _{j} &\sim &\pi \delta _{0}\left( \beta \right) +\left( 1-\pi \right) G
\\
G &\sim &DP\left( \alpha G_{0}\right) \\
G_{0} &\sim &N\left( \underline{\mu},\tau ^{2}\right) \\
\tau ^{2} &\sim & Inv-Gamma\left( \underline{a}_{1},\underline{a}_{2}\right) \\
\alpha &\sim &Gamma\left( \underline{\rho }_{1},\underline{\rho }_{2}\right)
\\
\pi &\sim & Beta\left( \underline{c},\underline{d}\right) ,\\
\sigma^{2} & \sim & \frac{1}{\sigma^{2}},
\end{eqnarray},
where $ \underline{\mu}, \underline{a}_{1},\underline{a}_{2}, \underline{\rho }_{1},\underline{\rho }_{2}, \underline{c},\underline{d}$ are parameters to be chosen by the researcher. The usual stick breaking representation for $\beta_j$ conditional on $\beta_{-j}$ and marginalized over $G$ is of the form
\begin{equation}
\left( \beta _{j}| \bm \beta _{-j}\right) \sim \frac{\alpha \left( 1-\pi \right) 
}{\alpha +K-p_{\beta _{1}}-1}N\left( \underline{\mu },\tau ^{2}\right) +\pi
\delta _{0}\left( \beta \right) +\sum\limits_{l=2}^{k_{\beta }}\frac{%
p_{\beta _{l}}\left( 1-\pi \right) }{\alpha +K-p_{\beta _{1}}-1}\delta
_{\beta _{l}}\left( \beta \right)  \label{conditional_prior_beta}
\end{equation}%
where $k_{\beta }$ is the number of atoms in the above equation (number of
mixture components plus the $\delta _{\beta }\left( 0\right) $ component),
and$\ p_{\beta _{n}}$ is the number of elements of the vector $\beta $ which
which are equal to $\delta _{\beta _{l}}\left( \beta \right) $, $%
n=1,2,...,k_{\beta }$, where it holds that $\delta _{\beta _{1}}\left( \beta
\right) =\delta _{0}\left( \beta \right) $. Additionally, for notational
convenience define the prior weights as%
\begin{eqnarray*}
w_{0} &=&\frac{\alpha \left( 1-\pi \right) }{\alpha +K-p_{\beta _{1}}-1} \\
w_{1} &=&\pi \\
w_{l} &=&\frac{p_{\beta _{l}}\left( 1-\pi \right) }{\alpha +K-p_{\beta
_{1}}-1},\text{ }l=2,...,k_{\beta }.
\end{eqnarray*}

\textit{Gibbs sampling from the conditional posterior:}

\begin{itemize}
\item Given $k_{\beta }$ number of mixture components, sample $ \bm \theta
=\left( \bm  \theta _{1},..., \bm \theta _{k_{\beta }}\right) $ from%
\[
\left(  \bm \theta |-\right) \sim N\left(  \bm E_{ \bm \theta }, \bm V_{ \bm \theta }\right) , 
\]%
with $ \bm E_{ \bm \theta }= \bm V_{\theta }\left(  \bm D^{-1}  \bm M + \sigma^{-2} \bm X_{\pi }^{\prime }%
\bm y\right) $ and $ \bm V_{\beta }=\left(  \bm D^{-1} + \sigma ^{-2} \bm X_{\pi}^{\prime} \bm X_{\pi }\right) ^{-1}$, where $ \bm D=\tau ^{2} \bm I_{k_{\beta }}$ and $ \bm M = \mu \mathbf{1}_{k_{\beta }}$. Here $ \bm X_{\pi }^{\prime }$ denotes the matrix $ \bm X$ with the columns corresponding to coefficients belonging to $%
\theta _{1}$ being replaced with zeros (or equivalently, with these columns
removed). Hence the remaining columns correspond to unrestricted
coefficients which belong to one of the remaining $k_{\beta }-1$ mixture
components.

\item Sample $\beta _{j}$ conditional on $\beta _{-j}$, data, and other
model parameters for $j=1,...,K$ from%
\[
\left( \beta _{j}| \bm \beta _{-j},-\right) \sim \overline{w}_{0}N\left(  E_{\beta
}, V_{\beta }\right) +\sum\limits_{l=1}^{k_{\beta }}\overline{w}_{l}  \bm \theta_{l}, 
\]%
so that with probability $\overline{w}_{l}$ we assign $\beta _{j}$ equal to
the atom of mixture component $l$ (i.e. $\beta _{j}=\theta _{l}$), while
with probability $\overline{w}_{0}$ we assign $\beta _{j}$ to a new $N\left(
E_{\beta },V_{\beta }\right) $ component. In the expression above it holds
that%
\begin{eqnarray*}
E_{\beta } &=&V_{\beta }\left( \tau ^{-2}\underline{\mu }+\sigma
^{-2}  \bm X^{\prime }  \bm y \right) \\
V_{\beta } &=&\left( \tau ^{-2}+\sigma ^{-2} \bm X^{\prime } \bm X\right) ^{-1},
\end{eqnarray*}%
and that%
\begin{eqnarray*}
\overline{w}_{0} &\propto &\frac{w_{0}N\left( 0;\underline{\mu },\tau
^{2}\right) \prod\nolimits_{i=1}^{n}N\left(  \widetilde{y}_{i};0,\sigma
^{2}\right) }{N\left( 0;E_{\beta },V_{\beta }\right) } \\
\overline{w}_{l} &\propto &w_{l}N\left( 0;\underline{\mu },\tau ^{2}\right)
\prod\nolimits_{i=1}^{n}N\left( \widetilde{y}_{i}; \bm X_{i,l} \bm \theta _{l},\sigma
^{2}\right) ,\text{ }l=1,...,k_{\beta },
\end{eqnarray*}%
where $\widetilde{y}_{i}=y_{i}-\sum\nolimits_{j^{\prime }\neq
j}X_{i,j^{\prime }}\beta _{j^{\prime }} = y_{i}-\left( \bm X_{\pi }\right)
_{i} \bm \theta + X_{j^{\prime },i}\beta_{j^{\prime }}$ for $j,j\prime =1,...,K$, $\left( X_{\pi }\right) _{i}$ is the $i$-th observation of the matrix $%
X_{\pi }$ constructed in step 1, and $N\left( a;b,c\right) $ denotes the
normal density with mean $b$ and variance $c$, evaluated at observation $a$.

\item Introduce an indicator variable $S_{\beta }=l$ if the coefficient $%
\beta _{j}$ belongs to cluster $l$, where $j=1,...,K$ and $l=1,...,k_{\beta
} $, in which case it holds that $\beta _{j}=\theta _{l}$. In addition, set $%
S_{\beta }=0$ if $\beta _{j}\neq \theta _{l}$, that is when $\beta _{j}$
does not belong to a preassigned cluster and a new cluster is introduced for
this coefficient. Then the conditional posterior of $S_{\beta }$ is%
\[
\left( S_{\beta }|-\right) \sim Multinomial\left( 0,1,...,k_{\beta };%
\overline{w}_{0},\overline{w}_{1},...,\overline{w}_{k_{\beta }}\right) . 
\]

\item Sample the restriction probability $\pi $ from the coniditional
distribution%
\[
\left( \pi |-\right) \sim Beta\left( \underline{c}+\sum\nolimits_{j=1}^{K}I%
\left( S_{\beta }=1\right) ,d+\sum\nolimits_{j=1}^{K}I\left( S_{\beta }\neq
1\right) \right) 
\]

\item Sample the latent variable $\eta $ from the posterior conditional%
\[
\left( \eta |-\right) \sim Beta\left( a+1,K-\sum\nolimits_{j=1}^{K}I\left(
S_{\beta }=1\right) \right) . 
\]

\item Sample the Dirichlet process precision coefficient $\alpha $ from the
conditional posterior%
\begin{eqnarray*}
\left( \alpha |-\right) &\sim &\pi _{\eta }Gamma\left( \underline{\rho }%
_{1}+k_{\beta }-n_{S_{\beta }=1},\underline{\rho }_{2}-\log \eta \right) + \\
&&\left( 1-\pi _{\eta }\right) Gamma\left( \underline{\rho }_{1}+k_{\beta
}-n_{S_{\beta }=1}-1,\underline{\rho }_{2}-\log \eta \right)
\end{eqnarray*}%
where the weight $\pi _{\eta }$ is given by%
\[
\frac{\pi _{\eta }}{1-\pi _{\eta }}=\frac{\underline{\rho }_{1}+k_{\beta
}-n_{S_{\beta }=1}-1}{\left( K-\sum\nolimits_{j=1}^{K}I\left( S_{\beta
}=1\right) \right) \left( \underline{\rho }_{2}-\log \eta \right) }, 
\]%
and $n_{S_{\beta }=1}=1$ if $\sum\nolimits_{j=1}^{K}I\left( S_{\beta
}=1\right) >0,$ and it is $0$ otherwise (i.e. when no coefficient $\beta
_{j} $ is restricted).\bigskip

\item Sample the variance $\tau ^{2}$ coefficient from the conditional
density%
\[
\left( \tau ^{2}|-\right) \sim iGamma\left( \underline{a}_{1}+\frac{1}{2}%
\left( k_{\beta }-1\right) ,\underline{a}_{2}^{-1}+\frac{1}{2}%
\sum_{l=2}^{k_{\beta }}\left(\bm \theta _{l}-\underline{\mu }\mathbf{1}\right)
^{2}\right) . 
\]
\end{itemize}

\end{appendix}

\end{document}